\documentstyle[preprint,epsfig,eqsecnum,aps,floats,tighten]{revtex}

\def\Missing#1#2{{\mbox{$#1\kern-0.57em\raise0.19ex\hbox{/}_{#2}$}}}

\def\vMissing#1#2{\ifmmode
            \vec{#1}\kern-0.57em\raise.19ex\hbox{/}_{#2}
         \else
            {{\mbox{$\vec{#1}\kern-0.57em\raise.19ex\hbox{/}_{#2}$}}}
         \fi}
\def\lsim{\mathrel{\rlap{\lower4pt\hbox{\hskip1pt$\sim$}}
    \raise1pt\hbox{$<$}}}        
\def\gsim{\mathrel{\rlap{\lower4pt\hbox{\hskip1pt$\sim$}}
    \raise1pt\hbox{$>$}}}        

\def\met{\mbox{$\Missing{E}{T}$}}
\def\vmet{\mbox{$\vMissing{E}{T}$}}

\def\pt{\mbox{$p_{T}$}}

\def\D0{D\O }

\newcommand{\Et}{\mbox{$E_{\rm T}$}}
\newcommand{\Etu}[1]{\mbox{$E^{\rm jet#1}_{T}$}}
\newcommand{\Etmax}{\mbox{$E_{T}^{\rm max}$}}
\newcommand{\Etjet}{\mbox{$E_{T}^{\rm jet}$}}

\newcommand{\Rc}{\mbox{${\mathcal R}$}}
\newcommand{\Rsep}{\mbox{${\mathcal{R}}_{\rm sep}$}}
\newcommand{\Exi}{E_{x}^{i}}
\newcommand{\Eyi}{E_{y}^{i}}
\newcommand{\Ezi}{E_{z}^{i}}

\newcommand{\ETi}{E_{T}^{i}}

\newcommand{\HT}{\mbox{${S}_{T}$}}

\newcommand{\DR}{\mbox{$\Delta \Rc$}}

\newcommand{\EA}{\mbox{$\eta$} }

\newcommand{\ipb}{\mbox{pb$^{-1}$}}
\newcommand{\inb}{\mbox{nb$^{-1}$}}

\def\simge{\mathrel{\rlap{\raise 0.53ex \hbox{$>$}}%
{\lower 0.53ex \hbox{$\sim$}}}}
\def\simle{\mathrel{\rlap{\raise 0.53ex \hbox{$<$}}%
{\lower 0.53ex \hbox{$\sim$}}}}

\def\inb{nb$^{-1}$}                     
\newcommand{\jjmass}{\mbox{$M$}}
\newcommand{\cteqthreem}{\mbox{CTEQ3M}}
\newcommand{\cteqfourm}{\mbox{CTEQ4M}}
\newcommand{\cteqfourhj}{\mbox{CTEQ4HJ}}
\newcommand{\cteqfoura}{\mbox{CTEQ4A}}
\newcommand{\mrsap}{\mbox{MRS{(A$^\prime$)}}}
\newcommand{\mrst}{\mbox{MRST}}
\newcommand{\mrstgu}{\mbox{MRST($g\!\uparrow$)}}
\newcommand{\mrstgd}{\mbox{MRST($g\!\downarrow$)}}
\newcommand{\modeta}{\mbox{$\mid \! \eta  \! \mid$}}
\newcommand{\modetaonetwo}{\mbox{$\mid \! \eta_{1,2}  \! \mid$}}

\newcommand{\modetajet}{\mbox{$\mid \! \eta^{\rm jet}  \! \mid$}}
\newcommand{\modetajetu}[1]{\mbox{$\mid \! \eta^{\rm jet#1}  \! \mid$}}
\newcommand{\chisq}{\mbox{$\chi^{2}$}}
\newcommand{\als}{\mbox{${\alpha_{{\scriptscriptstyle S}}}$}}

\def\ETmiss{\mbox{${\hbox{$E$\kern-0.5em\lower-.1ex\hbox{/}\kern+0.15em}}_T$ }}

\newcommand{\pbarp}{\mbox{$p\overline{p}$}}
\newcommand{\Emeas}{\mbox{$E_{\rm jet}^{\rm meas}$}}
\newcommand{\Eptcl}{\mbox{$E_{\rm jet}^{\rm ptcl}$}}
\newcommand{\gev}{\mbox{\rm GeV}}

\def\err#1#2#3 {{\it Erratum} {\bf#1},{\ #2} (19#3)}
\def\ib#1#2#3 {{\it ibid.} {\bf#1},{\ #2} (19#3)}
\def\nc#1#2#3 {Nuovo Cim. {\bf#1} ,#2(19#3)}
\def\nim#1#2#3 {Nucl. Instr. Meth. {\bf#1},{\ #2} (19#3)}
\def\np#1#2#3 {Nucl. Phys. {\bf#1},{\ #2} (19#3)}
\def\pl#1#2#3 {Phys. Lett. {\bf#1},{\ #2} (19#3)}
\def\prev#1#2#3 {Phys. Rev. {\bf#1},{\ #2} (19#3)}
\def\prl#1#2#3 {Phys. Rev. Lett. {\bf#1},{\ #2} (19#3)}
\def\rmp#1#2#3 {Rev. Mod. Phys. {\bf#1},{\ #2} (19#3)}
\def\zp#1#2#3 {Zeit. Phys. {\bf#1},{\ #2} (19#3)}     
\def\inb{nb$^{-1}$}                     
\newcommand{\gevcc}{\mbox{GeV/$c^2$}}                   
\def\tevcc{TeV/$c^2$}                   

\begin{document}
\pagestyle{myheadings}

\preprint{Fermilab-Pub-00/216-E, D\O\ Paper XXXX} 

\title{High-$\bbox{p_{T}}$ Jets in $\bbox{\bar{p}p}$ Collisions at
$\bbox{\sqrt{s}}$ = 630 and 1800 GeV}

%
\author{                                                                      
B.~Abbott,$^{56}$                                                             
A.~Abdesselam,$^{11}$                                                         
M.~Abolins,$^{49}$                                                            
V.~Abramov,$^{24}$                                                            
B.S.~Acharya,$^{16}$                                                          
D.L.~Adams,$^{58}$                                                            
M.~Adams,$^{36}$                                                              
G.A.~Alves,$^{2}$                                                             
N.~Amos,$^{48}$                                                               
E.W.~Anderson,$^{41}$
% ...
\\                                                                            
\vskip 0.30cm                                                                 
\centerline{(D\O\ Collaboration)}                                             
\vskip 0.30cm                                                                 
}                                                                             
\address{                                                                     
\centerline{$^{1}$Universidad de Buenos Aires, Buenos Aires, Argentina}       
\centerline{$^{2}$LAFEX, Centro Brasileiro de Pesquisas F{\'\i}sicas,         
                  Rio de Janeiro, Brazil}                                     
\centerline{$^{3}$Universidade do Estado do Rio de Janeiro,                   
                  Rio de Janeiro, Brazil}                                     
\centerline{$^{4}$Institute of High Energy Physics, Beijing,                  
                  People's Republic of China}                                 
\centerline{$^{5}$Universidad de los Andes, Bogot\'{a}, Colombia}             
\centerline{$^{6}$Charles University, Prague, Czech Republic}                 
\centerline{$^{7}$Institute of Physics, Academy of Sciences, Prague,          
                  Czech Republic}                                             
\centerline{$^{8}$Universidad San Francisco de Quito, Quito, Ecuador}         
\centerline{$^{9}$Institut des Sciences Nucl\'eaires, IN2P3-CNRS,             
                  Universite de Grenoble 1, Grenoble, France}                 
\centerline{$^{10}$CPPM, IN2P3-CNRS, Universit\'e de la M\'editerran\'ee,     
                  Marseille, France}                                          
\centerline{$^{11}$Laboratoire de l'Acc\'el\'erateur Lin\'eaire,              
                  IN2P3-CNRS, Orsay, France}                                  
\centerline{$^{12}$LPNHE, Universit\'es Paris VI and VII, IN2P3-CNRS,         
                  Paris, France}                                              
\centerline{$^{13}$DAPNIA/Service de Physique des Particules, CEA, Saclay,    
                  France}                                                     
\centerline{$^{14}$Panjab University, Chandigarh, India}                      
\centerline{$^{15}$Delhi University, Delhi, India}                            
\centerline{$^{16}$Tata Institute of Fundamental Research, Mumbai, India}     
\centerline{$^{17}$Seoul National University, Seoul, Korea}                   
\centerline{$^{18}$CINVESTAV, Mexico City, Mexico}                            
\centerline{$^{19}$FOM-Institute NIKHEF and University of                     
                  Amsterdam/NIKHEF, Amsterdam, The Netherlands}               
\centerline{$^{20}$University of Nijmegen/NIKHEF, Nijmegen, The               
                  Netherlands}                                                
\centerline{$^{21}$Institute of Nuclear Physics, Krak\'ow, Poland}            
\centerline{$^{22}$Institute for Theoretical and Experimental Physics,        
                   Moscow, Russia}                                            
\centerline{$^{23}$Moscow State University, Moscow, Russia}                   
\centerline{$^{24}$Institute for High Energy Physics, Protvino, Russia}       
\centerline{$^{25}$Lancaster University, Lancaster, United Kingdom}           
\centerline{$^{26}$Imperial College, London, United Kingdom}                  
\centerline{$^{27}$University of Arizona, Tucson, Arizona 85721}              
\centerline{$^{28}$Lawrence Berkeley National Laboratory and University of    
                  California, Berkeley, California 94720}                     
\centerline{$^{29}$University of California, Davis, California 95616}         
\centerline{$^{30}$California State University, Fresno, California 93740}     
\centerline{$^{31}$University of California, Irvine, California 92697}        
\centerline{$^{32}$University of California, Riverside, California 92521}     
\centerline{$^{33}$Florida State University, Tallahassee, Florida 32306}      
\centerline{$^{34}$University of Hawaii, Honolulu, Hawaii 96822}              
\centerline{$^{35}$Fermi National Accelerator Laboratory, Batavia,            
                   Illinois 60510}                                            
\centerline{$^{36}$University of Illinois at Chicago, Chicago,                
                   Illinois 60607}                                            
\centerline{$^{37}$Northern Illinois University, DeKalb, Illinois 60115}      
\centerline{$^{38}$Northwestern University, Evanston, Illinois 60208}         
\centerline{$^{39}$Indiana University, Bloomington, Indiana 47405}            
\centerline{$^{40}$University of Notre Dame, Notre Dame, Indiana 46556}       
\centerline{$^{41}$Iowa State University, Ames, Iowa 50011}                   
\centerline{$^{42}$University of Kansas, Lawrence, Kansas 66045}              
\centerline{$^{43}$Kansas State University, Manhattan, Kansas 66506}          
\centerline{$^{44}$Louisiana Tech University, Ruston, Louisiana 71272}        
\centerline{$^{45}$University of Maryland, College Park, Maryland 20742}      
\centerline{$^{46}$Boston University, Boston, Massachusetts 02215}            
\centerline{$^{47}$Northeastern University, Boston, Massachusetts 02115}      
\centerline{$^{48}$University of Michigan, Ann Arbor, Michigan 48109}         
\centerline{$^{49}$Michigan State University, East Lansing, Michigan 48824}   
\centerline{$^{50}$University of Nebraska, Lincoln, Nebraska 68588}           
\centerline{$^{51}$Columbia University, New York, New York 10027}             
\centerline{$^{52}$University of Rochester, Rochester, New York 14627}        
\centerline{$^{53}$State University of New York, Stony Brook,                 
                   New York 11794}                                            
\centerline{$^{54}$Brookhaven National Laboratory, Upton, New York 11973}     
\centerline{$^{55}$Langston University, Langston, Oklahoma 73050}             
\centerline{$^{56}$University of Oklahoma, Norman, Oklahoma 73019}            
\centerline{$^{57}$Brown University, Providence, Rhode Island 02912}          
\centerline{$^{58}$University of Texas, Arlington, Texas 76019}               
\centerline{$^{59}$Texas A\&M University, College Station, Texas 77843}       
\centerline{$^{60}$Rice University, Houston, Texas 77005}                     
\centerline{$^{61}$University of Virginia, Charlottesville, Virginia 22901}   
\centerline{$^{62}$University of Washington, Seattle, Washington 98195}       
}                                                                             

\date{17 December 2000}

\maketitle

\begin{abstract}
 Results are presented from analyses of jet data produced in \pbarp\
 collisions at $\sqrt{s}$ = 630 and 1800 GeV collected with the D\O\
 detector during the 1994--95 Fermilab Tevatron Collider run. We
 discuss details of detector calibration, and jet selection criteria
 in measurements of various jet production cross sections at
 $\sqrt{s}$ = 630 and 1800 GeV. The inclusive jet cross sections, the
 dijet mass spectrum, the dijet angular distributions, and the ratio
 of inclusive jet cross sections at $\sqrt{s}$ = 630 and 1800 GeV are
 compared to next-to-leading-order QCD predictions.  The order
 $\alpha_{\rm s}^{3}$ calculations are in good agreement with the
 data. We also use the data at $\sqrt{s}$ = 1800 GeV to rule out
 models of quark compositeness with a contact interaction scale less
 than $2.2$~TeV at the 95$\%$ confidence level.
\end{abstract}

\tableofcontents
\clearpage

\section{Introduction}
\label{sec:introduction}

 The quark model which suggested that hadrons are composite particles
 was first proposed in the early 1960's~\cite{first_theory_quark} and
 was confirmed as the quark-parton model in a series of experiments at
 the Stanford Linear Accelerator Center in the late 1960's and early
 1970's~\cite{SLAC_quarks}. The model developed during the 1970's into
 the theory of strong interactions, Quantum Chromodynamics
 (QCD)~\cite{jetsreview}, which describes the interactions of quarks
 and gluons (called partons).  Together with the theory of electroweak
 interactions, QCD forms the foundation of the standard model of
 particle physics (SM), which, thus far, describes accurately the
 interactions of all known elementary particles.

 Perturbative QCD (pQCD)~\cite{jetsreview} predicts the production
 cross sections at large transverse momentum ($p_{T}$) for
 parton-parton scattering in proton--antiproton (\pbarp )
 collisions~\cite{ua1_inc,ua2_inc,CDF_1,CDF_2,d0_inc}. The outgoing
 partons from the parton-parton scattering hadronize to form jets of
 particles. High-$p_{T}$ jets were observed clearly during
 experimentation at the CERN ISR~\cite{afs_jets} and the CERN \pbarp\
 collider~\cite{first_jets}. Significant deviations from predictions
 of pQCD can only be observed if the uncertainties in both
 experimental measurements and theoretical predictions are
 small. Calculations of high-\pt\ jet production involve the folding
 of parton scattering cross sections with experimentally determined
 parton distribution functions (PDFs).  These predictions have
 recently improved with next-to-leading-order (NLO) QCD
 calculations~\cite{aversa,eks,jetrad} and improved
 PDFs~\cite{cteq4m,mrst}. These $\cal{O}(\als^{{\rm 3}})$ predictions
 reduce theoretical uncertainties to
 $\sim$30$\%$~\cite{inc_jet_theory_uncertainties} (where \als\ is the
 strong coupling parameter).

\begin{figure}[htbp]
\begin{center}
\vbox{\centerline{\psfig{figure=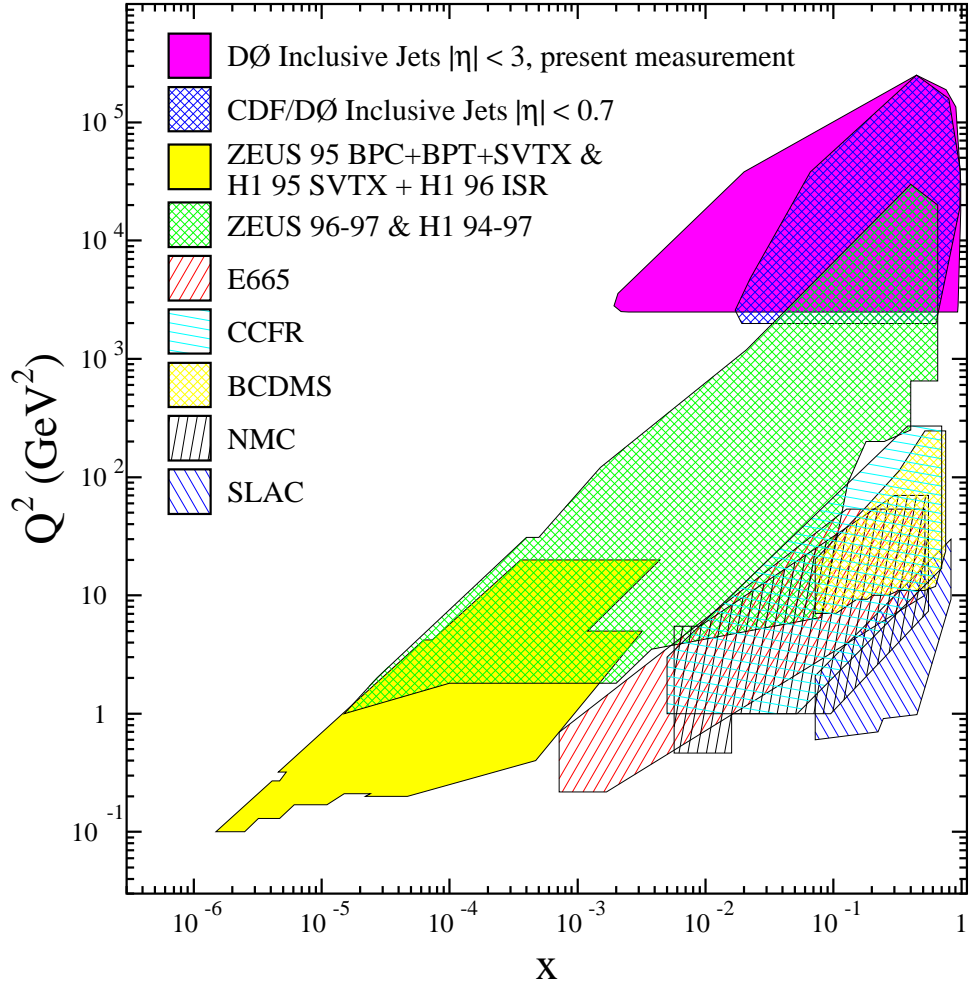,width=5.25in}}}
 \caption{The $x$ and $Q^2$ range of the data set analyzed by the D\O\
 experiment at $\sqrt{s} = 1.8$~TeV (D\O\ [8] and CDF [7] Inclusive
 Jets with $\modeta < 0.7$) compared with the data used to produce
 PDFs [27]. Also shown is the extended $x$ and $Q^2$ reach of the D\O\
 measurement of the inclusive jet cross section with $\modeta < 3$ as
 presented in Ref.~[26].  }
 \label{xq2jet_plot}
\end{center}
\end{figure}

 In this paper we describe the production of hadronic jets as observed
 with the D\O\ detector at the Fermilab Tevatron \pbarp\ collider at
 center-of-mass (CM) energies of 630 and 1800~GeV. High \pt\ jet
 production at $\sqrt{s} = 1800$~GeV probes the structure of the
 proton where the interacting partons carry a fraction of the proton
 momentum, $0.1 \lesssim x \lesssim 0.66$, for momentum transfers
 ($Q$) of $ 2.5 \times 10^3 \lesssim Q^2 \lesssim 2.3 \times
 10^5~\gev^2$, where $Q^2 = E_T^2$ and is equivalent to a distance
 scale of 10$^{-4}$~fm (see Fig.~\ref{xq2jet_plot}). Measurements of
 the inclusive jet cross section, the dijet angular distribution, and
 the dijet mass spectrum, can be used to test the predictions of
 pQCD. Additionally, new phenomena such as quark
 compositeness~\cite{compos} would reveal themselves as an excess of
 jet production at large transverse energy (\Et ) and dijet mass
 ($\jjmass$) relative to the predictions of QCD.

 Previous measurements by the CDF Collaboration of the inclusive jet
 cross section~\cite{CDF_1,CDF_2} and the inclusive dijet mass
 spectrum~\cite{cdf_dijet_mass} have reported an excess of jet
 production relative to a specific QCD prediction. More recent
 analysis of the dijet angular distribution by the \mbox{CDF}
 Collaboration~\cite{cdf_angular} has excluded models of quark
 compositeness in which the contact interaction scale is less than 1.6
 TeV at the 95$\%$ confidence level.

 This paper presents a detailed description of five measurements
 previously published by the D\O\ Collaboration: the inclusive jet
 cross sections at $\sqrt{s} = 1800$~GeV~\cite{d0_inc,Elvira} and
 630~GeV~\cite{d0_inc_630,jkrane}, the ratio of inclusive jet cross
 sections at $\sqrt{s} = 630$ and 1800~GeV~\cite{d0_inc_630,jkrane},
 the dijet angular distribution~\cite{d0_angular,fatyga}, and the
 dijet mass spectrum~\cite{d0_dijet_mass} at $\sqrt{s} = 1800$~GeV. In
 addition to the analyses presented in this paper, D\O\ has recently
 measured the inclusive jet cross section as a function of \Et\ and
 pseudorapidity, \modeta , at $\sqrt{s} = 1800$~GeV~\cite{levan} where
 $\eta \equiv -{\rm ln}[{\rm tan}(\theta/2)]$ and $\theta$ is the
 polar angle.

 For jet measurements, the most critical component of the D\O\
 detector is the calorimeter~\cite{d0_detector}.  A thorough
 understanding of the jet energy scale, jet resolutions, and knowledge
 of biases caused by jet triggering and reconstruction are necessary.
 After detector calibration, small experimental uncertainties can be
 achieved and precise statements can be made about the validity of QCD
 predictions. These results can then be used as the basis for searches
 for physics beyond the standard model.

 In this paper we discuss the theoretical predictions for the
 inclusive jet cross section, the inclusive dijet mass cross section,
 and the dijet angular distribution.  We describe the various
 measurements undertaken to understand and calibrate the D\O\ detector
 for jet measurements. Finally, four different physics measurements
 performed at D\O\ are presented: the inclusive jet cross section, the
 ratio of inclusive jet cross sections, the dijet angular
 distribution, and the dijet mass spectrum. The measurements presented
 here constitute a stringent test of QCD, with a total uncertainty
 substantially reduced relative to previous measurements.  Further,
 the results represent improved limits on the existence of phenomena
 not predicted by the standard model.
\clearpage

\section{Calorimeter}
\label{sec:Calorimeter}

 The D\O\ detector is described in detail elsewhere
 \cite{d0_detector,d0_cal_test_beam}.  The D\O\ uranium-liquid argon
 sampling calorimeters are uniform in structure and provide coverage
 for a pseudorapidity range $\modeta < 4.5$. They are nearly
 compensating with an $e/\pi$ ratio of less than 1.05 above 30 GeV.
 The central and end calorimeters are approximately 7 and 9
 interaction lengths deep respectively, ensuring containment of most
 particles except high-$p_{T}$ muons and neutrinos. The calorimeters
 are segmented into cells of size $\Delta \eta \times \Delta \phi =
 0.1 \times 0.1$, where $\phi$ is the azimuthal angle. These
 characteristics along with excellent energy resolution for electrons
 $(\sim 15\%/\sqrt{E[{\rm GeV}]})$ and pions $(\sim 50\%/\sqrt{E[{\rm
 GeV}]})$ make the D\O\ calorimeters especially well suited for jet
 measurements.

\begin{figure}[htbp]
\centerline{\psfig{figure=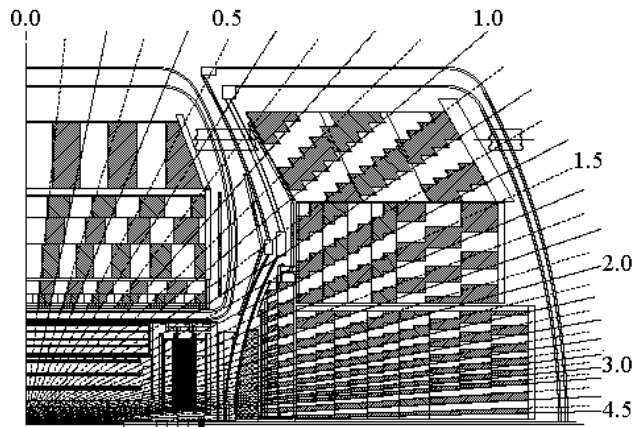,width=3.25in}}
\caption{A schematic view of one quarter of the D\O\ calorimeters. 
The shading pattern indicates the distinct readout cells. The rays
indicate the pseudorapidity intervals defined from the center of the
detector.}
\label{fig:cal_diagram}
\end{figure}

 The calorimeter is divided into three sections (see
 Fig.~\ref{fig:cal_diagram}): a central calorimeter (CC) covering low
 values of pseudorapidity, two end calorimeters (EC) covering forward
 and backward pseudorapidities, and the Intercryostat Detector (ICD)
 covering the gaps between the CC and EC ($0.8 \le \modeta \le
 1.6$). The CC and EC calorimeters each consist of an inner
 electromagnetic (EM) section, a fine hadronic (FH) section, and a
 coarse hadronic (CH) section. Each EM section is 21 radiation lengths
 deep and is divided into four longitudinal samples. The hadronic
 sections are divided into four (CC) and five (EC) layers. The ICD
 consists of scintillator tiles inserted in the space between the CC
 and EC cryostats. The Tevatron accelerator's Main Ring, which is used
 for preaccelerating protons, passes through the CH calorimeters.

\clearpage

\section {Jet Definitions}
\label{sec:jet_def}

 A jet is a collection of collimated particles produced by the
 hadronization of a high-\Et\ quark or gluon. In the measurements
 presented in this paper, we measure the energy and direction of the
 jets produced in \pbarp\ interactions and compare the measurements to
 various theoretical predictions.

 In addition to the jets produced by the high-\pt\ parton-parton
 scattering, there are many particles produced by the hadronization of
 the partons in the proton and antiproton that were not involved in
 the hard scattering process. Because of this there is no unequivocal
 method for experimentally selecting the particles that belong to a
 jet produced in high-\pt\ scattering. It is preferable to use a
 standard definition of a jet to facilitate comparisons of
 measurements from different experiments, and with theoretical
 predictions. In 1990 the so-called Snowmass Jet
 Algorithm~\cite{snowmass} was adopted as a standard definition.

 A jet algorithm can be run on several different input variables,
 calorimeter cells, and particles or partons produced by a Monte Carlo
 Simulation. To differentiate the results of the same algorithm being
 run on these different input we describe the resulting jets as
 follows: A jet (or calorimeter jet) is the result of the jet
 algorithm being run on calorimeter cells; A particle jet is created
 from particles produced by a Monte Carlo simulation after the
 hadronization step; Finally, a parton jet is formed from partons
 before hadronization takes place.
 
 \subsection{The Snowmass Accord}

 The Snowmass Jet Algorithm defines a jet as a collection of partons,
 particles, or calorimeter cells contained within a cone of opening
 angle ${\cal{R}}$. All objects in an event have a distance from the
 jet center, ${\cal{R}}_i \equiv \sqrt{\left(\eta_i - \eta_{\rm jet}
 \right)^2+\left(\phi_i - \phi_{\rm jet}\right)^2}$, where $\eta_{\rm
 jet}$ and $\phi_{\rm jet}$ define the direction of the jet and
 ($\eta_i$, $\phi_i$), are the coordinates of the parton, particle, or
 center of the calorimeter cell. If ${\cal{R}}_i \leq {\cal{R}}$ then
 the object is part of the jet.  The Snowmass suggested value of
 ${\cal{R}} = 0.7$ was used in these measurements. The \Et\ of the jet
 is given by
\begin{equation}
 \Et \equiv {\sum_{i \in {\cal{R}}_i \leq {\cal{R}}}} \ETi
 \label{snow_et},
\end{equation}
 where $i$ is an index for the $i^{\rm th}$ parton or cell. The
 direction of the jet is then given by
\begin{eqnarray}
\eta_{\rm jet} & = &
 { 1 \over \Et} 
 {\sum_{i \in {\cal{R}}_i \leq {\cal{R}}}} \ETi \eta^i,
 \label{snow_def} \\ 
\phi_{\rm jet} & = & 
 { 1 \over \Et} 
 {\sum_{i \in {\cal{R}}_i \leq {\cal{R}}}} \ETi \phi^i.
 \nonumber
\end{eqnarray}

 The Snowmass algorithm gives a procedure for finding the jets:
\begin{itemize}
\item Determine a list of jet ``seeds,'' each with a location
      $\eta_{\rm jet}$, 
      $\phi_{\rm jet}$.
\item Form a jet cone with direction $\eta_{\rm jet}$, 
      $\phi_{\rm jet}$.
\item Recalculate the \Et\ and direction of the jet.
\item Repeat steps 2 and 3 until the jet direction is stable.
\end{itemize}
 The definition of the jet seed is not given by the algorithm. At the
 parton level these seeds could be the partons, points lying between
 pairs of partons, or even a set of points randomly located in
 $\eta$-$\phi$ space. Experimentally, the seed could be defined as any
 cell above a given \Et\ threshold, all cells in the calorimeter, or
 clusters of calorimeter cells. It is up to each experimentalist or
 theorist to define a seed.

\subsection{The D\O\ Experiment's Jet Algorithm}

 At the calorimeter level in the D\O\ experiment, jets are defined in
 two sequential procedures.  In the first, or clustering, procedure
 all the energy depositions that belong to a jet are accumulated, and
 in the second the $\eta,\phi$, and $\Et$ of the jet are defined.  The
 clustering consists of the following steps:
\begin{itemize}
\item[1)] 
 Calorimeter towers (a set of four calorimeter cells of size $\Delta
 \eta \times \Delta \phi = 0.2 \times 0.2$) with $\Et>1$ GeV are
 ordered in \Et .  Starting with the highest-$\Et$ tower, preclusters
 are formed from contiguous towers around these seed towers.
\item[2)] The jet direction ($\eta_{\rm jet}$,$\phi_{\rm jet}$) is
 calculated using Eq.~\ref{snow_def} from the energy deposit pattern in
 a fixed cone of size \Rc\ around the precluster center.
\item[3)] The energy deposited in a cone of size $\Rc$ around the jet
axis is summed and the jet direction ($\eta_{\rm jet},\phi_{\rm jet}$)
is recalculated using the Snowmass algorithm (Eq.~\ref{snow_def}).
\item[4)] 
 Step 3 is iterated until the jet direction is stable.  This is
 typically achieved in two or three iterations.
\item[5)] 
 Only jets with $\Et>8$ GeV are retained.
\item[6)] Jets are merged or split according to the following
 criteria: two jets are merged into one jet if more than 50$\%$ of the
 $\Et$ of the jet with the smaller $\Et$ is contained in the overlap
 region.  If less than 50$\%$ of the $\Et$ is contained in the overlap
 region, the jets are split into two distinct jets and the energy of
 each calorimeter cell in the overlap region is assigned to the
 nearest jet.  The jet directions are recalculated using an
 alternative definition as given in Eq.~\ref{d0_def}.
\end{itemize}
  The D\O\ jet algorithm and the Snowmass algorithm calculate the
  final direction of the jet differently. In the D\O\ jet algorithm
  the final $\eta$ and $\phi$ of the jet are defined as:
\begin{eqnarray}
 \theta_{\rm jet} & = &
  \tan^{-1} \left[{ \sqrt { \left(\sum_{i} \Exi \right)^{2} 
    + \left(\sum_{i} \Eyi \right)^{2}} \over \sum_{i} \Ezi} \right] 
  \nonumber  \\
 \phi_{\rm jet} & = &
  \tan^{-1}\left({\sum_{i} \Eyi \over \sum_{i} \Exi}\right)
  \label{d0_def} \\
  \eta_{\rm jet} & = & 
   -\ln\left[\tan(\theta_{\rm jet}/2)\right] \nonumber
\end{eqnarray}

 {\raggedright{where $i$ corresponds to all towers whose centers are
 within the jet radius $\Rc$,
 $\displaystyle{E^{i}_{x}=E_{i}\sin\theta_{i}\cos\phi_{i}}$,
 $\displaystyle{E^{i}_{y}=E_{i}\sin\theta_{i}\sin\phi_{i}}$, and
 $\displaystyle{E^{i}_{z}=E_{i}\cos\theta_{i}}$.}}

 Applying the 8~GeV $\Et$ threshold to jets before merging and
 splitting has two important consequences.  The first is that two jets
 of $\Et<8$~GeV cannot be merged into a single jet to create a jet
 with $\Et>8$~GeV. The second is that jets may have $\Et<8$~GeV if
 they were involved in splitting.

\subsection{Corrected Jets}
\label{sec:corrected_jets}

 In this paper a ``true'' or corrected jet is the jet that would be
 found by the D\O\ jet algorithm if it was applied to the particles
 produced by the high-\pt\ parton-parton scattering before they hit
 the calorimeter. The jet does not include the particles produced by
 hadronization of the partons not involved in the hard scattering (the
 underlying event). The differences between jets observed in the
 calorimeter and the ``true'' jets are determined using Monte Carlo
 (MC) simulations of \pbarp\ interactions. The direction and \Et\ of
 the ``true'' jets are calculated using the Snowmass definition (see
 Eqs.~\ref{snow_et} and \ref{snow_def}) and are denoted by $E_T^{\rm
 ptcl}$, $\eta^{\rm ptcl}$, and $\phi^{\rm ptcl}$ (where ptcl denotes
 particle).

\subsection{Differences between the D\O\ and Snowmass Algorithms in Data}

\begin{figure}[htbp]
   \centerline {\psfig{figure=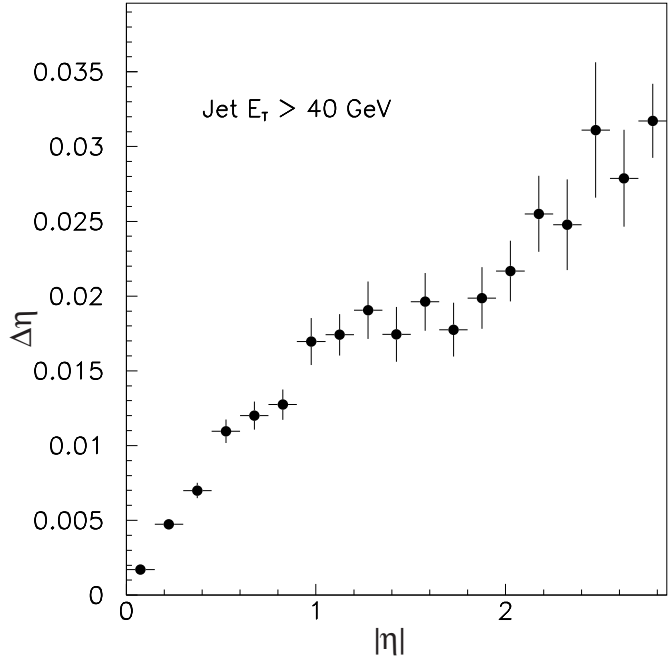,width=3.5in}}
   \caption{The average difference between the $\modeta$ of jets
   reconstructed using the D\O\ algorithm and the Snowmass algorithm
   for D\O\ data.}  \label{FIG:d0_vs_snowmass}
\end{figure}

 Because the D\O\ and Snowmass algorithms calculate the location, and
 hence angle, of the jet differently, a study to measure the
 differences was performed.  The same data events were reconstructed
 using the two different algorithms and the differences in location
 were compared. There were no differences in the \Et\ or $\phi$ of the
 jets. However, there were small differences in the jet $\modeta$,
 which increase as a function of the $\modeta$ of the jets and
 decrease as the transverse energy of the jets increases.  Figure
 \ref{FIG:d0_vs_snowmass} shows the average difference between the
 $\modeta$ of jets with $\Et >$~40 GeV reconstructed using the two
 different algorithms.  As can be seen, the difference is small, even
 at a large $\modeta$.

\subsection{Jet Algorithms at NLO}
 
\begin{figure}[htbp] 
\centerline{\psfig{figure=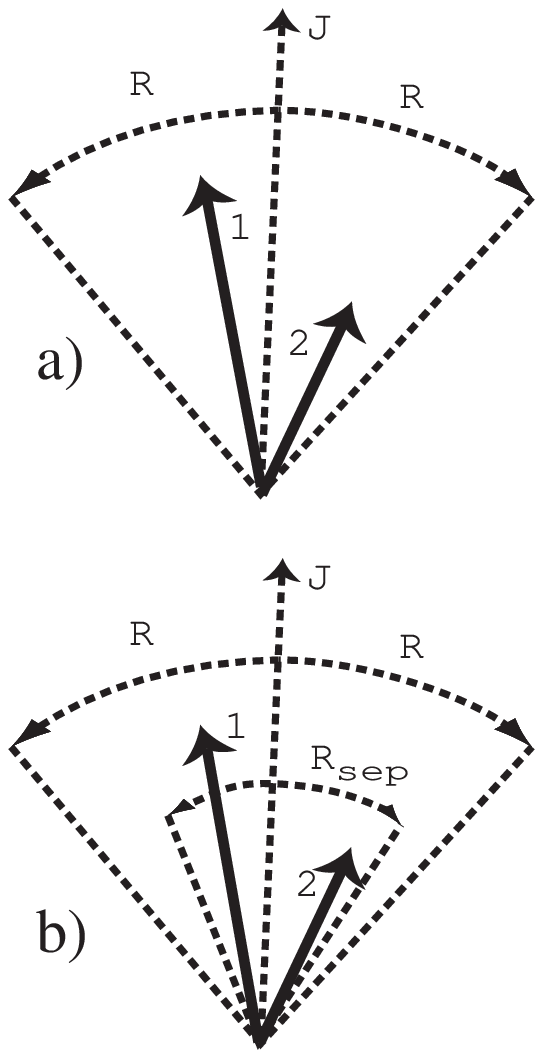,width=3.5in}}
\caption {Illustration and description of the jet definitions at NLO
 parton level as used by the D\O\ experiment. a) The jet definition in
 NLO according to Snowmas. Parton -1- and -2- are combined into jet
 -J-, if the parton distance to the jet axis is less than R. The jet
 axis is defined by partons 1 and 2, according to the Snowmass
 definition.  b) The jet definition in NLO according to the modified
 Snowmass with \Rsep . Use the standard Snowmass clustering, but in
 addition require the distance between the two partons to be less than
 \Rsep .}  \label{FIG:snowmass}
\end{figure}

 In pQCD calculations of parton-parton scattering at leading order
 (LO, ${\mathcal{O}}(\als ^{2})$) there can only be two partons in the
 final state. These partons are well separated and always form two
 jets when the Snowmass algorithm is applied. At next-to-leading order
 (NLO, ${\mathcal{O}}(\als ^{3})$), three partons can be produced in
 the final state. The Snowmass algorithm at the NLO parton level is
 illustrated in Fig.~\ref{FIG:snowmass}(a).  For any two partons in
 the final state, the seeds direction is given by applying
 Eq.~\ref{snow_def} to the two partons. If the partons lie within a
 distance \Rc\ from the seed's the two partons are combined to form a
 jet.
 
 In the Snowmass algorithm the partons contributing to a single jet
 can have a maximum separation of $2\Rc$. Consider a two parton final
 state with the partons separated by $2\Rc$. The experimentally
 observed energy pattern will be determined by the parton showering,
 hadronization, and calorimeter response.  Application of the D\O\ jet
 algorithm to the calorimeter energy deposition that results from the
 hadronization of the two partons will produce one or two jets
 depending on the splitting and merging criteria. The Snowmass
 algorithm is only capable of finding one jet, and hence cannot match
 the experimental measurement.

 This example illustrates the different treatment of jets at the
 parton and calorimeter level. To accommodate the differences between
 the jet definitions at the parton and calorimeter levels, an
 additional, purely phenomenological parameter has been suggested in
 Ref.~\cite{ellis}.  The variable is called $\Rsep$ and is the maximum
 allowed distance ($\Delta \Rc$) between two partons in a parton jet,
 divided by the cone size used: $\Rsep= \Delta \Rc / \Rc$.  This
 algorithm is illustrated in Fig.~\ref{FIG:snowmass}(b) and is
 referred to as the modified Snowmass algorithm.

 The value of $\Rsep$ depends on details of the jet algorithm used in
 each experiment. At the parton level $\Rsep$ is the distance between
 two partons when the clustering algorithm switches from a one-jet to
 a two-jet final state, even though both partons are contained within
 the jet defining cone.  The value of $\Rsep$ depends on the
 experimental splitting/merging scheme.  After several studies an
 $\Rsep$ value of 1.3 was found to best simulate the D\O\ merging and
 splitting criteria \cite{rsep}.

\subsection{Jet Reconstruction Efficiency}

\begin{figure}[htbp]
   \centerline{\psfig{figure=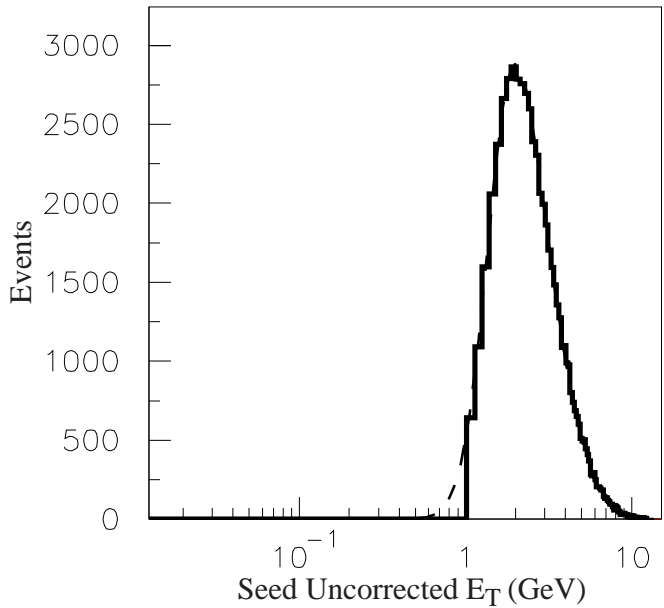,width=3.5in}}
   \caption{Seed tower distributions for ${\cal{R}}=0.7$ cone jets
   with an \Et\ range of 18--20 GeV. The data is represented by the
   solid histogram and the fit is given by the dashed curve. }
   \label{FIG:seed_towers}
\end{figure}

\begin{figure}[htbp]
   \centerline
   {\psfig{figure=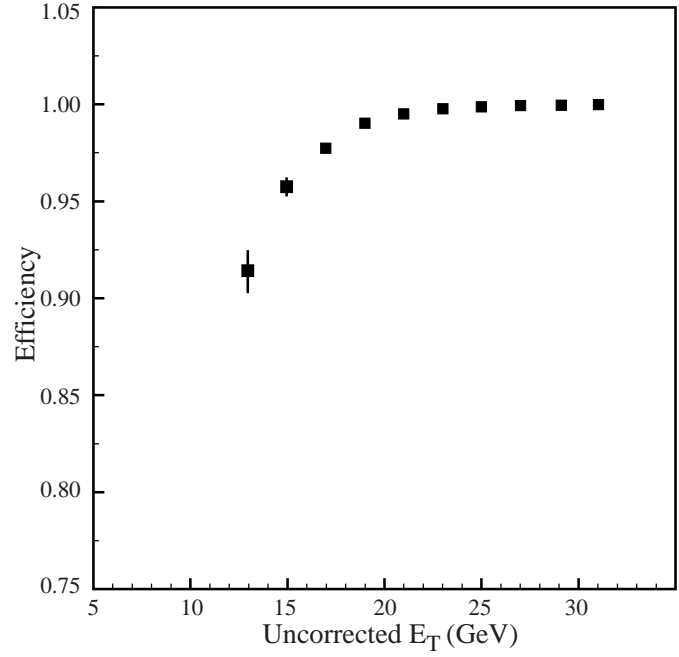,width=3.5in}}
   \caption{Reconstruction efficiency as a function of jet \Et .}
   \label{FIG:jet_efficiency}
\end{figure}

 The jet algorithm does not reconstruct all jets with the same
 efficiency.  Primarily this is due to calorimeter energy clusters not
 containing a seed tower of \Et\ greater than 1 GeV.  Since the jet
 algorithm explicitly depends on the \Et\ of a seed tower used to
 begin searching for a jet, the seed tower distributions are studied
 to determine if jets are likely to have seed towers below threshold.
 Figure~\ref{FIG:seed_towers} shows the seed-tower-\Et\ distribution
 of jets for an \Et\ range of 18 to 20 GeV (other \Et\ ranges have
 similar distributions). The distribution is fitted with
\begin{equation}
 A \cdot \exp \left[{\frac {-0.5 \cdot (-W+e^{W})}
               {\sqrt{2} \cdot \lambda} }\right] 
\end{equation}
 where $W = {(\Et-\chi)}/{\lambda}$, and $A$, $\chi$, and $\lambda$
 are free parameters in the fit.  Assuming that the seed towers are
 smoothly distributed in \Et , the fraction of jets not containing a
 seed tower exceeding 1 GeV is determined from the fit and used to
 calculate the jet reconstruction efficiency.
 Figure~\ref{FIG:jet_efficiency} shows the reconstruction efficiency
 for jets as a function of jet \Et .  For jets with an \Et\ of 20 GeV
 and $\Rc = 0.7$, the reconstruction efficiency is $99\%$.


\subsection{Biases in the Jet Algorithm}
\label{sec:eta_bias}

\begin{figure}[htbp]
   \centerline
   {\psfig{figure=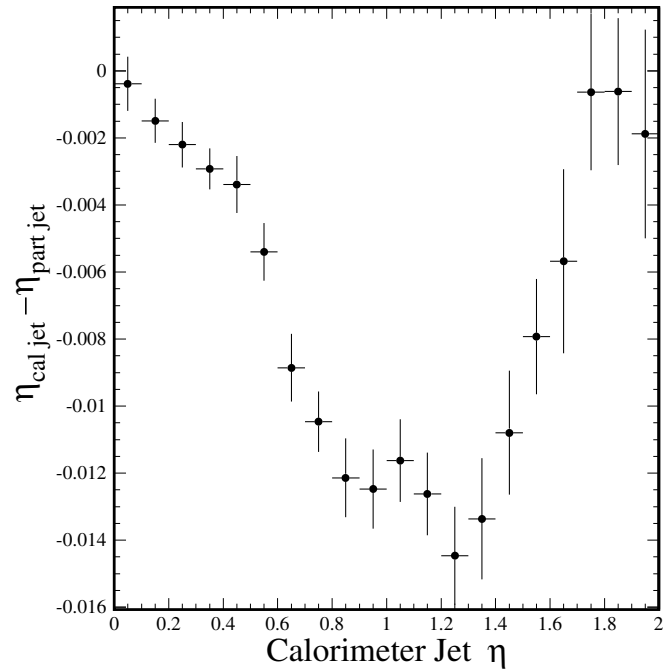,width=3.5in}}
   \caption{{\sc herwig} Monte Carlo simulation of the $\eta$ bias for
   all jet energies as a function of the reconstructed jet $\eta$.}
   \label{FIG:eta_bias}
\end{figure}

 The $\eta$ dependence of the calorimeter energy response together
 with algorithm related effects may cause a bias in the reconstructed
 jet $\eta$. The $\eta$ of the jet, assuming perfect position
 resolution, is:
\begin{equation}
\eta^{\rm ptcl} = \eta +\rho(E,\eta)
\end{equation}
 where $\rho(E,\eta)$ is the possible bias. To measure the bias, the
 {\sc herwig}~\cite{herwig} Monte Carlo event generator and the D\O\
 detector simulation, {\sc d\o geant}~\cite{geant}, are used. Jets are
 reconstructed at both the particle and calorimeter level.
 Statistically, \mbox{$\langle \rho(E,\eta) \rangle$} can be obtained
 as \mbox{$\langle \eta^{\rm ptcl} - \eta\rangle$} where a matching
 criterion is used to associate the particle jet to the reconstructed
 calorimeter jet.  Figure \ref{FIG:eta_bias} shows the $\eta$ bias for
 all jet energies as a function of $\eta$.  The bias in $\eta$ is less
 than 0.02 for all $\eta$. The magnitude of the bias is greatest when
 part of the jet falls into the Intercryostat Region ($0.8 < \eta <
 1.6$), which is the least instrumented region of the calorimeter.

  A similar study was performed to measure a possible bias in $\phi$
  (azimuth).  Since the calorimeter has a symmetric tower structure in
  $\phi$, no bias is expected.  The bias in $\phi$ (azimuth) was
  measured to be small --- much less than 0.01 radians.  Any bias
  introduced by this effect will be small for most physics analyses
  since the $\Delta \cal{R}$ between jets is typically used rather
  than the absolute $\phi$ or $\eta$ position. The analyses presented
  in this publication are not corrected for these effects.
\clearpage

\section{Theoretical Predictions}
\label{sec:theory_predictions}

\subsection{NLO QCD Predictions}

 Within the framework of pQCD, high \Et\ jet production can be
 described as an elastic collision between a single parton from the
 proton and a single parton from the
 antiproton~\cite{jetsreview}. After the collision, the outgoing
 partons form localized streams of particles called ``jets.''
 Predictions for the inclusive jet cross section, the dijet angular
 distribution, and the dijet mass spectrum are in general given
 by~\cite{jetsreview}:
\begin{eqnarray}
\sigma &=& \sum_{ij} \int dx_1 dx_2 
f_i\left(x_1,\mu_{F}^{2}\right)f_j\left(x_2,\mu_{F}^{2}\right)
\nonumber\\
& & {\hat{\sigma}}{\left[x_{1}P_1,x_{2}P_{2}, 
\als \left( \mu_{R}^{2}\right), \frac{Q^2}{\mu_F^2},
\frac{Q^2}{\mu_R^2} \right]}
\end{eqnarray}
 where $f_{i(j)}\left(x_{1(2)},\mu_{F}^{2}\right)$ represent the PDFs
 of the proton (antiproton) defined at factorization scale $\mu_F$,
 $\hat{\sigma}$ is the parton scattering cross section, $P_{1(2)}$ is
 the momentum of the proton (antiproton), $x_{1(2)}$ is the fraction
 of the proton (antiproton) momentum carried by the scattered parton,
 $Q$ is the hard scale that characterizes the parton scattering (which
 could be the \Et\ of a jet, the dijet mass of the event, etc.), and
 $\mu_R$ is the renormalization scale.

 The parton scattering cross sections have been calculated to
 next-to-leading order (NLO, ${\mathcal{O}}(\als ^{3})$). The
 perturbation series requires renormalization to remove ultraviolet
 divergences. This introduces a second scale to the problem, $\mu_R$.
 In addition, an arbitrary factorization scale, $\mu_F$, is introduced
 to remove the infrared divergences. Qualitatively, it represents the
 scale that separates the short- and long-range processes. A parton
 emitted with transverse momentum relative to the proton less than the
 scale $\mu_F$ will be included in the PDF, while a parton emitted at
 large transverse momentum will be included in $\hat{\sigma}$. The
 scales $\mu_R$ and $\mu_F$ should be chosen to be of the same order
 as the hard scale, $Q$, of the interaction. The larger the number of
 terms included in the perturbative expansion, the smaller the
 dependence on the values of $\mu_R$ and $\mu_F$. If all orders of the
 expansion could be included, the calculation should have no
 dependence on the choice of scales. In this article the
 renormalization scale is written as the product of a constant, $D$,
 and the hard scale of the interaction, $\mu = DQ$. Typically, the
 renormalization and factorization scales are set to the same value,
 $\mu = \mu_R = \mu_F$.

 Several pQCD NLO calculations have been
 performed~\cite{aversa,eks,jetrad}. In this paper we use the event
 generator {\sc jetrad}~\cite{jetrad} and a version of the analytic
 calculation {\sc eks}~\cite{eks} that integrates the cross section
 over bins. Both programs require the selection of a renormalization
 and factorization scale, a set of parton distribution functions, and
 a jet clustering algorithm.  Two partons are combined if they are
 contained within a cone of opening angle ${\cal{R}} = \sqrt{\Delta
 \eta^2 + \Delta \phi^2} = 0.7$, and are also within ${\cal{R}}_{\rm
 sep}=1.3$ (see Section~\ref{sec:jet_def}).  The authors of {\sc
 jetrad} have provided several choices for the renormalization
 scale. We have chosen a scale proportional to the \Et\ of the leading
 jet: $\mu = D\Etmax$, where $D$ is constant in the range $0.25 \le D
 \le 2.00$. An alternative scheme sets the scale to be proportional to
 the center-of-mass energy of the two outgoing partons: $\mu =
 C\sqrt{\hat{s}} = C\sqrt{x_{1}x_{2}s}$ where $C$ is constant in the
 range $0.25 \le C \le 1.00$, $x_{\rm 1} = \sum \Et_{\rm i}
 e^{\eta_{i}} / \sqrt{s}$, $x_{\rm 2} = \sum \Et_{\rm i} e^{-\eta_{i}}
 / \sqrt{s}$, and $i = 1, \dots, n$ where $n$ is the number of jets in
 the event. The authors of {\sc eks} prefer a different definition of
 the renormalization scale: the \Et\ of each jet in the event, $\mu =
 D\Etjet$ (a version of {\sc eks} that uses the renormalization scale
 $\mu = D\Etmax$ is also available).

 Several choices of PDF are considered. The \cteqthreem~\cite{cteq3m}
 and \mrsap~\cite{mrsap} PDFs are fits to collider and fixed target
 data sets published before 1994. \cteqfourm~\cite{cteq4m} updates
 these fits using data published before 1996, and \cteqfoura\ repeats the
 fits with values of $\als (M_Z)$ fixed in the range 0.110 to 0.122
 (\cteqfourm\ corresponds to an $\als (M_Z)$ of
 0.116). \cteqfourhj~\cite{cteq4m} adjusts the gluon distributions to
 fit a \mbox{CDF} inclusive jet cross section measurement~\cite{CDF_2}
 by increasing the effective weighting of the CDF data. \mrst\
 \cite{mrst} incorporates all data published before 1998. In addition
 to the standard \mrst\ PDF, two alternative PDFs are provided that
 vary the gluon distributions within the range allowed by experimental
 observations. The resulting distributions are called \mrstgu and
 \mrstgd .

 Since the publication of the \mrst\ and \mbox{CTEQ4} PDFs, problems
  were found in the implementations of the QCD evolution of the parton
  distributions in $Q^2$~\cite{pdf_problem}. This was caused by
  approximations to NLO QCD to reduce the time required for
  computation. The removal of these approximations could lead to
  changes of approximately $5\%$ in the theoretical predictions
  presented in this paper. Currently, PDFs calculated without the
  approximations are not available for use with {\sc jetrad} and {\sc
  eks}.
 
\subsubsection{Inclusive Jet Cross Section}

 The inclusive jet cross section may be expressed in several ways.
 Theoretical calculations are normally expressed in terms of the
 invariant cross section
\begin{equation}
 E \frac{d^{3} \sigma}{dp^3}.
\end{equation}
 In the D\O\ experiment the measured variables are the transverse
 energy (\Et) and pseudorapidity ($\eta$). In terms of these
 variables, the cross section is expressed as
\begin{equation}
\frac{d^{2}\sigma}{d\Et d\eta},
\end{equation}
 where the two are related by
\begin{equation}
  E \frac{d^{3} \sigma}{dp^3} \equiv  
\frac{d^{3} \sigma}{d^2p_T dy} \rightarrow 
\frac{1}{2 \pi E_{T}} \frac{d^{2}\sigma}{d\Et d\eta},
\end{equation}
 where $y$ is the rapidity of the jet.  The final expression follows
 if the jets are assumed to be massless.  For most measurements, the
 cross section is averaged over some range of pseudorapidity: in this
 paper $\modeta < 0.5$ and $0.1 < \modeta < 0.7$.

 The inclusive jet cross section measures the probability of observing
 a hadronic jet with a given \Et\ and $\eta$ in a \pbarp\
 collision. The term ``inclusive'' indicates that the presence or
 absence of additional objects in an event does not affect the
 selection of the data sample. An event which contains three jets that
 pass the selection criteria, for instance, will be entered into the
 cross section three times. The inclusive measurement is sometimes
 denoted $\sigma ( \pbarp \rightarrow {\rm jet} + X)$.

\begin{figure}[htbp]
\vbox{\centerline
{\psfig{figure=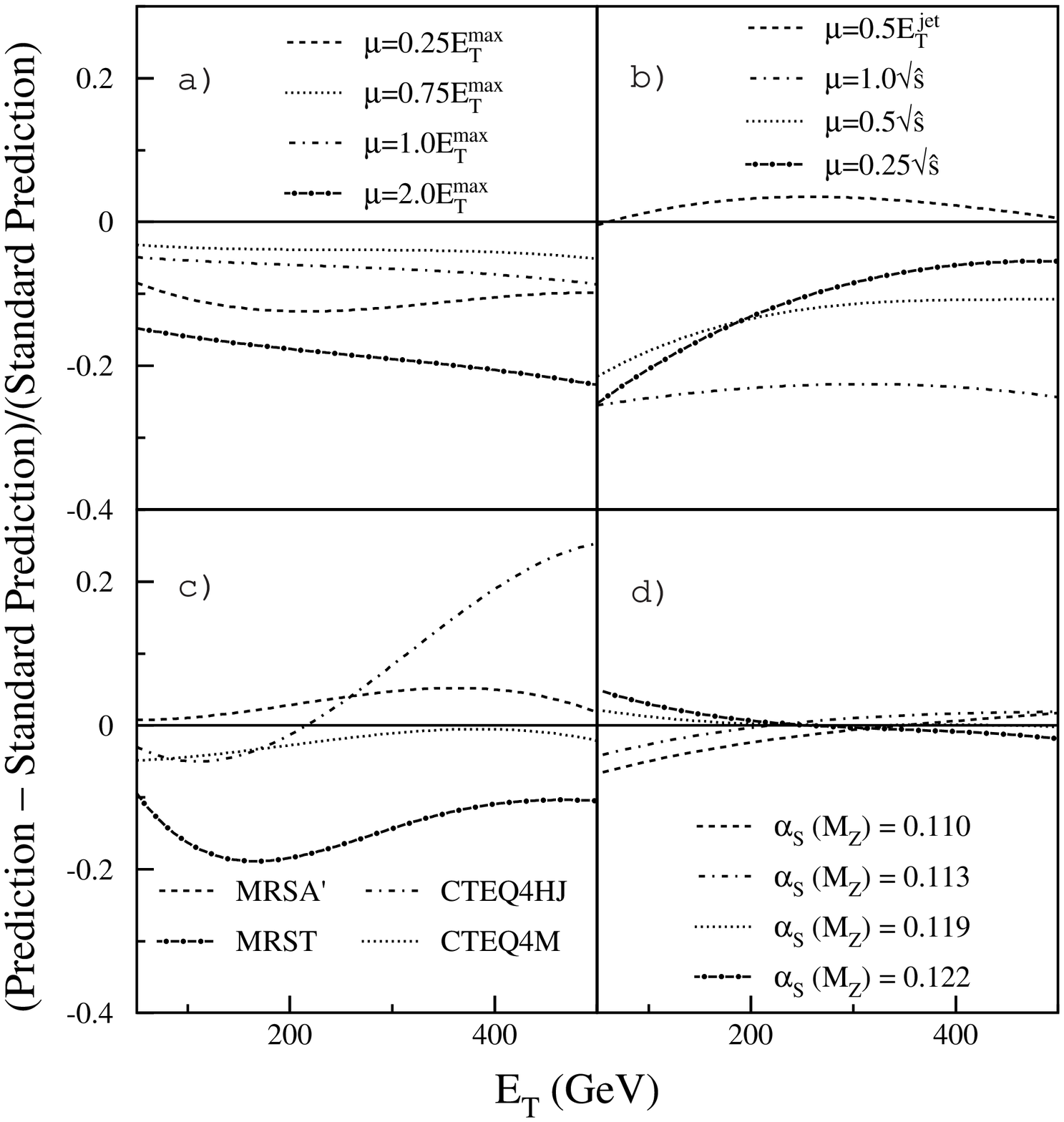,width=3.5in}}}
\caption{The difference between alternative predictions and the 
 reference prediction ($\mu =0.5\Etmax$, \cteqthreem ) for the
 inclusive jet cross section for $\modetajet < 0.5$ at $\sqrt{s}$ =
 1.8 TeV. Shown are the alternative predictions for the choices (a)
 $\mu =$ (0.25, 0.75, 1.0, 2.0)\Etmax , (b) $\mu =$ (0.25, 0.5,
 1.0)$\sqrt{\hat{s}}$ and 0.5\Etjet , (c) \cteqfourm , \cteqfourhj ,
 \mrsap , and \mrst , and (d) for $\als = 0.110 - 0.122$ using the
 \cteqfoura\ PDFs compared with the calculation using \cteqfourm ,
 for which $\als = 0.116$.}
\label{FIG:inc_cross_theory_uncertainties}
\end{figure}

 Theoretical predictions for the inclusive jet cross section are
 generated using the {\sc jetrad} and {\sc eks} programs.  Our
 reference prediction is the {\sc jetrad} calculation with $\mu =
 0.5\Etmax$, ${\mathcal{R}}_{\rm sep} =$ 1.3, and the \cteqthreem\
 PDF. The predictions are smoothed by fitting to the function
\begin{equation}
 A E_{T}^{-\alpha} \left(1 - \frac{2\Et}{\sqrt{s}} \right)^{\beta}
{\mathcal{P}}_{6}\left(\Et \right),
\end{equation}
 where ${\mathcal{P}}_{6}\left(\Et \right)$ is a sixth order
 polynomial. The resulting uncertainty due to smoothing is less than
 $2\%$ for a given \Et .  The uncertainty in the calculations
 resulting from the choices of different renormalization scales and
 PDFs is approximately 30$\%$ and varies as a function of \Et
 ~\cite{inc_jet_theory_uncertainties}.
 Figure~\ref{FIG:inc_cross_theory_uncertainties} shows the variations
 in the predictions for the inclusive jet cross section at $\sqrt{s} =
 1800$ GeV for {\sc jetrad}. The uncertainties in the inclusive jet
 cross section at $\sqrt{s} = 630$ GeV are of a similar size.
 
\subsubsection{Ratio of Inclusive Jet Cross Sections at
\protect{${\sqrt{s} =}$} 1800 and 630 GeV} \label{sec:ratio}

\begin{figure}[htbp]
\begin{center}
\vbox{\centerline{\psfig{figure=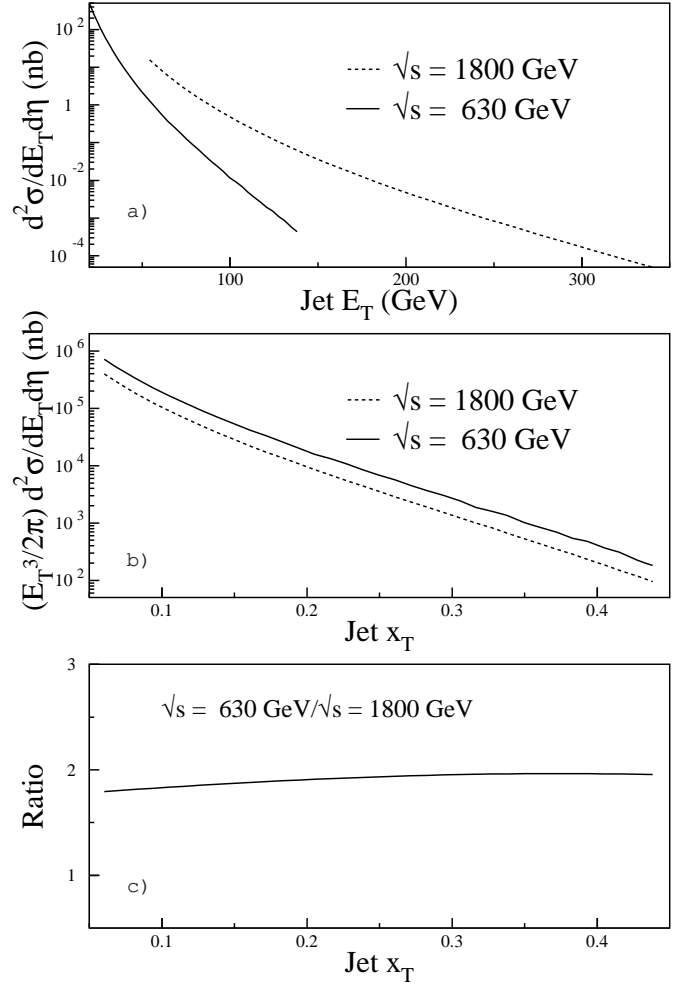,width=3.5in}}}
  \caption{Different presentations of the NLO inclusive jet cross
  sections at (dashed line) $\sqrt{s}=1800$ and (solid line) 630
  GeV. Theoretically, the scaled dimensionless cross sections (b)
  should be nearly exponential and close to one another. Panel (c)
  shows the ratio of dimensionless inclusive jet cross sections at
  $\sqrt{s} = 630$ and 1800 GeV for $\modetajet < 0.5$.}
  \label{xs_scale_example}
\end{center}
\end{figure}

 While it is possible to compare the inclusive jet cross sections as a
 function of \Et\ for both center-of-mass energies, the data will
 differ greatly in both magnitude and \Et\ range (see
 Fig.~\ref{xs_scale_example}(a)). If we express the cross section as a
 dimensionless quantity
\begin{equation}
E_{T}^4 E \frac{d^{3} \sigma}{dp^3} \equiv
\frac{E_{T}^{3}}{2\pi} \frac{d^{2}\sigma}{d\Et d\eta},
\end{equation}
 and calculate it as a function of $\displaystyle{x_{T} \equiv
 {2\Et}/{\sqrt{s}}}$, the ``scaling'' hypothesis, which is motivated
 by the Quark-Parton Model, predicts that it will be independent of
 the center-of-mass energy. However, QCD leads to scaling violation
 through the running of $\als$ and the evolution of the PDFs.
 
\begin{figure}[htbp]
\vbox{\centerline
{\psfig{figure=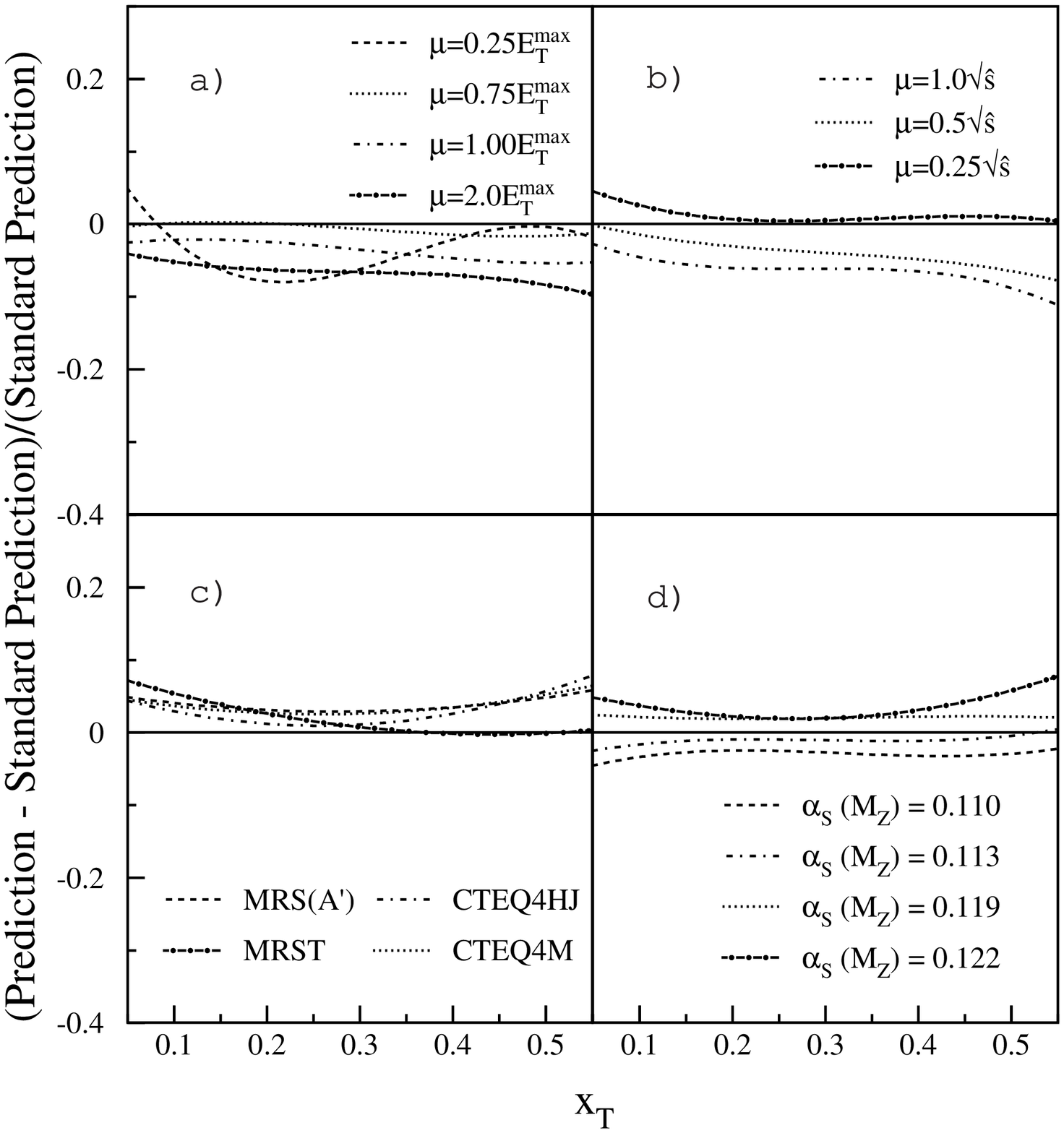,width=3.5in}}}
\caption{The difference between alternative predictions and the 
 reference prediction ($\mu =0.5\Etmax$, \cteqthreem ) of the ratio of
 inclusive jet cross sections at $\sqrt{s} = 630$ and 1800 GeV for
 $\modetajet < 0.5$. Shown are the alternative predictions for the
 choices (a) $\mu =$ (0.25, 0.75, 1.0, 2.0)\Etmax , (b) $\mu =$ (0.25,
 0.5, 1.0)$\sqrt{\hat{s}}$ and 0.5\Etjet , (c) \cteqfourm ,
 \cteqfourhj , \mrsap , and \mrst , and (d) for $\als = 0.110 - 0.122$
 using the \cteqfoura\ PDFs compared with the calculation using
 \cteqfourm , for which $\als = 0.116$.}
\label{FIG:ratio_theory_uncertainties}
\end{figure}

 By taking the ratio of the cross sections at $\sqrt{s} = 1800$ and
 630~GeV, many of the theoretical and experimental uncertainties are
 reduced. Variations in the prediction resulting from the choice of
 renormalization scale, factorization scale, and PDF are approximately
 10$\%$ and vary as a function of $x_T$. This is a significant
 reduction in the theoretical uncertainty compared to the
 uncertainties in the inclusive jet cross sections. The theoretical
 predictions for the ratio of the inclusive jet cross sections are
 calculated using the {\sc jetrad} and {\sc eks}
 programs. Figure~\ref{FIG:ratio_theory_uncertainties} shows the
 variations in the ratio between inclusive jet cross sections at
 $\sqrt{s} = 630$ and 1800 GeV for {\sc jetrad}.

 \subsubsection{Dijet Angular Distributions}

 At leading order two jets are produced. The invariant mass of the
 jets is given by
\begin{equation}
\jjmass^2 \equiv {\hat{s}} = 4 p_{T}^{2} \cosh^2 \left( \Delta y/2 \right)
\end{equation}
 where ${\hat{s}} = x_{1} x_{2} s$, is the CM energy squared of the
 interacting partons, and $\Delta y$ is the separation in rapidity of
 the two jets. If we assume that the jets are massless we can write
 the dijet invariant mass as
\begin{equation}
\jjmass^{2} = 2  \Etu{1}  \Etu{2}
  \left[ \cosh \left( \Delta \eta \right) - \cos \left( \Delta \phi
 \right) \right],
\end{equation}
 where $\phi$ is the azimuthal angle with respect to the beam. Since
 the dijet mass represents the center-of-mass energy of the
 parton-parton interaction, it directly probes the parton scattering
 cross section. The presence of higher-order processes can result in
 the production of additional jets. In this case the mass is
 calculated using the two highest-\Et\ jets in the event.

 If only two partons are produced in a parton-parton interaction, and
 we neglect the intrinsic transverse momentum of the scattering
 partons, the two jets will be back-to-back in azimuth and balance in
 transverse momentum.  The resulting two-jet inclusive cross section
 at LO can be written as a function of the $p_T$ and rapidity ($y_3$,
 $y_4$) of the jets~\cite{jetsreview}
\begin{equation}
{ \frac{d^{3} \sigma}{dy_3 dy_4 d p_T^2}}.
\end{equation}
This can be
 rewritten in terms of the dijet invariant mass and the center of mass
 scattering angle, $\theta^\star$, using the transformation~\cite{jetsreview}
\begin{equation}
{dp_T^2}dy_3 dy_4  \equiv 4 dx_1 dx_2 d \cos{\theta ^{\star}}
\end{equation}
resulting in
\begin{equation}
\frac{d^2 \sigma}{dM d\cos{\theta ^{\star}}} = \sum_{ij} \int^1_0 dx_1
dx_2 \delta \left( x_1 x_2 s - M^2 \right)
\frac{d{\hat{\sigma}}_{ij}}{d\cos{\theta ^{\star}}}. 
\end{equation}

\begin{figure}[htbp]
\vbox{\centerline{\psfig{figure=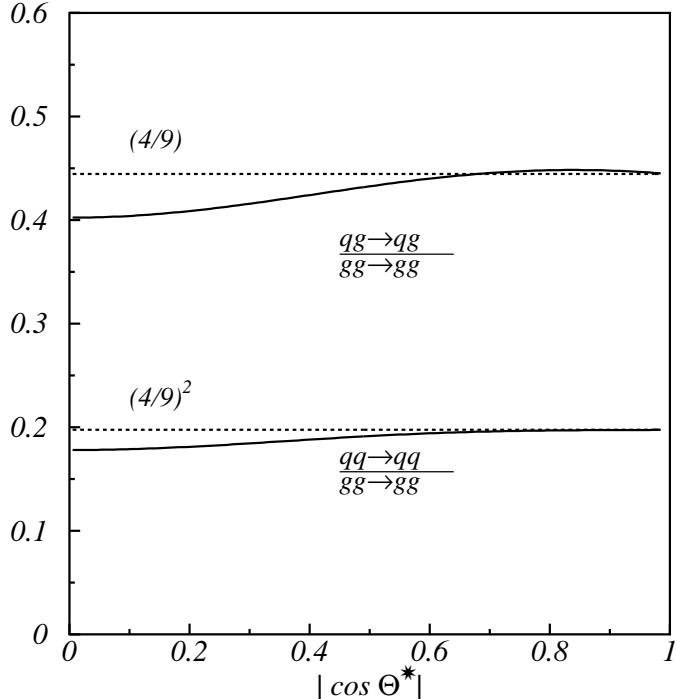,width=3.5in}}}
\caption{Quark-antiquark and quark-gluon angular distributions, 
         normalized to the angular distribution for gluon-gluon
	 scattering.}
\label{FIG:ang_dep}
\end{figure}

 The dijet angular distribution as measured in the dijet
 center-of-mass frame is sensitive to the QCD matrix elements.
 Angular distributions for the $qg \to qg$, $q \bar{q} \to q \bar{q}$,
 and $gg \to gg$ processes are similar. The properties of
 parton-parton scattering are almost independent of the partons
 involved (see Fig.~\ref{FIG:ang_dep}). The dominant process in QCD
 parton-parton scattering is $t$-channel exchange, which results in
 angular distributions peaked at small center-of-mass scattering
 angles. Many theoretical predictions for phenomenology beyond the SM
 have an isotropic angular distribution and could be detected using
 the measurement of the dijet angular distribution.

 At small center-of mass-scattering angles, $\theta^\star$, the dijet
 angular distribution predicted by leading order QCD is proportional
 to the Rutherford cross section:
\begin{equation}
{ {d\hat{\sigma} \over {d\cos \theta^{\star}}} \sim
{{1}\over{\sin ^4({{\theta ^{\star}}/{2}})}} }.
\end{equation}
 It is conventional to measure the angular distribution in the
 variable $\chi$, rather than $\cos \theta^{\star}$, where:
\begin{equation}
{{\chi} = {{1+|\cos \theta^{\star}|}\over{1-|\cos \theta^{\star}|}}} = 
\exp \left( \mid \! \Delta \eta  \! \mid \right). 
\end{equation}
 Plotting the dijet angular distribution in the variable $\chi$
 flattens out the distribution and facilitates comparison to
 theory~\cite{jetsreview} ($d\sigma/ d\chi$ is uniform for Rutherford
 scattering). The differential angular cross section measured in this
 analysis is:
\begin{equation}
{d^3 \sigma \over {dM \, d\chi \, d\eta_{\rm boost}}}, 
\end{equation}
 where $\eta_{\rm boost} = {0.5}(\eta_1 + \eta_2)$. The predictions
 are calculated using {\sc jetrad}.

 \subsubsection{Inclusive Differential Dijet Mass Cross Section}

\begin{figure}[htbp]
\vbox{\centerline
{\psfig{figure=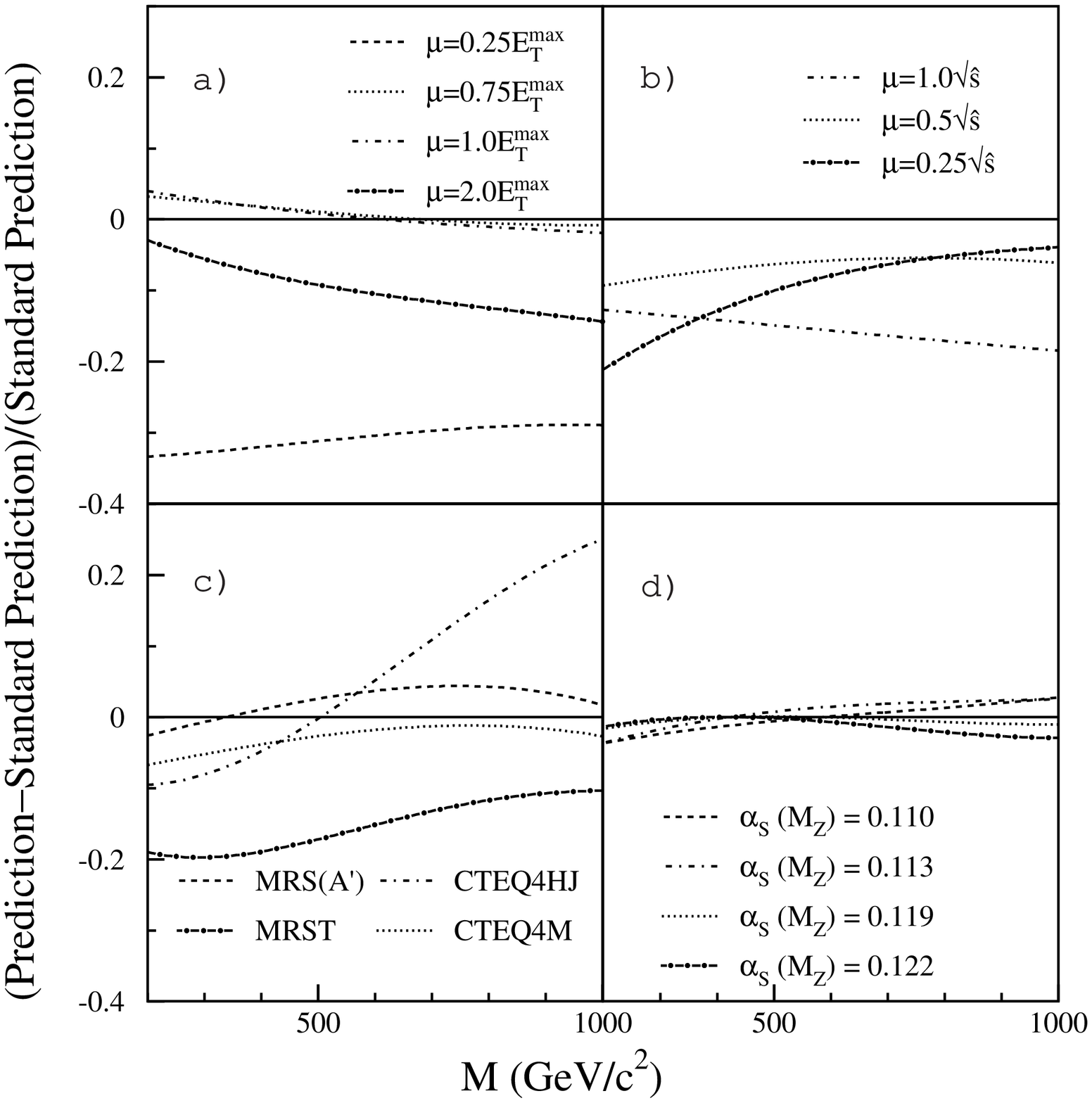,width=3.5in}}}
\caption{The differences between the alternative predictions and the
reference prediction ($\mu =0.5\Etmax$, \cteqthreem) of the inclusive
dijet mass cross section (Eq.~\ref{eq:dijet_mass_xsec}) at $\sqrt{s} =
1800$ GeV for $\modetajet < 1.0$. Shown are the alternative
predictions for the choices (a) $\mu =$ (0.25, 0.75, 1.0, 2.0)\Etmax ,
(b) $\mu =$ (0.25, 0.5, 1.0)$\sqrt{\hat{s}}$ and 0.5\Etjet , (c)
\cteqfourm , \cteqfourhj , \mrsap , and \mrst , and (d) for $\als =
0.110 - 0.122$ using the \cteqfoura\ PDFs compared with the
calculation using \cteqfourm , for which $\als = 0.116$.}
\label{FIG:mass_theory_uncertainties}
\end{figure}

 The inclusive triple differential dijet mass cross section is
 obtained by integrating over $\cos \theta^{\star}$ and is given by
\begin{equation}
\frac{d^{3} \sigma}{ d \jjmass d \eta^{{\rm jet1}} d \eta^{{\rm jet2}}}
\label{eq:dijet_mass_xsec}
\end{equation}
 where $\eta^{{\rm jet1,2}}$ are the pseudorapidities of the jets. We
 integrate the cross section over a range of pseudorapidity such that
 both jets satisfy $\modetajet < 1.0$.  The NLO predictions for this
 cross section are calculated using {\sc jetrad}. The {\sc jetrad}
 predictions were smoothed by fitting them to an ansatz function of
 the form
\begin{equation} A \cdot M^{- \alpha}
 \exp{\left[ -\beta \jjmass -\gamma M^{2} - \delta M^{3} \right]
 {\mathcal{P}}_{n}\left(\jjmass \right) }
\end{equation} 
 where ${\mathcal{P}}_{n}\left(\jjmass \right)$ is a polynomial of
 degree $n \le 6$ and $\alpha$, $\beta$, $\gamma$, and $\delta$ are
 fit parameters.  The uncertainty due to the form of the ansatz
 function not being quite right is estimated to be $< 2\%$. The
 uncertainties in the theoretical predictions are due to the choice of
 $\mu$ and PDF, and are approximately 40--50$\%$ with some dependence
 on \jjmass\ (see Fig.~\ref{FIG:mass_theory_uncertainties}).

\subsection{Quark Compositeness}
\label{sec:theory_compositeness}

 The existence of three generations of quarks and leptons suggests
 that they may not be fundamental particles. For example, it has been
 proposed~\cite{compos} that they could be composed of ``preons''
 which interact via a new strong interaction called metacolor. Below a
 characteristic energy scale $\Lambda$, the preons form metacolor
 singlets that are the quarks. The scale $\Lambda$ characterizes both
 the strength of the preon coupling and the physical size of the
 composite state ($\Lambda$ is defined so that $g^{2}/4\pi =
 1$). Limits are set assuming that all quarks are composite and
 \mbox{$\Lambda\gg\sqrt{\hat{s}}$} (where $\sqrt{\hat{s}}$ is the
 center of mass energy of the colliding partons), so that quarks
 appear to be point-like. Hence, the substructure coupling can be
 approximated by a four-fermion contact interaction described by an
 effective Lagrangian~\cite{compos}:
\begin{eqnarray}
{\mathcal L}& = \frac{g}{2 \Lambda^2} \biggl\{ \biggr. &
\eta^{0}_{LL} \left({\bar{q}}_{L} \gamma^\mu q_{L}\right) 
\left({\bar{q}}_{L} \gamma_\mu q_{L}\right) + \nonumber \\
	& &
\eta^{0}_{LR} \left({\bar{q}}_{L} \gamma^\mu q_{L}\right) 
\left({\bar{q}}_{R} \gamma_\mu q_{R}\right) + \nonumber \\
	& &
\eta^{0}_{RL} \left({\bar{q}}_{R} \gamma^\mu q_{R}\right) 
\left({\bar{q}}_{L} \gamma_\mu q_{L}\right) +\nonumber \\
	& &
\eta^{0}_{RR} \left({\bar{q}}_{R} \gamma^\mu q_{R}\right) 
\left({\bar{q}}_{R} \gamma_\mu q_{R}\right) + \nonumber \\
	& &
\eta^{1}_{LL} \left({\bar{q}}_{L} \gamma^\mu \frac{\lambda_a}{2}q_{L}\right) 
\left({\bar{q}}_{L} \gamma_\mu \frac{\lambda_a}{2}q_{L}\right)  +  \nonumber\\
	& &
\eta^{1}_{LR} \left({\bar{q}}_{L} \gamma^\mu \frac{\lambda_a}{2}q_{L}\right) 
\left({\bar{q}}_{R} \gamma_\mu \frac{\lambda_a}{2}q_{R}\right)  +   \nonumber\\
	&  &
\eta^{1}_{RL} \left({\bar{q}}_{R} \gamma^\mu \frac{\lambda_a}{2}q_{R}\right) 
\left({\bar{q}}_{L} \gamma_\mu \frac{\lambda_a}{2}q_{L}\right)  +  \nonumber\\
	&  &
\biggl.
\eta^{1}_{RR} \left({\bar{q}}_{R} \gamma^\mu \frac{\lambda_a}{2}q_{R}\right) 
\left({\bar{q}}_{R} \gamma_\mu \frac{\lambda_a}{2}q_{R}\right)
\biggr\}, \label{eq:compositeness}
\end{eqnarray}
 where $\eta^{0,1}_{HH}= 0, \pm 1$, and $H$= $L$, $R$ for left- or
 right-handed quarks. $\eta^{0(1)}_{HH}$ terms correspond to
 color-singlet (octet) contact interactions. These contact
 interactions modify the cross sections for quark-quark
 scattering. Limits are presented in
 Sections~\ref{sec:angular_distribution} and \ref{sec:dijet_mass} for
 the cases~\cite{compos,coloron_1}:
\begin{itemize}
\item $\Lambda_{LL}^{\pm}$, where $\eta^{0}_{LL}=\pm1$.
\item $\Lambda_{V}^{\pm}$, where 
 $\eta^{0}_{LL}= \eta^{0}_{RR} = \eta^{0}_{RL} = \eta^{0}_{LR} = \pm1$.
\item $\Lambda_{A}^{\pm}$, where
 $\eta^{0}_{LL}= \eta^{0}_{RR} = -\eta^{0}_{RL} = -\eta^{0}_{LR} = \pm1$.
\item $\Lambda_{(V-A)}^{\pm}$, where 
 $\eta^{0}_{LL}= \eta^{0}_{RR} = 0; \eta^{0}_{RL} = \eta^{0}_{LR} = \pm1$.
\item $\Lambda_{V_8}^{\pm}$, where 
 $\eta^{1}_{LL}= \eta^{1}_{RR} = \eta^{1}_{RL} = \eta^{1}_{LR} = \pm1$.
\item $\Lambda_{A_8}^{\pm}$, where
 $\eta^{1}_{LL}= \eta^{1}_{RR} = -\eta^{1}_{RL} = -\eta^{1}_{LR} = \pm1$.
\item $\Lambda_{{(V-A)}_8}^{\pm}$,
 where~$\eta^{1}_{LL}= \eta^{1}_{RR} = 0;
 \eta^{1}_{RL} = \eta^{1}_{LR} = \pm1$.
\end{itemize}
 Currently, there are no NLO compositeness calculations available;
 therefore LO calculations are used. The ratio of each LO prediction
 including compositeness to the LO prediction with no compositeness
 $(\Lambda = \infty)$ is used to scale the {\sc jetrad} NLO prediction:
\begin{equation}
\sigma \left( {\rm composite} \right)
= \frac{\sigma \left( \Lambda = X  \right)_{\rm LO}}{\sigma \left(
 \Lambda = \infty \right)_{\rm LO}}
 \sigma \left(\Lambda = \infty  \right)_{\rm NLO}
\end{equation}

\subsection{Coloron Limits}
\label{sec:theory_coloron}

 A flavor-universal coloron model~\cite{coloron_1} inspired by
 technicolor has been proposed to explain the nominal excess in the
 inclusive jet cross section as measured by CDF~\cite{CDF_2}. The
 model is minimal in its structure in that it involves the addition of
 one new interaction, one new scalar multiplet, and no new
 fermions. The QCD gauge group is extended to $SU(3)_{1} \times
 SU(3)_{2}$. At low energies, due to symmetry breaking, this results
 in the existence of ordinary massless gluons and an octet of heavy
 coloron bosons. Below the mass of the colorons ($M_c$),
 coloron-exchange can be approximated by the effective four-fermion
 interaction:
\begin{equation}
{\mathcal{L}}_{\rm eff} = - { {g_{3}^{2} \cot^{2}{\theta}} \over {2!
M_{c}^{2}}}
\left({\bar{q}} \gamma^\mu \frac{\lambda_a}{2}q\right) 
\left({\bar{q}} \gamma_\mu \frac{\lambda_a}{2}q\right)
\end{equation}
 where $\cot \theta$ represents the mixing between colorons and
 gluons, and $g_3^2 \equiv 4\pi\als$. If $\Lambda_{V_8}^{-}
 \sqrt{\alpha_{\rm s}} = M_{c}/\cot{\theta}$, this corresponds to
 Eq.~\ref{eq:compositeness} with $\eta^{1}_{LL} = \eta^{1}_{LR} =
 \eta^{1}_{RL} = \eta^{1}_{RR} = -1$ and would represent new
 color-octet vector current-current interactions.  Such interactions
 could arise from quark compositeness or from non-standard gluon
 interactions (e.g.  gluon compositeness)~\cite{chosimm}.

 The phenomenology of the coloron has been studied~\cite{coloron_2}
 and limits have been placed on $M_c$ and $\cot{\theta}$. Constraints
 on the size of the radiative corrections of the weak-interaction
 $\rho$ parameter require \mbox{$M_{c} / \cot{\theta} >
 450$~GeV}~\cite{coloron_1}, and a direct search for colorons in the
 dijet mass spectrum by the CDF Collaboration excludes colorons with mass
 below 1 TeV for $\cot{\theta} \lesssim 1.5$~\cite{cdfxjj}.
\clearpage

\section{Triggering}
\label{sec:triggers}

 The D\O\ trigger was based on a multi-level system.  The Level~0 (L\O
 ) trigger consisted of two scintillating hodoscopes, one on each side
 of the interaction region.  Coincident signals in the two hodoscopes
 indicate an inelastic collision and provide timing information for
 calculation of the position of the $z$-vertex of the interaction.
 The next level, Level 1 (L1), was a hardware trigger.  The L1 trigger
 required a specified number of calorimeter trigger tiles ($\Delta
 \eta \times \Delta \phi = 0.8 \times 1.6$) or towers ($\Delta \eta
 \times \Delta \phi = 0.2 \times 0.2$) above certain \Et\ thresholds.
 Different trigger versions with slightly different L1 requirements
 were instrumented during the run.  If a L1 rate was too large, a
 prescale was used to reduce the rate to an acceptable level. These
 prescale values were adjusted during the course of a beam store. A
 prescale of $P$ allows only 1 out of every $P$ events to be sent to
 the next level. Finally, Level 2 (L2) was a software trigger which
 selected the data to be written to tape.  A fast jet algorithm used
 at L2 defines jet \Et\ as the sum of the transverse energy within a
 cone of opening angle ${\cal R} =0.7$ centered on the \Et -weighted
 center of a L1 trigger tile or tower.

 The L2 triggers used in the QCD analyses at $\sqrt{s} =$ 1800 GeV are
 called JET\_30, JET\_50, JET\_85, and JET\_115. The names follow the
 nomenclature that a JET\_X trigger at L2 requires at least one jet
 with $E_{T}$ greater than X GeV. During the running at $\sqrt{s} =$
 630 GeV the L2 triggers were JET\_12, JET\_2\_12, and JET\_30. A
 complete description of the L1 and L2 trigger requirements is given
 in Table~\ref{TABLE:efficient}.
 
\begin{table*}[hbtp]
\begin{tabular}{ll@{ $>$ }rl@{ $>$ }lc} 
 Trigger & \multicolumn{2}{c}{Level 1 (GeV)}  & 
\multicolumn{2}{c}{Level 2 (GeV)}     & 98$\%$ efficient \\ 
\hline\hline
\multicolumn{6}{c}{$\sqrt{s} = 1800$~GeV}\\
\hline\hline
JET\_30  & 1 tile   & 15  & \multicolumn{2}{l}{1 jet with}  & 45 GeV     \\ 
         & \& 1 tile & 6  & \Et\       & ~30          &          \\
JET\_50  & 1 tile   & 15  & \multicolumn{2}{l}{1 jet with}  & 75  GeV   \\ 
         & \& 1 tile & 6  & \Et\       & ~50            &          \\
JET\_85  & 1 tile   & 35  & \multicolumn{2}{l}{1 jet with}  & 105 GeV    \\
         & \& 2 tiles & 6 & \Et\       & ~85            &          \\ 
JET\_115 & 1 tile   & 45  & \multicolumn{2}{l}{1 jet with}  & 170 GeV   \\ 
         & \& 1 tile & 6  & \Et\       & 115         &          \\%
\hline\hline
\multicolumn{6}{c}{$\sqrt{s} = 630$~GeV}\\
\hline\hline
JET\_12  & 1 tower  & 2   & \multicolumn{2}{l}{1 jet with}  &   20 GeV \\
              & \multicolumn{2}{c}{} & \Et\       & ~12          &          \\
JET\_2\_12\  & 2 towers  & 2   & \multicolumn{2}{l}{1 jet with}  & 30 GeV   \\
              & \multicolumn{2}{c}{} & \Et\       & ~12          &          \\
JET\_30\  & 1 tile   & 15  & \multicolumn{2}{l}{1 jet with}  & 45 GeV     \\ 
              & \multicolumn{2}{c}{} & \Et\       & ~30          &          \\
\end{tabular}
\caption{Typical trigger configurations used in inclusive analyses.  
 The L1 and L2 requirements are shown for each trigger.  Also shown
 are the leading uncorrected jet \Et\ at which the average event
 trigger efficiency exceeds 98$\%$. Redundant lower-\Et\ thresholds
 at L1 were used to provide extended lists of seeds for jet clustering
 at L2.}
\label{TABLE:efficient}
\end{table*}

 A study was performed to determine the trigger efficiency as a
 function of jet \Et\ for all triggers used in D\O\ QCD
 analyses. There is an efficiency for an event to pass the L1 trigger,
 and an efficiency for an event to pass the L2 trigger given that it
 passed L1. The combined efficiency to pass both L1 and L2 is :
\begin{equation}
 \epsilon_{\rm event}^{\rm total}=\epsilon_{\rm event}^{\rm L1} \times
 \epsilon_{\rm event}^{\rm L2|L1}
\end{equation}
 where $\epsilon_{\rm event}^{\rm L2|L1}$ is the efficiency for an
 event to pass L2 when it has passed L1. The L1 and L2 event
 efficiencies ($\epsilon_{\rm event}^{\rm L1}$ and $\epsilon_{\rm
 event}^{\rm L2|L1}$) depend on the event topology (\Et\ and $\eta$ of
 the jets in the event).  The event trigger efficiency as a function
 of single jet efficiencies for an event with $N_{\rm jets}$ is given
 by
\begin{equation}
\epsilon_{\rm event}=1-\prod_{i=1}^{N_{\rm jets}}\left[1-\epsilon_{i}
(E_{{T}i},\eta_{i})\right] ,
\end{equation}
 where $\epsilon_{i}$ is the single jet efficiency for the $i$th
 jet. The product represents the probability that none of the jets in
 the event pass the trigger requirements.

 The efficiency of the L1 trigger with the least restrictive
 requirements was measured using a data set that was required to pass
 only the L\O\ trigger. The single jet efficiency is given by the
 fraction of jets that satisfy the L1 requirements at a given \Et
 . The L1 efficiencies for more restrictive L1 triggers (MRT) were
 calculated using data samples that were required to pass a less
 restrictive L1 trigger (LRT). This allows the L1 efficiency of the
 more restrictive trigger to be calculated relative to the less
 restrictive trigger (given by $\epsilon_{\rm MRT,LRT}$). Hence the
 efficiency for a given L1 trigger is given by the product of the
 efficiencies of all less restrictive triggers at a given \Et . For
 example, the L1 efficiency for the Jet\_85 trigger is given by
\begin{equation}
\epsilon^{\rm L1}_{\rm Jet\_85} =  
\epsilon^{\rm L1}_{\rm Jet\_85,Jet\_50} \times 
\epsilon^{\rm L1}_{\rm Jet\_50,Jet\_30} \times 
\epsilon^{\rm L1}_{\rm Jet\_30,L\O }.
\end{equation}

\begin{figure}[htbp]
   \centerline
   {\psfig{figure=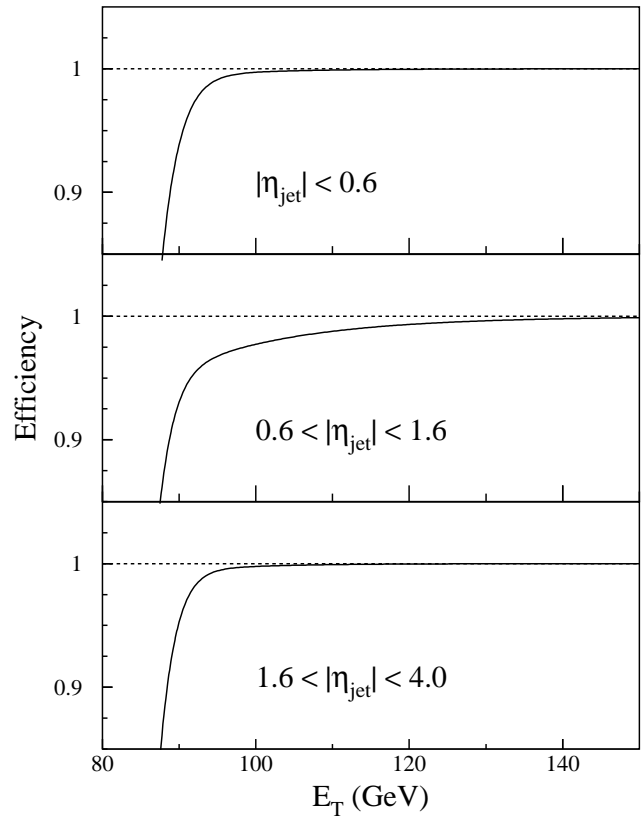,width=3.5in}}
   \caption{Average event efficiency for JET\_85 as a function of
   leading jet \Et\ and for three different pseudorapidity regions.}
   \label{FIG:sec03_event_efficiency}
\end{figure}

 The L2 trigger efficiencies for single jets are measured with respect
 to the L1 trigger. The fraction of these events which have a L2 jet
 above threshold determines the L2 single jet efficiency.
 Figure~\ref{FIG:sec03_event_efficiency} shows the event efficiency
 for Jet\_85 as a function of \Et .

 Table \ref{TABLE:efficient} shows the typical trigger requirements
 and the \Et\ value for the leading \Et\ jet at which each trigger
 averages an efficiency exceeding 98$\%$. The leading jet's \Et\ must
 be significantly higher than the L2 threshold in order for the
 trigger to be efficient.
\clearpage

\section{Luminosity}
\label{sec04:lumin}

 The beam luminosity was calculated from the counting rate of the L\O\
 counters and the cross section subtended by these counters. The cross
 section was determined using the geometric acceptance of the L\O\
 hodoscopes, the L\O\ hardware efficiency, and the world average (WA)
 of the \pbarp\ inelastic cross section measurements. The cross
 section of observed events in the L\O\ was found to be
 \mbox{$\sigma_{\rm L\O} = 43.1 \pm 1.9~{\rm mb}$} at $\sqrt{s} =$
 1800 GeV~\cite{d0_wcross} and \mbox{$\sigma_{\rm L\O} = 32.9 \pm
 1.1~{\rm mb}$} at $\sqrt{s} =$ 630 GeV (see \cite{luminosity_630} for
 a description of the method used). The effective luminosity was
 determined independently for each trigger on a run-by-run basis
 taking into account each trigger's prescale, the L\O\ inefficiency,
 and the detector deadtime. The WA \pbarp\ cross section at $\sqrt{s}
 =$ 1800 GeV used in this paper is based on measurements by the E710
 Collaboration~\cite{lumin_e710}, the CDF
 Collaboration~\cite{lumin_cdf}, and the E811
 Collaboration~\cite{E811_lumin}. At $\sqrt{s} = 630$~GeV there is no
 complete measurement of the \pbarp\ cross section. Hence, the \pbarp\
 cross section was obtained by interpolating between the WA \pbarp\
 cross sections measured at $\sqrt{s} = 546$ and 1800
 GeV~\cite{luminosity_630}. The WA \pbarp\ cross section at $\sqrt{s}
 = 546$~GeV is based on measurements by the UA4~\cite{ua4_lumin} and
 CDF~\cite{lumin_cdf} Collaborations. The CDF collaboration only uses
 its measurement of the \pbarp\ cross section to determine its
 luminosity. As a result the luminosities currently used by CDF are
 $6.1\%$ lower than those used by D\O . As a consequence of this all
 CDF cross sections are $6.1\%$ lower than D\O\ cross
 sections~\cite{d0_wcross}.

 The integrated luminosities at $\sqrt{s} =$ 1800 GeV as measured
 using L\O\ for the Jet\_30, Jet\_50, Jet\_85, and Jet\_115 triggers
 are 0.368, 4.89, 56.7, and 95.7 \ipb\ respectively, with an
 uncertainty of 5.1$\%$. The luminosities at $\sqrt{s} =$ 630 GeV for
 the JET\_12, JET\_2\_12, and, JET\_30 were 5.12, 31.9 and 538 \inb\
 respectively with an uncertainty of 4.4$\%$.

 The luminosity required corrections due to small discrepancies in the
 luminosity calculation during different running periods at $\sqrt{s}
 =$ 1800 GeV. The initial luminosities for triggers Jet\_85 and
 Jet\_115 were taken from the luminosity calculation exclusively
 determined with the L\O\ counters. The inclusive jet cross sections
 calculated with the first 7.3 \ipb\ of the data sample showed a
 10$\%$ difference for Jet\_115. The luminosity has been adjusted so
 that the dijet mass spectrum for the first 7.3 \ipb\ matches that of
 the remaining data. This adjustment was also applied to Jet\_85. Thus
 the luminosities for Jet\_85 and Jet\_115 are 56.5 and 94.9 \ipb\
 respectively, a change of 0.7$\%$ from the value obtained using the
 L\O\ counters.  This difference was added linearly to the 5.1$\%$
 error on the initial luminosity value for a total error of 5.8$\%$.

 In addition, for a part of the run the Jet\_30 and Jet\_50 triggers
 each required a single interaction at L\O . The luminosities of
 Jet\_30 and Jet\_50 from the L\O\ calculation are estimated to be
 accurate only to 10$\%$ due to uncertainties in the efficiency of the
 single interaction requirement.  The luminosity for the Jet\_50
 trigger was determined by matching the Jet\_50 cross section to the
 Jet\_85 cross section, and the Jet\_30 cross section was matched to
 the Jet\_50 cross section in regions of overlap. The trigger matching
 is analysis dependent; each analysis presented in this paper used the
 cross section of interest to match the triggers. The results obtained
 for the different measurements are consistent. The trigger matching
 errors are added in quadrature to the 5.8$\%$ error on Jet\_85.  The
 final luminosity and error for each trigger is shown in Table
 \ref{TABLE:luminosities}.

\begin{table}[htb]
\begin{center}
\begin{tabular}{llr}
\multicolumn{1}{c} {Trigger} &
\multicolumn{1}{c} {Luminosity} &
\multicolumn{1}{c} {Error} \\
\hline\hline
\multicolumn{3}{c}{$\sqrt{s} = 1800$~GeV}\\
\hline\hline
JET\_30  & 0.364 \ipb  & 7.8$\%$ \\ 	
JET\_50  & 4.84  \ipb  & 7.8$\%$ \\	
JET\_85  & 56.5  \ipb  & 5.8$\%$ \\	
JET\_115 & 94.9  \ipb  & 5.8$\%$ \\	
\hline\hline
\multicolumn{3}{c}{$\sqrt{s} = 630$~GeV}\\
\hline\hline
JET\_12	    & 5.12 \inb & 4.4$\%$ \\
JET\_2\_12  & 31.9 \inb & 4.4$\%$ \\
JET\_30     & 538. \inb & 4.4$\%$ \\
\end{tabular}
\caption{Corrected luminosity and errors for the inclusive jet 
triggers. The trigger matching for Jet\_30 and Jet\_50 at
$\sqrt{s}=$~1800~GeV was carried out using the dijet mass cross
section.}
\label{TABLE:luminosities}
\end{center}
\end{table}

 Since the analyses~\cite{d0_inc,d0_dijet_mass} were first presented,
 the E811 Collaboration measurement of the total inelastic cross
 section was published~\cite{E811_lumin}. Including this measurement
 in the WA changed the observed L\O\ cross section from
 \mbox{$\sigma_{\rm L\O} = 44.5 \pm 2.4~{\rm mb}$} to
 \mbox{$\sigma_{\rm L\O} = 43.1 \pm 1.9~{\rm mb}$} at $\sqrt{s} =$
 1800 GeV. This changed the integrated luminosity of Jet\_115 from
 $91.9 \pm 5.6~\ipb$ to $94.9 \pm 4.7~\ipb$, an increase of
 3.2$\%$. Hence all cross sections at $\sqrt{s} =$ 1800 GeV reported
 in this paper are reduced by $3.1\%$ from the previously published
 results. It is worth noting that the inclusion of the E811 result had
 no perceptible impact on the cross section interpolation to 630 GeV.

 The luminosity calculation consists of three distinct ingredients:
 the geometric acceptance of the L\O\ hodoscopes, the L\O\ hardware
 efficiency, and the \pbarp\ inelastic cross section. The luminosity
 uncertainties are listed in Table~\ref{t_lum_unc}.  The largest
 contribution to the luminosity uncertainty at $\sqrt{s}= 1800$~GeV
 derives from the World Average (WA) \pbarp\ total cross section. The
 \pbarp\ cross section at $\sqrt{s}= 630$~GeV was determined from a
 fit to the values at $\sqrt{s}= 546$ and 1800
 GeV~\cite{luminosity_630}.

\begin{table}[htbp]
\centering
\begin{tabular}{ccc}
Source of & \multicolumn{2}{c}{Uncertainty (Percent)} \\
\cline{2-3}
Uncertainty 		& 1800 GeV 	& 630 GeV \\
\hline\hline
World Average \pbarp\ Cross Sections
			& $3.70$ 	& $2.75$ \\
Hardware Efficiency 	& $2.32$ 	& $3.12$ \\
Geometric Acceptance 	& $2.73$ 	& $1.51$ \\
Time Dependencies	& $0.70$	& $0.00$\\
\hline\hline
All sources 		& $5.81$	& $4.43$ \\
\end{tabular}
\caption{Uncertainties in the luminosity calculation excluding trigger
matching. \label{t_lum_unc}}
\end{table}

\begin{figure}[htbp]
\begin{center}
\vbox{\centerline{\psfig{figure=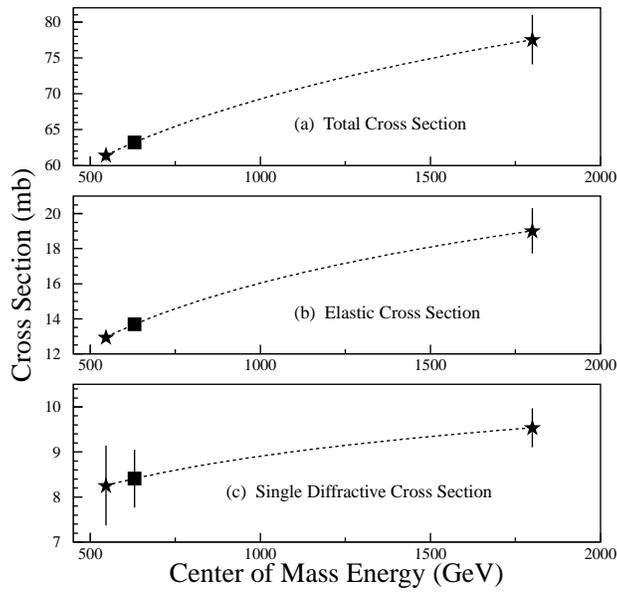,width=3.25in}}}
  \caption{The three fits to the World Average \pbarp\ cross
  sections. The stars depict the WA cross sections at $\sqrt{s} = 546$
  and 1800 GeV, and the closed square shows the interpolation to
  $\sqrt{s} = 630$~GeV. A fluctuation of the 1800 GeV point directly
  influences the interpolated value at 630 GeV, particularly in the
  case of the total cross section (a). } \label{lum_fits}
\end{center}
\end{figure}

 Two Monte Carlo minimum-bias event generators
 (\textsc{mbr}~\cite{Mbr} and \textsc{dtujet}~\cite{DtuJet}) were used
 to determine the geometric acceptance of the L\O\ hodoscopes. The
 difference in acceptance between the two MC results was taken as a
 source of systematic uncertainty for each $\sqrt{s}$. The consistent
 behavior of each generator relative to the other between
 center-of-mass energies indicates that the systematic uncertainty may
 be considered completely correlated.  Although the geometric
 acceptance of the L\O\ hodoscopes for diffractive processes must be
 considered in luminosity calculations, the uncertainty in the
 non-diffractive acceptance dominates.

 A study of zero-bias events (a random sampling of the detector during
 a beam-beam crossing) determined the hardware efficiency of L\O .
 Because the same estimation of the uncertainty appears in the
 calculation of the luminosities at both $\sqrt{s}$ values, the
 uncertainties are completely correlated. Table~\ref{t_lum_unc} lists
 the systematic uncertainty in the hardware efficiency for both
 center-of-mass energies.

\clearpage
\section{The Event Vertex}
\label{sec:vertex}

 The location of the event vertex was determined using the central
 tracking system~\cite{d0_detector}, which provides charged particle
 tracking over the region $\modeta < 3.2$. It measures the trajectory
 of charged particles with a resolution of 2.5~mrad in $\phi$ and
 28~mrad in $\theta$. From these measurements the position of the
 interaction vertex along the beam direction ($z$) can be determined
 with a resolution of 8~mm.

 As the instantaneous luminosity increases, the average number of
 \pbarp\ inelastic collisions per beam crossing increases. Hence there
 is the possibility of selecting the incorrect interaction vertex. If
 the incorrect vertex is chosen as the primary vertex, jet \Et\ and
 event missing transverse energy (\met ) will be miscalculated. This
 may result in a significant contribution to the jet spectra at very
 high-\Et\ since the high rate of jet production at lower \Et\ can
 cause contamination in the lower rate regions.  Visual scanning of
 the high-\Et\ jets shows that approximately 10$\%$ have misidentified
 interaction vertices.

 In order to study the effects of multiple interactions, a software
 tool called {\sc mitool}~\cite{mitool} was developed to provide
 information about the number of interactions.  This tool uses the
 L\O\ hodoscopes, the calorimeter, and the central tracker in order to
 evaluate the number of interactions. A sum-of-times inconsistent with
 a single interaction from the L\O\ hodoscopes indicates the
 possibility of the presence of more than one interaction.  The total
 energy in the calorimeter provides evidence of multiple
 interactions. If the total measured energy of an event is greater
 than 1.8 TeV, a multiple interaction is likely.  Additional
 information from the number of vertices found with the central
 tracker is also used. Using this information the most probable number
 of \pbarp\ interactions in the event is calculated.

 To a good approximation, the jet \Et\ and \EA can be calculated for
 the second \pbarp\ vertex using the measured vertex $z$-position and
 a simple geometric conversion.  Thus for all the jets in an event,
 the absolute magnitude of the vector sum of the jet \Et , denoted
 $\HT = \mid\!\!\Sigma \vec{E}_{T}^{\rm jet}\!\!\mid$ , can be
 calculated for each vertex.  Except for soft radiation falling below
 the jet reconstruction threshold, \HT\ will be equal in magnitude to
 the \met . Since QCD events should contain little \met , the correct
 vertex was selected by choosing the vertex with the minimum \HT .
\clearpage
\section{Jet and Event Selection}
\label{sec:quality_cuts}

 The existence of random spurious energy deposits in the calorimeter
 may either fake or modify a real jet.  Some sources of noise are
 electronic failures, cosmic ray showers, or accelerator losses due to
 Main Ring activity. A series of quality cuts was developed to remove
 this contamination.

\subsection{Removal of ``Hot'' Cells}
\label{section06:hot_cells}

 Before jet reconstruction, a cell suppression algorithm was
 implemented to suppress any cell with an unusually high deposition of
 energy relative to its longitudinal neighbors (a ``hot'' cell).
 Specifically, if a cell had more than 10 GeV of energy and more than
 20 times the average energy of its immediate longitudinal neighbors,
 the cell energy was set to zero.  This algorithm is successful in
 removing isolated high energy cells due to noise; however, the
 algorithm can also degrade the response to jets.

 Approximately 10$\%$ of the events have one or more suppressed cells.
 The rapidity distribution of the suppressed cells is very
 ``jet-like'' with a central plateau.  A cell was restored to a jet if
 it was within \mbox{$\DR = 0.7$} of the original jet direction and if
 the cumulative total of hot cell \Et\ was no more that 50$\%$ of the
 original jet \Et .  The jet rapidity and azimuth were then
 recalculated using the Snowmass definitions (Eq.~\ref{snow_def}). The
 event \met\ was also adjusted if a cell was restored to a jet.

 The restoration algorithm has been shown to be 99$\%$ efficient by
 fitting the \DR\ and restored cell fraction (the hot cell \Et\
 divided by the jet's original \Et ) distributions and estimating the
 inefficiency in the cut regions.  An event scan with restored jets
 (using relaxed restoration criteria) above 260 GeV showed no
 inefficiency. Less than 5$\%$ of these new jets are contaminated. For
 those events with a single suppressed cell, the \met\ is
 significantly reduced by the cell restoration.  The kinematic
 variables (\Et, $\eta$, and $\phi$) of the high-\Et\ jets which
 included a restored cell were compared to the kinematic variables
 calculated with a full reconstruction in which the suppression
 algorithm was disabled.  The differences were small and well within
 the characteristic resolutions of the variables.

\subsection{Quality Cuts}

 Even after the removal of isolated anomalously large cell energies,
 there still remain spurious jets.  Quality cuts were developed to
 remove these fake jets. The quality cuts were applied on either the
 jet or to the event.

 The jet quality cuts are based on the distribution of energy within
 the jet.  Three standard variables are used:
\begin{itemize}
\item[1)]
  Electromagnetic Fraction (EMF) --- the fraction of the jet energy
  contained in the electromagnetic section of the calorimeter.  Jets
  are retained if:
\begin{eqnarray}
          & \mbox{EMF}          & \leq 0.95 \hspace{0.5cm} 
          ( 1.2 < \mid \! \eta_{{\rm det}} \! \mid < 1.6 ), \nonumber \\
0.05 \leq & \mbox{EMF}          & \leq 0.95  \hspace{0.5cm} 
({\rm otherwise}) \label{EMFRcut}, 
\end{eqnarray}
  where $\eta_{{\rm det}}$ is the pseudorapidity of the jet calculated
  using a vertex position of $z = 0$.  The cut $\mbox{EMF} > 0.05$ is
  not applied for $1.2 < \mid \! \eta_{{\rm det}} \! \mid < 1.6$
  because of the gap between the CC and EC calorimeters
  (Section~\ref{sec:Calorimeter}).

\item[2)] 
 Coarse Hadronic Fraction (CHF) --- the fraction of the jet energy
 contained in the coarse hadronic section of the calorimeter.  This
 cut is designed to remove fake jets introduced by Main Ring particles
 depositing energy in the calorimeter. Jets are retained if:
\begin{equation}
\mbox{CHF}           < 0.4     \label{CHFRcut}.
\end{equation}
\item[3)] 
 Hot Cell Fraction (HCF) --- the ratio of the most energetic cell of a
 jet to the second most energetic cell. Jets are retained if:
\begin{equation}
\mbox{HCF} < 10.0.  \label{CELLcut}
\end{equation}
\end{itemize}

 A cut on \met\ is also used to remove bad events. Since QCD events
 are expected to have no intrinsic \met , a cut on events with large
 \met\ typically used:
\begin{equation}
 {\frac{\met}{\Etu{1}}}     < 0.7  \label{MISSETcut},
\end{equation}
 where $\Etu{1}$ is the transverse energy of the highest \Et\ jet in
 the event.  In the case of the inclusive jet analysis, the
 measurement is more susceptible to contamination from events in which
 the primary vertex is located outside of the tracking detector and a
 vertex due to an additional minimum bias event is identified as the
 primary vertex, leading to an overestimate of the jet \Et . In this
 case it is found that a \met\ cut of
\begin{eqnarray}
{\frac{\met}{\Etu{1}}}    <&  0.3  & 
 \hspace*{5mm} {\rm if~ }\Etu{1} > 100 {\rm ~GeV}, \nonumber \\
\met~~ 			  <&  30  {\rm ~GeV} &
 \hspace*{5mm}{\rm if~ }\Etu{1} \le 100 {\rm ~GeV} \label{MISSETcut_2}
\end{eqnarray}
 removes the contamination.

 Since the data collected at $\sqrt{s}=$ 630 GeV were taken at low
 instantaneous luminosity, there were fewer events with multiple
 interactions and incorrectly identified vertices. As a result, the
 quality cuts on the data were adjusted to maximize efficiency without
 increasing contamination. The resulting cuts are:
\begin{eqnarray}
\mbox{EMF} & <  & 0.90  \label{EMFRcut_630}, \\
\mbox{CHF} & <  & 0.4   \label{CHFRcut_630}, \\
\mbox{HCF} & <  & 20.0  \label{CELLcut_630}, \\
{\frac{\met}{\Etu{1}}}  &    < & 0.7.  \label{MISSETcut_630}
\end{eqnarray}

\subsection{Efficiency}

 The efficiencies of the quality cuts were measured.  The data sample
 used to calculate the efficiencies was selected by making cuts in
 $\eta$ and $\phi$.  We verified that the changes in shape of the EMF,
 HCF, and CHF distributions due to the \met\ were negligible.

\begin{figure}[htbp]
   \centerline
   {\psfig{figure=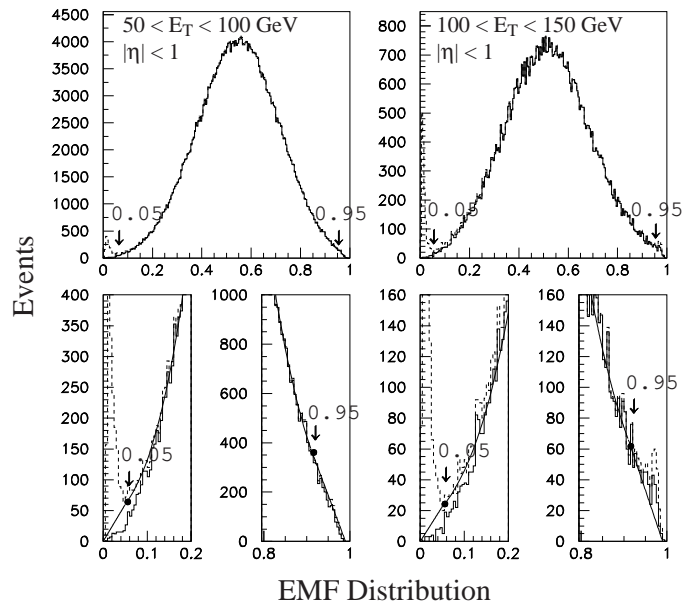,width=3.5in}} \caption{The
   measured EMF distributions for different \Et\ ranges. The lower
   plots show the cut values and the fit used to calculate the
   efficiency of the cut. The dashed histogram shows the full data
   sample and the solid histogram shows a data sample with minimal
   noise contamination. The arrows indicate the cut values. The peaks
   at EMF~$\approx 0$ or 1 are due to contamination.}
   \label{FIG:emf_distributions}
\end{figure}

\begin{figure}[htbp]
   \centerline
   {\psfig{figure=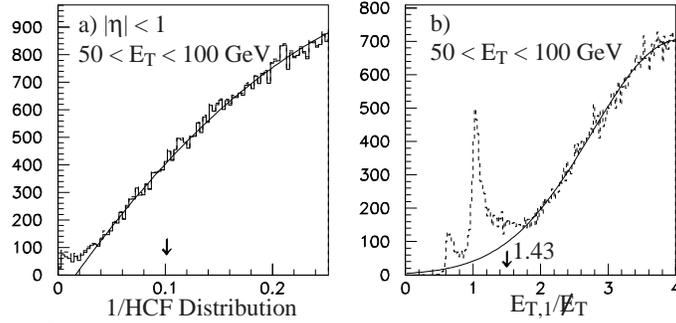,width=3.5in}} \caption{a)
   The 1/HCF distribution. The arrow shows the location of the cut for
   $\sqrt{s}=1800$~GeV. b) The distribution of ${{\Etu{1}}/\met}$. The
   arrow at ${{\Etu{1}}/{\met}} = 1.43$ corresponds to the
   $\met/\Etu{1}$ cut of 0.7. The peak at 1.0 is due to contamination
   from cosmic rays and the Main Ring. The dashed histograms show the
   distributions for the full data samples. }
   \label{FIG:chf_distributions}
\end{figure}

 To calculate the total efficiency, each individual efficiency is
 measured.  First the \met\ cut is applied to the data and the
 efficiency of the EMF cut is calculated.
 Figure~\ref{FIG:emf_distributions} illustrates the EMF distribution
 after the \met\ cut is applied. Contamination is visible as small
 peaks near EMF~$\approx 0$ or 1. A Gaussian-like curve is projected
 under the noise signal and used to estimate the data signal lost due
 to the cut. After the \met\ and EMF cuts are applied, the HCF
 efficiency is measured (Fig.~\ref{FIG:chf_distributions}(a) shows the
 HCF distribution). Then after both the EMF and HCF cuts were made,
 the efficiency of the CHF cut is measured. The total jet efficiency
 is calculated by multiplying the individual cut efficiencies
 together.

\begin{figure}[htbp]
   \centerline {\psfig{figure=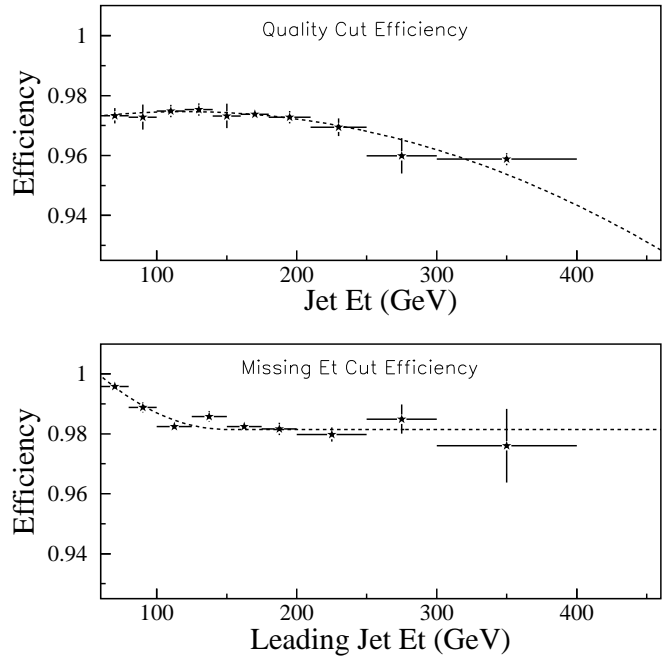,width=3.5in}}
   \caption{Top: The efficiency of the standard jet quality cuts for
   $\modeta < 0.5$ (Eqs.~\ref{EMFRcut}, \ref{CHFRcut}, and
   \ref{CELLcut}) at $\sqrt{s} = 1800$~GeV. Bottom: The efficiency of
   the \met\ cut used in the inclusive jet analysis
   (Eq.~\ref{MISSETcut_2}) at $\sqrt{s} = 1800$~GeV. The dotted curves
   show fits to the measured efficiencies.}
\label{FIG:efficiency}
\end{figure}

\begin{figure}[htbp]
   \centerline
   {\psfig{figure=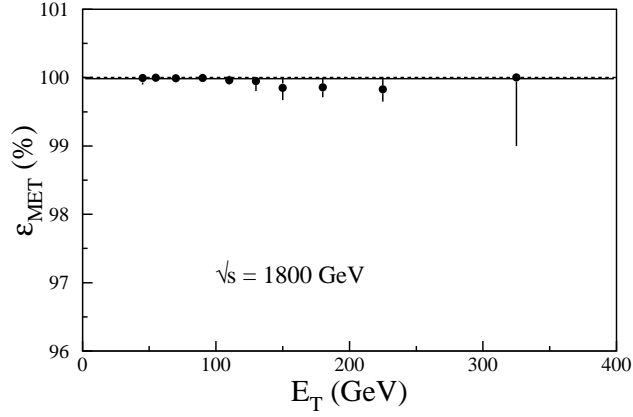,width=3.5in}}
   \caption{The efficiency of the \met\ cut used in the dijet analyses
   at $\sqrt{s} = 1800$~GeV (Eq.~\ref{MISSETcut}). }
\label{FIG:efficiency_2}
\end{figure}

\begin{figure}
\centerline
{\psfig{figure=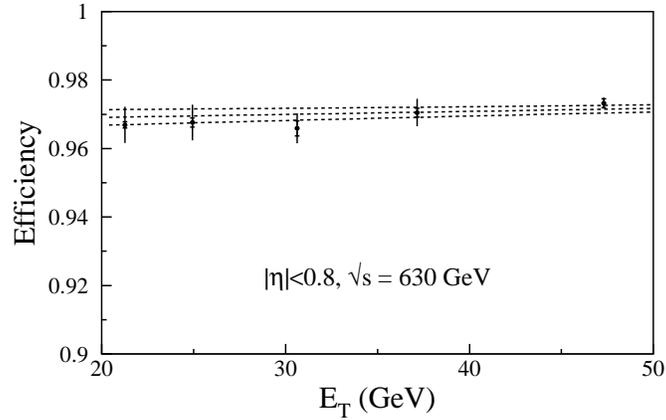,width=3.5in}}
\caption{The efficiency of the standard jet quality cuts for
   $\modeta < 0.8$ at $\sqrt{s} =$ 630 GeV (Eqs.~\ref{EMFRcut_630},
   \ref{CHFRcut_630}, and \ref{CELLcut_630}). The three curves show
   the fit to the efficiencies and the uncertainty in the fit.}
\label{fig:jet-eff-630}
\end{figure}

 The standard jet cuts remove most of the noise from the sample;
 however, there is still some contamination at high \Et\ due to cosmic
 rays and ``Main Ring Events''.  The \met\ cut removes this remaining
 noise.  By fitting this distribution in regions of the calorimeter
 where the noise effects are negligible, an extrapolation can be used
 to determine the efficiency.  Figure~\ref{FIG:chf_distributions}(b)
 shows the \met\ distribution used to calculate the efficiency.  The
 inclusive jet efficiencies as a function of jet \Et\ in the central
 region at $\sqrt{s}=1800$~GeV are shown in
 Fig.~\ref{FIG:efficiency}. Figure~\ref{FIG:efficiency_2} shows the
 efficiency of the \met\ cut for dijet events. The efficiencies of the
 quality cuts used at $\sqrt{s} =$ 630 GeV are given in
 Fig.~\ref{fig:jet-eff-630}.  The efficiency of the \met\ cut at
 $\sqrt{s} =$ 630~GeV is $> 99\%$.

\subsection{Contamination}

 In order to measure the remaining contamination after all quality
 cuts have been implemented, two separate studies were performed.
 Residual contamination was estimated by overlapping the observed hot
 cell distribution on a simulated inclusive jet sample.  The simulated
 cross section changed by less that 1$\%$ after imposition of the jet
 quality cuts.  The simulation also indicated that the jet quality
 cuts reject $>99\%$ of the ``fake'' jets with $\Et= 500$ GeV.
  
 To measure the contamination due to misvertexing, events at high-\Et\
 were visually inspected.  Misvertexing tends to cause lower \Et\ jets
 to migrate to higher \Et .  Since the cross section is steeply
 falling, this can corrupt the high-\Et\ cross section.  This study
 shows that after the vertex selection procedure has been applied,
 less than 1$\%$ of the events are contaminated at high \Et .
\clearpage

\section{Jet Energy Scale}
\label{sec:jet_energy_scale}

 The {\it in-situ} jet energy calibration uses reconstructed collider
 data, and is described in more detail in~\cite{energy_scale}.  The
 measured energy of a jet $\Emeas$ depends strongly on the jet
 definition. The particle-level (true) jet energy $\Eptcl$ is defined
 as the energy of a jet consisting of final-state particles produced
 by the high-\pt\ parton-parton scattering, and found using the
 Snowmass algorithm. The jet should not include the particles produced
 by the underlying event (Section~\ref{sec:corrected_jets}). The jet
 energy scale corrects the measured jet energy, on average, back to
 the energy of the final-state particle-level jet. $\Eptcl$ is
 determined as:
\begin{equation}
  \Eptcl = \frac{\Emeas- {E_{\rm O}}}{{R_{\rm jet}} \cdot
  {S_h}} \, \label{eq:escale_corr}
\end{equation}
where:
\begin{itemize}
\item ${E_{\rm O}}$ is an offset, which includes the underlying event,
noise from radioactive decays in the uranium absorber, the effects of
previous interactions (pile-up), and the contribution from additional
\pbarp\ interactions in the event.
\item ${R_{\rm jet}}$ is the calorimeter energy response 
 to jets. $R_{\rm jet}$ is typically less than unity due to energy
 deposited in uninstrumented regions of the detector, and differences
 in the response to electromagnetic and hadronic particles ($e/h >
 1$).
\item ${S_h}$ is the fraction of the jet energy that showered inside the 
 algorithm cone at the calorimeter level.
\end{itemize}
 The calibration is performed using data taken in \pbarp\ collisions
 at $\sqrt{s}=$1800 GeV and 630~GeV.

\subsection[Offset Correction, ${E_{\rm O}}$]
{Offset Correction, $\bbox{E_{\rm O}}$}

 The total offset correction is measured as a transverse energy
 density in $\eta$-$\phi$ space and subdivided as $D_{O} = D_{ue} +
 D_{\Theta}$. $D_{ue}$ represents the contribution due to the
 underlying event, i.e. energy associated with the spectator partons
 in a \pbarp\ event. $D_{\Theta}$ accounts for uranium noise, pile-up,
 and energy from additional \pbarp\ interactions. The offset
 correction ${E_O}$ is given by $D_O$ multiplied by the $\eta$-$\phi$
 area of the jet.

\begin{figure}[htbp]
\vbox{\centerline{\psfig{figure=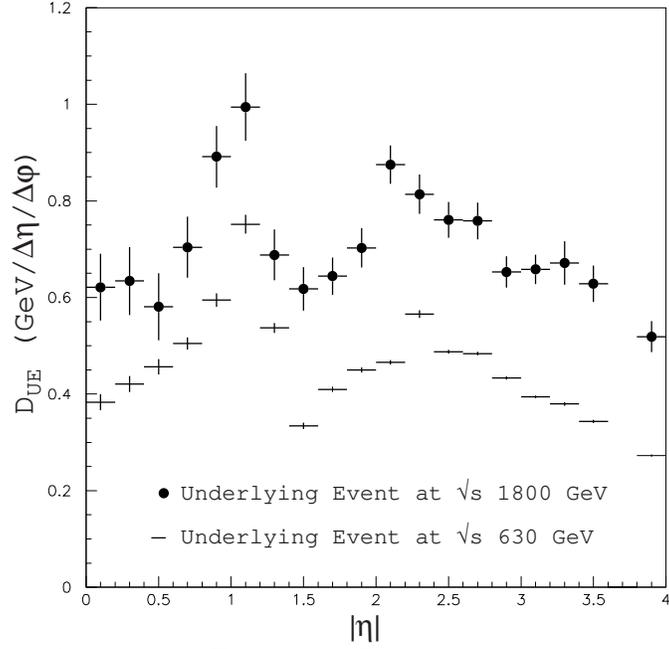,width=3.5in}}}
 \caption{Physics underlying event \Et\ density $D_{ue}$ versus $\eta$
         for events with $\sqrt{s}=1.8$~TeV and $\sqrt{s}=630$~GeV.}
\label{fig:offset_1}
\end{figure}

\begin{figure}[htbp]
\vbox{\centerline{\psfig{figure=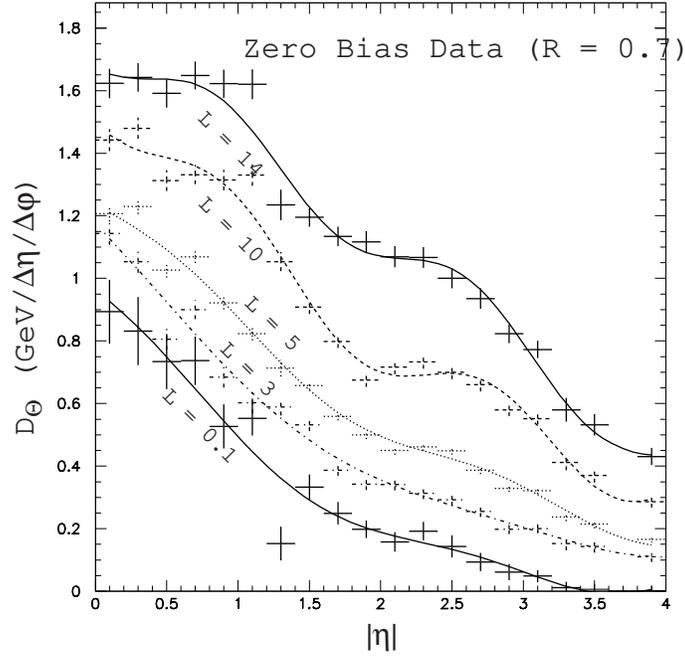,width=3.5in}}}
\caption{$D_{\Theta}$ versus $\eta$ for different luminosities in units
	 of 10$^{30}$ cm$^{-2}$sec$^{-1}$ at $\sqrt{s} = $~1.8~TeV.}
\label{fig:offset_2}
\end{figure}

 $D_{ue}$ is measured using the average transverse energy density in
 minimum-bias events (where a \pbarp\ interaction has occurred,
 usually inelastic scattering). $D_{\Theta}$ is determined from a
 zero-bias sample (a random sampling of the detector during a
 beam-beam crossing). The $\eta$ dependencies of both quantities and
 the luminosity dependence of $D_{\Theta}$ are shown in
 Figs.~\ref{fig:offset_1} and~\ref{fig:offset_2}.  The statistical and
 systematic errors of the offset correction are $8\%$ and 0.25 GeV
 respectively.

\subsection[Response Correction, ${R_{\rm jet}}$]
{Response Correction, $\bbox{R_{\rm jet}}$}\label{sec:mpf}

 D\O\ makes a direct measurement of the jet energy response using
 conservation of \pt\ in photon-jet ($\gamma$-jet)
 events~\cite{energy_scale}. The electromagnetic energy scale is
 determined from the D\O\ $Z (\rightarrow e^{+}e^{-})$, $J/\psi$, and
 $\pi^{0}$ data samples, using the masses of these known
 resonances. In the case of a $\gamma$-jet two body process, the jet
 response can be measured through:
\begin{equation}
R_{\rm jet} = 1 + \frac{\vmet \cdot \hat{n}_{T\gamma}}{E_{T\gamma}} \, , 
\end{equation}
 where $E_{T\gamma}$ and $\hat{n}_{T\gamma}$ are the transverse energy
 and direction of the photon. To avoid response and trigger biases,
 $R_{\rm jet}$ is binned in terms of $E^{\prime} = E_{T\gamma} \times
 {\rm cosh} (\eta_{\rm jet})$ and then mapped onto
 $\Emeas$. $E^{\prime}$ depends only on photon variables and jet
 pseudorapidity, which are both measured with very good resolution.

\subsubsection{$\eta$-Dependent Corrections}

 Most measurements need a high degree of accuracy in the jet energy
 scale at all rapidities. An $\eta$-dependent correction becomes
 necessary. The cryostat factor $F_{\rm cry}$ is defined as the ratio
 $R_{\rm jet}^{\rm EC}/R_{\rm jet}^{\rm CC}$.  The measured factor
 0.977 $\pm$ 0.005 is constant as a function of $E^{\prime}$. This was
 expected because the CC and the EC calorimeters use the same
 technology.

\begin{figure}[htbp]
\centerline{\psfig{figure=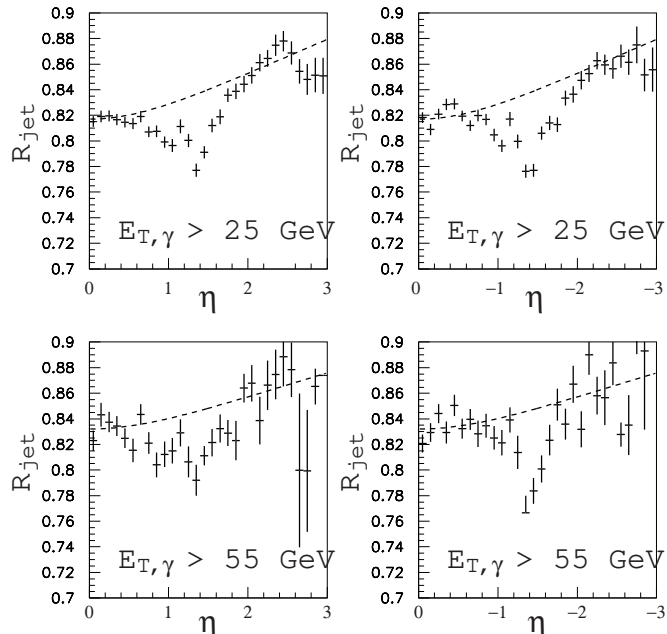,width=3.5in}}
\caption{Response versus $\eta$ for $\gamma$-jet data before the
$\eta$-dependent correction.  The dashed line is the fit to the
expected IC response.}
\label{fig:eta_dep}
\end{figure}

 The intercryostat region (IC), which covers the pseudorapidity range
 \mbox{$0.8<\modeta<1.6$}, is the least well-instrumented region of
 the calorimeter system. A substantial amount of energy is lost in the
 cryostat walls, module end plates, and support structures. An IC
 correction is performed after the $F_{\rm cry}$ correction and before
 the energy-dependent response correction.  Because the energy
 dependence of $R_{\rm jet}$ is included in $R_{\rm jet}$ as a
 function of $\eta$, this function is not a constant, but should be
 smooth. The IC correction is set so that the response as a function
 of $\eta$ agrees with the fit to the functional form, $R_{\rm jet} =
 a + b \cdot \ln \left[ \cosh (\eta) \right]$, of the CC and EC
 response, as shown in Fig.~\ref{fig:eta_dep}.

\subsubsection{Energy-Dependent Correction}

\begin{figure}[htbp]
\vbox{\centerline
{\psfig{figure=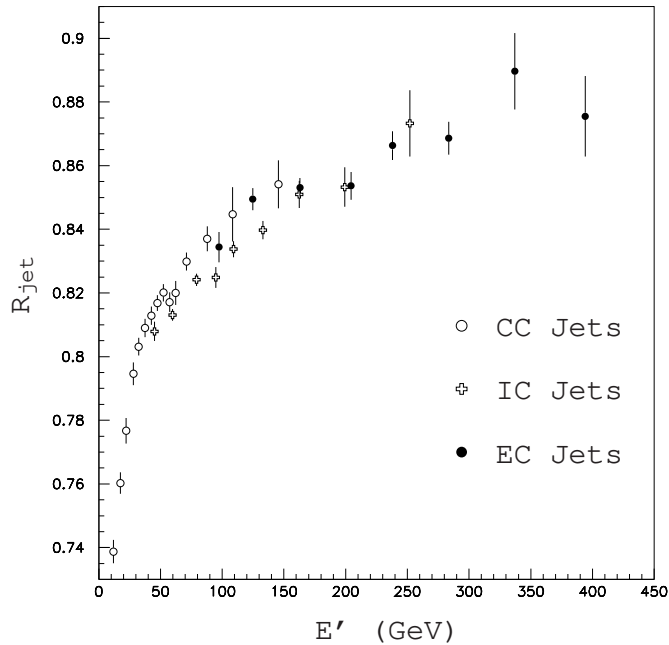,width=3.5in}}}
 \caption{$R_{\rm jet}$ versus $E^{\prime}$ measured in the CC, IC and EC
calorimeter regions after $\eta$ dependent corrections.}
\label{fig:ene_dep}
\end{figure}

 Following the above procedure, the energy dependence of $R_{\rm jet}$
 is then determined as a function of $E^\prime$ as illustrated in
 Fig.~\ref{fig:ene_dep}.  Uniformity of the calorimeters allows the
 use of data from both the CC and the EC to measure $R_{\rm jet}$. The
 rapidly falling photon cross sections limit the use of CC data to
 energies $\lsim$120 GeV.  EC data are used to extend the energy reach
 to $\sim$300 GeV.  We exploit the fact that jet energy in the EC is
 larger than in the CC for the same $E_{T}$.  Monte Carlo data are
 also included at the highest energy to constrain the extrapolation. A
 set of $\gamma$-jet events is generated using {\sc
 herwig}~\cite{herwig}, processed through the {\sc d\o
 geant}~\cite{geant} detector simulation, and reconstructed with the
 standard photon and jet algorithms.  The Monte Carlo simulation is
 improved by incorporating the single particle response of the
 calorimeter as measured in test beam.

\begin{figure}[htbp]
\vbox{\centerline
{\psfig{figure=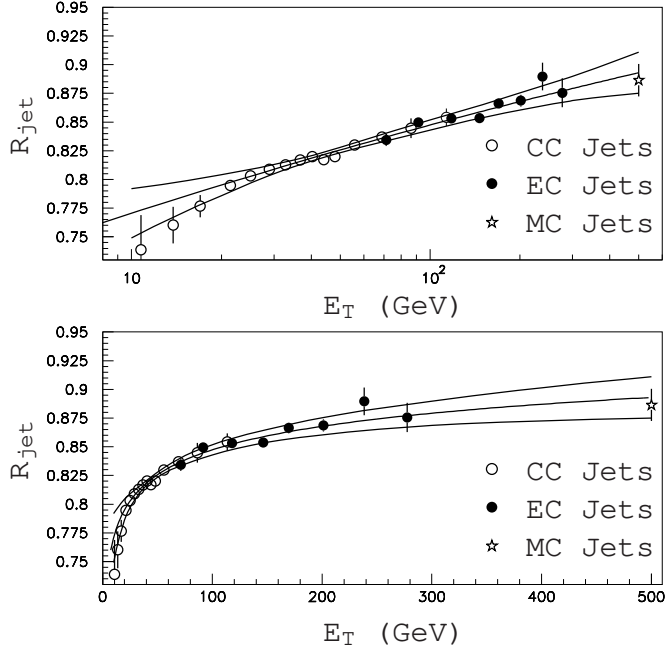,width=3.5in}}}
\caption{$R_{\rm jet}$ versus energy for the ${\cal{R}}=0.7$ cone jet algorithm. 
 The solid lines are the fit and the associated uncertainty band.}
\label{fig:ene_dep_2}
\end{figure}

 The response versus energy for the $\Rc = 0.7$ cone algorithm is
 shown in Fig.~\ref{fig:ene_dep}.  The CC and EC data, and the
 expected response from MC at $\Et = 500$~GeV are fit with the
 functional form \mbox{$R_{\rm jet}(E) = a + b \cdot \ln(E) + c \cdot
 \ln(E)^2$} (see Fig.~\ref{fig:ene_dep_2}).  This function is
 motivated by the hadronic shower becoming gradually more
 ``electromagnetic'' (EM) with increasing energy~\cite{ferbel}.  If
 $e$ and $h$ are the responses of the calorimeter to the EM and non-EM
 components of a hadronic shower, and $\pi$ is the response to charged
 pions, then $e/\pi = 1/ [\frac{h}{e} \, - \langle f_{EM} \rangle
 (\frac{h}{e}-1)]$.  The functional form for the mean electromagnetic
 fraction of the jet $\langle f_{EM} \rangle$ is $\sim \alpha \cdot
 \ln(E)$, giving the expected logarithmic dependence for energy
 carried by the charged pions and, therefore, jets.

 In addition to the uncertainty from the fit (1.5$\%$, 0.5$\%$,
1.6$\%$ for 20, 100, 450~GeV jets respectively), there is also a
$\sim$0.5$\%$ uncertainty from the $W$ boson background in the photon
sample. Some of the events in the $\gamma$-jet sample are not two-body
processes. In the IC region, the $\eta$-dependent corrections
contribute an additional $\sim$1$\%$ uncertainty.

\subsection[Showering Correction, ${S_h}$]
{Showering Correction, $\bbox{S_h}$}

 As a jet of particles strikes the detector, it interacts with the
 calorimeter material producing a wide shower of particles. Some
 particles directed inside the cone deposit a fraction of their energy
 outside the cone (and vice versa) as the shower develops inside the
 calorimeter. We do not correct for any QCD radiation or particles
 that are radiated from the cone; we only correct for the effects of
 the detector.

 The correction for this showering is determined using jet energy
 density profiles from data and particle-level {\sc
 herwig}~\cite{herwig} Monte Carlo.  The data contains the
 contributions of both gluon radiation and showering effects outside
 the cone. The former contribution is subtracted using the
 particle-level Monte Carlo profiles. $S_h$ is defined as the inverse
 of the measured correction factor; that means $S_h$ is the fraction
 of the jet energy showered inside the algorithm cone in the
 calorimeter (Eq.~\ref{eq:escale_corr}). The showering correction is
 negligible for ${\cal{R}} = 0.7$ cone jets above $\sim$100 GeV in the
 central region ($\modeta < 1.0$) with an uncertainty of
 $\sim1\%$. Both the correction and uncertainty are larger for lower
 energies, higher $\eta$, and smaller cone sizes.

\subsection{Correlations of the Uncertainties}
\label{sec:energy_scale_correlations}

 The uncertainties in the jet energy scale can be separated into five
 sources: offset, $\eta$-dependent corrections, response corrections,
 method, and showering corrections. The correlations of these
 uncertainties as a function of \Et\ and $\eta$ have been studied:
\begin{enumerate}
 \item {\bf Offset:} This is the dominant uncertainty at low \Et\, but
 is unimportant at high \Et . The uncertainty due to the offset
 correction is divided into two parts: a systematic error related to
 uncertainties in the method which is correlated as a function of \Et
 , and a statistical error that is uncorrelated as a function of
 pseudorapidity due to the finite size of the data sample used to
 determine the offset.
 \item {\bf $\bbox{\eta}$-Dependent Correction:} The uncertainty due
 to this correction was separated into two parts. The first is due to
 the cryostat factor and is correlated as a function of \Et\ and
 $\eta$.  The second is the IC correction, which is uncorrelated as a
 function of \Et\ and $\eta$.
 \item{\bf Response Correction:} The uncertainty associated with the
 hadronic response is unimportant at low \Et\ but dominant at high \Et
 .  As a result of using a fit, the uncertainty is partially
 correlated as a function of \Et .  The correlation matrices for
 various jet cone sizes can be found in Ref.~\cite{energy_scale}.
 \item{\bf Method:} The uncertainty in the method used to determine
 the energy scale correction arises from the data selection
 requirements, and punch-through at very high energies.  The method
 uncertainty is correlated as a function of \Et .
 \item {\bf Showering Correction:} The uncertainty due to this
 correction is small except at very low \Et\ and is considered to be
 fully correlated as a function of \Et .
\end{enumerate}
 
\subsection{Summary and Verification Studies}

\begin{figure}[htbp]
\vbox{\centerline
{\psfig{figure=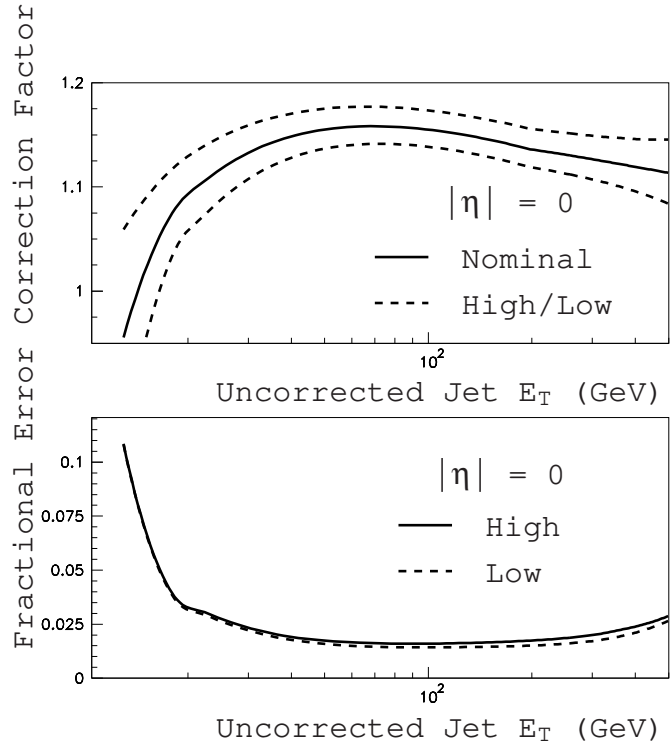,width=3.5in}}}
 \caption{ Corrections and errors for $\eta_{\rm jet}=$0.0, ${\mathcal
 R}=$0.7. The high (low) curve depicts the $+(-)1\sigma$
 uncertainties.}
\label{fig:summ}
\end{figure}

 Figure~\ref{fig:summ} shows the magnitude of the correction and
 uncertainty for ${\cal{R}} = 0.7$ cone jets with $\eta=$0.  The
 overall correction factor to jet energy in the central calorimeter is
 1.160 $\pm$ 0.018 and 1.120 $\pm$ 0.025 at 70 GeV and 400 GeV,
 respectively.  Point-to-point correlations in the energy uncertainty
 are very high for jets with $200 < \Et < 450$~GeV.

\begin{figure}[htbp]
\vbox{\centerline
{\psfig{figure=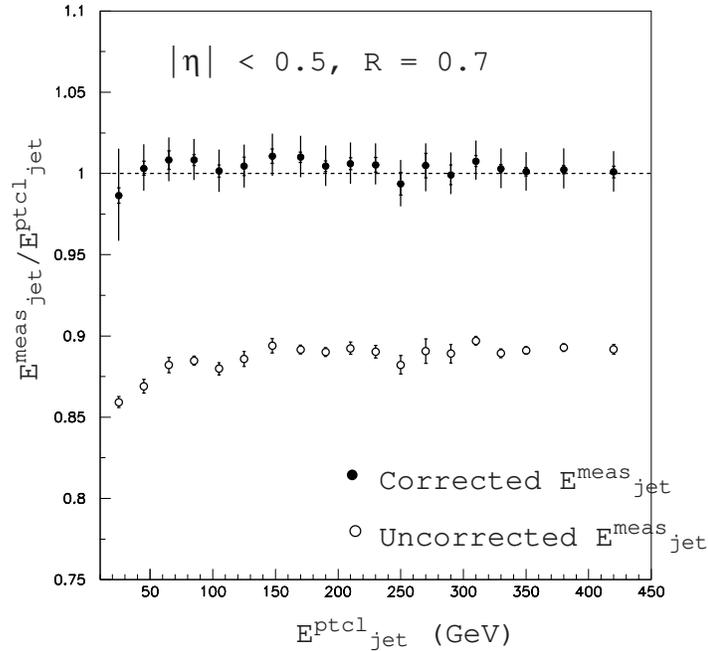,width=3.5in}}}
 \caption{ Monte Carlo verification test. Corrected $\Emeas/\Eptcl$
 ratio is consistent with 1.0 within errors. The inner error bars
 depict the statistical error due to the size of the Monte Carlo
 sample, and the outer error bars represent the systematic uncertainty
 on the energy scale.}
\label{fig:summ_2}
\end{figure}

 The accuracy of the jet energy scale correction is verified using a
 {\sc herwig} $\gamma$-jet sample and the {\sc d\o geant} detector
 simulation.  A Monte Carlo jet energy scale is derived and the
 corrected jet energy is compared directly to the energy of the
 associated particle jet.  Figure~\ref{fig:summ_2} shows the ratio of
 calorimeter to particle jet energy before (open circles) and after
 (full circles) the jet scale correction in the CC. The ratio is
 consistent with unity to within $\sim0.5\%$.

\clearpage

\section{Jet Resolutions} 
\label{SEC:jet_resolutions}
	
 The observed energy distributions are smeared due to resolution
 effects. The fractional energy resolution $\sigma_{E}/E$ may be
 parameterized as:
\begin{equation}
{\sigma_{E} \over {E}} = \sqrt{ {{N^{2}}\over{E^{2}}} + 
 {{S^{2}}\over{E}} + C^{2}}.
\end{equation}
 The nature of the incident particles, sampling fluctuations, and
 showering fluctuations, contribute mostly to the sampling term,
 $S$. Detector imperfections and deviations from an electron/hadron
 single particle response of unity, limit the resolution at high
 energies and are described by the constant term, $C$. Noise
 fluctuations (including the effects of multiple interactions) affect
 the low energy range and are given by the noise term, $N$.

 In the analyses reported here, we measure the \Et\ of the jets; hence
 we need to measure the resolution of \Et , which will have the same
 form:
\begin{equation}
{\sigma_{E_{T}} \over {\Et}} = \sqrt{ {{N^{2}}\over{\Et^{2}}} +
 {{S^{2}}\over{\Et}} + C^{2}}. \label{reso_param}
\end{equation}
 The relationship between ${\sigma_{E}/ {E}}$ and ${\sigma_{E_T}/
 {\Et}}$ depends on the $\eta$ resolution, $\sigma_{\eta}$.  Using
 $\Et = E/\cosh{\eta}$ and assuming that $\sigma_{E_T}$ and
 $\sigma_{\eta}$ are uncorrelated then:
\begin{equation}
\left( {{\sigma_{E_T}}\over{\Et}} \right)^{2} \approx 
\left( {{\sigma_{E}}\over{E}} \right)^{2} + 
\mid \! \tanh{\eta} \! \mid^{2} \sigma_{\eta}^{2}. 
\end{equation}

 In addition to the detector resolution, other contributions must be
 folded into the resolutions used for physics analyses. These are, for
 example, fluctuations of the out-of-cone losses, and the fluctuations
 of the vertex $z$-position about its measured value.

 Using D\O\ dijet data we derive the energy resolutions using energy
 conservation in the transverse plane.  The following criteria are
 applied to dijet events in order to eliminate sources of
 contamination due to additional low-\Et\ jets:
\begin{itemize}
\item 
 The $z$-coordinate of the interaction vertex must be within 50 cm of
 the center of the detector.
\item 
 The two leading-\Et\ jets must be back-to-back ($\Delta \phi >
 175^{\circ}$).
\item 
 If there is a third jet in the event, it must have \Etu{3}\ less than
 a specified value.
\item 
 All jets in the event must satisfy the jet quality cuts.
\item 
 Both leading jets are required to be in the same $\eta$ region so
 that their resolutions are similar, i.e. $\modetajetu{1} \approx
 \modetajetu{2}$.
\end{itemize}

\begin{figure}[htbp] \centerline
   {\psfig{figure=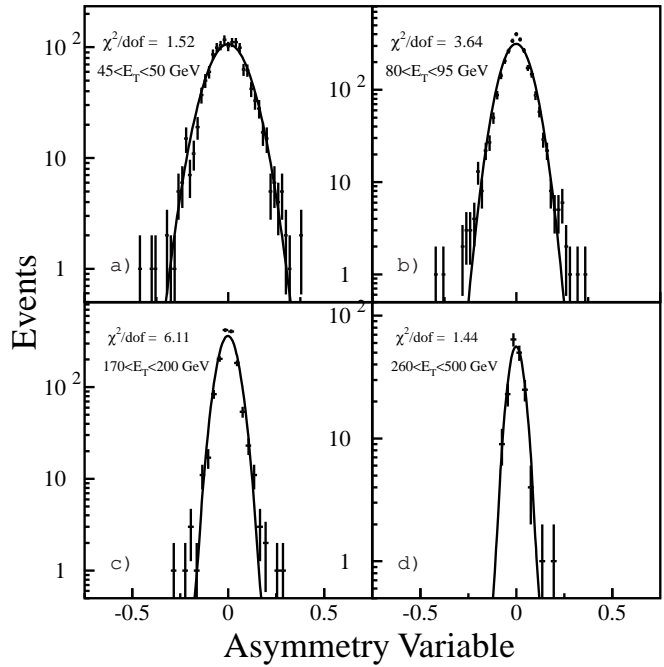,width=3.5in}}
   \caption{Asymmetry distribution in several $E_{T}$ bins for jets
   with $\modeta < 0.5$ and $\Etu{3} < 8$~GeV. } \label{FIG:asymmetry}
\end{figure}

 The dijet balance method is based on the asymmetry variable $A$, which
 is defined as:
\begin{equation}
A = { {\Etu{1} - \Etu{2}}\over{\Etu{1} + \Etu{2}}},
\label{eq_jetasym}
\end{equation}
 where $\Etu{1}$ and $\Etu{2}$ are the randomly ordered transverse
 energies of the two leading-\Et\ jets in an event. The variance of
 the asymmetry distribution can be written as:
\begin{equation}
\sigma_{A}^{2} = 
\left\bracevert { {\partial A} \over {\partial \Etu{1} }} \right\bracevert^{2}
{\sigma^{2}_{E_T^{\rm jet1}}} +
\left\bracevert { {\partial A} \over {\partial \Etu{2} }} \right\bracevert^{2}
{\sigma^{2}_{E_T^{\rm jet2}}}.
\end{equation}
 Assuming $\Et \equiv \Etu{1}=\Etu{2}$ and $\sigma_{E_T} \equiv
 \sigma_{E_T^{\rm jet1}} = \sigma_{E_T^{\rm jet2}}$, the fractional
 transverse energy can be expressed as a function of $\sigma_{A}$ in
 the following way:
\begin{equation}
 \left( {{\sigma_{E_T}}\over{\Et}} \right) = \sqrt{2}\sigma_{A}.
\end{equation}
 Figure \ref{FIG:asymmetry} shows the asymmetry distributions $A$ for
 different \Et\ bins.  The asymmetry distributions show minimal tails
 ($\ll 1\%$) and are well-described by a Gaussian distribution.

\subsection{Soft Radiation Correction}

\begin{figure}[htbp]
\centerline
 {\psfig{figure=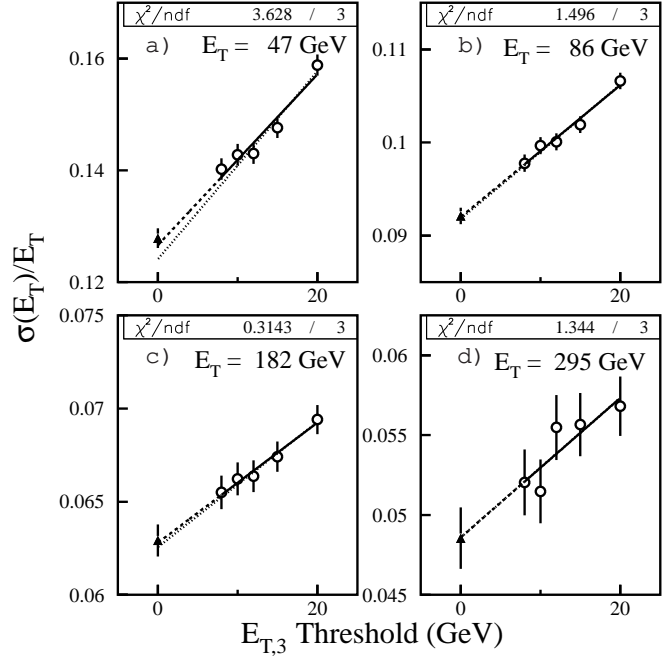,width=3.5in}}
   \caption{Resolutions as a function of the cut on $\Etu{3}$ for
   different \Et\ bins ($\modeta < 0.5$). The solid line shows the fit
   to the data points, the dashed line shows the extrapolation to
   $\Etu{3} = 0$, and the dotted line shows the fit excluding the
   $\Etu{3} < 8$~GeV point.}  \label{FIG:k_factor}
\end{figure}

 Although the $\Delta \phi$ and third-jet \Et\ cuts ($\Delta \phi >
 175^{\circ}$ and $\Etu{3} < 8$ \gev) are designed to remove events
 with more than two reconstructed jets, events may still contain soft
 radiation that prevents the two leading-\Et\ jets from balancing in
 the transverse plane; therefore the measured resolutions are
 overestimates of the hypothetical ``true resolutions.'' To evaluate
 this bias, the resolutions were determined from samples with
 different \Etu{3}\ cuts: 8, 10, 12, 15, and 20 \gev .  The
 resolutions are then extrapolated to a ``true'' dijet system with
 $\Etu{3}$ = 0.  Figure \ref{FIG:k_factor} shows the fractional jet
 resolutions as a function of \Etu{3}\ cut for several \Et\ bins.

\begin{figure}[hbtp]
\centerline{\psfig{figure=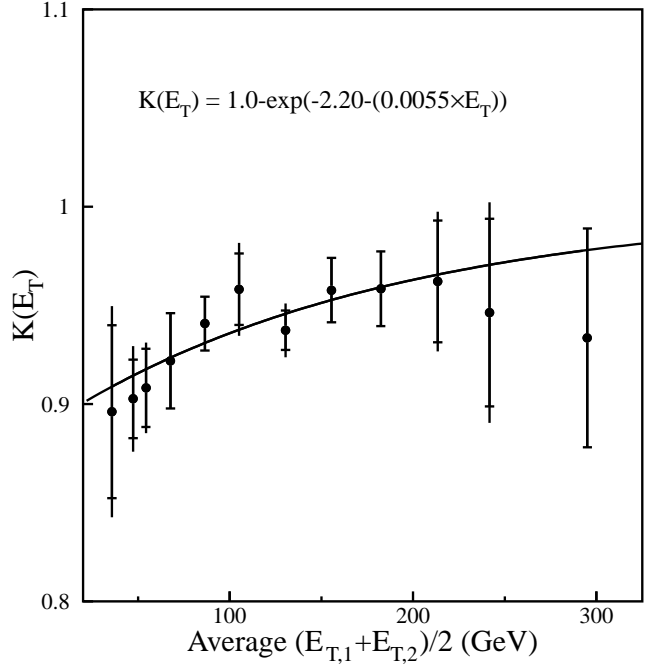,width=3.5in}}
   \caption{The soft radiation correction, $K(\Et)$, as a function of
  \Et\ ($\modeta < 0.5$). The error bars show the total uncertainty in
  the point-to-point correlations. The inner error bars show the
  uncertainty in the resolutions measured with $\Etu{3} > 8 \gev$. }
  \label{FIG:k_factor_2}
\end{figure}

 This procedure is repeated for every \Et\ bin. We expect the
 correction for additional radiation in the event to be continuous as
 a function of jet \Et\ and to be given by a function $K(\Et
 )$. Because the soft radiation bias should primarily affect small
 values of \Et\ but be negligible at high \Et , we parameterize the
 soft radiation correction $K(\Et)$ with the function:
\begin{equation}
K\left(\Et\right) = 1 - \exp{\left(-{a_{0}} - {a_{1}}\Et \right)}.
\end{equation}
 For the pseudorapidity bin $\modeta < 0.5$, $a_{0} = 2.20$ and $a_{1}
 = 0.0055$ (Fig.~\ref{FIG:k_factor_2}). This parameterization corrects
 the resolutions of each \Et\ bin for the effects of soft radiation.

 Note that the point-to-point correlations in
 Fig.~\ref{FIG:k_factor_2} are very large because each data point
 represents a subsample of the data point to its immediate right. In
 addition, it is not clear that the linear trend continues down to
 $\Etu{3} = 0$; hence we do not use the errors obtained from the fits
 to calculate the error on the corrected resolutions. The uncertainty
 in the extrapolation is the sum in quadrature of the following: the
 uncertainty in the resolution at $\Etu{3} > 8$~GeV, the difference in
 the extrapolation to $\Etu{3} = 0$ including and excluding the sample
 with the \Etu{3}\ cut of 8~GeV, and the uncertainty in the fit to the
 point-to-point correlations.

\subsection{Particle Jet Imbalance}

 Since we are correcting our measurements to the particle level, we
 must not include the effects of hadronization of the quarks and
 gluons in the resolutions.  The energy carried by particles emitted
 outside the particle-level cone does not belong to the particle
 jet. In other words, at LO the total $\vec{p_{\rm T}}$ of a dijet
 event at the particle level is zero, but the two reconstructed
 particle jets do not necessarily balance, since there could be
 particles emitted outside the cones. The asymmetry distribution
 measures the detector resolution convoluted with the contribution of
 the dijet imbalance at the particle level. The latter must be
 removed.

\begin{figure}[htbp]
   \centerline
   {\psfig{figure=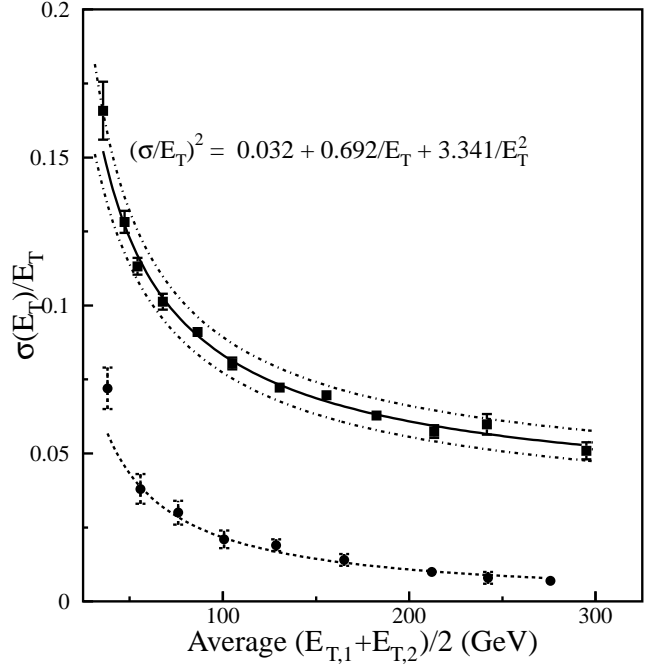,width=3.5in}}
   \caption{$\sigma_{\rm \Et/E_{T}}$ as a function of average \Et\ for
   $\modeta < 0.5$.  The data points (squares) indicate the
   resolutions after the soft radiation correction and the solid curve
   shows the fit to the resolutions. The dash-dot lines show the
   systematic uncertainty due to the method.  The dashed line is a fit
   to the particle-level resolutions obtained from MC points
   (circles).}  \label{FIG:resolutions}
\end{figure}

 The particle-level resolution is obtained by applying the same
 techniques as used on the data to a {\sc herwig}~\cite{herwig} Monte
 Carlo sample, e.g. no energy fluctuations.  The calorimeter
 resolution is obtained by removing the particle-level resolution
 using:
\begin{equation}
\left(\frac{\sigma_{\Et} }{\Et} \right)^{2} = 
\left(\frac{\sigma_{\Et} }{\Et}\right)^{2}_{\rm data}
-\left(\frac{\sigma_{\Et} }{\Et}\right)^{2}_{\rm MC}.
\end{equation}
 The fractional $\Et$ resolutions before the particle jet imbalance
 correction are shown in Fig.~\ref{FIG:resolutions} along with the MC
 data used to calculate the particle jet imbalance correction. The
 fully-corrected resolutions are given in
 Table~\ref{tab:1800_resolution_numbers}.

\begin{table}[htb]
\raggedright
\caption{\mbox{The measured jet resolutions at $\sqrt{s}= 1800$~GeV} and
 their uncertainties. }
\label{tab:1800_resolution_numbers}
\begin{tabular}{ddd}
  $\langle \Et \rangle$~(GeV) & $\sigma(\Et ) / \Et$ & $\Delta \left(
  \sigma(\Et ) / \Et \right)$ \\
\hline\hline
  35.75 &     0.154 &     0.009 \\
  47.32 &     0.120 &     0.004 \\
  54.25 &     0.106 &     0.003 \\
  67.70 &     0.096 &     0.003 \\
  86.43 &     0.088 &     0.001 \\
 105.08 &     0.078 &     0.002 \\
 130.42 &     0.070 &     0.001 \\
 155.54 &     0.068 &     0.001 \\
 182.40 &     0.062 &     0.001 \\
 213.44 &     0.056 &     0.002 \\
 241.69 &     0.059 &     0.003 \\
 295.10 &     0.050 &     0.003 \\
\end{tabular}
\end{table}

 \subsection{Studies of Systematic Uncertainties}

\begin{figure}[htbp]
\vbox{\centerline
{\psfig{figure=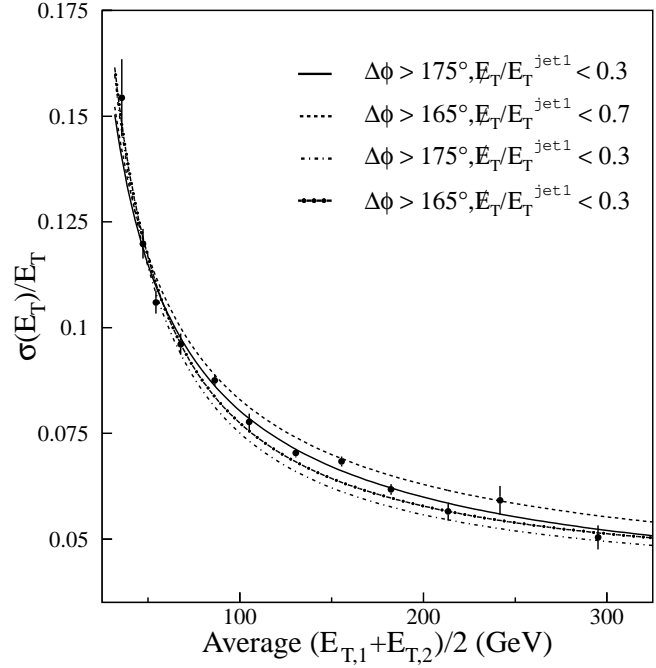,width=3.5in}}}
\caption{Fully corrected $\sigma_{E_T}/E_{T}$ as a function of
 average \Et\ for $\modeta < 0.5$ ({\rm i.e.} the soft radiation
 correction and the particle-level dijet imbalance corrections have
 been applied). The data points (solid curve) show the resolution as
 calculated with cuts $\met/ \Etu{1} < 0.7$ and $\Delta \phi >
 175^{\circ}$. The dashed line shows the effect of using a cut of
 $\Delta \phi > 165^{\circ}$. In addition, the effects of using a
 \met\ cut of $\met/ \Etu{1} < 0.3$ when $\Etu{1} > 100$~GeV, or $\met
 < 30$ GeV when $\Etu{1} < 100$ GeV are shown (dash-dot and solid-dots
 lines).}
\label{sec_2:fig_7}
\end{figure}

 In principle, the soft radiation correction should remove the effects
 of additional gluon radiation in the data sample; however, this may
 not be the case because not all the particles present in the detector
 appear in reconstructed jets. It is also possible that the
 requirement that jets be back-to-back ($\Delta \phi > 175^{\circ}$)
 preferentially selects events with better-than-average
 resolution. The possible size of these effects is studied by changing
 the back-to-back requirement to $\Delta \phi > 165^{\circ}$ and
 repeating the determination of the resolutions. The result of this
 study is shown in Fig.~\ref{sec_2:fig_7}. The resulting resolutions
 are slightly higher than the resolutions calculated with a cut of
 $\Delta \phi > 175^{\circ}$ and this difference is included in the
 overall systematic error.
 
 Some analyses require a tighter cut on the \met\ than the standard
 cut. In particular, the measurement of the inclusive jet cross
 section requires a \met\ cut of $\met/ \Etu{1} < 0.3$ when $\Etu{1} >
 100$ GeV, or $\met < 30$ GeV when $\Etu{1} < 100$ GeV. Any
 strengthening of the \met\ requirement will implicitly reduce the
 difference between the \Et 's of the two jets selected and also
 reduce the amount of soft gluon radiation; hence the resolutions
 should improve. The resolution parameterization using this \met\ cut
 is depicted in Fig.~\ref{sec_2:fig_7}.
 
\begin{figure}[htbp]
\vbox{\centerline
{\psfig{figure=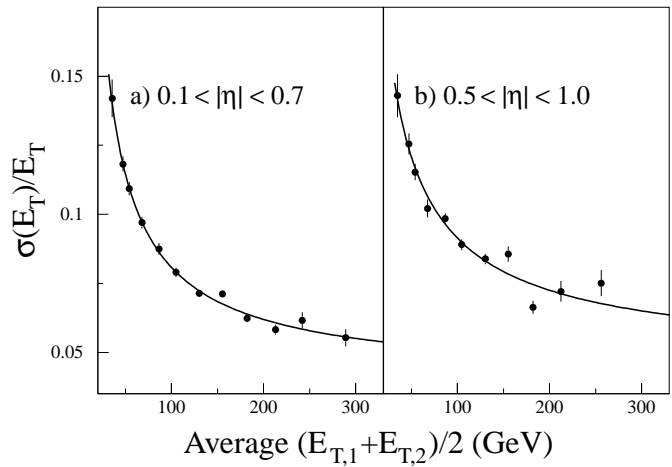,width=3.5in}}}
\caption{Fully corrected $\sigma_{E_T}$ as a function of \Et\  
for different rapidity regions.
}
\label{FIG:sec_2:fig_8}
\end{figure}

 The fractional \Et\ resolutions are parameterized using
 Eq.~\ref{reso_param} for all rapidities ($\modeta < 1$) and are given
 in Table~\ref{Table:resolution_parameters} and are plotted in
 Figs.~\ref{sec_2:fig_7} and \ref{FIG:sec_2:fig_8}.

\begin{table*}[hbt]
 \begin{center} \caption{The resolution fit parameters at $\sqrt{s} =
 1800$ GeV.}
 \label{Table:resolution_parameters} \vspace*{2mm}
 \begin{tabular}{cr@{ $\pm$ }lr@{ $\pm$ }lr@{ $\pm$ }l} 
 &\multicolumn{6}{c}{Fit Variables for a \met\
 cut of $\met/\Etu{1} < 0.7$.} \\
\cline{2-7}
 $\eta$ & \multicolumn{2}{c}{$C$} & \multicolumn{2}{c}{$S$} & 
\multicolumn{2}{c}{$N$} \\
\hline\hline
$\modeta < 0.5$        &  0.033 & 0.006 & 0.686 & 0.065 &2.621 & 0.810 \\
$ 0.5 < \modeta < 1.0$ &  0.047 & 0.008 & 0.783 & 0.137 &0.590 & 9.334 \\
$ 0.1 < \modeta < 0.7$ &  0.040 & 0.013 & 0.641 & 0.160 &2.891 & 1.413  \\
\hline\hline
 &\multicolumn{6}{c}{Fit Variables for a \met\ cut of $\met/\Etu{1} < 0.3$.}\\
\cline{2-7}
 $\eta$ & \multicolumn{2}{c}{$C$} & \multicolumn{2}{c}{$S$} & 
\multicolumn{2}{c}{$N$} \\ 
\hline\hline
$\modeta < 0.5$        & 0.037 &0.002 &  0.514 & 0.027 & 4.009 & 0.202 \\
$ 0.5 < \modeta < 1.0$ & 0.036 &0.006 &  0.736 & 0.059 & 1.972 & 0.904 \\
$ 0.1 < \modeta < 0.7$ & 0.038 &0.005 &  0.550 & 0.074 & 3.654 & 0.487 \\ 
   \end{tabular}
 \end{center}
\end{table*}

\subsection{Jet Resolutions at $\boldmath \sqrt{s}$ = 630 GeV}
\label{sec:630_resolutions}

 The jet resolutions at $\sqrt{s} =$~630 GeV are measured using the
 same techniques as the resolutions at $\sqrt{s} =$~1800 GeV. These
 resolutions are supplemented at low values of jet \Et\ by resolutions
 measured using photon-jet events.

\begin{figure}[htbp]
   \centerline {\psfig{figure=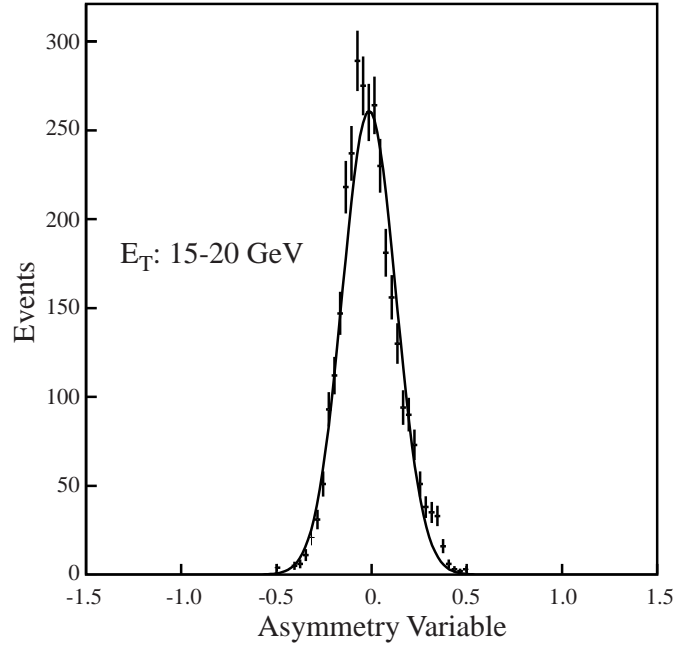,width=3.5in}}
   \caption{Distribution of photon-jet asymmetry for jet \Et\ between
   15 and 20 GeV in the central region. } \label{photasym}
\end{figure}

 The energy resolution for photons is approximately 10 times better
 than that for a jet, allowing a convenient redefinition of
 Eq.~\ref{eq_jetasym}. The photon-jet asymmetry is defined as
\begin{equation}
A_{\gamma,{\rm jet}} =  { {E^{\gamma}_{T} - E^{\rm jet}_{T}  }
\over{E^{\gamma}_{T}}},  \label{eq_photasym}
\end{equation}
 where $E^{\gamma}_{T}$ and $E^{\rm jet}_{T}$ are the fully corrected
 photon and jet transverse energies, respectively.  If one
 approximates $E^{\gamma}_{T} \approx E^{\rm jet}_{T} \equiv
 E_{\mathrm{T}}$ as before, and takes $\delta E^{\gamma}_{T} \approx
 0$, the standard deviation of the photon-jet asymmetry identically
 becomes the fractional jet resolution:
\begin{equation}
\left(\frac{\sigma _{E_T}}{\Et}\right)= \sigma_{A_{\gamma,{\rm jet}}}.  \label{eq_photres}
\end{equation}
 Figure \ref{photasym} displays a typical distribution of photon-jet
 asymmetry.

\begin{figure}[htbp]
\centerline
{\psfig{figure=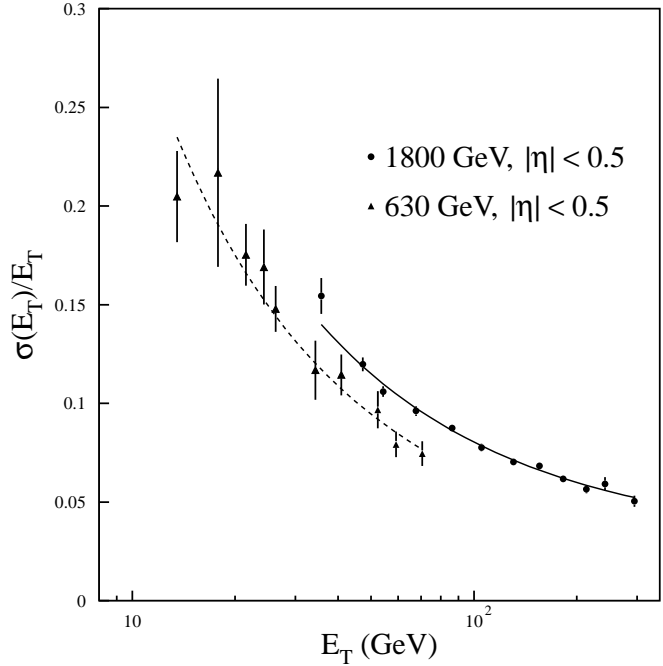,width=3.5in}}
\caption{ The single jet resolutions at $\sqrt{s}=630$~GeV (triangles)
and 1800~GeV (circles).  The resolutions at the two center-of-mass
energies have been fitted separately to Eq.~\ref{reso_param}. The fit
to the $\sqrt{s}=1800$~GeV data is the solid line, and the fit to the
$\sqrt{s}=630$~GeV data is the dashed line.}
\label{ccssnn}
\end{figure}

 As described in previous sections, the measured resolution is
 adjusted to reflect third-jet biases and the particle-jet
 asymmetry. The results bolster the low--statistics dijet results at
 $\sqrt{s}=630$ GeV. The resulting resolutions are given in
 Table~\ref{tab:630_resolution_numbers} and are compared with the
 resolutions at $\sqrt{s}=1800$~GeV in Fig.~\ref{ccssnn}. It is clear
 that the resolutions at the different center-of-mass energies are
 significantly different (a probability of agreement of 0.0007).

\begin{table}[htbp]
\begin{center}
\caption{The measured jet resolutions (and uncertainties) at
 $\sqrt{s}= 630$~GeV. }
\label{tab:630_resolution_numbers}
\begin{tabular}{lddd}
  Data Set &
  $\langle \Et \rangle$~(GeV) & $\sigma(\Et) / \Et$ & 
  $\Delta \left(  \sigma(\Et ) / \Et \right)$ \\
\hline\hline
$\gamma$-jet &  13.51 &     0.205 &     0.023 \\
$\gamma$-jet &  17.81 &     0.217 &     0.048 \\
$\gamma$-jet &  21.52 &     0.175 &     0.016 \\
$\gamma$-jet &  24.27 &     0.169 &     0.019 \\
jet-jet      &  26.28 &     0.148 &     0.012 \\
jet-jet      &  34.35 &     0.117 &     0.015 \\
jet-jet      &  40.87 &     0.114 &     0.010 \\
jet-jet      &  52.27 &     0.097 &     0.009 \\
jet-jet      &  59.12 &     0.079 &     0.007 \\
jet-jet      &  70.53 &     0.075 &     0.006 \\
\end{tabular}
\end{center}
\end{table}

\subsubsection{Parameterization of the Jet Resolutions.}

\begin{figure}[htbp]
\centerline
{\psfig{figure=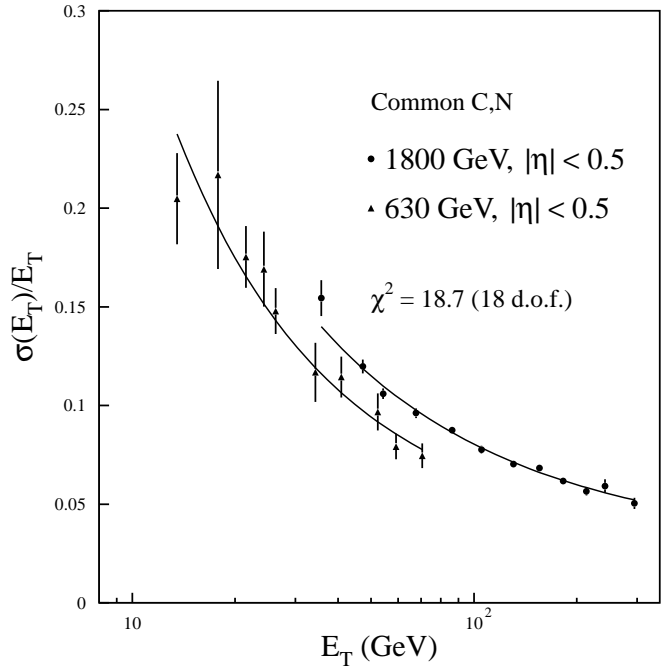,width=3.5in}}
\caption{ The single jet resolutions at $\sqrt{s}=630$~GeV (triangles)
and 1800~GeV (circles).  The resolutions at the two center of mass
energies have been fitted using the {\bf CSSN} model (solid lines).}
\label{cssn}
\end{figure}

 There are several parameterization choices that can be used to fit
 the data at both center-of-mass energies. We considered five
 alternative parameterizations of the resolutions:
\begin{enumerate}

 \item Fit the data simultaneously with Eq.~\ref{reso_param}: the {\bf
 CSN} model.
 \item Fit the data with common $C$ and $S$ terms and different noise
 terms ($N_{1800}$, $N_{630}$) at the two CM energies: the {\bf CSNN}
 model.
 \item Fit the data with common $C$ and $N$ terms and different
 sampling terms ($S_{1800}$, $S_{630}$) at the two CM energies: the
 {\bf CSSN} model (Fig.~\ref{cssn}).
 \item Fit the data with a common $C$ term and different sampling and
 noise terms at the two CM energies: the {\bf CSSNN} model.
 \item Fit the data with no common terms: the {\bf CCSSNN} model.
\end{enumerate}
 A model where only the $C$ term was allowed to vary between the two
 CM energies was not considered because $C$ depends on the physical
 structure of the calorimeter, and hence should not change. The
 \chisq\ and numbers of degrees of freedom for these five models are
 calculated and compared in Table~\ref{models}. The fit parameters are
 given in Table~\ref{fit_parameters}.

\begin{table}[htbp]
\caption{\chisq\ for the different models that can be used to
parameterize the single jet resolutions. \label{models}}
 \centering
\begin{tabular}{cddd}
Model & \chisq\ & Degrees of freedom & Probability \\
\hline \hline
{\bf CSN} 	& 44.9 & 19 & 0.0007 \\ 
{\bf CSNN} 	& 25.5 & 18 & 0.11   \\ 
{\bf CSSN} 	& 18.7 & 18 & 0.41   \\ 
{\bf CSSNN}	& 17.9 & 17 & 0.35   \\ 
{\bf CCSSNN}	& 17.9 & 16 & 0.33   \\ 
\end{tabular}
\end{table}

\begin{table}[htbp] 
\caption{The fit parameters for all models used to fit the resolution
 data. The correlation matrix for the {\bf CSSN} model is also
 given.}\label{fit_parameters}
\centering
\begin{tabular}{crrr}
 Model	 & Parameter & \hspace*{1cm}Value & Statistical Error \\ 
\hline\hline
		& $N$		& 1.098    &  1.128 \\
{\bf CSN}	& $S$		& 0.745    &  0.038 \\
	 	& $C$		& 0.028    &  0.004 \\
\hline\hline
		& $N$		& 2.571    &  0.309 \\ 
{\bf CSSN}	& $S_{1800}$	& 0.691    &  0.027 \\
          	& $S_{630}$	& 0.510    &  0.057 \\
	 	& $C$		& 0.032    &  0.003 \\
\hline 
\multicolumn{4}{c}{Correlation Matrix}\\
\hline 
 1.000&$-$0.812&$-$0.838& 0.575\\
$-$0.812& 1.000& 0.751&$-$0.902\\
$-$0.838& 0.751& 1.000&$-$0.589\\
 0.575&$-$0.902&$-$0.589& 1.000\\  
\hline\hline  
          	& $N_{1800}$	& 3.543    &  0.399 \\
{\bf CSNN}	& $N_{630}$	& 1.907    &  0.437 \\ 
		& $S$		& 0.590    &  0.049 \\
	 	& $C$		& 0.040    &  0.003 \\
\hline\hline  
          	& $N_{1800}$	& 2.510    &  0.893 \\
		& $N_{630}$	& 2.587    &  0.374 \\ 
{\bf CSSNN}	& $S_{1800}$	& 0.696    &  0.068 \\
		& $S_{630}$	& 0.509    &  0.063 \\
	 	& $C$		& 0.031    &  0.007 \\
\end{tabular}
\end{table}

\begin{figure}[htb]
\centerline
{\psfig{figure=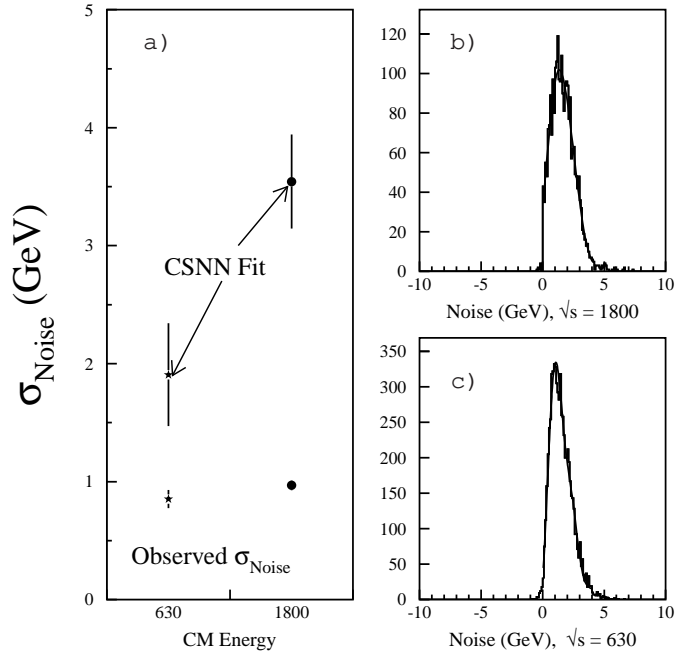,width=3.5in}}
\caption{ a) The average value and RMS width of the calorimeter noise
distributions is given by the two lower points. The two upper points
are the values of the $N$ parameter obtained in the fit to the
resolution data using the {\bf CSNN} model. b) and c): The noise
distribution found within a standard jet cone at each center-of-mass
energy (the cone used to measure the noise is required to be at least
$90^\circ$ in $\phi$ from any other jet in the event).}
\label{fig:noise}
\end{figure}

 It is clear from the \chisq\ of the parameterizations that the data
 cannot be represented by a single fit with common $C$, $S$, and $N$
 ({\bf CSN} model). Of the other models, the {\bf CSSN} model gives
 the best fit to the data. If we allow additional parameters to be
 included in the fit, the \chisq\ does not improve significantly.  The
 {\bf CSNN} model does not fit the data as well.  The noise
 distribution in the calorimeter is similar at the two different CM
 energies (Fig.~\ref{fig:noise}); hence the {\bf CSSN}
 parameterization model was chosen to fit the resolutions.  The cause
 of the change in sampling term as a function of CM energy is not
 known. One possible explanation is that a jet sample at $\sqrt{s} =
 630$~GeV will have relatively more ``quark'' jets than a sample at
 $\sqrt{s} = 1800$~GeV with the same mean \Et ~\cite{jkrane,snihur}.

\subsection{Monte Carlo Consistency Tests}

\begin{figure}[htbp]
   \centerline
   {\psfig{figure=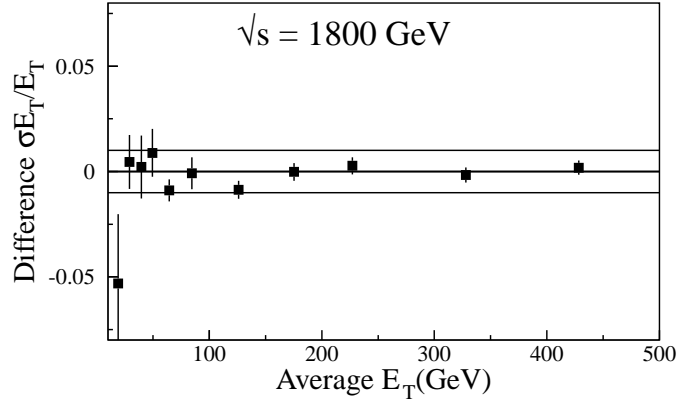,width=3.5in}}
   \caption{Resolution closure from {\sc herwig} Monte Carlo
   simulation; the difference in resolution obtained using the two
   techniques. The degree of closure is within $1\%$ for all data
   points above 25 GeV.}  \label{FIG:resolution_consistency}
\end{figure}

 To verify the resolution extraction methods, a Monte Carlo study
 compared events with and without the detector simulation.  The jet
 resolutions of the MC sample are measured in two ways; the first is
 the asymmetry method, and the second is a direct measurement of the
 resolutions.  If the \Et\ of a jet as measured by the calorimeter is
 simply denoted by $E_{T}$, and the \Et\ as measured at the
 particle-level is denoted by $E_{T}^{\rm ptcl}$, then the jet
 resolution can be derived from the ratio:
\begin{equation}
\frac{E_{T}^{\rm ptcl}-\Et}{E_{T}^{\rm ptcl}}\text{.}  \label{eq_pjet_res}
\end{equation}
 Figure~\ref{FIG:resolution_consistency} shows the differences in the
 resolutions as measured by the two methods. The differences between
 the two are scattered about zero, indicating lack of bias in the
 method. The differences between the two methods, less than $1\%$,
 indicates the magnitude of the systematic uncertainty, which can be
 parameterized as
\begin{equation}
\Delta \left( \frac{\sigma _{E_T}}{{E}_{T}}\right) =
\frac{2.0}{\Et^{2}}+0.01.  \label{eq_reso_sys}
\end{equation}

\begin{figure}[htb]
\centerline
{\psfig{figure=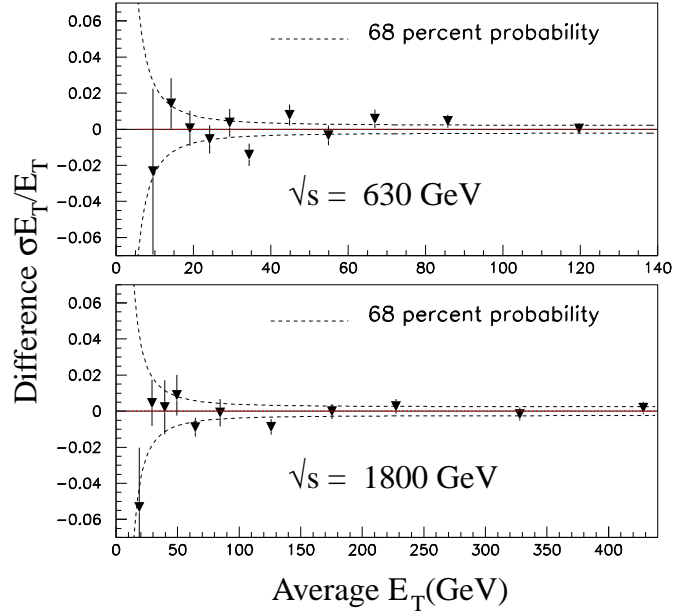,width=3.5in}}
\caption{ The improved resolution closure obtained using the {\sc
herwig} Monte Carlo simulation, for both center-of-mass energies. For
most of the kinematic range, the degree of closure lies within a
fraction of a percent.}
\label{mc_closure_630}
\end{figure}

 Subsequent to the publication by D\O\ of the inclusive jet cross
 section~\cite{d0_inc} (Section~\ref{sec:inclusive_jet}) and the dijet
 mass spectrum~\cite{d0_dijet_mass} (Section~\ref{sec:dijet_mass}) at
 $\sqrt{s} = 1800$~GeV the MC closure data for the resolutions were
 reexamined (see Fig.~\ref{mc_closure_630}). As a result of this, the
 MC closure error at $\sqrt{s} = 1800$~GeV was reduced for $\Et >
 40$~GeV:
\begin{eqnarray}
\Delta \left( \frac{\sigma_{E_T}}{{E}_{\mathrm{T}
}}\right)_{\sqrt{s} = 630~\mbox{\footnotesize  GeV}} & = &
  \frac{2.23}{\Et^{2}}+0.0021. \nonumber \\
\Delta \left( \frac{\sigma_{E_T}}{{E}_{\mathrm{T}
}}\right)_{\sqrt{s} = 1800~\mbox{\footnotesize  GeV}} & = &
 \frac{14.1}{\Et^{2}}+0.0024. \nonumber \\
 \label{eq_reso_sys_new}
\end{eqnarray}
 The effect of reducing the error on the inclusive jet cross section
 and dijet mass spectrum was negligible, and hence the results were
 not updated. The reduced errors are important for the analysis of the
 ratio of inclusive jet cross sections at $\sqrt{s} = 630$ and
 1800~GeV~\cite{d0_inc_630} (Section~\ref{sec:inclusive_jet_ratio}).

\begin{figure}[htbp]
\centerline
{\psfig{figure=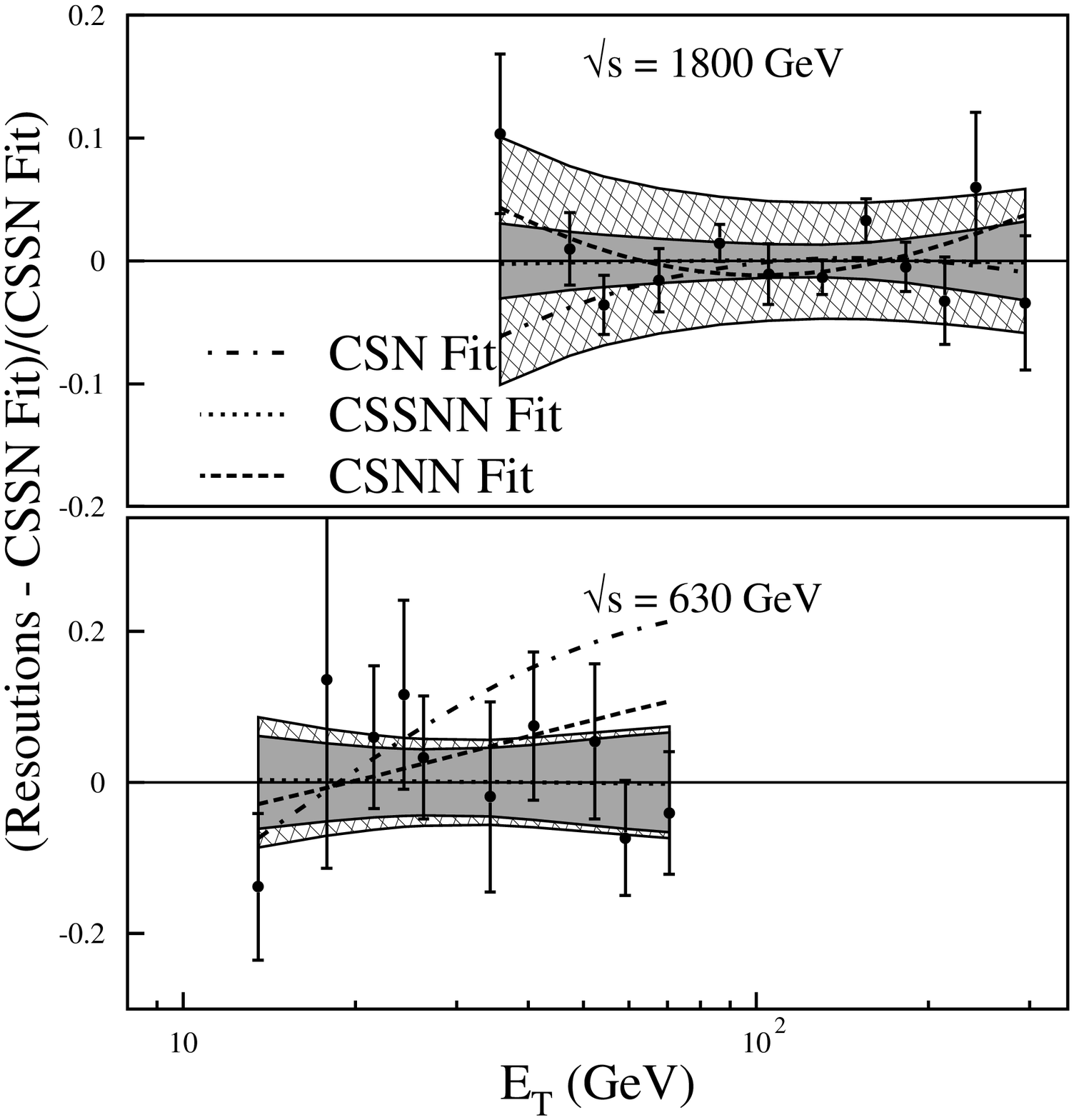,width=3.5in}}
\caption{A comparison of the measured jet resolutions and the fit
using the {\bf CSSN} model. Also shown are curves representing
comparisons between the different models and {\bf CSSN}. The shaded
regions show the uncertainty in the fit. The hatched region shows the
magnitude of the fit and MC closure uncertainties added in
quadrature.}
\label{fig:comp_fits}
\end{figure}

 In Fig.~\ref{fig:comp_fits} the measured resolutions are compared
 with the {\bf CSSN} fit. The shaded region shows the size of the fit
 uncertainty, and the hatched region shows the size of the fit and MC
 closure uncertainties added in quadrature. Also shown are the other
 models. It is clear that the combined fit and MC closure
 uncertainties are of reasonable size and that the total uncertainties
 are not underestimated.

\subsection{$\boldmath {\eta}$ and $\boldmath {\phi}$ Resolutions}

\begin{figure}[htbp]
   \centerline
   {\psfig{figure=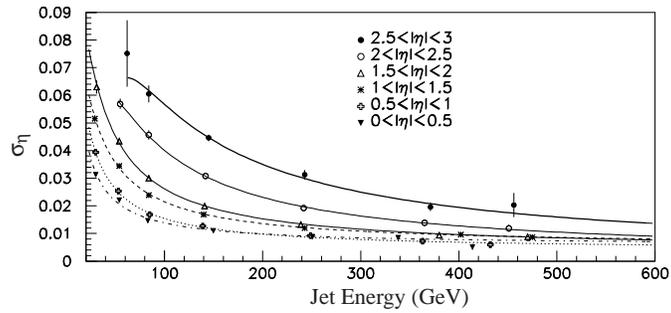,width=3.5in}}
   \caption{$\eta$ resolution as a function of the particle-level jet
   energy using a {\sc herwig} simulation.  }
   \label{FIG:eta_resolution}
\end{figure}

\begin{figure}[htbp]
   \centerline
   {\psfig{figure=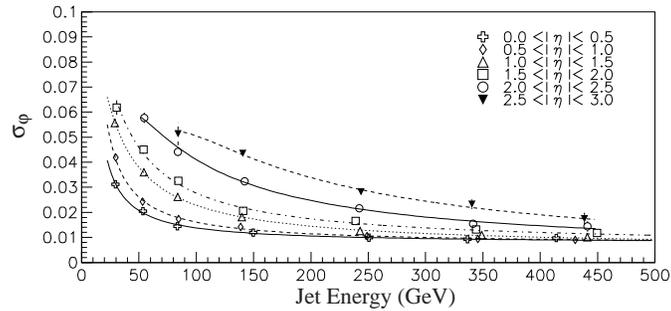,width=3.5in}}
   \caption{$\phi$ resolution as a function of the particle-level jet
   energy using a {\sc herwig} simulation.  }
   \label{FIG:phi_resolution}
\end{figure}

 After the $\eta$-bias correction is applied, the average $\eta$ of
 the reconstructed jet is equal to the $\eta$ of a particle-level jet,
 but due to calorimeter showering effects, both $\eta$ and $\phi$
 resolutions remain non-zero.  The $\eta$ resolution is obtained by
 using {\sc herwig} Monte Carlo and studying $\eta_{\rm ptcl}-\eta$ as
 a function of jet energy and $\eta$.  Figure \ref{FIG:eta_resolution}
 shows the $\eta$-resolution as a function of jet energy for different
 energy regions.  The distributions show no tails and are
 well-described by a Gaussians.  The $\phi$ resolution is determined
 by measuring $\phi_{\rm ptcl}-\phi$ as a function of jet energy and
 $\eta$.  Figure \ref{FIG:phi_resolution} shows the $\phi$ resolutions
 which are similar in magnitude to the $\eta$ resolutions.

\clearpage

\section[Inclusive Jet Cross Section at 
$\protect\bbox{\sqrt{s}}$~=~1800~GeV] {Inclusive Jet Cross Section at
$\protect\bbox{\sqrt{\lowercase{s}}}$~=~1800~G{\lowercase{e}}V}
\label{sec:inclusive_jet}

 In this section we describe the measurement of the inclusive jet
 cross section in the pseudorapidity ranges $\modeta < 0.5$ and $0.1 <
 \modeta < 0.7$. The inclusive jet cross section is given by:
\begin{equation}
\frac{d^{2}\sigma}{d\Et d\eta} = 
\frac{N_{i} C_{i}}{{\cal{L}}_{i} \epsilon_{i} \Delta \Et \Delta \eta}
\label{eq:inc_cross_section}
\end{equation}
 where $N_i$ is the number of accepted jets in \Et\ bin $i$ of width
 $\Delta \Et$; $C_i$ is the resolution unsmearing correction; ${\cal
 L}_{i}$ is the integrated luminosity; $\epsilon_{i}$ is the
 efficiency of the trigger, vertex selection, and the jet quality
 cuts; and $\Delta \eta$ is the width of the pseudorapidity bin.

\subsection{Data Selection} 
 
 The selected data are events with one or more jets which satisfy the
 requirements of the inclusive jet triggers. Jets are required to pass
 the standard jet quality criteria to be included in the cross section
 sample (Section~\ref{sec:quality_cuts}). The \met\ of the event is
 required to satisfy Eq.~\ref{MISSETcut_2}. The vertex of the event
 must be within 50 cm of $z = 0$. The efficiency for each jet is then
 given by the product of the efficiencies of the jet quality cuts
 ($\epsilon_{{\rm jet}}$), the efficiency of the cut on $\met$
 ($\epsilon_{\rm met}$), the efficiency for an event to pass the
 trigger ($\epsilon_{\rm trigger}$), and the efficiency for passing
 the vertex cut ($\epsilon_{\rm vertex}$):
\begin{equation}
\label{inc_weight}
\epsilon_i = {{ \epsilon_{\rm jet} \epsilon_{\rm met} \epsilon_{\rm
trigger} \epsilon_{\rm vertex}}}.
\end{equation}
 The values of $\epsilon_{{\rm jet}}$ and $\epsilon_{\rm met}$ are
 plotted in Fig.~\ref{FIG:efficiency}. The efficiency of the vertex
 requirement is $90 \pm 1\%$.

\subsection{Filter Efficiency and Luminosity Matching}

\begin{figure}[htbp]
\begin{center}
\vbox{\centerline{\psfig{figure=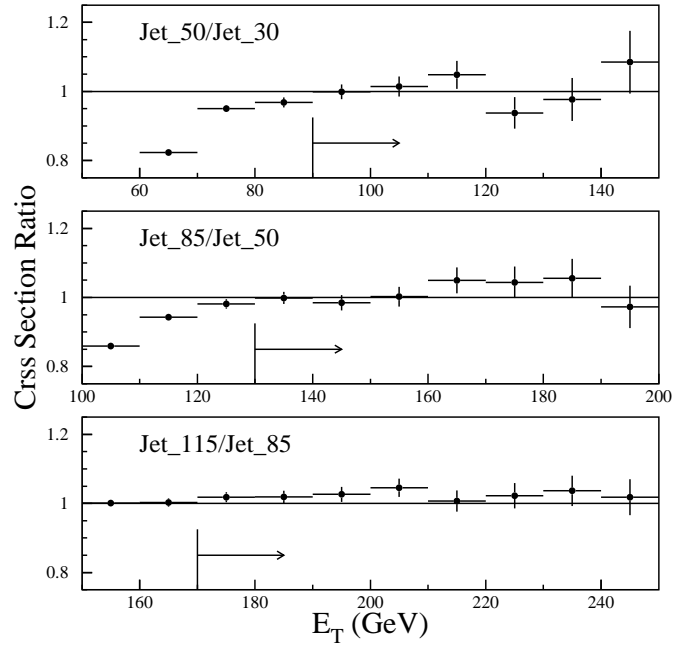,width=3.5in}}}
  \caption{Inclusive cross section ratios. The arrows signify the \Et\
  above which the higher threshold trigger is used.} \label{Ratios}
\end{center}
\end{figure}

 Figure~\ref{Ratios} shows the cross section ratios for
 Jet\_50/Jet\_30, Jet\_85/Jet\_50 and Jet\_115/Jet\_85.  Since the
 denominator in each ratio represents a less restrictive trigger than
 the numerator, the numerator trigger is efficient where the ratio
 stabilizes at a constant value. Thus Jet\_50, Jet\_85, and Jet\_115
 are efficient above 90, 130, and 170 GeV, respectively.  The
 efficiency for Jet\_30 was determined to be $100\%$ at 50 GeV
 (Section~\ref{sec:triggers}).

 The determination of the integrated luminosity for each of the jet
 triggers is described in detail in Section~\ref{sec04:lumin}. The
 luminosity used for Jet\_50 is determined by matching the Jet\_50
 inclusive jet cross section to the Jet\_85 cross section above
 130~GeV, introducing a $1.1\%$ statistical error. The Jet\_30
 luminosity is determined by matching to the Jet\_50 cross section
 above 90~GeV, which results in a $1.4\%$ statistical error. Hence the
 matching error for Jet\_30 is given by $1.1\%$ and $1.4\%$ added in
 quadrature, or $1.7\%$. These errors are added to the $5.8\%$ error
 on Jet\_85.  The final Jet\_30 and Jet\_50 luminosities are then
 0.350 \ipb\ and 4.76 \ipb\ with errors of $6.1\%$ and $5.9\%$,
 respectively.

\begin{figure}[htbp]
 \begin{center}
\vbox{\centerline{\psfig{figure=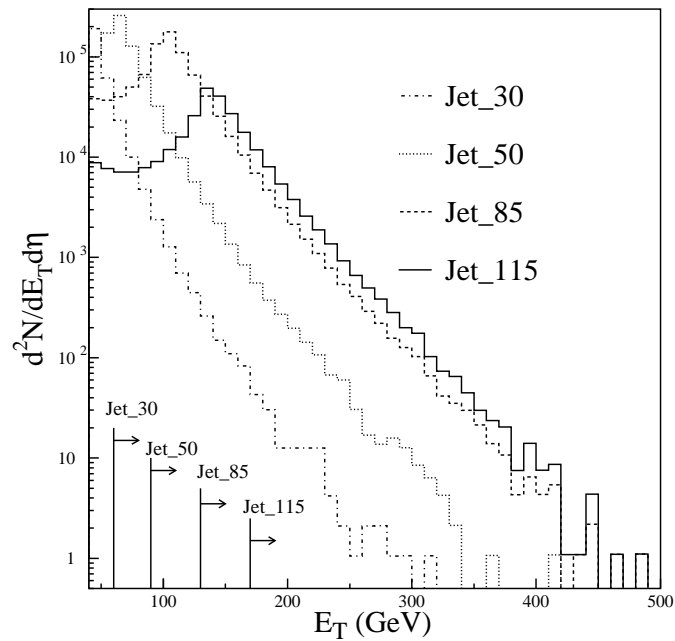,width=3.5in}}}
\caption{Energy-corrected \Et\ spectra for Jet\_115 (solid line),
 Jet\_85 (dashed), Jet\_50 (dotted), and Jet\_30 (dot-dashed). The
 arrows signify the \Et\ range in which each trigger's spectrum is
 used.}
\label{Four}
\end{center}
\end{figure}

 Figure~\ref{Four} shows \Et\ spectra for the four jet
 triggers, without luminosity normalization, in the central rapidity
 region ($\modeta < 0.5$) after efficiency and energy corrections.

\subsection{Observed Cross Section}

\begin{figure}[htbp]
\begin{center}
\vbox{\centerline{\psfig{figure=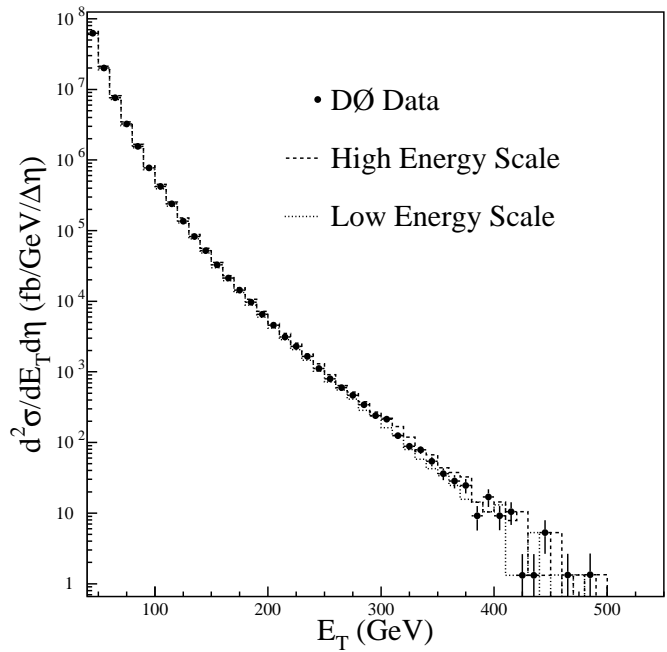,width=3.5in}}} 
\caption{Energy-corrected and luminosity-normalized \Et\ spectra.  The
points with error bars correspond to the nominal energy scale
correction. The dashed (dotted) histogram corresponds to the high
(low) energy scales corrections.} \label{observed}
\end{center}
\end{figure}

 Figure~\ref{observed} shows the central cross section compiled from
 the four triggers.  As suggested by the cross section ratios, and in
 order to maximize statistics, the spectrum from $60 \leq \Et \leq 90$
 GeV is taken from the Jet\_30 data, $90$--$130$ GeV from Jet\_50,
 $130$--$170$ GeV from Jet\_85, and above $170$ GeV from Jet\_115.
 The three data sets in Fig.~\ref{observed} correspond to the low,
 nominal, and high energy scale corrections. The differences can be
 considered to be an error estimate on the cross section which
 dominates all other sources of error (luminosity, jet, missing \Et ,
 and vertex cuts).

\subsection{Highest $\bbox{\Et}$ Event Scanning}

 Since the cross section decreases rapidly as the \Et\ increases, a
 small amount of contamination can have a significant effect on the
 measured cross sections at large \Et .  The data set included 46
 events that passed selection cuts and contained a central jet
 ($\modeta \leq$ 0.7) with transverse energy greater than 375
 GeV. These events were visually scanned for contamination. We defined
 an event to be ``good'' if it had at least two jets with
 well-contained energy, if there were no isolated cells forming jets,
 and if there was no activity in the muon chambers consistent with
 cosmic ray interactions associated with the event. These conditions
 were intended to reject high-\Et\ jets arising from noisy calorimeter
 cells, cosmic rays, or beam halo from the main ring, which passes
 through the D\O\ detector~\cite{d0_detector}.  The 46 events
 contained 62 jets with \Et\ greater than 375 GeV. Seven of these jets
 included restored cells and seven of the events preferred the second
 vertex.  All of the events passed visual inspection.

\subsection{Resolution Unfolding}
\label{sec:1800_unfolding}

\begin{figure}[htbp]
\begin{center}
\vbox{\centerline
{\psfig{figure=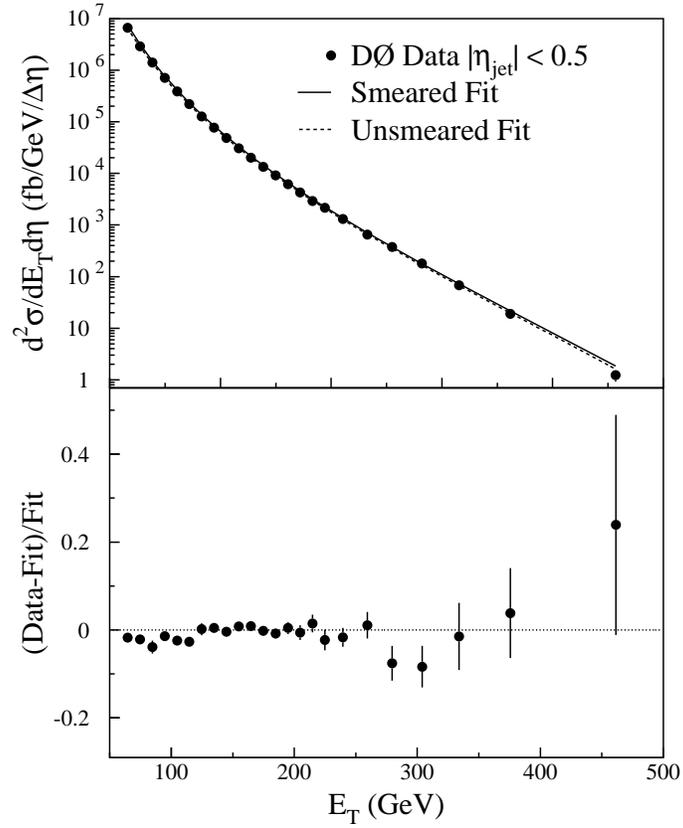,width=3.5in}}}
\caption[]{Data with smeared and unsmeared fit hypotheses
(Eq.~\ref{1800_ansatz}). The lower pane shows the smeared fit
residuals, (data $-$ smeared fit)/smeared fit.}
\label{observed_hypothesis}
\end{center}
\end{figure}

 The steep \Et\ spectrum is distorted by jet energy resolution. The
 distortion was corrected by using an ansatz function for the cross
 section,
\begin{equation}
 \exp\left(A \right) E_{T}^{\displaystyle \alpha}
 \left(1 - \frac{2\Et}{\sqrt{s}} \right)^{\displaystyle \beta} ,
\label{1800_ansatz}
\end{equation}
 smearing it with the measured resolution
 (Table~\ref{Table:resolution_parameters}), and comparing the smeared
 result with the measured cross section. The parameters $A, \alpha,$
 and $\beta$ were varied until the best fit was found between the
 observed cross section and the smeared trial spectrum. The \chisq\
 for the fit is 21.2 for 24 bins and three parameters, corresponding
 to 21 degrees of freedom
 (Table~\ref{tab:inc_ans_1800}). Figure~\ref{observed_hypothesis}
 shows an example of the energy scale corrected data with the best-fit
 smeared and unsmeared ansatz functions.  Simulations have shown that
 $\eta$-smearing causes negligible changes in the inclusive cross
 sections \cite{Elvira}.

\begin{table}[htbp]
\caption{Unsmearing ansatz function parameters for the inclusive jet
cross section (in fb) at $\sqrt{s}=1800$ GeV.}
\label{tab:inc_ans_1800}
\begin{tabular}{lld}
Rapidity Range	& Parameter 		& Value  	\\
\hline\hline 
		& $A$	 		& 37.28		\\	
$\modeta < 0.5$	& $\alpha$		& $-$5.04 	\\
		& $\beta$		&  8.23		\\
\hline 
		& $A$	 		& 37.30		\\	
$0.1 < \modeta < 0.7$	& $\alpha$	& $-$5.05	\\
		& $\beta$		&  8.37 	\\
\end{tabular}
\end{table}

\begin{figure}[htbp]
\begin{center}
\vbox{\centerline
{\psfig{figure=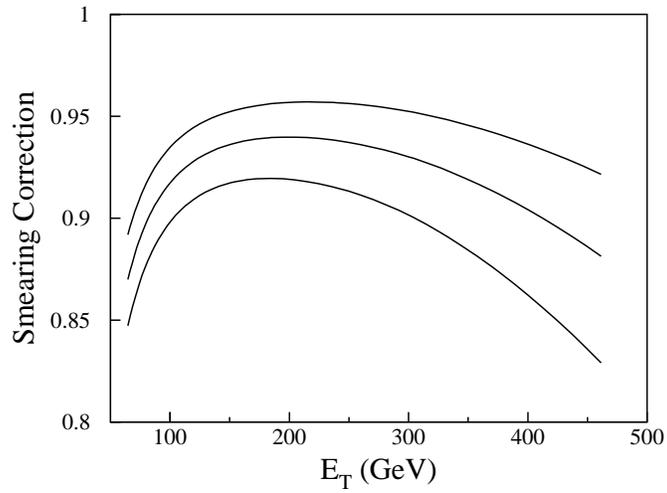,width=3.5in}}} 
\caption{ The nominal unsmearing correction is given by the central
line. See the text for an explanation of the other curves.}
\label{unsmearing_corr}
\end{center}
\end{figure}

 Figure~\ref{unsmearing_corr} shows the unsmearing correction as a
 function of transverse energy. The observed cross section is
 multiplied by this correction.  The central curve shows the nominal
 correction. The change in cross section is greatest at low \Et\ due
 to the steepness of the inclusive spectra and the relatively poor,
 rapidly changing jet resolution. The magnitude of the correction is
 $-13\%$ at 64.6~GeV, drops to $\approx -6\%$ at 205~GeV, and then
 rises to $-12\%$ at 461~GeV.

 The two outer curves of Fig.~\ref{unsmearing_corr} show the extent of
 the uncertainties in the nominal correction due to the resolution
 uncertainties.  This error was estimated directly with the data by
 unfolding with the upper and lower estimates of the resolution
 curves. For $\modeta < 0.5$ the maximum error is $3\%$. Varying the
 fit parameters by up to 3 standard deviations results in negligible
 changes in the resolution correction.

 The resolution correction errors due to the fit procedure and
 statistical fluctuations of the data were estimated by performing the
 unfolding procedure on distributions simulated with {\sc jetrad}.  A
 generated theoretical distribution was smeared with a resolution
 function. The ratio of the generated theoretical distribution to the
 smeared theoretical distribution was taken as the ``true'' unsmearing
 correction. Next, the previously described unfolding procedure was
 applied to the ``smeared theory'' and the resulting unsmearing
 correction was compared with the ``true'' unsmearing correction. The
 difference between the two corrections provided a measurement of the
 unfolding error. Above $\Et = 50$~GeV, the differences were less than
 $1\%$.  The error due to statistical fluctuation was estimated by
 simulating many jet samples containing the same total number of jets
 as the data sample. The statistical fluctuations between the
 different simulated samples lead to an error below 0.25$\%$ in any
 \Et\ bin.  A detailed description of this unfolding, and the
 unfolding error estimation procedures can be found in
 Ref.~\cite{Elvira}.
 
\subsection{Unfolded Cross Section}

\begin{figure}[htbp]
\begin{center}
\vbox{\centerline
{\psfig{figure=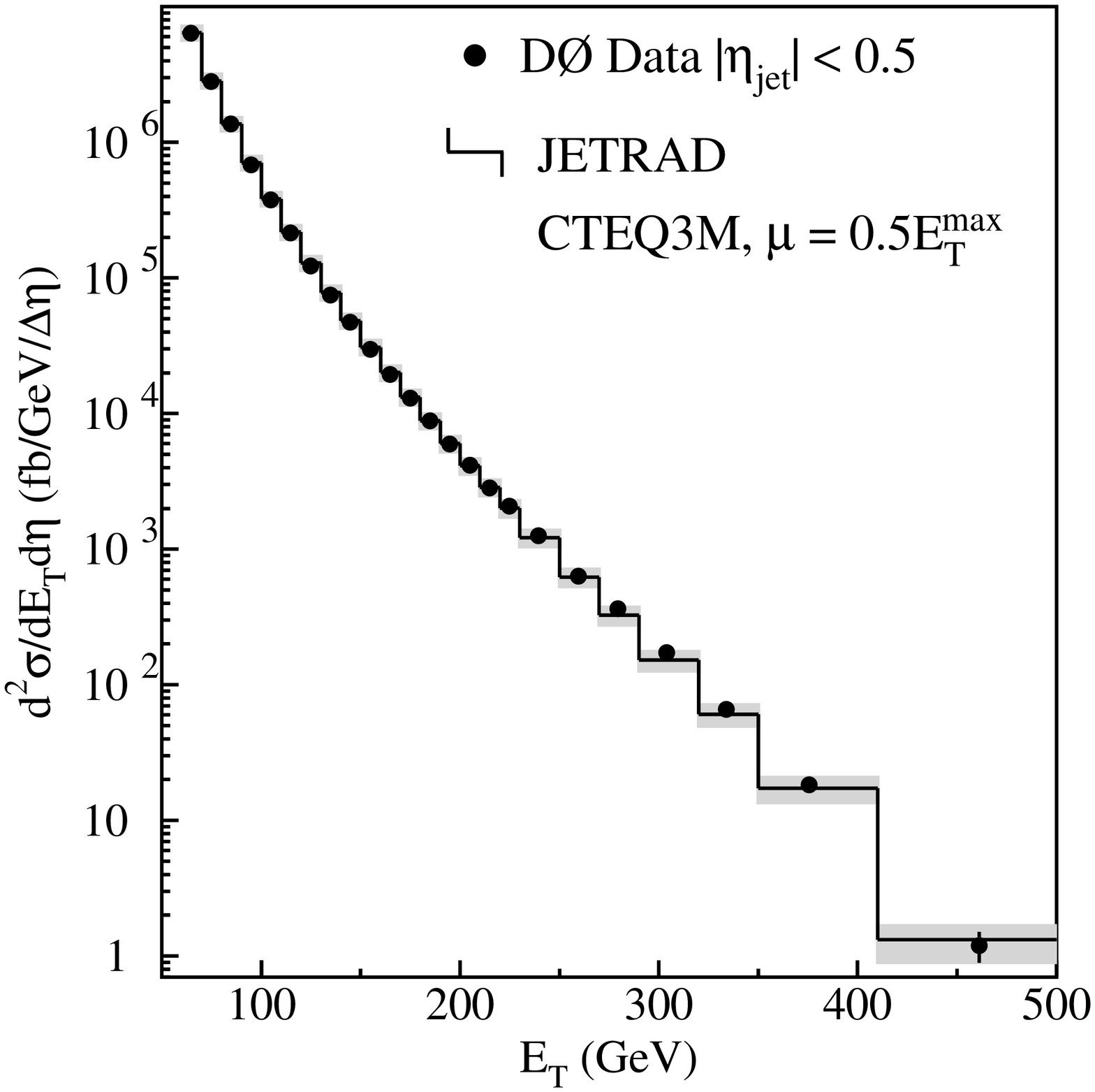,width=3.5in}}}
\caption{The $ \modeta < 0.5 $ inclusive jet cross section.
Statistical uncertainties are invisible on this scale except for the
highest \Et\ bin.  The histogram represents the {\sc jetrad}
prediction and the shaded band represents the $\pm 1\sigma$ systematic
uncertainty band about the prediction excluding the $5.8\%$ luminosity
uncertainty.}
\label{Cross_section}
\end{center}
\end{figure}

 The central inclusive jet cross section is shown in
 Fig.~\ref{Cross_section}. The cross section values are plotted in
 each bin at the \Et\ value for which the average integrated cross
 section is equal to the value of the analytical function
 (Eq.~\ref{1800_ansatz}) fitted to the
 cross-section~\cite{Ppoint}. The error bars are purely statistical
 and are visible only for the highest \Et\ value.  The error band
 indicates a one standard deviation variation of all systematic
 uncertainties, except the $5.8\%$ uncertainty on the absolute
 normalization. The measured cross-section is compared to the
 inclusive cross section for the same \Et\ values calculated with the
 {\sc jetrad} program using the \cteqthreem\ PDF and the scale $\mu
 =0.5\Etmax$. This prediction lies within the error band for all \Et\
 bins.  Table~\ref{inc_tab:table1} lists the cross sections for
 $\modeta < 0.5$ and $0.1 < \modeta < 0.7$.

\begin{table*}[htbp]
\squeezetable
\caption{  The $\modeta < 0.5$ and $0.1 <\modeta < 0.7$ cross sections 
 (Eq.~\ref{eq:inc_cross_section}).  Also given is the value of the
 fit to the cross section using Eq.~\ref{1800_ansatz}.}
\begin{tabular}{ccccccccc}
  & \multicolumn{4}{c}{$\modeta < 0.5$}
 & \multicolumn{4}{c}{$0.1 <\modeta < 0.7$} \\
\cline{2-5}\cline{6-9}
 Bin Range   &Plotted \Et\ &  Cross Sec.           &  Sys.         & Fitted &
              Plotted \Et\ &  Cross Sec.           &  Sys.         & Fitted \\
  (GeV)      &   (GeV)     &  $\pm$ Stat.          &
  Uncer.($\%$) &   Cross Sec. &
	         (GeV)     &  $\pm$ Stat.          &
  Uncer.($\%$) &   Cross Sec. \\ 
             &             & (fb/GeV/$\Delta\eta$) &               &
    (fb/GeV/$\Delta\eta$)  &
	                   & (fb/GeV/$\Delta\eta$) &               &
    (fb/GeV/$\Delta\eta$)  \\ \hline
 ~60~--~~70 & ~64.6 & $( 6.39\pm 0.04 )\times 10^{6}$ & $\pm 10$ & $
 6.27 \times 10^{6}$ & ~64.6& $( 6.26\pm 0.04 )\times 10^{6}$ & $\pm
 10$ &$ 6.13 \times 10^{6}$ \\ ~70~--~~80 & ~74.6 & $( 2.80\pm 0.03
 )\times 10^{6}$ & $\pm 10$ & $ 2.74 \times 10^{6}$ & ~74.6& $(
 2.74\pm 0.03 )\times 10^{6}$ & $\pm 10$ &$ 2.67 \times 10^{6}$ \\
 ~80~--~~90 & ~84.7 & $( 1.36\pm 0.02 )\times 10^{6}$ & $\pm 10$ & $
 1.31 \times 10^{6}$ & ~84.7& $( 1.34\pm 0.02 )\times 10^{6}$ & $\pm
 10$ &$ 1.28 \times 10^{6}$ \\ ~90~--~100 & ~94.7 & $( 6.84\pm 0.04
 )\times 10^{5}$ & $\pm 10$ & $ 6.74 \times 10^{5}$ & ~94.7& $(
 6.66\pm 0.04 )\times 10^{5}$ & $\pm 10$ &$ 6.53 \times 10^{5}$ \\
 100~--~110 & 104.7 & $( 3.76\pm 0.03 )\times 10^{5}$ & $\pm 10$ & $
 3.67 \times 10^{5}$ & 104.7& $( 3.63\pm 0.03 )\times 10^{5}$ & $\pm
 10$ &$ 3.54 \times 10^{5}$ \\ 110~--~120 & 114.8 & $( 2.14\pm 0.02
 )\times 10^{5}$ & $\pm 10$ & $ 2.08 \times 10^{5}$ & 114.8& $(
 2.07\pm 0.02 )\times 10^{5}$ & $\pm 10$ &$ 2.01 \times 10^{5}$ \\
 120~--~130 & 124.8 & $( 1.23\pm 0.02 )\times 10^{5}$ & $\pm 10$ & $
 1.23 \times 10^{5}$ & 124.8& $( 1.19\pm 0.01 )\times 10^{5}$ & $\pm
 10$ &$ 1.18 \times 10^{5}$ \\ 130~--~140 & 134.8 & $( 7.46\pm 0.04
 )\times 10^{4}$ & $\pm 10$ & $ 7.49 \times 10^{4}$ & 134.8& $(
 7.16\pm 0.03 )\times 10^{4}$ & $\pm 10$ &$ 7.18 \times 10^{4}$ \\
 140~--~150 & 144.8 & $( 4.71\pm 0.03 )\times 10^{4}$ & $\pm 10$ & $
 4.69 \times 10^{4}$ & 144.8& $( 4.51\pm 0.03 )\times 10^{4}$ & $\pm
 10$ &$ 4.48 \times 10^{4}$ \\ 150~--~160 & 154.8 & $( 2.97\pm 0.02
 )\times 10^{4}$ & $\pm 10$ & $ 3.00 \times 10^{4}$ & 154.8& $(
 2.83\pm 0.02 )\times 10^{4}$ & $+11,-10$&$ 2.86 \times 10^{4}$ \\
 160~--~170 & 164.8 & $( 1.94\pm 0.02 )\times 10^{4}$ & $+11,-10$ & $
 1.96 \times 10^{4}$ & 164.8& $( 1.83\pm 0.02 )\times 10^{4}$ & $\pm
 11$ &$ 1.86 \times 10^{4}$ \\ 170~--~180 & 174.8 & $( 1.30\pm 0.01
 )\times 10^{4}$ & $\pm 11$ & $ 1.30 \times 10^{4}$ & 174.8& $(
 1.23\pm 0.01 )\times 10^{4}$ & $\pm 11$ &$ 1.23 \times 10^{4}$ \\
 180~--~190 & 184.8 & $( 8.83\pm 0.10 )\times 10^{3}$ & $\pm 11$ & $
 8.75 \times 10^{3}$ & 184.8& $( 8.38\pm 0.09 )\times 10^{3}$ & $\pm
 11$ &$ 8.28 \times 10^{3}$ \\ 190~--~200 & 194.8 & $( 5.95\pm 0.08
 )\times 10^{3}$ & $\pm 11$ & $ 5.98 \times 10^{3}$ & 194.8& $(
 5.64\pm 0.07 )\times 10^{3}$ & $+12,-11$&$ 5.64 \times 10^{3}$ \\
 200~--~210 & 204.8 & $( 4.15\pm 0.07 )\times 10^{3}$ & $+12,-11$ & $
 4.13 \times 10^{3}$ & 204.8& $( 3.93\pm 0.06 )\times 10^{3}$ &
 $+12,-11$&$ 3.88 \times 10^{3}$ \\ 210~--~220 & 214.8 & $( 2.84\pm
 0.06 )\times 10^{3}$ & $+12,-11$ & $ 2.88 \times 10^{3}$ & 214.8& $(
 2.67\pm 0.05 )\times 10^{3}$ & $\pm 12$ &$ 2.70 \times 10^{3}$ \\
 220~--~230 & 224.8 & $( 2.08\pm 0.05 )\times 10^{3}$ & $\pm 12$ & $
 2.03 \times 10^{3}$ & 224.8& $( 1.95\pm 0.04 )\times 10^{3}$ &
 $+13,-12$&$ 1.90 \times 10^{3}$ \\ 230~--~250 & 239.4 & $( 1.26\pm
 0.03 )\times 10^{3}$ & $+13,-12$ & $ 1.24 \times 10^{3}$ & 239.4& $(
 1.17\pm 0.02 )\times 10^{3}$ & $+13,-12$&$ 1.15 \times 10^{3}$ \\
 250~--~270 & 259.4 & $( 6.34\pm 0.19 )\times 10^{2}$ & $+14,-13$ & $
 6.40 \times 10^{2}$ & 259.4& $( 5.84\pm 0.17 )\times 10^{2}$ &
 $+14,-13$&$ 5.94 \times 10^{2}$ \\ 270~--~290 & 279.5 & $( 3.65\pm
 0.15 )\times 10^{2}$ & $+14,-13$ & $ 3.39 \times 10^{2}$ & 279.5& $(
 3.21\pm 0.12 )\times 10^{2}$ & $+15,-14$&$ 3.13 \times 10^{2}$ \\
 290~--~320 & 303.9 & $( 1.73\pm 0.08 )\times 10^{2}$ & $+16,-14$ & $
 1.60 \times 10^{2}$ & 303.9& $( 1.56\pm 0.07 )\times 10^{2}$ &
 $+16,-15$&$ 1.47 \times 10^{2}$ \\ 320~--~350 & 333.9 & $( 6.60\pm
 0.50 )\times 10^{1}$ & $+17,-16$ & $ 6.50 \times 10^{1}$ & 333.9& $(
 6.06\pm 0.44 )\times 10^{1}$ & $+18,-16$&$ 5.91 \times 10^{1}$ \\
 350~--~410 & 375.7 & $( 1.83\pm 0.19 )\times 10^{1}$ & $+21,-18$ & $
 1.91 \times 10^{1}$ & 375.7& $( 1.48\pm 0.15 )\times 10^{1}$ &
 $+22,-19$&$ 1.72 \times 10^{1}$ \\ 410~--~560 & 461.1 & $( 1.20\pm
 0.30 )\times 10^{0}$ & $+30,-25$ & $ 1.57 \times 10^{0}$ & 460.9& $(
 1.05\pm 0.25 )\times 10^{0}$ & $+31,-26$&$ 1.39 \times 10^{0}$ \\
\end{tabular}
\label{inc_tab:table1}
\end{table*}

\subsection{Cross Section Uncertainties}

\begin{figure}[htbp]
\begin{center}
\vbox{\centerline
{\psfig{figure=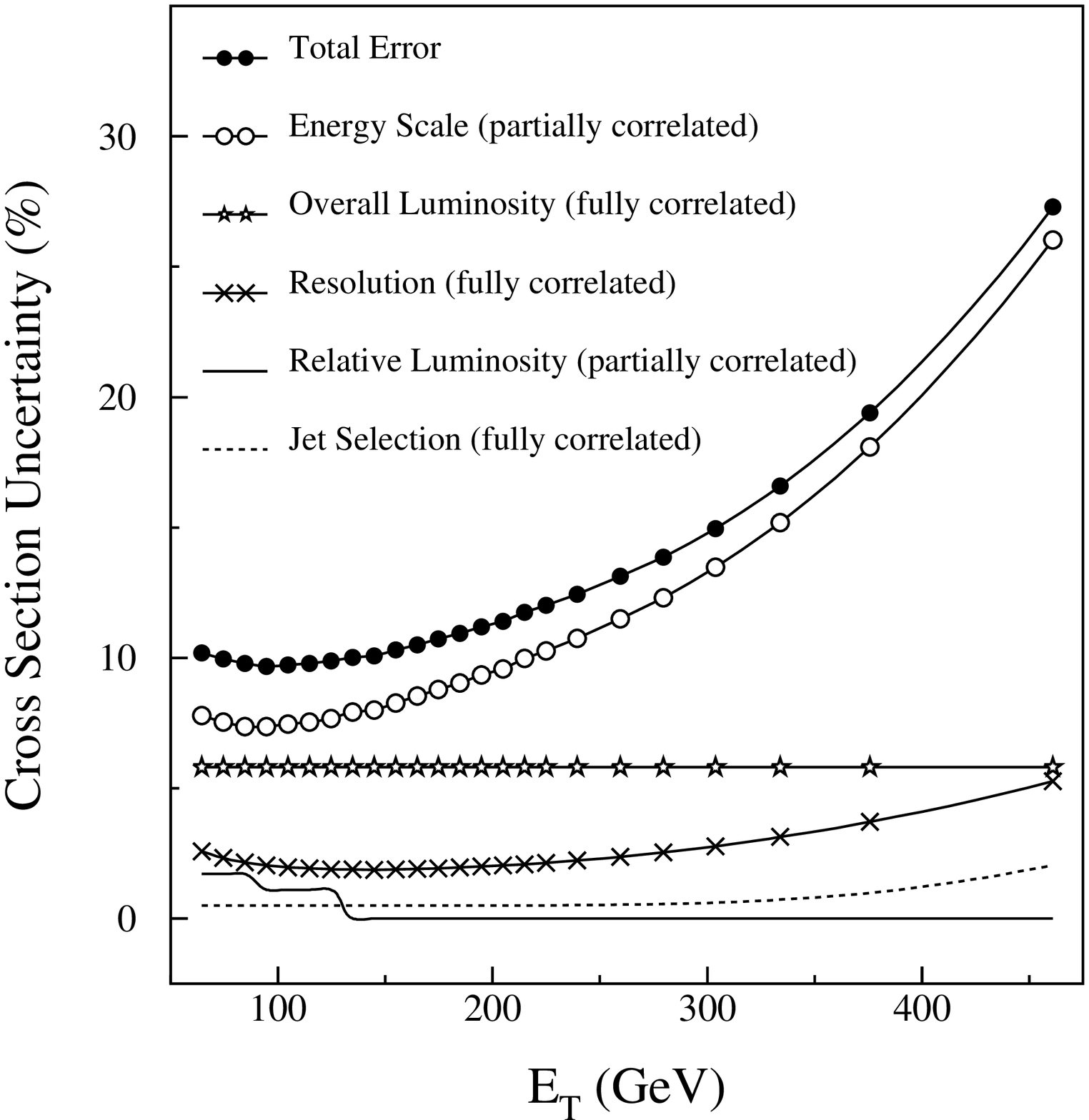,width=3.5in}}}
\caption{Contributions to the $\modeta < 0.5$ inclusive jet cross
section uncertainty plotted by component.}
\label{error_components}
\end{center}
\end{figure}

 The cross section uncertainties are dominated by the uncertainties in
 the energy scale correction. Table~\ref{1800_cross_section_errors}
 summarizes the uncertainties in the unfolded cross section. A
 detailed list of the uncertainties and their magnitudes is given in
 Tables~\ref{1800_cross_section_errors_a} and
 \ref{1800_cross_section_errors_b}.  Figure~\ref{error_components}
 shows the various uncertainties for the $\modeta < 0.5$ cross
 section.  The second uppermost curve shows the uncertainty in the
 energy scale, which varies from $8\%$ at low \Et\ to $30\%$ at 450
 GeV.  Clearly, this contribution dominates all other sources of error
 except at low \Et\ where the $5.8\%$ luminosity error is of
 comparable magnitude.  The other sources of error (jet and event
 selection, trigger matching, and jet resolution) are relatively
 small.

\begin{table}[htbp]
\begin{center}
\caption{Unfolded cross section errors}
\label{1800_cross_section_errors}
\begin{tabular}{lcl}
Source & Percentage & Comment \\ \hline Jet and Event Selection & $<2$
& Correlated
\\
\hline Luminosity      &   5.8             &   Correlated \\
\hline Luminosity Match&                   &                   \\
~~~60--~90 GeV  &   1.7             &   Statistical, Correlated \\
~~~90--130 GeV  &   1.1             & Trigger-to-Trigger\\ \hline
Energy Scale    & 15--30             & Mostly Correlated\\ \hline
Unfolding 	&                   &                    \\ 
~~Resolution Function 	&  1--3        	& Correlated         \\ 
~~Closure       &  1--2 	&Correlated           \\
\end{tabular}
\end{center}
\end{table}

\begin{figure}[htbp]
\begin{center}
\vbox{\centerline{\psfig{figure=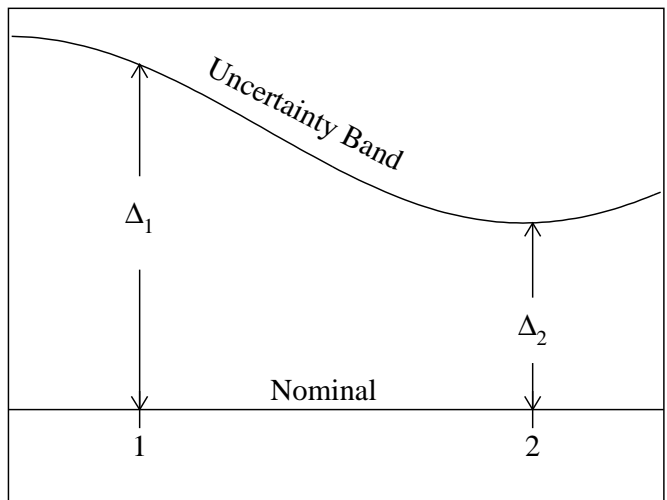,width=3.5in}}}
  \caption{Example of an error band relative to some nominal
  distribution (illustrated here with a flat line). If the errors at
  points \textbf{1} and \textbf{2} are completely correlated, then a
  one standard deviation $\left( 1\sigma \right)$ $\Delta
  _{\mathrm{1}}$ at the first position necessarily results in a
  $1\sigma$ $\Delta _{\mathrm{2}}$ at the second position.}
\label{diagrams_corr}
\end{center}
\end{figure}

\begin{figure}[hbtp]
\begin{center}
\vbox{\centerline{\psfig{figure=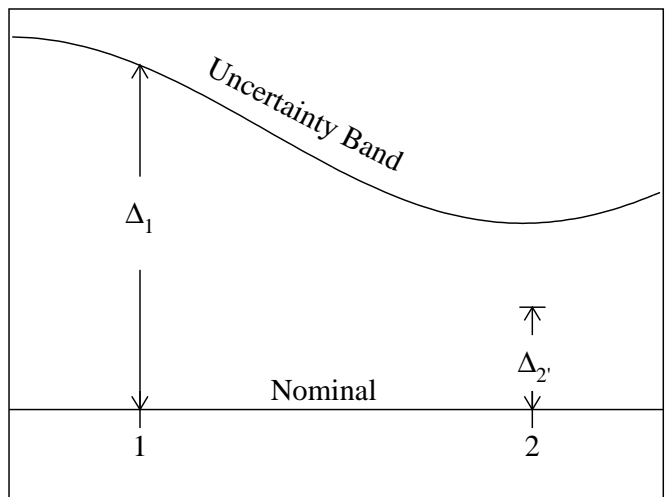,width=3.5in}}}
  \caption{If the errors at points \textbf{1} and \textbf{2} are
  partially correlated, then a full $1\sigma$ $\Delta _{\mathrm{1}}$
  at the first position results in a smaller than $1\sigma $ $\Delta
  _{\mathrm{2}^{\prime }}$ at the second position.  The correlation
  factor illustrated here is 0.55.}
\label{diagrams_partial}
\end{center}
\end{figure}

 Most of the systematic uncertainties in the inclusive jet cross
 section are highly correlated as a function of \Et . The
 uncertainties are separated into three ``types,'' depending on the
 correlation $\left( \rho \right)$ between two bins:
 
\begin{itemize}
\item $\rho =1:$ ``Completely correlated,'' indicating that a
 $1\sigma $ fluctuation in an error at a particular \Et\ bin is
 accompanied by a $1\sigma $ fluctuation at all other \Et\ bins
 (Fig.~\ref{diagrams_corr}).
\item $\rho =\rho \left(\Et_{1}, \Et_{2}
 \right) =\left[ -1,1\right]:$ ``Partially correlated,'' possessing a
 varying degree of correlation in \Et . A $1\sigma $ fluctuation thus
 implies a less than $1\sigma$ fluctuation elsewhere
 (Fig.~\ref{diagrams_partial}); negative $\rho$ indicates the shifts
 will have opposite directions at the two points.  This type of error
 is the most complicated to calculate and propagate.
\item $\rho =0:$ ``Uncorrelated,''
 statistical in nature or otherwise independent of one another. Some
 small errors with unknown (but probably positive) \Et\ correlation
 are treated as uncorrelated for simplicity. Such treatment is
 conservative. 
\end{itemize}

 The uncertainties due to jet selection are correlated as a function
 of \Et .  The uncertainties due to unsmearing are also correlated.
 The luminosity uncertainty is correlated as a function of \Et . The
 trigger matching uncertainties are correlated as a function of \Et\
 for bins that are derived from the same trigger sample and
 uncorrelated for all other bins.  The energy scale errors are
 partially correlated as a function of \Et\ and are discussed below.

\begin{figure}[htbp]
\begin{center}
\vbox{\centerline
{\psfig{figure=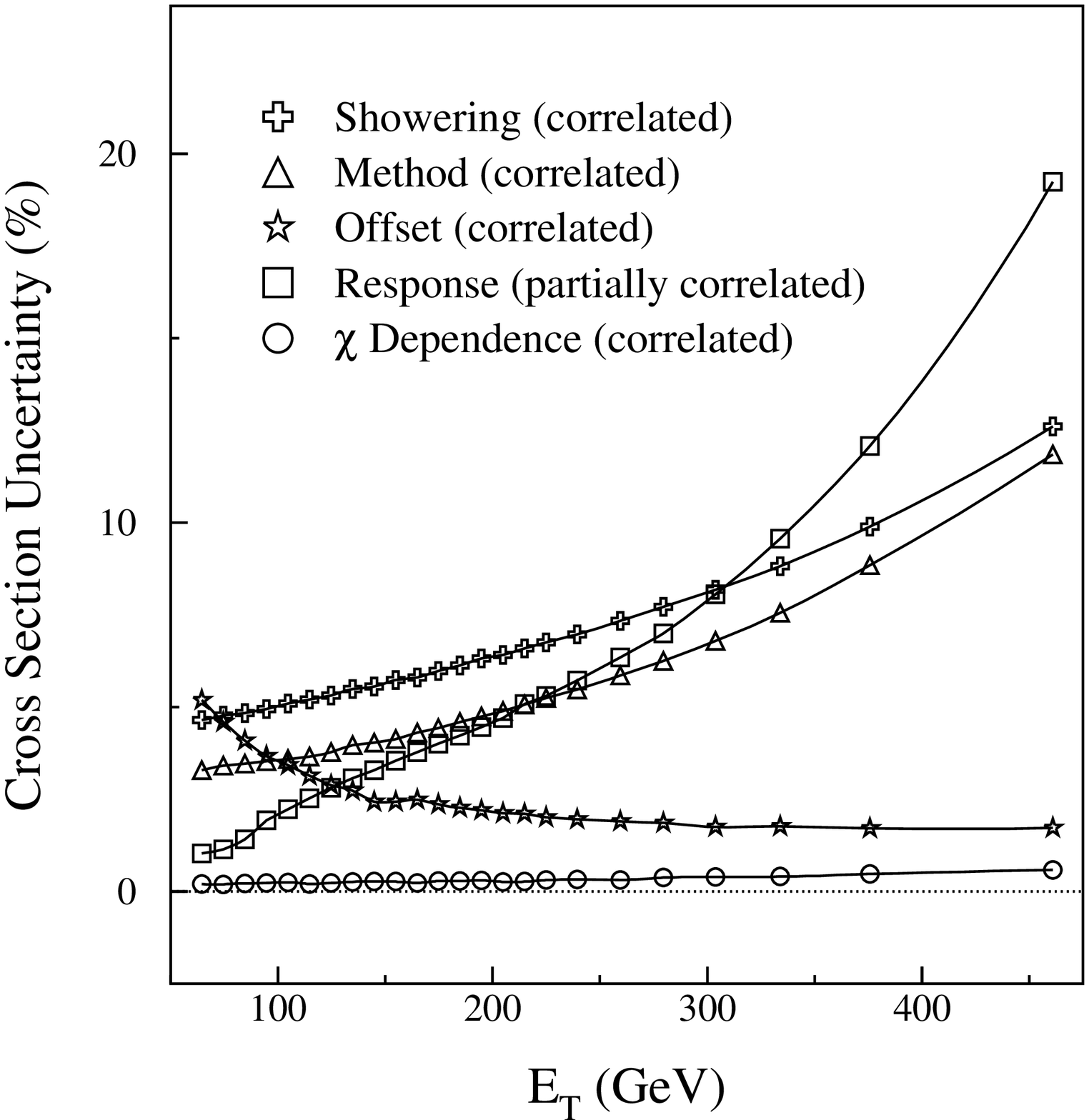,width=3.5in}}}
\caption[]{Percentage cross section errors for
 $\modeta < 0.5$ associated with the components of energy scale
 correction.}
\label{escale_components}
\end{center}
\end{figure}

 The energy scale calibration is implemented as a series of
 corrections, each with its own uncertainty
 (Section~\ref{sec:jet_energy_scale}).  The uncertainty due to the
 energy scale is separated into several components so that the
 correlations as a function of \Et\ can be studied
 (Fig.~\ref{escale_components}).  The energy scale uncertainties were
 calculated with a Monte Carlo simulation of the inclusive jet cross
 section. At each uncorrected \Et\ the simulation generated an
 ensemble of jets with rapidity, vertex position, luminosity, and
 variable correlations derived from the
 data. Figure~\ref{escale_components} shows the components of energy
 scale uncertainty as a function of \Et . The \Et\ of each of the
 simulated jets was then corrected and the resulting uncertainty in
 the jet cross section calculated. These uncertainties are in good
 agreement with the uncertainties derived from the data.

\begin{figure}[htbp]
\begin{center}
\vbox{\centerline
{\psfig{figure=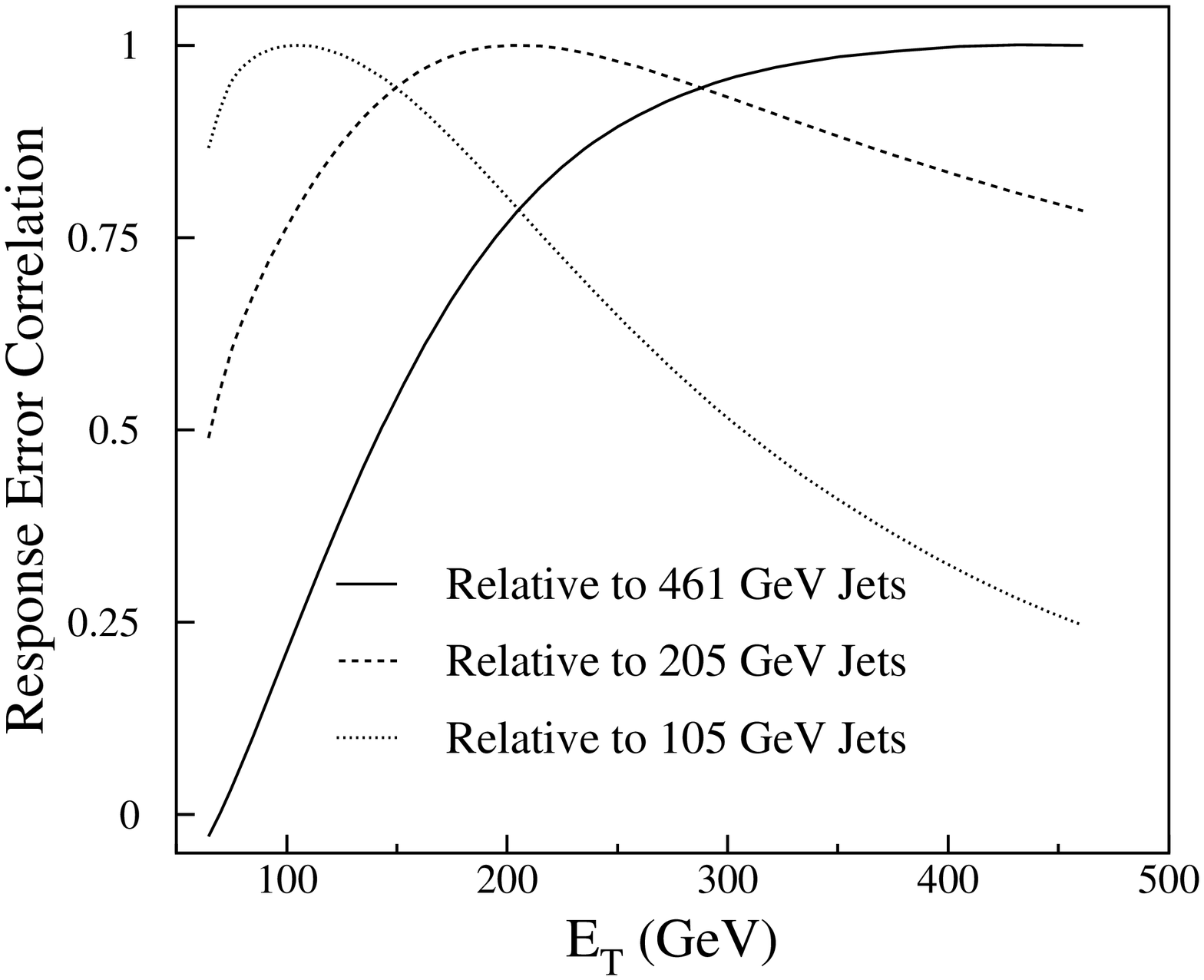,width=3.25in}}}
\caption{The correlations of the uncertainty due to the hadronic
response correction as a function of \Et. The solid curve shows the
correlations relative to the 461 GeV bin, the dashed curve with
respect to the 205 GeV bin, and the dotted curve with respect to the
105 GeV bin. }
\label{escale_err}
\end{center}
\end{figure}

 The uncertainties due to the offset correction, the $\eta$-dependent
 correction, the showering correction, and the method are all
 correlated as a function of \Et . The hadronic response uncertainty
 is partially correlated as a function of \Et\
 (Section~\ref{sec:jet_energy_scale}).  The hadronic response
 correlations are illustrated in Fig.~\ref{escale_err} and are given
 in Table~\ref{response_correlations}. In addition, the response
 uncertainties are only approximated by a Gaussian uncertainty
 distribution. Tables~\ref{TABLE:hadronic_sigma} and
 \ref{TABLE:hadronic_sigma_17} give the actual uncertainties for a
 given percentage confidence level (C.L.). i.e. if one has a $+20\%$
 error in the cross section at a given \Et\ corresponding to $95\%$
 C.L., then with $95\%$ probability the response errors will cause a
 deviation in the cross section of $\le 20\%$.  The correlations for
 the total systematic uncertainties are given in
 Table~\ref{inclusive_correlations}.

\begin{table*}[htbp]
\squeezetable
\caption{Percentage $\modeta < 0.5$ cross section uncertainties.  The
last row gives the nature of the \Et\ bin-to-bin correlations: 0
signifies uncorrelated uncertainties, 1 correlated, and p partially
correlated.  }
\label{1800_cross_section_errors_a}
\begin{tabular}{ddddddddddddddddd}
\Et   & Stat & Jet  & Lumin  & Lumin  &  \multicolumn{2}{c}{Unsmearing} 
&\multicolumn{10}{c}{Energy Scale} \\
\cline{6-7}\cline{8-17}
GeV & Error & Sel & & Match & High & Low &
\multicolumn{2}{c}{Underlying} & \multicolumn{2}{c}{$\eta$} &
\multicolumn{2}{c}{Method} &\multicolumn{2}{c}{Shower} 
&\multicolumn{2}{c}{Response}\\
\cline{8-9}\cline{10-11}\cline{12-13}\cline{14-15}\cline{16-17}
      &       &     &        &        &       &       
&  High & Low &  High & Low &  High & Low &  High & Low  &  High & Low \\
\hline\hline
    64.6 & 0.7 & 0.5 & 5.8 & 1.8 & 2.5 & -2.6 & 5.3 & -5.0 & 0.2 &
-0.2 & 3.4 & -3.2 & 4.7 & -4.6 & 1.1 & -1.0 \\ 74.6 & 1.0 & 0.5 & 5.8
& 1.8 & 2.3 & -2.4 & 4.7 & -4.5 & 0.2 & -0.2 & 3.5 & -3.3 & 4.8 & -4.7
& 1.2 & -1.1 \\ 84.7 & 1.5 & 0.5 & 5.8 & 1.8 & 2.1 & -2.2 & 4.2 & -4.1
& 0.2 & -0.2 & 3.5 & -3.5 & 4.9 & -4.8 & 1.5 & -1.3 \\ 94.7 & 0.6 &
0.5 & 5.8 & 1.1 & 1.9 & -2.1 & 3.8 & -3.7 & 0.2 & -0.2 & 3.6 & -3.5 &
5.0 & -4.9 & 1.9 & -1.9 \\ 104.7 & 0.8 & 0.5 & 5.8 & 1.1 & 1.9 & -2.1
& 3.5 & -3.4 & 0.2 & -0.2 & 3.6 & -3.6 & 5.1 & -5.1 & 2.2 & -2.2 \\
114.8 & 1.0 & 0.5 & 5.8 & 1.1 & 1.8 & -2.0 & 3.2 & -3.1 & 0.2 & -0.2 &
3.7 & -3.6 & 5.3 & -5.1 & 2.6 & -2.5 \\ 124.8 & 1.4 & 0.5 & 5.8 & 1.1
& 1.8 & -2.0 & 3.0 & -2.9 & 0.2 & -0.2 & 3.9 & -3.7 & 5.4 & -5.2 & 2.8
& -2.8 \\ 134.8 & 0.5 & 0.5 & 5.8 & 0.0 & 1.7 & -2.0 & 2.9 & -2.8 &
0.2 & -0.2 & 4.1 & -3.9 & 5.6 & -5.4 & 3.1 & -3.0 \\ 144.8 & 0.6 & 0.5
& 5.8 & 0.0 & 1.7 & -2.0 & 2.7 & -2.6 & 0.2 & -0.2 & 4.1 & -3.9 & 5.6
& -5.5 & 3.3 & -3.2 \\ 154.8 & 0.8 & 0.5 & 5.8 & 0.0 & 1.7 & -2.0 &
2.5 & -2.5 & 0.3 & -0.3 & 4.3 & -3.9 & 5.9 & -5.6 & 3.7 & -3.4 \\
164.8 & 1.0 & 0.5 & 5.8 & 0.0 & 1.7 & -2.1 & 2.4 & -2.4 & 0.3 & -0.3 &
4.5 & -4.1 & 5.9 & -5.7 & 3.9 & -3.6 \\ 174.8 & 0.9 & 0.5 & 5.8 & 0.0
& 1.7 & -2.1 & 2.3 & -2.3 & 0.3 & -0.3 & 4.7 & -4.2 & 6.1 & -5.9 & 4.1
& -3.9 \\ 184.8 & 1.1 & 0.5 & 5.8 & 0.0 & 1.8 & -2.1 & 2.2 & -2.2 &
0.3 & -0.3 & 4.9 & -4.3 & 6.2 & -6.0 & 4.3 & -4.2 \\ 194.8 & 1.4 & 0.5
& 5.8 & 0.0 & 1.8 & -2.2 & 2.1 & -2.1 & 0.3 & -0.3 & 5.1 & -4.4 & 6.4
& -6.2 & 4.5 & -4.4 \\ 204.8 & 1.7 & 0.5 & 5.8 & 0.0 & 1.8 & -2.2 &
2.1 & -2.0 & 0.3 & -0.3 & 5.3 & -4.5 & 6.5 & -6.3 & 4.8 & -4.6 \\
214.8 & 2.0 & 0.5 & 5.8 & 0.0 & 1.9 & -2.3 & 2.0 & -2.0 & 0.3 & -0.3 &
5.6 & -4.5 & 6.7 & -6.4 & 5.2 & -5.0 \\ 224.8 & 2.4 & 0.5 & 5.8 & 0.0
& 1.9 & -2.4 & 2.0 & -1.9 & 0.3 & -0.3 & 5.8 & -4.7 & 6.9 & -6.6 & 5.4
& -5.2 \\ 239.4 & 2.1 & 0.5 & 5.8 & 0.0 & 2.0 & -2.5 & 1.9 & -1.9 &
0.3 & -0.3 & 6.1 & -4.9 & 7.1 & -6.8 & 5.8 & -5.7 \\ 259.4 & 3.0 & 0.5
& 5.8 & 0.0 & 2.1 & -2.6 & 1.8 & -1.8 & 0.3 & -0.3 & 6.6 & -5.1 & 7.5
& -7.2 & 6.4 & -6.3 \\ 279.5 & 4.0 & 0.6 & 5.8 & 0.0 & 2.2 & -2.8 &
1.8 & -1.8 & 0.4 & -0.3 & 7.1 & -5.4 & 7.9 & -7.5 & 7.1 & -6.9 \\
303.9 & 4.7 & 0.6 & 5.8 & 0.0 & 2.4 & -3.1 & 1.8 & -1.8 & 0.4 & -0.4 &
7.9 & -5.7 & 8.4 & -7.9 & 8.3 & -7.8 \\ 333.9 & 7.6 & 0.7 & 5.8 & 0.0
& 2.7 & -3.5 & 1.8 & -1.7 & 0.4 & -0.4 & 9.0 & -6.1 & 9.1 & -8.6 & 9.8
& -9.3 \\ 375.7 & 10.2 & 1.0 & 5.8 & 0.0 & 3.2 & -4.2 & 1.7 & -1.7 &
0.5 & -0.5 & 10.8 & -6.8 & 10.2 & -9.6 & 12.4 & -11.7 \\ 461.1 & 25.0
& 2.1 & 5.8 & 0.0 & 4.6 & -5.9 & 1.7 & -1.7 & 0.6 & -0.6 & 15.0 & -8.6
& 13.2 & -12.0 & 20.3 & -18.2 \\
\hline
\multicolumn{1}{c}{Correl.} & 
\multicolumn{1}{c}{0} &   	
\multicolumn{1}{c}{1} &		
\multicolumn{1}{c}{1} &		
\multicolumn{1}{c}{p} &		
\multicolumn{1}{c}{1} &    	
\multicolumn{1}{c}{1} &    	
\multicolumn{1}{c}{1} &    	
\multicolumn{1}{c}{1} &    	
\multicolumn{1}{c}{1} &    	
\multicolumn{1}{c}{1} &    	
\multicolumn{1}{c}{1} &    	
\multicolumn{1}{c}{1} &    	
\multicolumn{1}{c}{1} &    	
\multicolumn{1}{c}{1} &    	
\multicolumn{1}{c}{p} &    	
\multicolumn{1}{c}{p} \\   	
\end{tabular}
\end{table*}

\begin{table*}[htbp]
\squeezetable
\caption{Percentage $0.1 < \modeta < 0.7$ cross section uncertainties.
The last row gives the nature of the \Et\ bin-to-bin correlations: 0
signifies uncorrelated uncertainties, 1 correlated, and p partially
correlated.  }
\label{1800_cross_section_errors_b}
\begin{tabular}{ddddddddddddddddd}
\Et   & Stat & Jet  & Lumin  & Lumin  &  \multicolumn{2}{c}{Unsmearing}
 &\multicolumn{10}{c}{Energy Scale} \\
\cline{6-7}\cline{8-17}
GeV   & Error & Sel &        & Match  &  High & Low  
 &  \multicolumn{2}{c}{Underlying} & \multicolumn{2}{c}{$\eta$} & 
\multicolumn{2}{c}{Method} &\multicolumn{2}{c}{Shower}
 &\multicolumn{2}{c}{Response}\\
\cline{8-9}\cline{10-11}\cline{12-13}\cline{14-15}\cline{16-17}
      &       &     &        &        &       &       &
  High & Low &  High & Low &  High & Low &  High & Low  &  High & Low \\
\hline\hline
  64.6 & 0.6 & 0.5 & 5.8 & 1.8 & 2.3 & -2.6 & 5.3 & -5.0 & 0.6 & -0.6
& 3.4 & -3.2 & 4.8 & -4.8 & 1.1 & -1.1\\ 74.6 & 0.9 & 0.5 & 5.8 & 1.8
& 2.1 & -2.4 & 4.7 & -4.5 & 0.6 & -0.6 & 3.5 & -3.4 & 4.9 & -4.7 & 1.3
& -1.1\\ 84.7 & 1.4 & 0.5 & 5.8 & 1.8 & 2.0 & -2.3 & 4.2 & -4.1 & 0.6
& -0.6 & 3.5 & -3.5 & 5.0 & -4.9 & 1.6 & -1.5\\ 94.7 & 0.5 & 0.5 & 5.8
& 1.1 & 1.9 & -2.2 & 3.8 & -3.7 & 0.6 & -0.6 & 3.6 & -3.5 & 5.1 & -5.0
& 2.1 & -2.1\\ 104.7 & 0.7 & 0.5 & 5.8 & 1.1 & 1.8 & -2.2 & 3.5 & -3.4
& 0.7 & -0.7 & 3.7 & -3.6 & 5.2 & -5.2 & 2.3 & -2.3\\ 114.7 & 1.0 &
0.5 & 5.8 & 1.1 & 1.8 & -2.1 & 3.2 & -3.1 & 0.7 & -0.7 & 3.8 & -3.6 &
5.5 & -5.2 & 2.7 & -2.6\\ 124.8 & 1.3 & 0.5 & 5.8 & 1.1 & 1.8 & -2.1 &
3.0 & -2.9 & 0.7 & -0.7 & 3.9 & -3.7 & 5.5 & -5.4 & 2.9 & -2.9\\ 134.8
& 0.5 & 0.5 & 5.8 & 0.0 & 1.8 & -2.1 & 2.8 & -2.8 & 0.7 & -0.7 & 4.1 &
-3.9 & 5.7 & -5.5 & 3.2 & -3.1\\ 144.8 & 0.6 & 0.5 & 5.8 & 0.0 & 1.8 &
-2.2 & 2.7 & -2.6 & 0.7 & -0.7 & 4.2 & -4.0 & 5.8 & -5.6 & 3.6 &
-3.3\\ 154.8 & 0.8 & 0.5 & 5.8 & 0.0 & 1.8 & -2.2 & 2.5 & -2.5 & 0.8 &
-0.8 & 4.4 & -4.0 & 5.9 & -5.8 & 3.7 & -3.5\\ 164.8 & 0.9 & 0.5 & 5.8
& 0.0 & 1.8 & -2.2 & 2.4 & -2.4 & 0.8 & -0.8 & 4.6 & -4.2 & 6.0 & -5.9
& 4.0 & -3.8\\ 174.8 & 0.9 & 0.5 & 5.8 & 0.0 & 1.8 & -2.3 & 2.3 & -2.3
& 0.8 & -0.8 & 4.8 & -4.2 & 6.2 & -6.0 & 4.2 & -4.0\\ 184.8 & 1.1 &
0.5 & 5.8 & 0.0 & 1.9 & -2.4 & 2.2 & -2.2 & 0.8 & -0.8 & 5.0 & -4.3 &
6.4 & -6.2 & 4.4 & -4.4\\ 194.8 & 1.3 & 0.5 & 5.8 & 0.0 & 1.9 & -2.4 &
2.1 & -2.1 & 0.8 & -0.8 & 5.2 & -4.4 & 6.5 & -6.3 & 4.7 & -4.4\\ 204.8
& 1.6 & 0.5 & 5.8 & 0.0 & 1.9 & -2.5 & 2.1 & -2.0 & 0.9 & -0.9 & 5.4 &
-4.5 & 6.7 & -6.5 & 5.0 & -4.8\\ 214.8 & 1.9 & 0.5 & 5.8 & 0.0 & 2.0 &
-2.6 & 2.0 & -2.0 & 0.9 & -0.9 & 5.7 & -4.6 & 6.9 & -6.6 & 5.3 &
-5.0\\ 224.8 & 2.2 & 0.5 & 5.8 & 0.0 & 2.0 & -2.7 & 2.0 & -2.0 & 0.9 &
-0.9 & 6.0 & -4.8 & 7.0 & -6.7 & 5.6 & -5.5\\ 239.4 & 2.0 & 0.5 & 5.8
& 0.0 & 2.1 & -2.8 & 1.9 & -1.9 & 1.0 & -0.9 & 6.2 & -4.9 & 7.3 & -6.9
& 6.0 & -5.8\\ 259.4 & 2.9 & 0.5 & 5.8 & 0.0 & 2.3 & -3.0 & 1.9 & -1.9
& 1.0 & -1.0 & 6.8 & -5.1 & 7.7 & -7.2 & 6.7 & -6.4\\ 279.5 & 3.9 &
0.6 & 5.8 & 0.0 & 2.4 & -3.3 & 1.8 & -1.8 & 1.1 & -1.0 & 7.4 & -5.4 &
8.0 & -7.6 & 7.5 & -7.2\\ 303.9 & 4.5 & 0.6 & 5.8 & 0.0 & 2.7 & -3.6 &
1.8 & -1.8 & 1.1 & -1.1 & 8.1 & -5.7 & 8.5 & -8.0 & 8.7 & -8.3\\ 333.9
& 7.2 & 0.7 & 5.8 & 0.0 & 3.0 & -4.1 & 1.8 & -1.8 & 1.2 & -1.2 & 9.3 &
-6.2 & 9.2 & -8.6 & 10.2 & -9.8\\ 375.7 & 10.3 & 1.0 & 5.8 & 0.0 & 3.6
& -4.9 & 1.8 & -1.7 & 1.3 & -1.3 & 11.1 & -6.9 & 10.3 & -9.7 & 13.5 &
-12.5\\ 460.9 & 24.3 & 2.1 & 5.8 & 0.0 & 5.2 & -7.0 & 1.7 & -1.7 & 1.6
& -1.6 & 15.1 & -8.7 & 13.3 & -12.2 & 21.8 & -19.4\\
\hline
\multicolumn{1}{c}{Correl.} & 
\multicolumn{1}{c}{0} &   	
\multicolumn{1}{c}{1} &		
\multicolumn{1}{c}{1} &		
\multicolumn{1}{c}{p} &		
\multicolumn{1}{c}{1} &    	
\multicolumn{1}{c}{1} &    	
\multicolumn{1}{c}{1} &    	
\multicolumn{1}{c}{1} &    	
\multicolumn{1}{c}{1} &    	
\multicolumn{1}{c}{1} &    	
\multicolumn{1}{c}{1} &    	
\multicolumn{1}{c}{1} &    	
\multicolumn{1}{c}{1} &    	
\multicolumn{1}{c}{1} &    	
\multicolumn{1}{c}{p} &    	
\multicolumn{1}{c}{p} \\   	
\end{tabular}
\end{table*}

\begin{table*}[htbp]
\squeezetable
\caption{The correlations for the uncertainty due to the energy scale response 
 for $\modetajet < 0.5$, and $0.1 <\modetajet < 0.7$. The correlation
 values above the diagonal are the correlations for $\modetajet < 0.5$
 and the correlations below the diagonal correspond to $0.1
 <\modetajet < 0.7$.  In both cases the correlation matrices are
 symmetric.}  \label{response_correlations}
 \begin{tabular}{rccccccccccccccccccccccccl} & 1 & 2 & 3 & 4 & 5 & 6 &
 7 & 8 & 9 & 10 & 11 & 12 & 13 & 14 & 15 & 16 & 17 & 18 & 19 & 20 & 21
 & 22 & 23 & 24 & \\ \hline\hline & 1.00& 0.97& 0.94& 0.90& 0.87&
 0.83& 0.80& 0.76& 0.72& 0.68& 0.64& 0.60& 0.56& 0.53& 0.49& 0.45&
 0.42& 0.37& 0.31& 0.26& 0.20& 0.14& 0.07&-0.03& 1 \\ & & 1.00& 0.99&
 0.97& 0.95& 0.92& 0.90& 0.86& 0.83& 0.79& 0.75& 0.72& 0.68& 0.64&
 0.60& 0.56& 0.53& 0.48& 0.41& 0.35& 0.29& 0.22& 0.15& 0.03& 2 \\ 1 &
 1.00 & & 1.00& 0.99& 0.98& 0.96& 0.94& 0.92& 0.89& 0.85& 0.82& 0.78&
 0.75& 0.71& 0.67& 0.64& 0.60& 0.55& 0.49& 0.43& 0.37& 0.30& 0.22&
 0.10& 3 \\ 2 & 0.97& 1.00 & & 1.00& 1.00& 0.99& 0.97& 0.95& 0.93&
 0.90& 0.87& 0.84& 0.80& 0.77& 0.74& 0.70& 0.67& 0.62& 0.56& 0.50&
 0.44& 0.37& 0.29& 0.17& 4 \\ 3 & 0.94& 0.99& 1.00 & & 1.00& 1.00&
 0.99& 0.97& 0.95& 0.93& 0.91& 0.88& 0.85& 0.82& 0.79& 0.76& 0.72&
 0.68& 0.62& 0.57& 0.51& 0.44& 0.36& 0.25& 5 \\ 4 & 0.90& 0.97& 0.99&
 1.00 & & 1.00& 1.00& 0.99& 0.98& 0.96& 0.94& 0.91& 0.89& 0.86& 0.83&
 0.80& 0.78& 0.73& 0.68& 0.63& 0.57& 0.51& 0.43& 0.32& 6 \\ 5 & 0.87&
 0.95& 0.98& 1.00& 1.00 & & 1.00& 1.00& 0.99& 0.98& 0.96& 0.94& 0.92&
 0.90& 0.87& 0.85& 0.82& 0.78& 0.73& 0.68& 0.63& 0.57& 0.50& 0.39& 7
 \\ 6 & 0.83& 0.92& 0.96& 0.99& 1.00& 1.00 & & 1.00& 1.00& 0.99& 0.98&
 0.96& 0.95& 0.93& 0.91& 0.88& 0.86& 0.83& 0.78& 0.74& 0.68& 0.63&
 0.56& 0.45& 8 \\ 7 & 0.80& 0.90& 0.94& 0.97& 0.99& 1.00& 1.00 & &
 1.00& 1.00& 0.99& 0.98& 0.97& 0.95& 0.93& 0.91& 0.89& 0.86& 0.82&
 0.78& 0.73& 0.68& 0.62& 0.51& 9 \\ 8 & 0.76& 0.86& 0.92& 0.95& 0.97&
 0.99& 1.00& 1.00 & & 1.00& 1.00& 0.99& 0.98& 0.97& 0.96& 0.94& 0.92&
 0.90& 0.86& 0.82& 0.78& 0.73& 0.67& 0.57& 10 \\ 9 & 0.72& 0.83& 0.89&
 0.93& 0.95& 0.98& 0.99& 1.00& 1.00 & & 1.00& 1.00& 0.99& 0.98& 0.97&
 0.96& 0.95& 0.92& 0.89& 0.86& 0.82& 0.77& 0.72& 0.62& 11 \\ 10 &
 0.68& 0.79& 0.85& 0.90& 0.93& 0.96& 0.98& 0.99& 1.00& 1.00 & & 1.00&
 1.00& 0.99& 0.99& 0.98& 0.96& 0.95& 0.92& 0.89& 0.85& 0.81& 0.76&
 0.67& 12 \\ 11 & 0.64& 0.75& 0.82& 0.87& 0.91& 0.94& 0.96& 0.98&
 0.99& 1.00& 1.00 & & 1.00& 1.00& 0.99& 0.99& 0.98& 0.96& 0.94& 0.91&
 0.88& 0.84& 0.79& 0.71& 13 \\ 12 & 0.60& 0.72& 0.78& 0.84& 0.88&
 0.91& 0.94& 0.96& 0.98& 0.99& 1.00& 1.00 & & 1.00& 1.00& 0.99& 0.99&
 0.98& 0.96& 0.94& 0.91& 0.87& 0.83& 0.75& 14 \\ 13 & 0.56& 0.68&
 0.75& 0.80& 0.85& 0.89& 0.92& 0.95& 0.97& 0.98& 0.99& 1.00& 1.00 & &
 1.00& 1.00& 1.00& 0.99& 0.97& 0.95& 0.93& 0.90& 0.86& 0.78& 15 \\ 14
 & 0.53& 0.64& 0.71& 0.77& 0.82& 0.86& 0.90& 0.93& 0.95& 0.97& 0.98&
 0.99& 1.00& 1.00 & & 1.00& 1.00& 0.99& 0.98& 0.97& 0.95& 0.92& 0.88&
 0.81& 16 \\ 15 & 0.49& 0.60& 0.67& 0.74& 0.79& 0.83& 0.87& 0.91&
 0.93& 0.96& 0.97& 0.99& 0.99& 1.00& 1.00 & & 1.00& 1.00& 0.99& 0.98&
 0.96& 0.94& 0.90& 0.84& 17 \\ 16 & 0.45& 0.56& 0.64& 0.70& 0.76&
 0.80& 0.85& 0.88& 0.91& 0.94& 0.96& 0.98& 0.99& 0.99& 1.00 & 1.00 & &
 1.00& 1.00& 0.99& 0.98& 0.96& 0.93& 0.87& 18 \\ 17 & 0.42& 0.53&
 0.60& 0.67& 0.72& 0.78& 0.82& 0.86& 0.89& 0.92& 0.95& 0.96& 0.98&
 0.99& 1.00 & 1.00& 1.00 & & 1.00& 1.00& 0.99& 0.98& 0.95& 0.91& 19 \\
 18 & 0.37& 0.48& 0.55& 0.62& 0.68& 0.73& 0.78& 0.83& 0.86& 0.90&
 0.92& 0.95& 0.96& 0.98& 0.99 & 0.99& 1.00& 1.00 & & 1.00& 1.00& 0.99&
 0.97& 0.94& 20 \\ 19 & 0.31& 0.41& 0.49& 0.56& 0.62& 0.68& 0.73&
 0.78& 0.82& 0.86& 0.89& 0.92& 0.94& 0.96& 0.97 & 0.98& 0.99& 1.00&
 1.00 & & 1.00& 1.00& 0.99& 0.96& 21 \\ 20 & 0.26& 0.35& 0.43& 0.50&
 0.57& 0.63& 0.68& 0.74& 0.78& 0.82& 0.86& 0.89& 0.91& 0.94& 0.95 &
 0.97& 0.98& 0.99& 1.00& 1.00 & & 1.00& 1.00& 0.98& 22 \\ 21 & 0.20&
 0.29& 0.37& 0.44& 0.51& 0.57& 0.63& 0.68& 0.73& 0.78& 0.82& 0.85&
 0.88& 0.91& 0.93 & 0.95& 0.96& 0.98& 0.99& 1.00& 1.00 & & 1.00& 0.99&
 23 \\ 22 & 0.14& 0.22& 0.30& 0.37& 0.44& 0.51& 0.57& 0.63& 0.68&
 0.73& 0.77& 0.81& 0.84& 0.87& 0.90 & 0.92& 0.94& 0.96& 0.98& 0.99&
 1.00& 1.00 & & 1.00& 24 \\ 23 & 0.07& 0.15& 0.22& 0.29& 0.36& 0.43&
 0.50& 0.56& 0.62& 0.67& 0.72& 0.76& 0.79& 0.83& 0.86 & 0.88& 0.90&
 0.93& 0.95& 0.97& 0.99& 1.00& 1.00& & \\ 24 &-0.03& 0.03& 0.10& 0.17&
 0.25& 0.32& 0.39& 0.45& 0.51& 0.57& 0.62& 0.67& 0.71& 0.75& 0.78 &
 0.81& 0.84& 0.87& 0.91& 0.94& 0.96& 0.98& 0.99& 1.00& \\
 \end{tabular}
\end{table*}

\begin{table*}[htbp]
\squeezetable
\caption{The percentage cross section uncertainties due to the energy scale
 response correction that correspond to a given percentage confidence
 level for $\modeta < 0.5$.}
\label{TABLE:hadronic_sigma}
\begin{tabular}{rdddddddddddd}
    & \multicolumn{6}{c}{Upper} &  \multicolumn{6}{c}{Lower}\\ \cline{2-7}\cline{8-13}
Bin & 
\multicolumn{1}{c}{$40\%$} &  \multicolumn{1}{c}{$68.3\%$} &  
\multicolumn{1}{c}{$86\%$} &  \multicolumn{1}{c}{$90\%$} & 
\multicolumn{1}{c}{$95\%$} &  \multicolumn{1}{c}{$99\%$} &
\multicolumn{1}{c}{$40\%$} &  \multicolumn{1}{c}{$68.3\%$} &  
\multicolumn{1}{c}{$86\%$} &  \multicolumn{1}{c}{$90\%$} & 
\multicolumn{1}{c}{$95\%$} &  \multicolumn{1}{c}{$99\%$} \\
\hline \hline 
1 & 0.7 & 1.1 & 1.3 & 1.3 & 1.6 & 1.9 &-0.7 & -1.0 & -1.3 & -1.3 &
-1.6 & -1.8 \\ 2 & 0.7 & 1.1 & 1.5 & 1.4 & 1.8 & 2.2 &-0.7 & -1.1 &
-1.5 & -1.4 & -1.7 & -2.1 \\ 3 & 1.0 & 1.5 & 1.9 & 1.9 & 2.2 & 2.6
&-0.9 & -1.3 & -1.8 & -1.8 & -2.2 & -2.6 \\ 4 & 1.3 & 1.9 & 2.5 & 2.6
& 3.0 & 3.6 &-1.3 & -1.8 & -2.4 & -2.6 & -3.0 & -3.5 \\ 5 & 1.4 & 2.2
& 3.0 & 3.1 & 3.4 & 4.0 &-1.4 & -2.2 & -2.9 & -3.0 & -3.3 & -4.0 \\ 6
& 1.6 & 2.6 & 3.3 & 3.4 & 3.8 & 4.6 &-1.5 & -2.5 & -3.1 & -3.3 & -3.6
& -4.4 \\ 7 & 1.6 & 2.9 & 3.6 & 3.8 & 4.1 & 5.0 &-1.6 & -2.8 & -3.5 &
-3.7 & -4.0 & -4.8 \\ 8 & 1.6 & 3.0 & 3.9 & 4.0 & 4.2 & 5.3 &-1.7 &
-3.0 & -3.8 & -3.0 & -4.2 & -5.3 \\ 9 & 1.7 & 3.3 & 4.1 & 4.2 & 4.6 &
5.7 &-1.7 & -3.2 & -4.1 & -4.2 & -4.5 & -5.6 \\ 10 & 1.9 & 3.6 & 4.5 &
4.6 & 5.0 & 6.2 &-1.8 & -3.4 & -4.3 & -4.4 & -4.7 & -5.9 \\ 11 & 2.0 &
3.9 & 4.8 & 4.9 & 5.1 & 6.7 &-1.8 & -3.6 & -4.6 & -4.6 & -4.9 & -6.3
\\ 12 & 2.1 & 4.1 & 5.1 & 5.0 & 5.4 & 7.0 &-2.0 & -3.9 & -5.0 & -4.9 &
-5.2 & -6.8 \\ 13 & 2.1 & 4.3 & 5.3 & 5.2 & 5.7 & 7.4 &-2.1 & -4.1 &
-5.2 & -5.2 & -5.6 & -7.2 \\ 14 & 2.4 & 4.5 & 5.6 & 5.6 & 6.1 & 7.8
&-2.3 & -4.5 & -5.5 & -5.5 & -6.0 & -7.5 \\ 15 & 2.6 & 4.8 & 5.9 & 5.9
& 6.5 & 8.3 &-2.5 & -4.6 & -5.7 & -5.8 & -6.4 & -7.9 \\ 16 & 3.1 & 5.2
& 6.4 & 6.5 & 7.3 & 9.0 &-2.9 & -5.0 & -6.2 & -6.3 & -7.0 & -8.6 \\ 17
& 3.3 & 5.4 & 6.7 & 6.9 & 7.7 & 9.4 &-3.1 & -5.2 & -6.4 & -6.6 & -7.5
& -8.9 \\ 18 & 3.8 & 5.8 & 7.1 & 7.5 & 8.8 & 10.2 &-3.7 & -5.6 & -7.0
& -7.4 & -8.4 & -9.8 \\ 19 & 4.4 & 6.4 & 8.2 & 8.5 & 10.0 & 11.7 &-4.3
& -6.3 & -7.8 & -8.1 & -9.4 & -11.0 \\ 20 & 5.0 & 7.1 & 9.0 & 9.4 &
11.1 & 13.0 &-4.9 & -6.9 & -8.8 & -9.0 &-10.6 & -12.3 \\ 21 & 6.0 &
8.3 & 10.8 & 11.1 & 13.3 & 15.6 &-5.7 & -7.8 &-10.0 &-10.2 &-12.0 &
-14.0 \\ 22 & 7.4 & 10.0 & 13.3 & 13.3 & 15.8 & 18.9 &-7.0 & -9.4
&-12.3 &-12.2 &-14.5 & -17.0 \\ 23 & 9.4 & 12.6 & 17.1 & 16.8 & 20.0 &
24.4 &-9.0 & -12.0 &-15.7 &-15.6 &-18.2 & -21.4 \\ 24 &16.2 & 22.4 &
29.6 & 30.2 & 35.3 & 43.1 &-14.6 & -20.1 &-25.5 &-26.2 &-29.6 & -34.6
\\
\end{tabular}

\end{table*}
\begin{table*}[htbp]
\squeezetable
\caption{The percentage cross section uncertainties due to the energy scale
 response correction that correspond to a given percentage confidence
 level for $0.1 < \modeta < 0.7$.}
\label{TABLE:hadronic_sigma_17}
\begin{tabular}{rdddddddddddd}
    & \multicolumn{6}{c}{Upper} &  \multicolumn{6}{c}{Lower}\\ 
\cline{2-7}\cline{8-13} 
Bin & 
\multicolumn{1}{c}{$40\%$} &  \multicolumn{1}{c}{$68.3\%$} &  
\multicolumn{1}{c}{$86\%$} &  \multicolumn{1}{c}{$90\%$} & 
\multicolumn{1}{c}{$95\%$} &  \multicolumn{1}{c}{$99\%$} &
\multicolumn{1}{c}{$40\%$} &  \multicolumn{1}{c}{$68.3\%$} &  
\multicolumn{1}{c}{$86\%$} &  \multicolumn{1}{c}{$90\%$} & 
\multicolumn{1}{c}{$95\%$} &  \multicolumn{1}{c}{$99\%$} \\
\hline \hline 
1 & 0.7 & 1.1 & 1.4 & 1.4 & 1.6 & 2.0 & -0.8 & -1.1 & -1.4 & -1.4 &
-1.7 & -1.9 \\ 2 & 0.8 & 1.2 & 1.6 & 1.6 & 1.9 & 2.3 & -0.8 & -1.2 &
-1.5 & -1.5 & -1.9 & -2.2 \\ 3 & 1.1 & 1.6 & 2.1 & 2.1 & 2.5 & 2.9 &
-1.0 & -1.5 & -2.1 & -2.2 & -2.5 & -2.9 \\ 4 & 1.3 & 2.1 & 2.7 & 2.9 &
3.2 & 3.9 & -1.3 & -2.0 & -2.7 & -2.8 & -3.2 & -3.8 \\ 5 & 1.5 & 2.4 &
3.1 & 3.2 & 3.6 & 4.2 & -1.4 & -2.3 & -3.0 & -3.1 & -3.4 & -4.2 \\ 6 &
1.6 & 2.7 & 3.4 & 3.6 & 3.9 & 4.9 & -1.5 & -2.6 & -3.3 & -3.5 & -3.8 &
-4.6 \\ 7 & 1.6 & 3.0 & 3.7 & 3.9 & 4.1 & 5.1 & -1.6 & -2.9 & -3.7 &
-3.9 & -4.1 & -5.0 \\ 8 & 1.7 & 3.2 & 4.0 & 4.1 & 4.4 & 5.5 & -1.7 &
-3.1 & -3.9 & -4.0 & -4.3 & -5.4 \\ 9 & 1.8 & 3.6 & 4.4 & 4.5 & 4.8 &
6.1 & -1.7 & -3.4 & -4.2 & -4.3 & -4.6 & -5.8 \\ 10 & 1.9 & 3.8 & 4.7
& 4.7 & 5.0 & 6.4 & -1.8 & -3.5 & -4.4 & -4.5 & -4.8 & -6.2 \\ 11 &
2.1 & 4.0 & 5.0 & 5.0 & 5.3 & 6.9 & -1.9 & -3.8 & -4.8 & -4.8 & -5.1 &
-6.6 \\ 12 & 2.1 & 4.2 & 5.3 & 5.1 & 5.6 & 7.2 & -2.0 & -4.0 & -5.1 &
-5.0 & -5.4 & -6.9 \\ 13 & 2.3 & 4.4 & 5.5 & 5.5 & 6.0 & 7.7 & -2.4 &
-4.3 & -5.5 & -5.5 & -6.1 & -7.6 \\ 14 & 2.6 & 4.7 & 5.8 & 5.9 & 6.4 &
8.1 & -2.4 & -4.5 & -5.6 & -5.6 & -6.2 & -7.7 \\ 15 & 2.9 & 5.0 & 6.2
& 6.3 & 7.2 & 8.8 & -2.8 & -4.8 & -6.0 & -6.1 & -6.7 & -8.2 \\ 16 &
3.2 & 5.3 & 6.6 & 6.8 & 7.7 & 9.2 & -3.0 & -5.1 & -6.3 & -6.5 & -7.4 &
-8.8 \\ 17 & 3.6 & 5.7 & 7.0 & 7.3 & 8.3 & 9.8 & -3.6 & -5.5 & -6.9 &
-7.2 & -8.2 & -9.6 \\ 18 & 4.0 & 6.0 & 7.5 & 7.9 & 9.3 & 10.9 & -4.0 &
-5.9 & -7.3 & -7.6 & -8.8 &-10.3 \\ 19 & 4.7 & 6.7 & 8.6 & 8.9 & 10.4
& 12.2 & -4.5 & -6.5 & -8.1 & -8.4 & -9.8 &-11.4 \\ 20 & 5.4 & 7.6 &
9.9 & 10.1 & 12.1 & 14.2 & -5.2 & -7.3 & -9.4 & -9.6 &-11.3 &-13.2 \\
21 & 6.5 & 8.8 & 11.8 & 11.9 & 14.2 & 16.9 & -6.2 & -8.2 &-10.8 &
-10.9 &-12.8 &-15.0 \\ 22 & 7.8 & 10.5 & 14.1 & 13.9 & 16.5 & 19.8 &
-7.4 & -9.7 &-12.8 & -12.6 &-14.9 &-17.8 \\ 23 & 9.9 & 13.4 & 18.2 &
18.0 & 21.4 & 26.5 & -9.7 &-13.0 &-17.0 & -17.1 &-19.8 &-23.5 \\ 24 &
17.2 & 24.4 & 32.2 & 33.0 & 38.3 & 47.3 & -15.9 &-22.1 &-28.0 & -28.8
&-32.2 &-37.1 \\
\end{tabular}
\end{table*}

\begin{table*}[htbp]
\squeezetable
\caption{The systematic error correlations for the inclusive jet cross
 section for $\modetajet < 0.5$, and $0.1 <\modetajet < 0.7$. The
 correlation values above the diagonal are the correlations for
 $\modetajet < 0.5$ and the correlations below the diagonal correspond
 to $0.1 <\modetajet < 0.7$.  In both cases the correlation matrices
 are symmetric.}  \label{inclusive_correlations}
 \begin{tabular}{rccccccccccccccccccccccccl} & 1 & 2 & 3 & 4 & 5 & 6 &
 7 & 8 & 9 & 10 & 11 & 12 & 13 & 14 & 15 & 16 & 17 & 18 & 19 & 20 & 21
 & 22 & 23 & 24 & \\ \hline\hline & 1.00& 1.00& 0.99& 0.98& 0.97&
 0.96& 0.95& 0.93& 0.92& 0.91& 0.90& 0.89& 0.88& 0.86& 0.86& 0.84&
 0.83& 0.82& 0.80& 0.78& 0.75& 0.71& 0.66& 0.53& 1 \\ & & 1.00& 1.00&
 0.99& 0.98& 0.97& 0.96& 0.95& 0.94& 0.93& 0.92& 0.91& 0.90& 0.89&
 0.88& 0.86& 0.86& 0.84& 0.82& 0.80& 0.77& 0.73& 0.68& 0.55& 2 \\ 1 &
 1.00 & & 1.00& 0.99& 0.99& 0.98& 0.97& 0.96& 0.95& 0.94& 0.93& 0.92&
 0.92& 0.91& 0.90& 0.88& 0.88& 0.86& 0.84& 0.82& 0.79& 0.75& 0.70&
 0.57& 3 \\ 2 & 1.00& 1.00 & & 1.00& 1.00& 0.99& 0.99& 0.98& 0.97&
 0.96& 0.96& 0.95& 0.94& 0.93& 0.92& 0.91& 0.90& 0.89& 0.86& 0.84&
 0.81& 0.77& 0.72& 0.59& 4 \\ 3 & 0.99& 1.00& 1.00 & & 1.00& 1.00&
 1.00& 0.99& 0.98& 0.97& 0.97& 0.96& 0.95& 0.94& 0.94& 0.92& 0.92&
 0.90& 0.88& 0.86& 0.83& 0.79& 0.74& 0.61& 5 \\ 4 & 0.98& 0.99& 0.99&
 1.00 & & 1.00& 1.00& 0.99& 0.99& 0.98& 0.98& 0.97& 0.96& 0.96& 0.95&
 0.94& 0.93& 0.92& 0.90& 0.88& 0.85& 0.81& 0.76& 0.63& 6 \\ 5 & 0.97&
 0.98& 0.99& 1.00& 1.00 & & 1.00& 0.99& 0.99& 0.99& 0.98& 0.98& 0.97&
 0.97& 0.96& 0.95& 0.94& 0.93& 0.91& 0.89& 0.87& 0.83& 0.78& 0.65& 7
 \\ 6 & 0.96& 0.97& 0.98& 1.00& 1.00& 1.00 & & 1.00& 1.00& 1.00& 0.99&
 0.99& 0.99& 0.98& 0.97& 0.97& 0.96& 0.95& 0.93& 0.91& 0.89& 0.85&
 0.80& 0.68& 8 \\ 7 & 0.95& 0.96& 0.98& 0.99& 1.00& 1.00& 1.00 & &
 1.00& 1.00& 1.00& 0.99& 0.99& 0.99& 0.98& 0.97& 0.97& 0.96& 0.94&
 0.93& 0.90& 0.87& 0.82& 0.70& 9 \\ 8 & 0.93& 0.95& 0.96& 0.98& 0.99&
 0.99& 0.99& 1.00 & & 1.00& 1.00& 1.00& 1.00& 0.99& 0.99& 0.98& 0.98&
 0.97& 0.95& 0.94& 0.92& 0.88& 0.84& 0.72& 10 \\ 9 & 0.92& 0.94& 0.95&
 0.97& 0.98& 0.99& 0.99& 1.00& 1.00 & & 1.00& 1.00& 1.00& 1.00& 0.99&
 0.99& 0.98& 0.98& 0.96& 0.95& 0.93& 0.90& 0.85& 0.74& 11 \\ 10 &
 0.91& 0.93& 0.95& 0.97& 0.98& 0.98& 0.99& 1.00& 1.00& 1.00 & & 1.00&
 1.00& 1.00& 1.00& 0.99& 0.99& 0.98& 0.97& 0.96& 0.94& 0.91& 0.87&
 0.76& 12 \\ 11 & 0.90& 0.92& 0.94& 0.96& 0.97& 0.98& 0.98& 0.99&
 1.00& 1.00& 1.00 & & 1.00& 1.00& 1.00& 1.00& 0.99& 0.99& 0.98& 0.97&
 0.95& 0.92& 0.88& 0.78& 13 \\ 12 & 0.89& 0.91& 0.93& 0.95& 0.96&
 0.97& 0.98& 0.99& 1.00& 1.00& 1.00& 1.00 & & 1.00& 1.00& 1.00& 1.00&
 0.99& 0.98& 0.97& 0.96& 0.93& 0.90& 0.80& 14 \\ 13 & 0.88& 0.90&
 0.92& 0.94& 0.95& 0.96& 0.97& 0.99& 0.99& 0.99& 1.00& 1.00& 1.00 & &
 1.00& 1.00& 1.00& 0.99& 0.99& 0.98& 0.97& 0.94& 0.91& 0.81& 15 \\ 14
 & 0.87& 0.89& 0.91& 0.93& 0.94& 0.96& 0.97& 0.98& 0.99& 0.99& 1.00&
 1.00& 1.00& 1.00 & & 1.00& 1.00& 1.00& 0.99& 0.99& 0.97& 0.95& 0.92&
 0.83& 16 \\ 15 & 0.85& 0.88& 0.90& 0.92& 0.94& 0.95& 0.96& 0.97&
 0.98& 0.99& 0.99& 1.00& 1.00& 1.00& 1.00 & & 1.00& 1.00& 1.00& 0.99&
 0.98& 0.96& 0.93& 0.85& 17 \\ 16 & 0.84& 0.86& 0.89& 0.91& 0.93&
 0.94& 0.95& 0.97& 0.98& 0.98& 0.99& 0.99& 1.00& 1.00& 1.00 & 1.00 & &
 1.00& 1.00& 1.00& 0.99& 0.97& 0.95& 0.87& 18 \\ 17 & 0.83& 0.85&
 0.87& 0.90& 0.91& 0.93& 0.94& 0.96& 0.97& 0.98& 0.98& 0.99& 0.99&
 1.00& 1.00 & 1.00& 1.00 & & 1.00& 1.00& 0.99& 0.98& 0.96& 0.89& 19 \\
 18 & 0.82& 0.84& 0.86& 0.89& 0.90& 0.92& 0.93& 0.95& 0.96& 0.97&
 0.98& 0.98& 0.99& 0.99& 1.00 & 1.00& 1.00& 1.00 & & 1.00& 1.00& 0.99&
 0.97& 0.91& 20 \\ 19 & 0.80& 0.82& 0.84& 0.86& 0.88& 0.90& 0.91&
 0.93& 0.94& 0.95& 0.96& 0.97& 0.98& 0.98& 0.99 & 0.99& 1.00& 1.00&
 1.00 & & 1.00& 1.00& 0.99& 0.94& 21 \\ 20 & 0.77& 0.79& 0.81& 0.84&
 0.86& 0.87& 0.89& 0.91& 0.92& 0.93& 0.95& 0.96& 0.96& 0.97& 0.98 &
 0.98& 0.99& 0.99& 1.00& 1.00 & & 1.00& 1.00& 0.96& 22 \\ 21 & 0.74&
 0.76& 0.78& 0.81& 0.83& 0.85& 0.86& 0.88& 0.90& 0.91& 0.93& 0.94&
 0.95& 0.96& 0.96 & 0.97& 0.98& 0.99& 0.99& 1.00& 1.00 & & 1.00& 0.98&
 23 \\ 22 & 0.71& 0.73& 0.75& 0.77& 0.79& 0.81& 0.83& 0.85& 0.87&
 0.88& 0.90& 0.91& 0.92& 0.93& 0.94 & 0.95& 0.96& 0.97& 0.98& 0.99&
 1.00& 1.00 & & 1.00& 24 \\ 23 & 0.65& 0.66& 0.68& 0.71& 0.73& 0.75&
 0.77& 0.79& 0.81& 0.83& 0.85& 0.86& 0.88& 0.89& 0.90 & 0.92& 0.93&
 0.94& 0.96& 0.97& 0.98& 0.99& 1.00& & \\ 24 & 0.52& 0.53& 0.55& 0.57&
 0.59& 0.61& 0.63& 0.66& 0.68& 0.70& 0.73& 0.75& 0.77& 0.79& 0.80 &
 0.82& 0.84& 0.86& 0.88& 0.91& 0.93& 0.95& 0.98& 1.00& \\
 \end{tabular}
\end{table*}

\subsection{Comparison of the Data to Theory }
\label{sec:inc_comparison}

\begin{figure}[htbp]
\begin{center}
\vbox{\centerline
{\psfig{figure=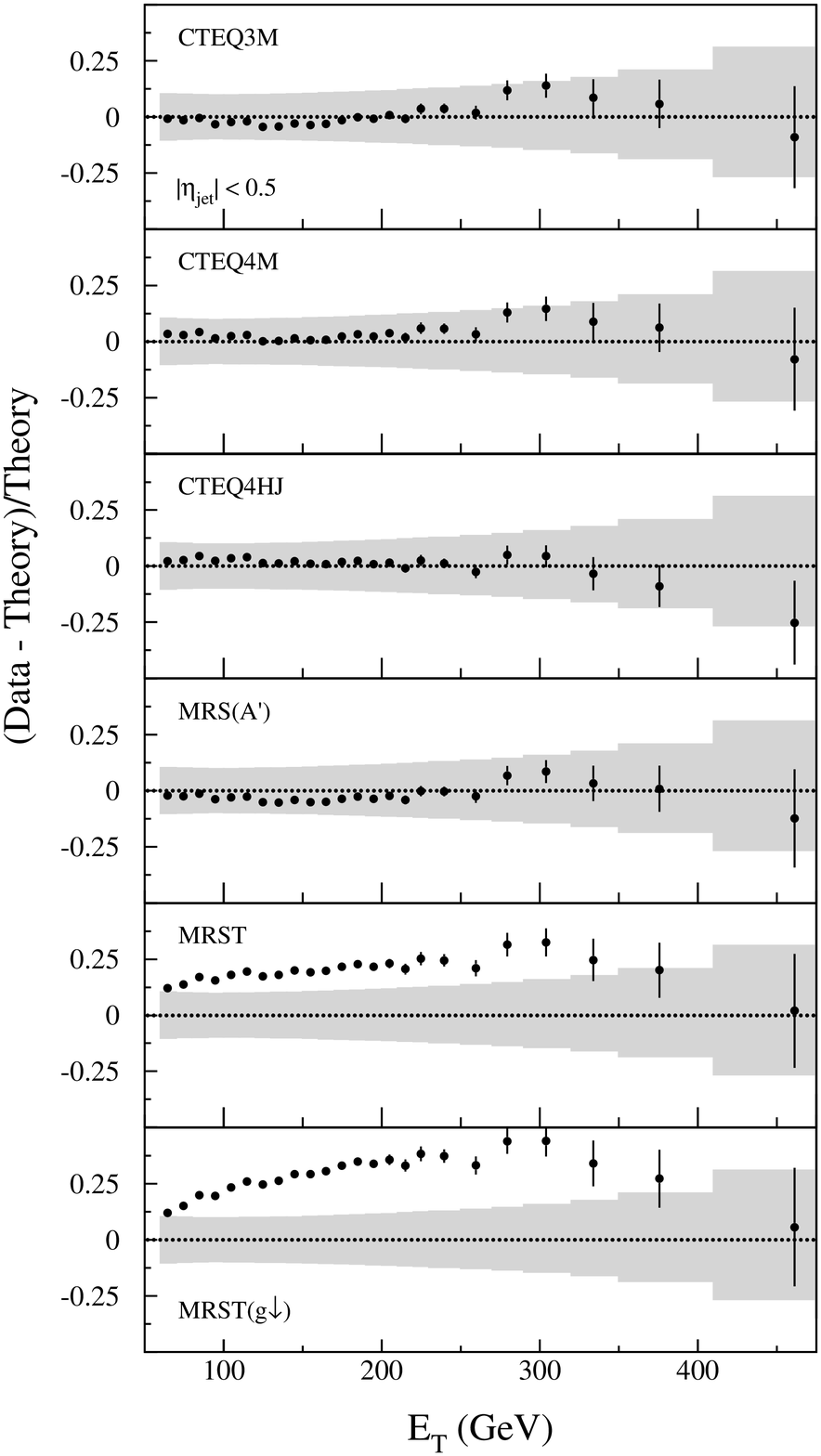,width=3.5in}}}
\caption{The difference between data and {\sc jetrad} QCD predictions
normalized to predictions for $\modeta < 0.5$.  The shaded region
represents the $\pm1\sigma$ systematic uncertainties about the
prediction.}
\label{data_cteq3m}
\end{center}
\end{figure}

\begin{figure}[hbtp]
\begin{center}
\vbox{\centerline
{\psfig{figure=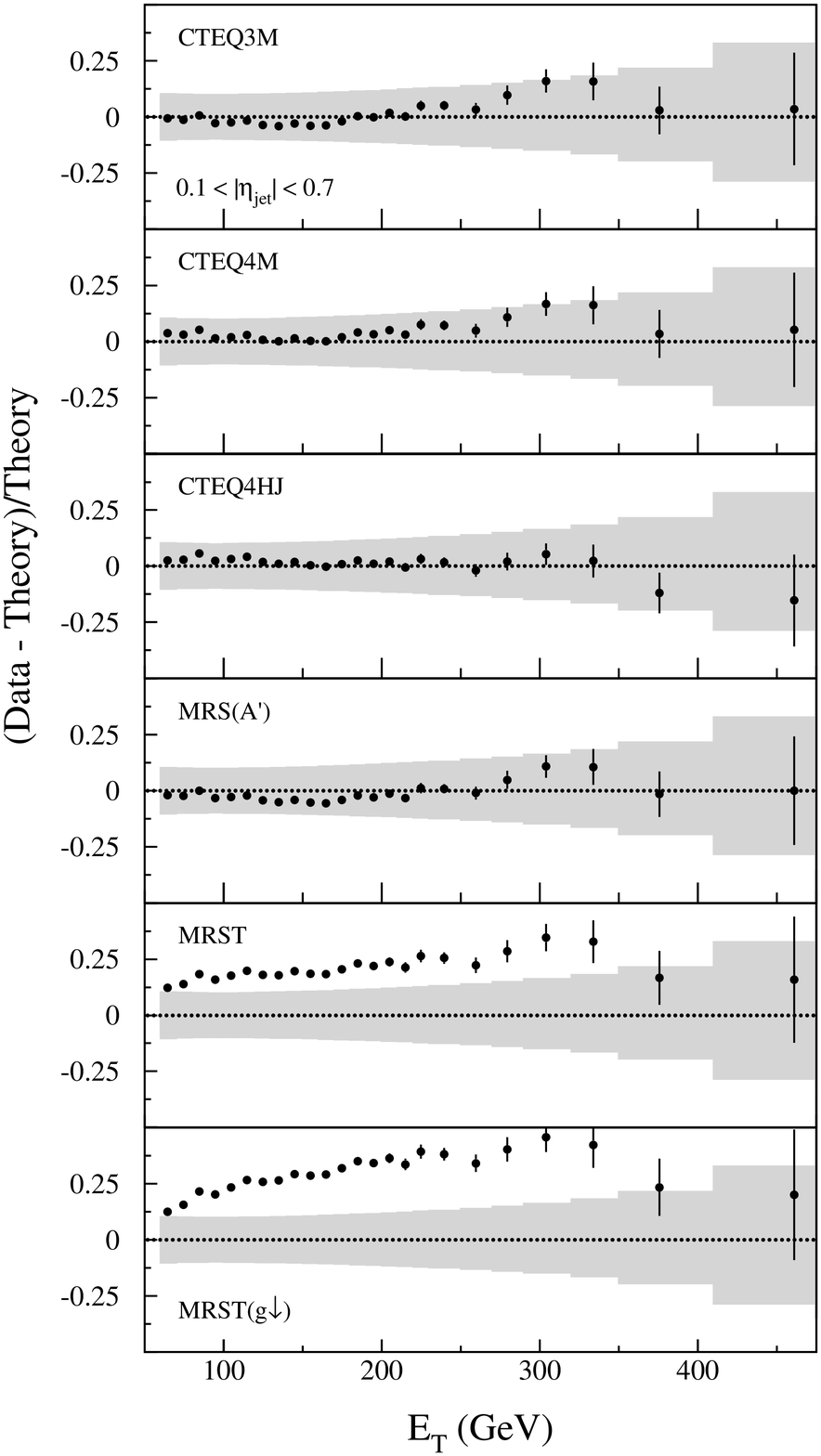,width=3.5in}}}
\caption{The difference between data and {\sc jetrad} QCD predictions
normalized to predictions for $0.1 < \modeta < 0.7$. The error bars
show the statistical uncertainties.  The shaded region represents the
$\pm1\sigma$ systematic uncertainties about the prediction. }
\label{data_cteq3m_07}
\end{center}
\end{figure}

 Figures~\ref{data_cteq3m} and \ref{data_cteq3m_07} show the
 fractional difference between the data, $D$, and a {\sc jetrad}
 theoretical prediction, $T$, normalized by the prediction,
 ($(D-T)/T$), for $\modeta < 0.5$ and $0.1 < \modeta < 0.7 $
 respectively. The {\sc jetrad} prediction was generated with
 $\mu=0.5\Etmax$, ${\cal{R}}_{\rm sep}=1.3$, and several different
 choices of PDF.  The error bars represent statistical errors only.
 The outer bands represent the total cross section error excluding the
 $5.8\%$ luminosity uncertainty. Given the experimental and
 theoretical uncertainties, the predictions are in agreement with the
 data; in particular, the data above $\Et = 350$~GeV show no
 indication of an excess relative to QCD.

 The data and theory can be compared quantitatively with a \chisq\
 test incorporating the uncertainty covariance matrix
 (Table~\ref{inclusive_correlations},~\cite{inclusive_matrix}). The
 \chisq\ is given by:
\begin{equation}
\chisq = \sum_{i,j} \delta_{i} V_{ij}^{-1} \delta_{j}
\end{equation}
 where $\displaystyle{\delta_{i}}$ is the difference between the data
 and theory for a given \Et\ bin, and $\displaystyle{V_{ij}}$ is
 element $i,j$ of the covariance matrix:
\begin{equation}
V_{ij} = \rho_{ij} \cdot \Delta \sigma_{i} \cdot \Delta \sigma_{j}.
\end{equation} 
 where $\displaystyle{\Delta \sigma}$ is the sum of the systematic
 error and the statistical error added in quadrature if $i=j$ and the
 systematic error if $i \neq j$, and $\displaystyle{\rho_{ij}}$ is the
 correlation between the systematic uncertainties of \Et\ bins as
 given in Table~\ref{inclusive_correlations}. The systematic
 uncertainty is given by the percentage uncertainty times the
 theoretical prediction (see Appendix for a discussion of the \chisq
 ).  The resulting \chisq\ values are given in
 Table~\ref{inclusive_chi_squares} for all of the theoretical choices
 described above. The choice of PDF and renormalization scale is
 varied.  Each comparison has 24 degrees of freedom.

 All but one of the {\sc jetrad} predictions adequately describe the
 $\modeta < 0.5$ and $0.1 < \modeta < 0.7$ cross sections. For these,
 the probabilities for $\chi^{2}$ to exceed the listed values are
 between $10\%$ and $86\%$. The prediction using \cteqfourhj\ and $\mu
 = 0.5 \Etmax$ produces the highest probability for both measurements.
 The prediction with the \mrstgd\ PDF has a probability of agreement
 with the data of $0.3\%$, and thus is incompatible with our data.

 Comparisons between the data and {\sc eks} calculations using various
 PDFs, ${\cal{R}}_{\rm{sep}}=1.3$, and with renormalization scales
 $\mu =$ (0.25, 0.50, 0.75, 1.00, 1.25, 1.50, 1.75, 2.00)$\Et$
 (where \Et\ = \Etmax\ and \Etjet ) are also made
 (Table~\ref{inclusive_chi_squares_eks}).  The {\sc eks} predictions
 give a reasonable description of the $\modeta < 0.5$ cross
 section. However, unlike the {\sc jetrad} predictions, the {\sc eks}
 calculation using \cteqfourm\ and $\mu = 0.5 \Etmax$ has the highest
 probability of agreement. The {\sc eks} predictions for $0.1 <
 \modeta < 0.7$ all give \chisq\ values with probabilities below
 $10\%$ for the choices examined.

\begin{table}
\begin{center}
\caption{ $\chi^{2}$ comparisons between {\sc jetrad} and $\modeta <
0.5 $ and $0.1 < \modeta < 0.7 $ data for $\mu = 0.5\Etmax$,
${\cal{R}}_{\rm{sep}}=1.3$, and various PDFs.  There are 24 degrees
of freedom.}
\begin{tabular}{lldddd}
  PDF        & $\displaystyle{\mu}$ 	  &
  \multicolumn{2}{c}{$\modeta \leq 0.5$}  &  
  \multicolumn{2}{c}{$0.1 \leq \modeta \leq 0.7$} \\
\cline{3-4}\cline{5-6}
	     &                          
& \chisq\ & \multicolumn{1}{c}{Prob.} & \chisq\ 
& \multicolumn{1}{c}{Prob.}
\\ \hline\hline

  \cteqthreem\ 	& $0.50\Etmax$   & 25.3 & 0.39  	&32.7   &  0.11  \\
  \cteqfourm\   & $0.50\Etmax$   & 20.1 & 0.69  	&26.8   &  0.31  \\
  \cteqfourhj\  & $0.50\Etmax$   & 16.8 & 0.86  	&22.4   &  0.56  \\
  \mrsap\     	& $0.50\Etmax$   & 20.4 & 0.67  	&28.5   &  0.24  \\
  \mrst\       	& $0.50\Etmax$   & 25.3 & 0.39  	&29.6   &  0.20  \\
 \mrstgu\    	& $0.50\Etmax$   & 21.6 & 0.60  	&30.1   &  0.18  \\
 \mrstgd\    	& $0.50\Etmax$   & 47.5 & 0.003  	&47.9   &  0.003 \\
\hline\hline				                   	                 
  \cteqthreem\  & $0.25\Etmax$      & 21.4 & 0.61  	&28.1   &  0.26  \\
  \cteqthreem\  & $0.50\Etmax$      & 25.3 & 0.39  	&32.7   &  0.11  \\
  \cteqthreem\  & $0.75\Etmax$      & 25.8 & 0.37  	&32.5   &  0.11  \\
  \cteqthreem\  & $1.00\Etmax$      & 24.8 & 0.42  	&31.7   &  0.14  \\
\end{tabular}

\label{inclusive_chi_squares}
\end{center}
\end{table}

\begin{table*}
\caption{ $\chi^{2}$ comparisons between {\sc eks} and the data for
$\modeta < 0.5 $ and $0.1 < \modeta < 0.7 $ with $\mu = D\Etmax$ or
$D\Etjet$, ${\cal{R}}_{\rm{sep}}=1.3$, and various PDFs.  There are
24 degrees of freedom.}
\begin{center}
\begin{tabular}{ccdddddddd}
 PDF & D where & \multicolumn{4}{c}{$\modeta < 0.5$}
 & \multicolumn{4}{c}{$0.1 < \modeta < 0.7$}\\
\cline{3-6}\cline{7-10}
     & $\mu =D E_{T}^{\rm xxx}$  & 
\multicolumn{2}{c}{$\Etjet$} & \multicolumn{2}{c}{$\Etmax$} & 
\multicolumn{2}{c}{$\Etjet$} & \multicolumn{2}{c}{$\Etmax$} \\
\cline{3-4}\cline{5-6}\cline{7-8}\cline{9-10}
     &         &  \chisq\ & \multicolumn{1}{c}{Prob.}
 &  \chisq\ & \multicolumn{1}{c}{Prob.} &
 \chisq\ & \multicolumn{1}{c}{Prob.}
 &  \chisq\ & \multicolumn{1}{c}{Prob.} \\   \hline\hline
\cteqthreem\ & 0.25 & 21.1 & 0.63 & 17.9 & 0.81 & 32.3 & 0.12 &  ---   &  --- \\
\cteqthreem\ & 0.50 & 20.7 & 0.66 & 19.3 & 0.74 & 33.7 & 0.09 & 33.3   & 0.10 \\
\cteqthreem\ & 0.75 & 20.4 & 0.67 & 19.4 & 0.73 & 33.3 & 0.10 & 33.0   & 0.10 \\
\cteqthreem\ & 1.00 & 20.2 & 0.68 & 19.5 & 0.73 & 32.9 & 0.11 & 32.7   & 0.11 \\
\cteqthreem\ & 1.25 & 20.4 & 0.68 & 19.8 & 0.71 & 32.8 & 0.11 & 32.8   & 0.11 \\
\cteqthreem\ & 1.50 & 20.8 & 0.65 & 20.3 & 0.68 & 33.1 & 0.10 & 33.1   & 0.10 \\
\cteqthreem\ & 1.75 & 21.5 & 0.61 & 21.2 & 0.63 & 33.5 & 0.09 & 33.6   & 0.09 \\
\cteqthreem\ & 2.00 & 22.4 & 0.55 & 22.1 & 0.57 & 34.2 & 0.08 & 34.3   & 0.08 \\
\hline\hline
\cteqfourm\   & 0.50 & 19.4 & 0.73 & 18.2 & 0.80 & 33.8 & 0.09 & 34.3 & 0.08 \\
\cteqfourhj\  & 0.50 & ---  & ---  & 23.3 & 0.50 & ---  & ---  & ---  &  --- \\
\cteqfoura 1  & 0.50 & ---  & ---  & 18.4 & 0.78 & ---  & ---  & ---  &  --- \\
\cteqfoura 2  & 0.50 & ---  & ---  & 18.3 & 0.79 & ---  & ---  & ---  &  --- \\
\cteqfoura 4  & 0.50 & ---  & ---  & 18.4 & 0.78 & ---  & ---  & ---  &  --- \\
\cteqfoura 5  & 0.50 & ---  & ---  & 19.2 & 0.74 & ---  & ---  & ---  &  --- \\
\mrsap\       & 0.50 & ---  & ---  & 19.3 & 0.74 & ---  & ---  & 36.8 & 0.05 \\
\end{tabular}
\label{inclusive_chi_squares_eks}
\end{center}
\end{table*}

\subsection{Comparison with Previously Published Results}

\begin{figure}[htbp]
\begin{center}
\vbox{\centerline
{\psfig{figure=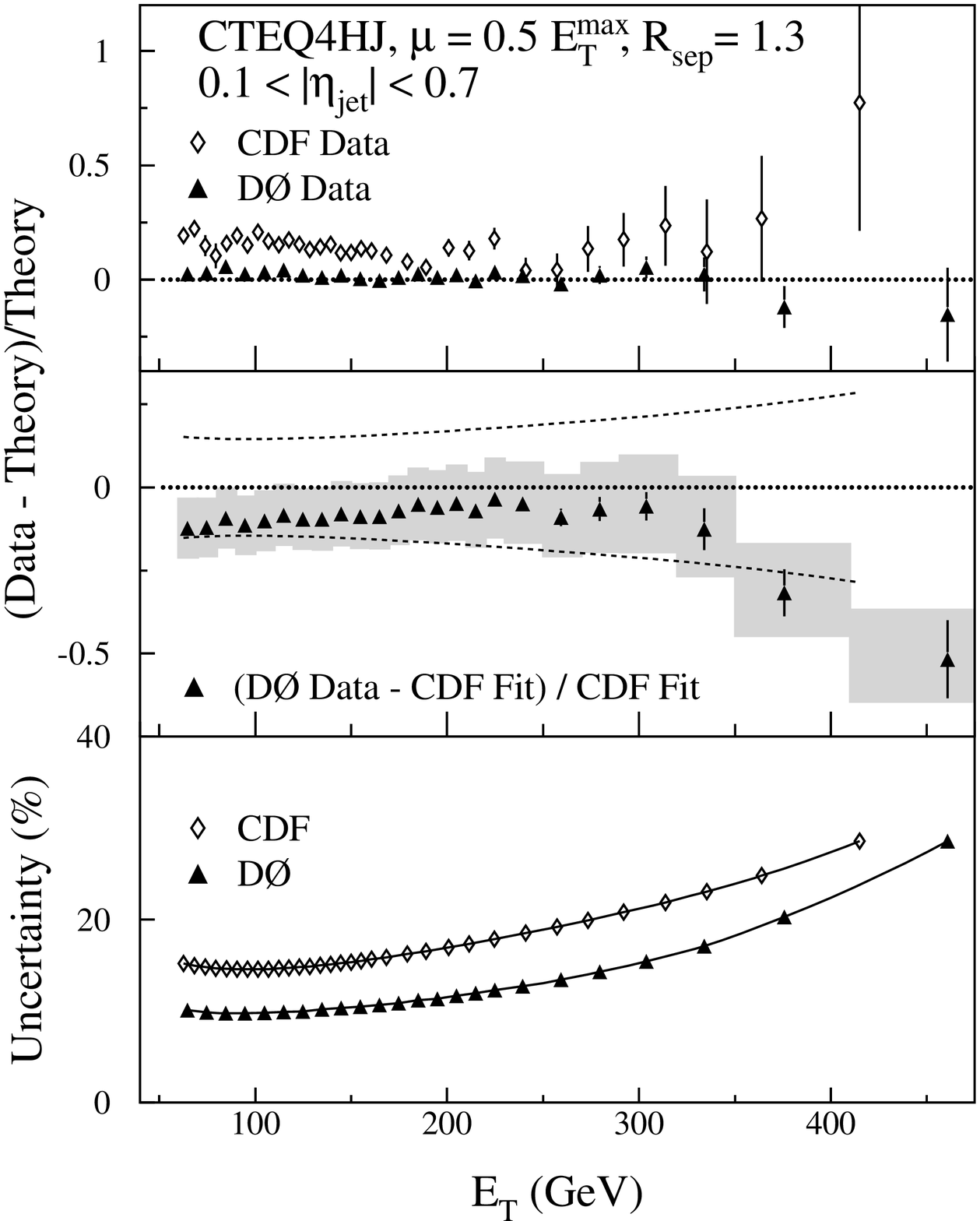,width=3.5in}}}
  \caption{Top: Normalized comparisons of our data and of the CDF data
  in Ref.~[7] to the {\sc jetrad} prediction using \cteqfourhj\ with
  $\mu = 0.5 \Etmax$.  Middle: Difference between our data and
  smoothed results of CDF normalized to the latter.  The shaded region
  represents the $\pm1\sigma$ systematic uncertainties about the D\O\
  data. The dashed curves show the $\pm1\sigma$ systematic
  uncertainties about the smoothed CDF data. Bottom: A comparison of
  the systematic uncertainties of the D\O\ measurement and the CDF
  measurement.}  \label{inclusive:Fig_cdf}
\end{center}
\end{figure}

 The top panel in Fig.~\ref{inclusive:Fig_cdf} shows $(D-T)/T$ for our
 data in the $0.1 \leq \modeta \leq 0.7$ region relative to a {\sc
 jetrad} calculation using the \cteqfourhj\ PDF, $\mu=0.5\Etmax$, and
 ${\cal{R}}_{\rm{sep}}=1.3$. Also shown are the previously published
 CDF data from the 1992--1993 Fermilab Tevatron running
 period~\cite{CDF_2} relative to the same {\sc jetrad} calculation.
 For this rapidity region, we have carried out a $\chi^{2}$ comparison
 between our data and the nominal curve describing the central values
 of the data of Ref.~\cite{CDF_2}.  Comparing our data to the nominal
 curve, as though it were theory, we obtain a \chisq\ of 56.5 for 24
 degrees of freedom (probability of $0.02\%$).  Thus our data cannot
 be described with this parameterization.  As illustrated in the
 middle panel of Fig.~\ref{inclusive:Fig_cdf}, our data and the curve
 differ at low and high \Et ; such differences cannot be accommodated
 by the highly correlated uncertainties of our data.  If we include
 the systematic uncertainties of the data of Ref.~\cite{CDF_2} in the
 covariance matrix, the $\chi^{2}$ is reduced to 30.8 (probability of
 $16\%$), representing acceptable agreement.

\subsection{Rapidity Dependence of the Inclusive Jet Cross Section}

 D\O\ has subsequently extended the measurement of the inclusive jet
 cross section as a function of \Et\ to $\modeta <3$ in several bins
 of pseudorapidity~\cite{levan}. In this analysis the details of the
 jet energy scale corrections, single jet resolutions, and vertex
 selection were updated to minimize uncertainties for jets at large
 pseudorapidity ($\modeta > 1.5$). These cross sections are compared
 with {\sc jetrad} predictions generated with $\mu=0.5\Etmax$,
 ${\cal{R}}_{\rm sep}=1.3$, and similar choices of PDF given in
 Table~\ref{inclusive_chi_squares}. The data and theory were also
 compared using the same \chisq\ test as used in this paper
 (Section~\ref{sec:inc_comparison}). The data indicate an preference
 for the \cteqfourhj , \mrstgu , and \cteqfourm\ PDFs~\cite{levan}.

\subsection{Conclusions}

 We have made the most precise measurement to date of the inclusive
 jet cross section for $\Et \geq 60$~GeV at $\sqrt{s}=1800$~GeV. No
 excess production of high-\Et\ jets is observed. QCD predictions are
 in good agreement with the observed cross section for most standard
 parton distribution functions and different renormalization scales
 ($\mu =$ 0.25--2.00$\Et$ where \Et\ = \Etmax\ or \Etjet ).
\clearpage

\section[Ratio of Inclusive Jet Cross Sections at 
$\protect\bbox{\sqrt{s}}$~=~1800 and 630 GeV] {Ratio of Inclusive Jet
Cross Sections at $\protect\bbox{\sqrt{\lowercase{s}}}$~=~1800 and 630
G\lowercase{e}V}
\label{sec:inclusive_jet_ratio}

\subsection[Inclusive Jet Cross Section at ${\sqrt{s} =}$ 630 GeV]
{Inclusive Jet Cross Section at $\bbox{\sqrt{s} =}$ 630 GeV}

 The inclusive jet cross section for $\modeta < 0.5$ at $\sqrt{s}=630$
 GeV consists of data collected with three triggers: Jet\_12,
 Jet\_2\_12, and Jet\_30. To form the inclusive jet cross section, an
 \Et\ region of each trigger is selected to maximize statistical power
 while maintaining full trigger efficiency. Any given cross section
 bin receives contributions from one and only one trigger. The
 luminosity in any given bin is the luminosity exposure for that
 trigger (given in Table~\ref{TABLE:luminosities}).
 
 The inclusive jet cross section at $\sqrt{s}=1800$ GeV was determined
 prior to the 630 GeV analysis.  To facilitate the ratio calculation
 as a function of $x_T \equiv {2\Et}/{\sqrt{s}}$, the bin boundaries
 for the 630 GeV analysis were selected such that
\begin{equation} 
 E_{T}^{630}=\frac{630}{1800}\cdot E_{T}^{1800}, \label{eq_bin_select} 
\end{equation} 
 i.e., such that the bin edges match in $x_{T}$--space.  Most of the
 resulting bins are 3.5 GeV wide, but some bins have a width of
 7.0~GeV, 10.5~GeV, or more.

\begin{figure}[htbp]
\begin{center}
\vbox{\centerline
{\psfig{figure=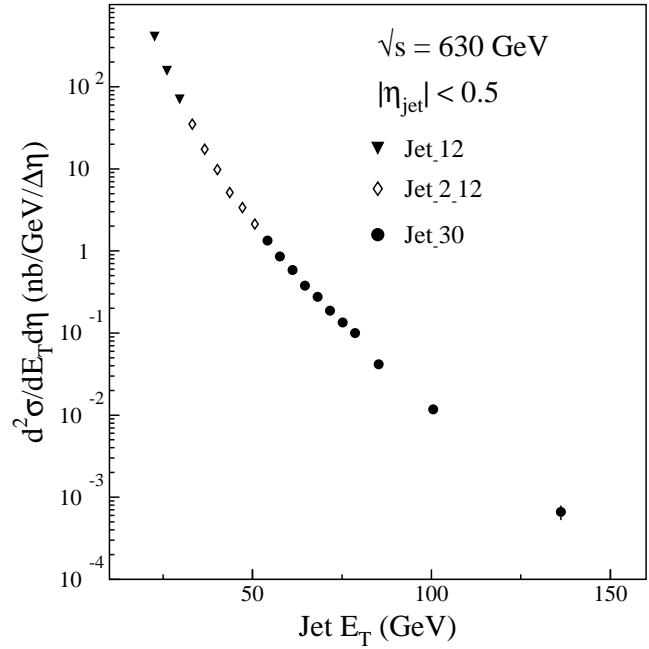,width=3.5in}}}
  \caption{The observed inclusive jet cross section for
  $\protect\sqrt{s}=630$ GeV. Symbols indicate the three jet triggers
  (shaded triangles: Jet\_12; hollow diamonds: Jet\_2\_12; shaded
  circles: Jet\_30).}
\label{raw_xs}
\end{center}
\end{figure}

 Figure \ref{raw_xs} displays the observed cross section at
 $\sqrt{s}=630$ GeV. The three different symbols indicate the \Et\
 region for each jet trigger. Vertical lines (mostly hidden by the
 symbols) indicate the statistical uncertainty on each point.

\begin{figure}[htbp]
\begin{center}
\vbox{\centerline
{\psfig{figure=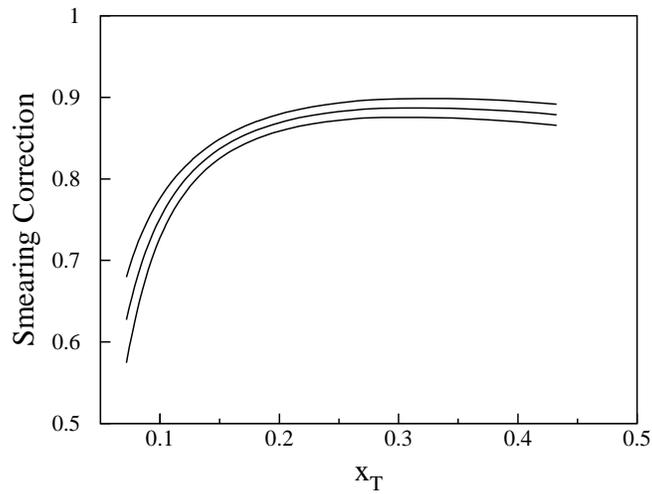,width=3.5in}}}
  \caption{ The nominal unsmearing correction at $\sqrt{s}= 630$ GeV
  is given by the central line. The outer curves depict the
  uncertainty in the unsmearing due to the uncertainties in the
  measurement of the resolution of the jet energy. }
\label{corr_fac}
\end{center}
\end{figure}

 The cross section is corrected for the effects of jet resolution
 using the same method as used for the $\sqrt{s}=1800$~GeV cross
 section (Section~\ref{sec:1800_unfolding}). The single jet
 resolutions at $\sqrt{s}=630$~GeV are given in
 Section~\ref{sec:630_resolutions}.  The resulting ansatz fit
 parameters are given in Table~\ref{t_corr_fac} and the unsmearing
 correction is plotted in Fig.~\ref{corr_fac}.

\begin{table*}[htbp] 
\centering
\caption{Unsmearing ansatz function parameters (see
Eq.~\ref{1800_ansatz}) for the inclusive jet cross section for
$\modeta < 0.5$ (in nb) at $\sqrt{s}=630$ and 1800~ GeV and their
uncertainties.}
\label{t_corr_fac}
\begin{tabular}{cccrrr}
$\sqrt{s}$ (GeV) & Parameter & Value &  \multicolumn{3}{c}{Error Matrix} \\
 \hline\hline
	  & $A$ & $23.43$ & $ 4.48825\times 10^{-3}$ & $-1.15352\times
10^{-3}$ & $-6.70083\times 10^{-3}$\\ 1800 & $\alpha$ & $-5.04$ &
$-1.15352\times 10^{-3}$ & $ 2.98882\times 10^{-4}$ & $ 1.79044\times
10^{-3}$\\ & $\beta$ & $8.23$ & $-6.70083\times 10^{-3}$ & $
1.79044\times 10^{-3}$ & $ 1.21005\times 10^{-2}$ \\ \hline\hline &
$A$ & $22.7$ & $ 2.28649\times 10^{-2}$ & $-8.62822\times 10^{-3}$ &
$-5.76500\times 10^{-2}$ \\ 630 & $\alpha$ & $-5.33$ & $-8.62822\times
10^{-3}$ & $ 3.28592\times 10^{-3}$ & $ 2.24650\times 10^{-2}$ \\ &
$\beta$ & $6.58$ & $-5.76500\times 10^{-2}$ & $ 2.24650\times 10^{-2}$
& $ 0.16449\times 10^{-1}$ \\
\end{tabular}
\end{table*}

\begin{figure}[htbp]
\begin{center}
\vbox{\centerline
{\psfig{figure=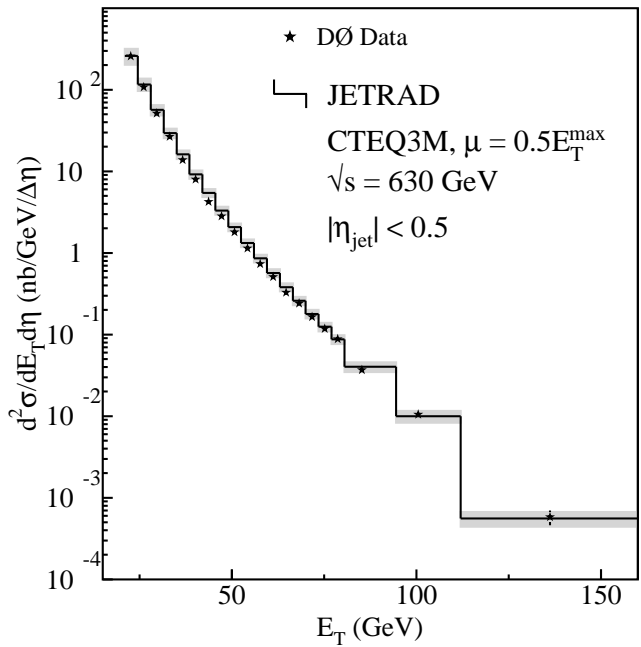,width=3.5in}}}
 \caption{The $ \modeta < 0.5 $ inclusive cross section at $\sqrt{s} =
 630$~GeV.  Statistical uncertainties are not visible on this scale
 (except for the last point).  The histogram represents the {\sc
 jetrad} prediction and the shaded band represents the $\pm 1\sigma$
 systematic uncertainty band about the prediction.  } \label{xsec_630}
\end{center}
\end{figure}

\begin{figure}[htbp]
\begin{center}
\vbox{\centerline
{\psfig{figure=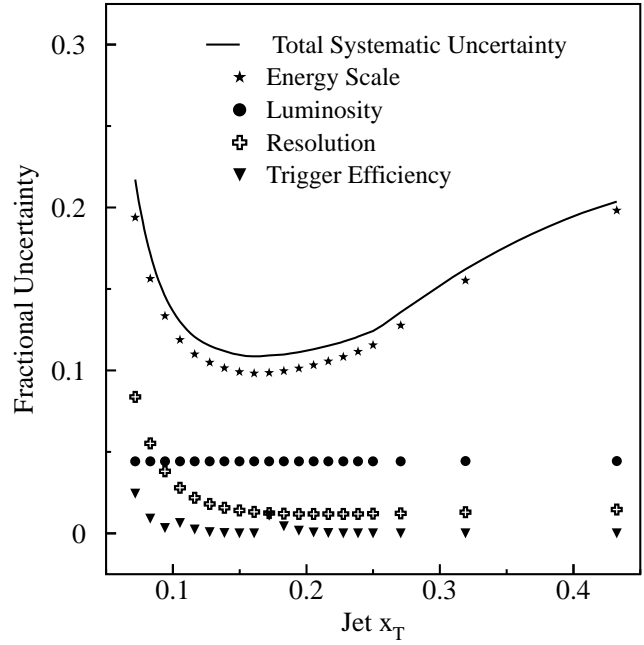,width=3.5in}}}
\caption{Contributions to the $\modeta < 0.5$ at $\sqrt{s} = 630$~GeV
cross section uncertainty plotted by component.}
\label{error_components_630_xsec}
\end{center}
\end{figure}

\begin{figure}[htbp]
\vbox{\centerline
{\psfig{figure=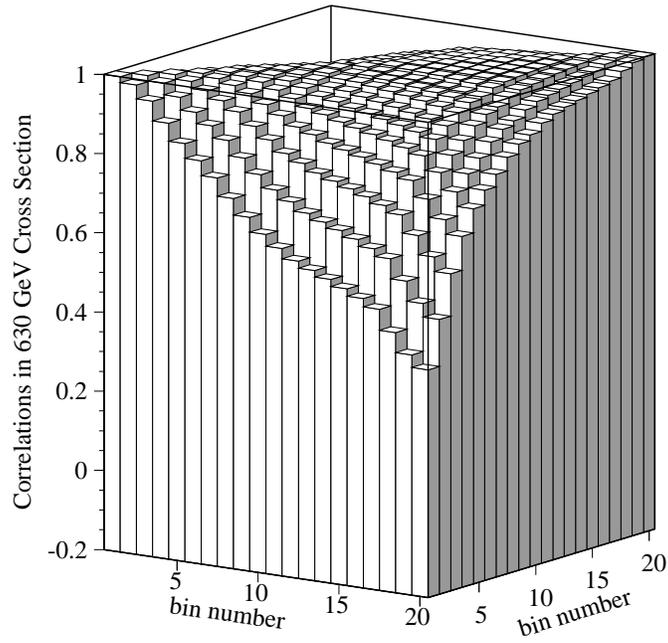,width=3.5in}}}
 \caption{The correlations for the total systematic uncertainty for
 the inclusive jet cross section for $\modeta < 0.5$ at $\sqrt{s} =
 630$~GeV (Table~\ref{response_correlations_630}).}
\label{630_correl_fig}
\end{figure}

 The resulting inclusive jet cross section at $\sqrt{s}=630$~GeV is
 given in Table~\ref{t_xsec_values} and is plotted in
 Fig.~\ref{xsec_630}. The uncertainties in the cross section are given
 in Table~\ref{630_cross_section_errors} and are also plotted in
 Fig.~\ref{error_components_630_xsec}.  The bin-to-bin correlations of
 the uncertainties are shown in Fig.~\ref{630_correl_fig} and are
 given in Table~\ref{response_correlations_630}.

 The magnitude of the energy scale uncertainties are larger for the
 cross section at $\sqrt{s}=630$~GeV than at $\sqrt{s}=1800$~GeV
 (Table~\ref{1800_cross_section_errors_a}). This is caused by several
 factors.  The cross section at 630 GeV begins with jet $\Et > 20$~GeV
 compared with $60$~GeV at $\sqrt{s}=1800$~GeV. The uncertainty in the
 energy scale offset correction (which is additive) has a much larger
 effect on 20~GeV jets than on 60~GeV jets. For $\Et > 60$~GeV the
 cross section at $\sqrt{s}=630$~GeV is much steeper than the cross
 section at 1800 GeV, hence the same uncertainty in the energy scale
 will lead to a larger uncertainty in the cross section.

\begin{figure}[htbp]
\begin{center}
\vbox{\centerline{\psfig{figure=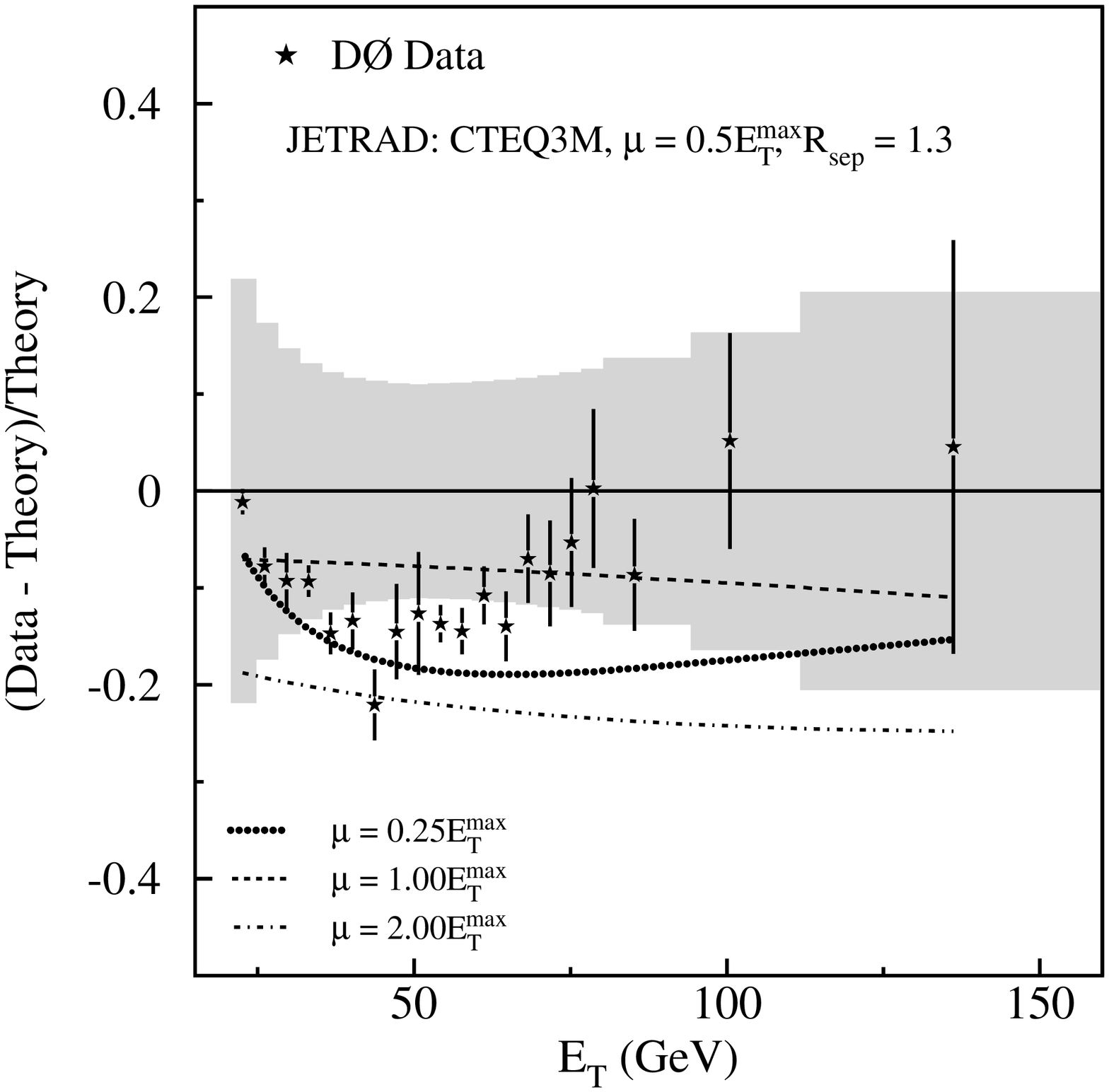,width=3.5in}}}
  \caption{The difference between the data and the prediction ({\sc
  jetrad}), divided by the prediction for $\modeta < 0.5$ at $\sqrt{s}
  = 630$~GeV. The solid stars represent the comparison to the
  calculation using \cteqthreem\ with $\mu = 0.5 \Etmax$. The shaded
  region represents the $\pm1\sigma$ systematic uncertainties about
  the prediction. The effects of changing the renormalization scale
  are also shown (each curve shows the difference between the
  alternative prediction and the standard prediction).}
\label{dtt_4hj_et2}
\end{center}
\end{figure}

\begin{figure}[htbp]
\begin{center}
\vbox{\centerline
{\psfig{figure=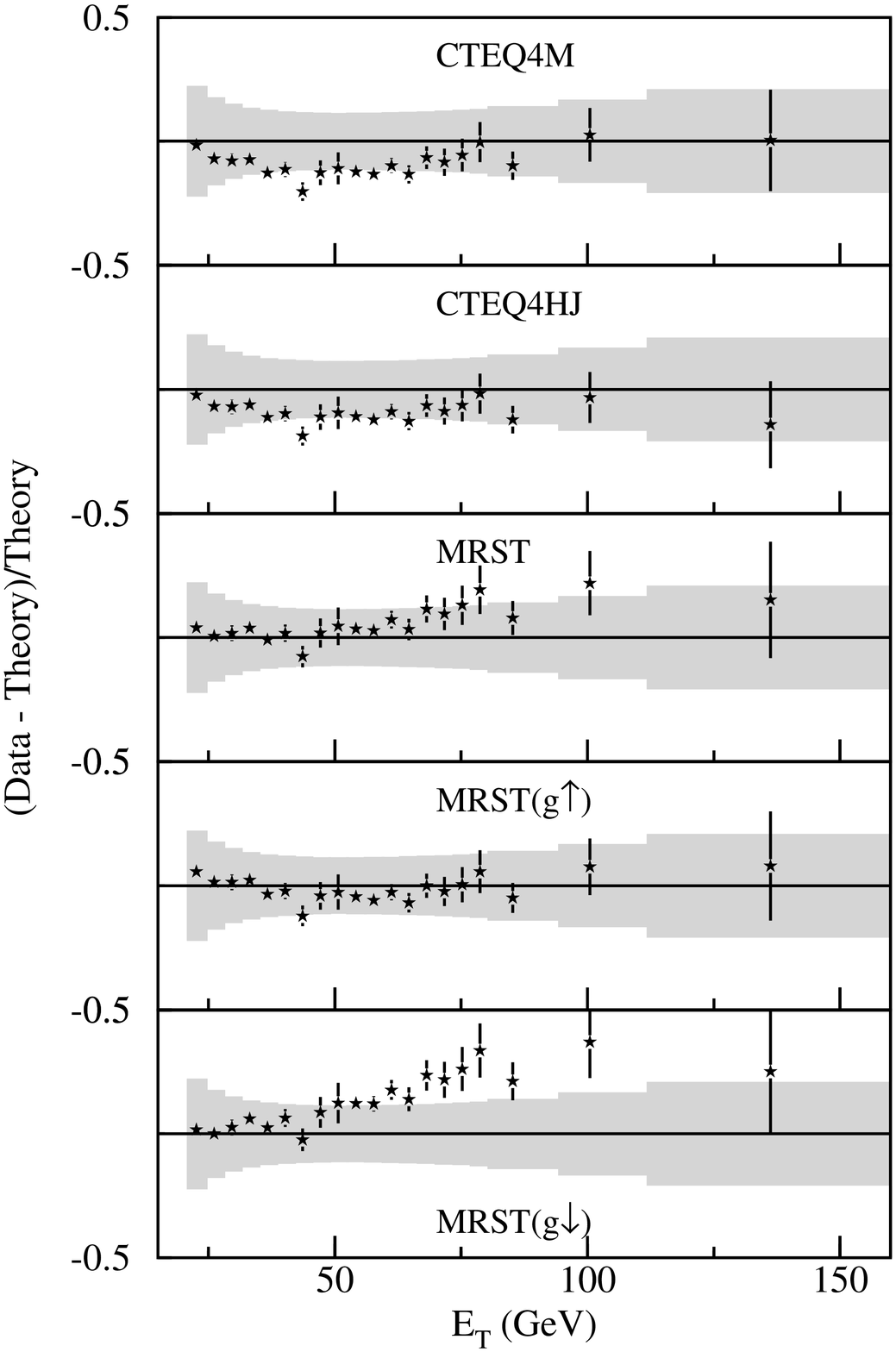,width=3.5in}}}
  \caption{The difference between the data and the prediction ({\sc
  jetrad}), divided by the prediction for $\modeta < 0.5$. The solid
  stars represent the comparison to the calculation using $\mu = 0.5
  \Etmax$ and the PDFs \cteqfourm , \cteqfourhj , \mrst , \mrstgu , and
  \mrstgd . The shaded region represents the $\pm1\sigma$ systematic
  uncertainties about the prediction.}
\label{dtt_4hj_et2_2}
\end{center}
\end{figure}

 Figures~\ref{dtt_4hj_et2} and \ref{dtt_4hj_et2_2} show the fractional
 differences between the data and several {\sc jetrad} predictions
 using different choices of renormalization scale and PDF. These NLO
 QCD predictions are in reasonable agreement with the data.  The data
 and predictions are compared quantitatively using a \chisq\ test
 (Section~\ref{sec:inc_comparison}). The resulting \chisq\ values are
 given in Table~\ref{t_dtt_630}; each comparison has 20 degrees of
 freedom. All but two of the {\sc jetrad} predictions adequately
 describe the cross section at $\sqrt{s} = 630$~GeV. For these, the
 probabilities for $\chi^{2}$ to exceed the listed values are between
 $6.4\%$ and $78\%$. The prediction using \mrstgu\ and $\mu = 0.5
 \Etmax$ produces the highest probability. The predictions using
 \mrstgd\ with $\mu = 0.5 \Etmax$, and \cteqthreem\ with $\mu =
 2\Etmax$, thus are inconsistent with our measurements with
 probabilities $\le 0.4\%$.

\begin{table*}[htbp] 
\caption{Inclusive jet cross section for $\modeta < 0.5$ at 
$\sqrt{s} = 630$~GeV.\label{t_xsec_values}}
\begin{tabular}{d@{--}dddcd}
 \multicolumn{2}{c}{Bin Range} & \multicolumn{1}{c}{Plotted} &
  \multicolumn{1}{c}{Plotted} & \multicolumn{1}{c}{Cross Section} &
  \multicolumn{1}{c}{Systematic} \\ \multicolumn{2}{c}{\Et } &
  \multicolumn{1}{c}{\Et } & \multicolumn{1}{c}{$x_T$} &
  \multicolumn{1}{c}{$\pm$ statistical error} &
  \multicolumn{1}{c}{Uncertainty} \\ \multicolumn{2}{c}{(GeV)} &
  \multicolumn{1}{c}{(GeV) } & & \multicolumn{1}{c}{(nb)} &
  \multicolumn{1}{c}{($\%$)}\\
 \hline\hline
 21.0 & 24.5 &   22.6 &   0.07 & $(  2.56 \pm  0.03)\times 10^{ 2}$ &21.7  \\
 24.5 & 28.0 &   26.1 &   0.08 & $(  1.07 \pm  0.02)\times 10^{ 2}$ &17.2  \\
 28.0 & 31.5 &   29.6 &   0.09 & $(  5.14 \pm  0.16)\times 10^{ 1}$ &14.6  \\
 31.5 & 35.0 &   33.1 &   0.11 & $(  2.67 \pm  0.05)\times 10^{ 1}$ &13.0  \\
 35.0 & 38.5 &   36.7 &   0.12 & $(  1.37 \pm  0.04)\times 10^{ 1}$ &12.1  \\
 38.5 & 42.0 &   40.2 &   0.13 & $(  7.96 \pm  0.27)\times 10^{ 0}$ &11.5  \\
 42.0 & 45.5 &   43.7 &   0.14 & $(  4.24 \pm  0.20)\times 10^{ 0}$ &11.2  \\
 45.5 & 49.0 &   47.2 &   0.15 & $(  2.83 \pm  0.16)\times 10^{ 0}$ &11.0  \\
 49.0 & 52.5 &   50.7 &   0.16 & $(  1.81 \pm  0.13)\times 10^{ 0}$ &10.9  \\
 52.5 & 56.0 &   54.2 &   0.17 & $(  1.14 \pm  0.03)\times 10^{ 0}$ &10.9  \\
 56.0 & 59.5 &   57.7 &   0.18 & $(  7.35 \pm  0.21)\times 10^{-1}$ &11.0  \\
 59.5 & 63.0 &   61.2 &   0.19 & $(  5.07 \pm  0.17)\times 10^{-1}$ &11.1  \\
 63.0 & 66.5 &   64.7 &   0.21 & $(  3.29 \pm  0.14)\times 10^{-1}$ &11.3  \\
 66.5 & 70.0 &   68.2 &   0.22 & $(  2.42 \pm  0.12)\times 10^{-1}$ &11.5  \\
 70.0 & 73.5 &   71.7 &   0.23 & $(  1.64 \pm  0.10)\times 10^{-1}$ &11.8  \\
 73.5 & 77.0 &   75.2 &   0.24 & $(  1.18 \pm  0.08)\times 10^{-1}$ &12.1  \\
 77.0 & 80.5 &   78.7 &   0.25 & $(  8.79 \pm  0.72)\times 10^{-2}$ &12.4  \\
 80.5 & 94.5 &   85.2 &   0.27 & $(  3.69 \pm  0.23)\times 10^{-2}$ &13.6  \\
 94.5 &112.0 &  100.5 &   0.32 & $(  1.05 \pm  0.11)\times 10^{-2}$ &16.2  \\
112.0 &196.0 &  136.2 &   0.43 & $(  5.81 \pm  1.19)\times 10^{-4}$ &20.4  \\
\end{tabular}
\end{table*}

\begin{table*}[htbp]
\squeezetable
\caption{Percentage cross section uncertainties for $\modeta < 0.5$ at
$\sqrt{s} = 630$~GeV.  The last row gives the nature of the bin-to-bin
$x_T$ correlations: 0 signifies uncorrelated uncertainties; 1
correlated; and p partially correlated.  }
\label{630_cross_section_errors}
\begin{tabular}{dddddddd}
$x_T$ & Statistical & Jet Selection  & Luminosity  &
 Trigger & Unsmearing & Energy Scale & Total \\
\hline\hline
    0.07 &    1.3 &    0.2 &    4.4 &    2.4 &    8.4 &   19.4 &   21.7 \\
    0.08 &    2.1 &    0.2 &    4.4 &    0.9 &    5.5 &   15.6 &   17.2 \\
    0.09 &    3.2 &    0.2 &    4.4 &    0.3 &    3.8 &   13.3 &   14.6 \\
    0.11 &    1.8 &    0.2 &    4.4 &    0.6 &    2.8 &   11.9 &   13.0 \\
    0.12 &    2.5 &    0.2 &    4.4 &    0.2 &    2.2 &   11.0 &   12.1 \\
    0.13 &    3.4 &    0.2 &    4.4 &    0.1 &    1.8 &   10.5 &   11.5 \\
    0.14 &    4.7 &    0.1 &    4.4 &    0.0 &    1.6 &   10.2 &   11.2 \\
    0.15 &    5.8 &    0.1 &    4.4 &    0.0 &    1.4 &    9.9 &   11.0 \\
    0.16 &    7.3 &    0.1 &    4.4 &    0.0 &    1.3 &    9.8 &   10.9 \\
    0.17 &    2.2 &    0.1 &    4.4 &    1.1 &    1.3 &    9.9 &   10.9 \\
    0.18 &    2.8 &    0.1 &    4.4 &    0.4 &    1.2 &   10.0 &   11.0 \\
    0.19 &    3.4 &    0.1 &    4.4 &    0.2 &    1.2 &   10.1 &   11.1 \\
    0.21 &    4.2 &    0.2 &    4.4 &    0.1 &    1.2 &   10.3 &   11.3 \\
    0.22 &    4.9 &    0.2 &    4.4 &    0.0 &    1.2 &   10.6 &   11.5 \\
    0.23 &    6.0 &    0.2 &    4.4 &    0.0 &    1.2 &   10.8 &   11.8 \\
    0.24 &    7.0 &    0.2 &    4.4 &    0.0 &    1.2 &   11.2 &   12.1 \\
    0.25 &    8.2 &    0.2 &    4.4 &    0.0 &    1.2 &   11.6 &   12.4 \\
    0.27 &    6.3 &    0.2 &    4.4 &    0.0 &    1.2 &   12.8 &   13.6 \\
    0.32 &   10.6 &    0.3 &    4.4 &    0.0 &    1.3 &   15.5 &   16.2 \\
    0.43 &   20.4 &    0.3 &    4.4 &    0.0 &    1.5 &   19.8 &   20.4 \\
\hline 
Correlation & 0 & 0 & 1 & 0 & 1 & p & p \\
\end{tabular}
\end{table*}

\begin{table*}[htbp]
\squeezetable
\caption{The correlations for the total systematic uncertainty for the
inclusive jet cross section for $\modeta < 0.5$ at $\sqrt{s} =
630$~GeV and the ratio of cross sections. The correlation values above
the diagonal are for the cross section at $\sqrt{s} = 630$~GeV and the
values below the diagonal correspond to the ratio of cross sections.
In both cases the correlation matrices are symmetric.}
\label{response_correlations_630}
\begin{tabular}{rcccccccccccccccccccccl}
   & & 1.00& 0.99& 0.96& 0.91& 0.86& 0.82& 0.79& 0.74& 0.69& 0.66&
   0.62& 0.60& 0.58& 0.56& 0.54& 0.52& 0.50& 0.44& 0.39& 0.36& 1\\ & &
   & 1.00& 0.99& 0.96& 0.93& 0.90& 0.87& 0.83& 0.79& 0.76& 0.73& 0.70&
   0.69& 0.67& 0.65& 0.63& 0.61& 0.56& 0.51& 0.48& 2\\ 1 & 1.00& & &
   1.00& 0.99& 0.97& 0.95& 0.93& 0.90& 0.87& 0.84& 0.82& 0.79& 0.78&
   0.77& 0.75& 0.73& 0.72& 0.67& 0.62& 0.59& 3\\ 2 & 0.98& 1.00& & &
   1.00& 0.99& 0.98& 0.97& 0.94& 0.92& 0.90& 0.88& 0.86& 0.85& 0.84&
   0.83& 0.81& 0.80& 0.76& 0.71& 0.68& 4\\ 3 & 0.95& 0.99& 1.00& & &
   1.00& 1.00& 0.99& 0.97& 0.96& 0.94& 0.93& 0.91& 0.90& 0.89& 0.88&
   0.87& 0.85& 0.82& 0.77& 0.74& 5\\ 4 & 0.90& 0.95& 0.98& 1.00& & &
   1.00& 1.00& 0.99& 0.97& 0.96& 0.95& 0.94& 0.93& 0.92& 0.91& 0.90&
   0.89& 0.85& 0.81& 0.78& 6\\ 5 & 0.84& 0.91& 0.95& 0.99& 1.00& & &
   1.00& 0.99& 0.99& 0.98& 0.97& 0.96& 0.95& 0.94& 0.94& 0.93& 0.92&
   0.88& 0.85& 0.82& 7\\ 6 & 0.79& 0.87& 0.93& 0.97& 0.99& 1.00& & &
   1.00& 1.00& 0.99& 0.98& 0.98& 0.97& 0.97& 0.96& 0.95& 0.94& 0.91&
   0.88& 0.85& 8\\ 7 & 0.74& 0.83& 0.89& 0.95& 0.98& 0.99& 1.00& & &
   1.00& 1.00& 0.99& 0.99& 0.98& 0.98& 0.97& 0.97& 0.96& 0.94& 0.91&
   0.88& 9\\ 8 & 0.68& 0.77& 0.84& 0.91& 0.94& 0.96& 0.97& 1.00& & &
   1.00& 1.00& 0.99& 0.99& 0.99& 0.98& 0.98& 0.97& 0.95& 0.93&
   0.90&10\\ 9 & 0.62& 0.72& 0.80& 0.87& 0.91& 0.94& 0.96& 0.99& 1.00&
   & & 1.00& 1.00& 1.00& 0.99& 0.99& 0.99& 0.98& 0.97& 0.94& 0.92&11\\
   10 & 0.57& 0.67& 0.76& 0.84& 0.89& 0.92& 0.95& 0.98& 0.99& 1.00& &
   & 1.00& 1.00& 1.00& 0.99& 0.99& 0.99& 0.97& 0.95& 0.93&12\\ 11 &
   0.52& 0.63& 0.72& 0.80& 0.86& 0.90& 0.93& 0.97& 0.98& 0.99& 1.00& &
   & 1.00& 1.00& 1.00& 0.99& 0.99& 0.98& 0.96& 0.94&13\\ 12 & 0.47&
   0.58& 0.68& 0.76& 0.83& 0.87& 0.90& 0.95& 0.97& 0.98& 0.99& 1.00& &
   & 1.00& 1.00& 1.00& 0.99& 0.98& 0.97& 0.95&14\\ 13 & 0.43& 0.54&
   0.64& 0.73& 0.80& 0.84& 0.88& 0.93& 0.95& 0.97& 0.99& 0.99& 1.00& &
   & 1.00& 1.00& 1.00& 0.99& 0.97& 0.95&15\\ 14 & 0.39& 0.50& 0.60&
   0.70& 0.76& 0.82& 0.85& 0.91& 0.94& 0.96& 0.98& 0.99& 0.99& 1.00& &
   & 1.00& 1.00& 0.99& 0.98& 0.96&16\\ 15 & 0.34& 0.46& 0.56& 0.66&
   0.73& 0.78& 0.83& 0.88& 0.91& 0.94& 0.96& 0.98& 0.99& 0.99& 1.00& &
   & 1.00& 0.99& 0.98& 0.97&17\\ 16 & 0.30& 0.41& 0.52& 0.62& 0.69&
   0.75& 0.79& 0.85& 0.89& 0.92& 0.95& 0.97& 0.98& 0.99& 0.99& 1.00& &
   & 1.00& 0.99& 0.98&18\\ 17 & 0.25& 0.37& 0.47& 0.57& 0.65& 0.71&
   0.76& 0.82& 0.86& 0.89& 0.93& 0.95& 0.96& 0.98& 0.99& 0.99& 1.00& &
   & 1.00& 0.99&19\\ 18 & 0.14& 0.25& 0.36& 0.46& 0.55& 0.61& 0.66&
   0.73& 0.78& 0.82& 0.86& 0.90& 0.92& 0.94& 0.96& 0.97& 0.98& 1.00& &
   & 1.00&20\\ 19 &-0.01& 0.09& 0.20& 0.30& 0.38& 0.45& 0.50& 0.58&
   0.63& 0.69& 0.74& 0.79& 0.82& 0.85& 0.88& 0.91& 0.93& 0.97& 1.00& &
   & \\ 20 &-0.13&-0.04& 0.06& 0.15& 0.24& 0.30& 0.36& 0.44& 0.50&
   0.56& 0.62& 0.68& 0.72& 0.75& 0.79& 0.83& 0.86& 0.93& 0.98& 1.00& &
   \\
\end{tabular}
\end{table*}

\begin{table}[htbp]
 \centering
\caption{\chisq\ comparisons of the  inclusive jet cross section 
 for $\modeta < 0.5$ at $\sqrt{s} = 630$~GeV with several theoretical
 predictions (20 degrees of freedom). \label{t_dtt_630}}
\begin{tabular}{cccc}
PDF	     & {$\mu$} 		& $\chi^{2}$ 	& prob. \\ 
\hline \hline 
	     & $2\Etmax$ 	& 40.5 		& $0.4\%$ \\ 
\cteqthreem\ & \Etmax\		& 25.9 		& $17\%$ \\ 
	     & $0.5\Etmax$ 	& 30.4 		& $6.4\%$ \\ 
	     & $0.25\Etmax$ 	& 27.5 		& $12\%$ \\ 
\hline
\cteqfourm\  & 0.5\Etmax\ 	& 24.2 		& $24\%$ \\ 
\cteqfourhj\ & 0.5\Etmax\ 	& 19.0 		& $53\%$ \\ 
\mrst\ 	     & 0.5\Etmax\ 	& 22.6 		& $31\%$ \\ 
\mrstgu\     & 0.5\Etmax\ 	& 14.9 		& $78\%$ \\ 
\mrstgd\     & 0.5\Etmax\ 	& 51.8 		& $0.01\%$ \\ 
\end{tabular}
\end{table}

\subsection{The Ratio of Jet Cross Sections}

 The dimensionless inclusive jet cross section
 (Section~\ref{sec:ratio}) is given by
\begin{equation}
\sigma_{\sqrt{s ({\rm GeV})}} = 
\frac{E_{T}^{3}}{2\pi} \frac{d^{2}\sigma}{d\Et d\eta}, \label{eq_r}
\end{equation}
 where $\displaystyle{{d^{2}\sigma}/{d\Et d\eta}}$ is given by
 Eq.~\ref{eq:inc_cross_section}. The ratio of inclusive jet cross
 sections for $\modeta < 0.5$ is calculated in bins of identical
 $x_T$:
\begin{equation}
R\left(x_T \right) = \frac{\sigma_{630}\left(x_T \right)
}{\sigma_{1800}\left(x_T \right) }.
\end{equation}

\subsection{Uncertainties in the Ratio of Jet Cross Sections}

 Most of the systematic uncertainties in the inclusive jet cross
 section are highly correlated as a function of \Et\ and
 center-of-mass energy, and cancel when the ratio of the two cross
 sections is calculated. To determine the uncertainty in the ratio,
 all uncertainties are separated into three categories, depending on
 the correlation $\left( \rho \right)$ as a function of \Et\ and CM
 energies.  In most cases, complete correlation in \Et\ at one CM
 energy implies complete correlation between CM energies, but
 exceptions exist and are highlighted in the following sections.

\subsubsection{Luminosity Uncertainties}

 The luminosity calculation at $\sqrt{s}=630$ GeV shares many common
 uncertainties with the calculation at 1800 GeV
 (Section~\ref{sec04:lumin}). The uncertainty from the fit to the
 world average (WA) \pbarp\ total cross section determines the
 uncertainty in the luminosity at $\sqrt{s} = 630$~GeV
 (Fig.~\ref{lum_fits}). A $1\sigma $ shift in the mean value of the
 cross section at $\sqrt{s} = 1800$~GeV directly impacts the central
 value of the cross section at $\sqrt{s} = 630$~GeV, resulting in a
 shift of unequal magnitude but like direction. The magnitude of the
 shift at 630 GeV, subtracted in quadrature from the interpolation
 uncertainty, defines two uncertainty components: the shift, which is
 completely correlated with the 1800~GeV cross section uncertainty,
 and the remainder, which is added in quadrature with the other
 independent luminosity uncertainties. The uncertainty components in
 the WA elastic and single-diffractive \pbarp\ cross sections are
 handled with the same procedure.  Table~\ref{t_lum} lists the
 systematic uncertainties due to the luminosity for the ratio.

\begin{table}[htbp] 
\caption{The uncertainties in the ratio of cross sections due to the
 luminosity calculation. *Includes trigger matching
 uncertainty.\label{t_lum}}
\centering
\begin{tabular}{cd}
Source 				&  \multicolumn{1}{c}{Uncertainty (Percent)}\\ 
\hline\hline		
World Average \pbarp\ cross section	& $\pm$3.2  \\
Hardware efficiency 			& $\pm$3.6 \\
Geometric acceptance 			& $\pm$0.8  \\
Uncorrelated$^{*}$ 	 		& $\pm$2.6 \\
\hline\hline		
All sources$^{*}$ 			& $\pm$4.2 \\
\end{tabular}
\end{table}

\subsubsection{Jet and Event Selection Uncertainties}

 At 1800 GeV, the total uncertainty for jet cut efficiencies, the
 \met\ cut efficiency, and the vertex cut efficiency is $1\%$. An
 independent study at 630 GeV determined cut uncertainties that were
 smaller (Table~\ref{t_eff}). Despite some similarities in
 methodology, these uncertainties are all considered to be independent
 of one another in the ratio.

\begin{table}[htbp] \centering
\caption{Uncertainty from jet and event selection.\label{t_eff}}
\begin{tabular}{ccd}
  	 	& {Uncertainty Source}	& \multicolumn{1}{c}{Uncertainty}\\ 
\hline \hline 
1800 GeV 	& all selection & \multicolumn{1}{c}{$1\%$ below 350 GeV}\\ 
		& cut efficiencies & \multicolumn{1}{c}{$2\%$ above 350 GeV}\\
\hline\hline
 		&Jet cuts 	& \multicolumn{1}{c}{$0.12$ to $0.53\%$} \\ 
630 GeV 	& \met\ cut 	& 0.03$\%$ \\ 
		& vertex cut 	& 0.006$\%$ \\ 
\end{tabular}
\end{table}

\subsubsection{Resolution and Unsmearing Uncertainties}

\begin{figure}[htbp]
\begin{center}
\vbox{\centerline
{\psfig{figure=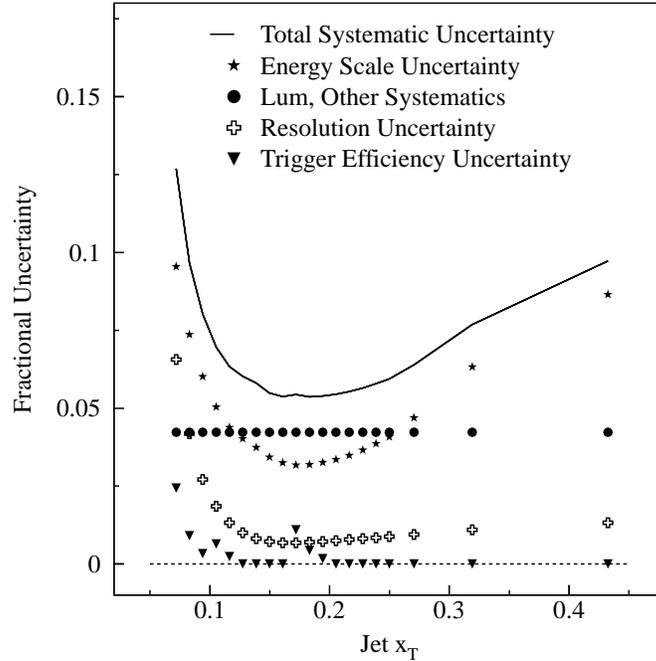,width=3.5in}}}
  \caption{The uncertainty components in the ratio of inclusive jet
  cross sections as a function of $x_T$ plotted by component.}
\label{all_errors}
\end{center}
\end{figure}

 Uncertainty in the unsmearing correction is dominated by the
 uncertainty in the jet resolution measurement. In the case of
 $\sqrt{s}=1800$ GeV, the systematic uncertainty dominates; for
 $\sqrt{s}=630$ GeV, poor statistics result in a fit uncertainty that
 is larger in magnitude than the systematic uncertainty. The
 systematic uncertainties in the unsmearing correction are assumed to
 be uncorrelated between the CM energies, as are the fitting
 uncertainties. The magnitudes of the resolution and unsmearing
 uncertainties are illustrated in Fig.~\ref{all_errors}.

\subsubsection{Energy Scale Uncertainties}

 The uncertainty in the inclusive jet cross section and in the ratio
 of cross sections is calculated using a Monte Carlo Simulation.  The
 event generator performs several steps for each $\sqrt{s}$ and each
 cross section bin in $x_T$.  First, it generates a sample of jets
 with an $x_T$ spectrum which matches that observed in data. Second, it
 closely imitates true running conditions by simulating luminosity,
 vertexing, and smearing effects; thus the energy scale corrections
 of each Monte Carlo jet will closely match the corrections in real
 data. Third, the uncertainties from the energy scale corrections are
 calculated. Finally, the weighted average uncertainties and
 correlations in each bin are combined to form a covariance matrix.

 The jet \Et\ distribution must be identical to the observed (smeared)
 jet cross section in data. The Monte Carlo simulation:
\begin{enumerate}
\item Randomly generates the initial parton momenta $x_{1}$ and
 $x_{2}$.
\item Generates the corresponding $p_T$ and other kinematic quantities
 for both of the final-state partons (which result in jets).
\item Smears the jets according to the known resolution functions and
 then selects one jet at random.
\item  Checks that the selected jet falls within the desired $x_T$
 bin and has $\modetajet <0.5$ (or starts over).
\item Generates a weight for the jet, to reproduce the steeply
 falling spectrum of the inclusive jet cross section, using either a
 theoretical weight based on \cteqfourm\ and the scale of the collision, or
 an experimental weight based on the ansatz from unsmearing.
\end{enumerate}

 Because the generated jet distribution already represents the
 energy-scale-corrected jet \Et , and because the response correlation
 is given in terms of the energy before the response correction, the
 energy scale algorithm must be run ``in reverse'' to find the
 uncertainties and their correlations as a function of jet \Et\ and CM
 energy.

\begin{figure}[htbp]
 \begin{center}
 \vbox{\centerline{\psfig{figure=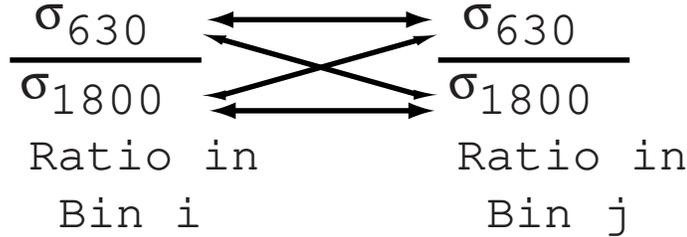,width=3.25in}}}
   \caption{Correlations between two ratio bins $i$ and $j$. Arrows
   indicate the four possible correlation ($\rho $) terms. The
   uppermost arrow is $\rho _{ij}^{630-630}$, while the
   ``$\searrow $'' arrow is $\rho _{ij}^{630-1800}$.}
 \label{six_cor}
 \end{center}
 \end{figure}

 The ratio of inclusive jet cross sections is given in
 Eq.~\ref{eq_r}. The elements of the covariance matrix are
 \begin{equation}
 C_{ij}=\left\langle \hat{\rho}_{ij}\;\delta R_{i}\;\delta R_{j}\right\rangle
 ,  \label{eq_cov}
 \end{equation}
  where $\hat{\rho}$ expresses the correlation between the $x_{T}$ bins
  $i$ and $j$, and the uncertainties in the ratio $\delta R$ may be expressed
  as
 \begin{eqnarray}
 \delta R_{i} & = & \frac{\partial R_{i}}{\partial \sigma _{i}^{630}}\;\delta
 \sigma _{i}^{630}+\frac{\partial R_{i}}{\partial \sigma _{i}^{1800}}\;\delta
 \sigma _{i}^{1800},  \label{eq_del_r}
 \end{eqnarray}
  where the two partial derivatives possess opposite signs:
 \begin{eqnarray}
 \frac{\partial R_{i}}{\partial \sigma _{i}^{630}} & = &\
  frac{1}{\sigma _{i}^{1800} }=\frac{R_{i}}{\sigma _{i}^{630}}
  \hspace*{0.25in} \nonumber \\
 \frac{\partial R_{i}}{\partial \sigma _{i}^{1800}} & = &
  \frac{-\sigma _{i}^{630}}{\left( \sigma _{i}^{1800}\right) ^{2}}
  =-\frac{R_{i}}{\sigma _{i}^{1800}}.
  \label{eq_partials}
 \end{eqnarray}
 Defining $x\equiv x_{T}$, the dependence of $\delta \sigma $ on jet
 energy is given by:
\begin{equation}
 \delta \sigma _{i}^{a}=\frac{\partial \sigma _{i}^{a}}{\partial x_{i}}\delta
 x_{i}^{a}=\frac{2}{a}\sin \theta _{i}\cdot \frac{\partial \sigma _{i}^{a}}{
 \partial x_{i}}\;\delta E_{i}^{a}.
\end{equation}
  The  cross section uncertainty is now expressed in terms of
  jet energy, the jet angle, the CM energy ($a$), and the slope of the
  dimensionless cross section. The final expression for the covariance
  matrix elements becomes
\begin{eqnarray}
 C_{ij} & = & \sum_{a,b}\sum_{k,l}q \frac{2}{a}\sin \theta _{k} \frac{2}{b}
 \sin \theta _{l} \frac{R_{k}}{\sigma _{k}^{a}}\frac{\partial \sigma
 _{k}^{a}}{\partial x_{k}} \frac{R_{l}}{\sigma _{l}^{b}}\frac{\partial
 \sigma _{l}^{b}}{\partial x_{l}}  \nonumber \\
 & & 
 \left\langle \rho _{kl}^{ab}\;\delta
 E_{k}^{a}\;\delta E_{l}^{b}\right\rangle ,  \label{eq_cov2}
\end{eqnarray}
  where $a$ and $b$ indicate CM energies; $\rho _{kl}^{ab}$ is the
  correlation between the uncertainties of the two jets whose energies
  fall in bins $k$ and $l$, originating from the data sets at
  $\sqrt{s}=a$ and $b$; and $q$ is a factor that accounts for the
  negative sign in Eq.~\ref{eq_partials}: $q=1$ when $a=b $, and
  $q=-1$ otherwise. The bracket notation indicates the average. The
  summations indicate the four relevant correlations, visually
  described in Fig.~\ref{six_cor}.

  As mentioned previously, interpolation of the correlation matrix
  determines the values of $\rho _{kl}^{ab}$ for the response
  uncertainty. For the completely correlated uncertainties, all $\rho
  $'s take the value of unity; for the uncorrelated uncertainties, all
  $\rho $'s are zero.  The major contribution originates from the
  partially correlated response uncertainty.

\subsubsection{Combined Uncertainty in the Ratio}

\begin{figure}[htbp]
\begin{center}
\vbox{\centerline{\psfig{figure=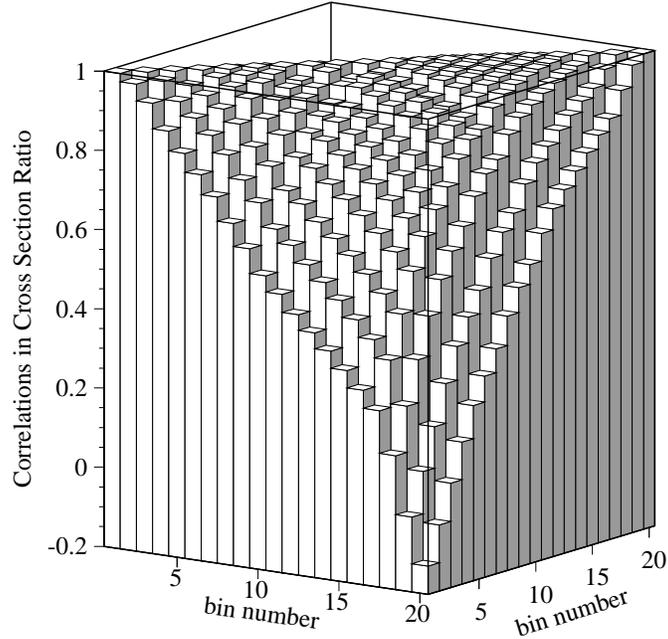,width=3.5in}}}
  \caption{The correlation matrix for the ratio of cross
  sections. Axes indicate the bin numbers.  }
\label{corr_total}
\end{center}
\end{figure}

 The individual uncertainties of the earlier sections fall into
 several classifications, summarized in Table
 \ref{t_correl_classif}. Complete cancellation of uncertainties occurs
 when the uncertainties are completely correlated between CM energies.
 The components of the systematic uncertainties for the ratio of cross
 sections are plotted in Fig.~\ref{all_errors} and given in
 Table~\ref{ratio_errors}. The uncertainty in the energy scale
 correction dominates at each end of the spectrum; resolution and
 contributions from other sources (primarily the luminosity
 uncertainty) become important at intermediate values of $x_T$.
 Figure~\ref{corr_total} plots the point-to-point uncertainty
 correlations between data points.

\begin{table}[htbp]
\begin{center}
\caption{Uncertainty correlations in the ratio of cross sections.
``0'' indicates no correlation, ``1'' indicates complete
correlation.\label{t_correl_classif}}
\begin{tabular}{lccc}
Uncertainty Source & \multicolumn{2}{c}{Correlation in} & Comments \\ 
	& $\sqrt{s}$ & Jet \Et\  \\ \hline\hline
Luminosity & partial & 1 \\
Filter Match & 0 & 1 & 1800 GeV Only \\
Event Cuts   & 0 & 0 & \\
Jet Cuts   & 0 & 0 & \\
Resolution \\
~~Fits & partial & 1 &  \\
~~Closure & 1 & 1 & \\
Unsmearing Fits & 0 & 1 & \\
Energy Scale & \\
~~Offset & partial & 1 & \\
~~Response fit & 1 & partial & \\
~~Response at 630 GeV & 0 & 1 &\\
~~Showering & 1 & 1 & \\
\end{tabular}
\end{center}
\end{table}

\begin{table*}[htbp]
\squeezetable
\caption{Percentage uncertainties in the ratio of inclusive jet cross
sections at $\sqrt{s} = 630$ and 1800~GeV for $\modeta < 0.5$. }
\label{ratio_errors}
\begin{tabular}{ddddddddd}
$x_T$ & Statistical & Jet Selection &Trigger Match  & Luminosity  & 
Trigger Efficiency & Unsmearing & Energy Scale & Total \\
\hline\hline
 0.07 &   1.5 & 1.1 &    1.7 &    4.1 &    2.4 &    6.6 &    9.5 &   12.7 \\
 0.08 &   2.4 & 1.1 &    1.7 &    4.1 &    0.9 &    4.2 &    7.4 &    9.7 \\
 0.09 &   3.5 & 1.1 &    1.7 &    4.1 &    0.3 &    2.7 &    6.0 &    8.0 \\
 0.11 &   1.9 & 1.1 &    1.1 &    4.1 &    0.6 &    1.9 &    5.0 &    7.0 \\
 0.12 &   2.7 & 1.1 &    1.1 &    4.1 &    0.2 &    1.3 &    4.4 &    6.3 \\
 0.13 &   3.5 & 1.1 &    1.1 &    4.1 &    0.1 &    1.0 &    4.0 &    6.0 \\
 0.14 &   4.9 & 1.1 &    1.1 &    4.1 &    0.0 &    0.8 &    3.7 &    5.8 \\
 0.15 &   5.8 & 1.1 &    0.0 &    4.1 &    0.0 &    0.7 &    3.4 &    5.5 \\
 0.16 &   7.3 & 1.1 &    0.0 &    4.1 &    0.0 &    0.7 &    3.2 &    5.4 \\
 0.17 &   2.4 & 1.1 &    0.0 &    4.1 &    1.1 &    0.7 &    3.2 &    5.4 \\
 0.18 &   3.0 & 1.1 &    0.0 &    4.1 &    0.4 &    0.7 &    3.2 &    5.4 \\
 0.19 &   3.5 & 1.1 &    0.0 &    4.1 &    0.2 &    0.7 &    3.3 &    5.4 \\
 0.21 &   4.4 & 1.1 &    0.0 &    4.1 &    0.1 &    0.8 &    3.4 &    5.5 \\
 0.22 &   5.1 & 1.1 &    0.0 &    4.1 &    0.0 &    0.8 &    3.5 &    5.5 \\
 0.23 &   6.2 & 1.1 &    0.0 &    4.1 &    0.0 &    0.8 &    3.7 &    5.6 \\
 0.24 &   7.3 & 1.1 &    0.0 &    4.1 &    0.0 &    0.8 &    3.9 &    5.8 \\
 0.25 &   8.5 & 1.1 &    0.0 &    4.1 &    0.0 &    0.9 &    4.1 &    5.9 \\
 0.27 &   6.6 & 1.1 &    0.0 &    4.1 &    0.0 &    1.0 &    4.7 &    6.4 \\
 0.32 &  11.0 & 1.1 &    0.0 &    4.1 &    0.0 &    1.1 &    6.3 &    7.7 \\
 0.43 &  20.5 & 1.1 &    0.0 &    4.1 &    0.0 &    1.3 &    8.7 &    9.7 \\
\end{tabular}
\end{table*}

\subsection{Results and Comparison to Theoretical Predictions}

\begin{figure}[htbp]
\begin{center}
\vbox{\centerline
{\psfig{figure=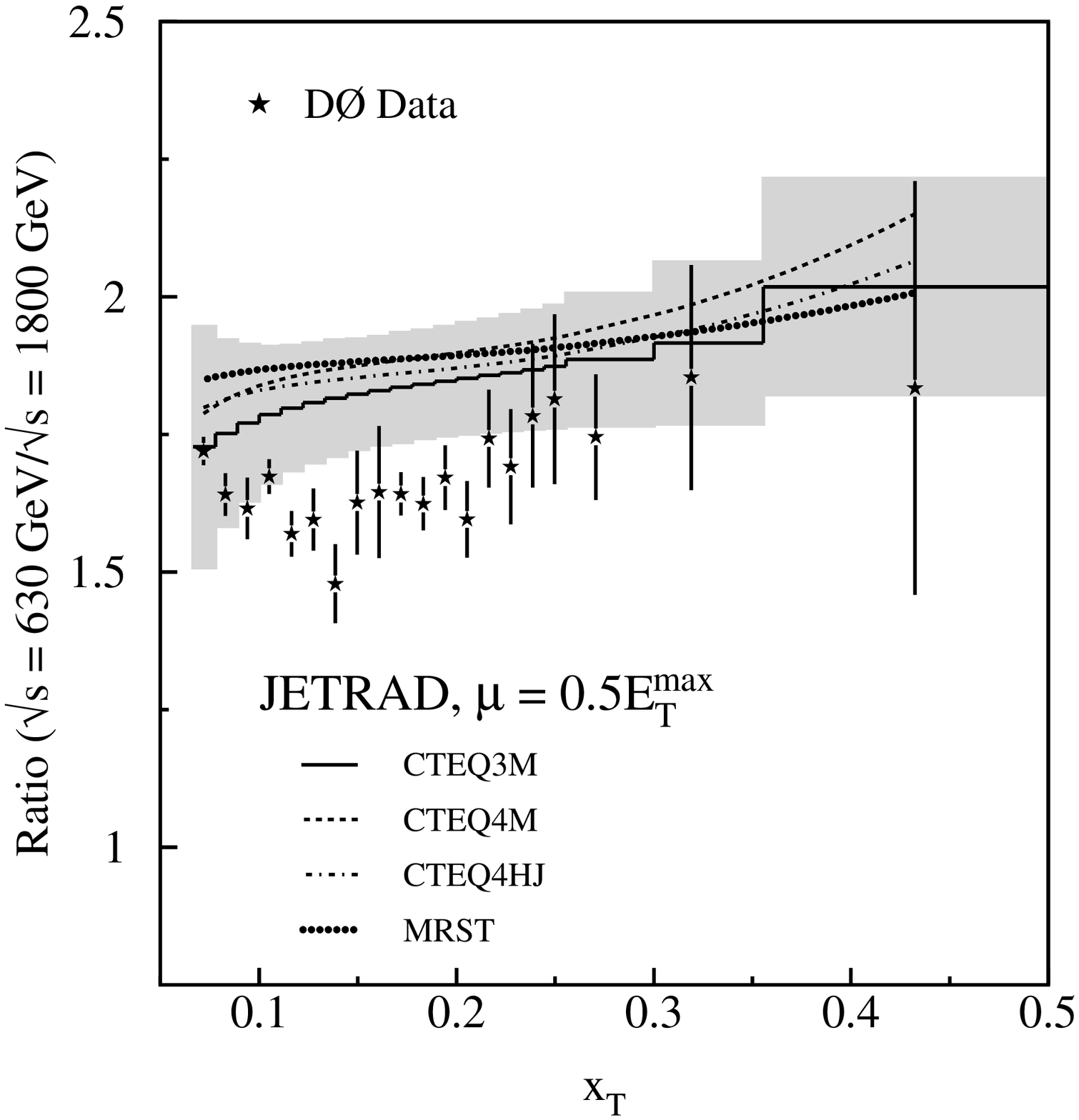,width=3.5in}}}
 \caption{The ratio of dimensionless cross sections for $\modeta <
 0.5$ compared with {\sc jetrad} predictions with $\mu = 0.5 \Etmax$
 and the \cteqthreem , \cteqfourm , \cteqfourhj , and \mrst\ PDFs. The
 shaded band represents the $\pm 1\sigma$ systematic uncertainty band
 about the prediction. }
\label{rat_pdf}
\end{center}
\end{figure}

\begin{figure}[hbtp]
\begin{center}
\vbox{\centerline
{\psfig{figure=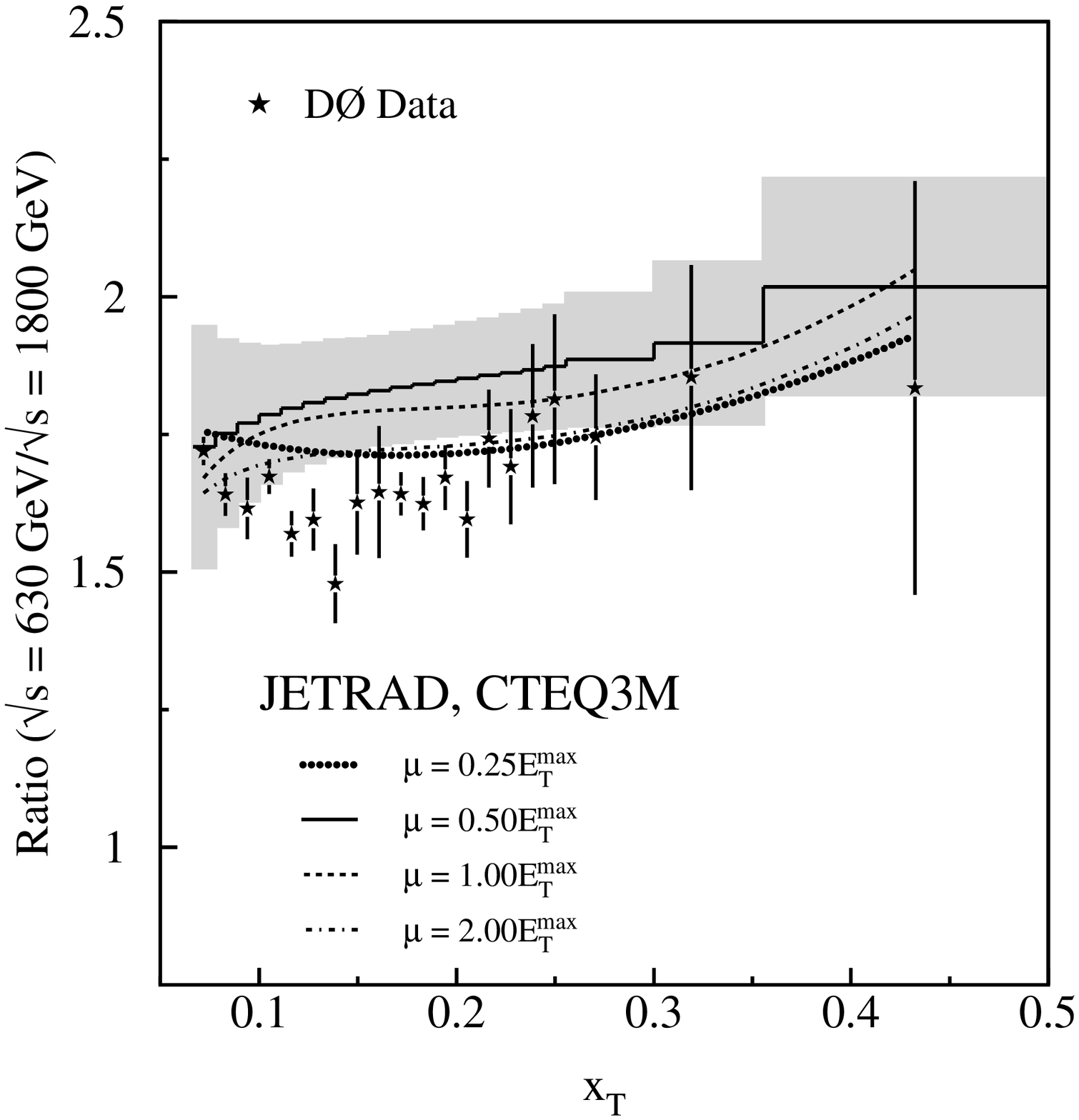,width=3.5in}}}
 \caption{The ratio of dimensionless cross sections for $\modeta <
 0.5$ compared with {\sc jetrad} predictions with various values of
 $\mu$ and the \cteqthreem\ PDF.  The shaded band represents the $\pm
 1\sigma$ systematic uncertainty band about the prediction.  }
\label{rat_mu}
\end{center}
\end{figure}

 The ratio between the inclusive jet cross sections at $\sqrt{s}
 =$~630 and 1800 GeV is given in Table~\ref{ratio_values}.
 Figures~\ref{rat_pdf} and \ref{rat_mu} show the ratios of cross
 sections compared with {\sc jetrad} predictions using different
 choices of PDF and renormalization scale. The measured ratios lie
 approximately $10\%$ below the theoretical predictions, which have an
 uncertainty of approximately $10\%$
 (Section~\ref{sec:ratio}). Table~\ref{t_chi6} lists the $\chi^{2}$
 distributions for the ratio of cross sections compared to selected
 theoretical predictions.  The \chisq\ values lie in the range
 15.1--24 for 20 degrees of freedom (corresponding to probabilities in
 the range $28\%$ to $77\%$). The best agreement occurs for extreme
 choices of renormalization scales: $\mu = (0.25, 2.00) \Etmax$. As
 expected, there is very little dependence on the choice of PDF.

\begin{table}[htbp]
\begin{center}
\caption{The ratio of the inclusive jet cross sections for 
$\modeta < 0.5$ at $\sqrt{s}=$ 630 and 1800 GeV.}
\label{ratio_values}
\begin{tabular}{cccd}
   \multicolumn{1}{c}{($x_T$)} & \multicolumn{1}{c}{($x_T$)} &
 \multicolumn{1}{c}{Ratio of Cross Sections} & Systematic  \\ 
 \multicolumn{1}{c}{Bin Range}     &  \multicolumn{1}{c}{Plotted}  &
 $\pm$ statistical error & Uncertainty ($\%$)\\
 \hline\hline
  0.067 --  0.078 &  0.072 & $  1.72\pm  0.03 $ &   12.7  \\
  0.078 --  0.089 &  0.083 & $  1.64\pm  0.04 $ &    9.7  \\
  0.089 --  0.100 &  0.094 & $  1.62\pm  0.06 $ &    8.0  \\
  0.100 --  0.111 &  0.105 & $  1.67\pm  0.03 $ &    7.0  \\
  0.111 --  0.122 &  0.116 & $  1.57\pm  0.04 $ &    6.3  \\
  0.122 --  0.133 &  0.127 & $  1.59\pm  0.06 $ &    6.0  \\
  0.133 --  0.144 &  0.139 & $  1.48\pm  0.07 $ &    5.8  \\
  0.144 --  0.156 &  0.150 & $  1.63\pm  0.09 $ &    5.5  \\
  0.156 --  0.167 &  0.161 & $  1.64\pm  0.12 $ &    5.4  \\
  0.167 --  0.178 &  0.172 & $  1.64\pm  0.04 $ &    5.4  \\
  0.178 --  0.189 &  0.183 & $  1.62\pm  0.05 $ &    5.4  \\
  0.189 --  0.200 &  0.194 & $  1.67\pm  0.06 $ &    5.4  \\
  0.200 --  0.211 &  0.205 & $  1.60\pm  0.07 $ &    5.5  \\
  0.211 --  0.222 &  0.216 & $  1.74\pm  0.09 $ &    5.5  \\
  0.222 --  0.233 &  0.228 & $  1.69\pm  0.10 $ &    5.6  \\
  0.233 --  0.244 &  0.239 & $  1.78\pm  0.13 $ &    5.8  \\
  0.244 --  0.256 &  0.250 & $  1.81\pm  0.15 $ &    5.9  \\
  0.256 --  0.300 &  0.271 & $  1.74\pm  0.11 $ &    6.4  \\
  0.300 --  0.356 &  0.319 & $  1.85\pm  0.20 $ &    7.7  \\
  0.356 --  0.622 &  0.432 & $  1.83\pm  0.38 $ &    9.7  \\
\end{tabular}
\end{center}
\end{table}

\begin{table}[htbp] 
\centering
\caption{The calculated $\chi^{2}$ for the ratio of cross sections 
 (20 degrees of freedom).   \label{t_chi6}}
\begin{tabular}{ccdd}
{PDF} & {Renormalization Scale} &  \chisq\ &  \multicolumn{1}{c}{Prob.} \\ 
\hline\hline
 		& $2\Etmax$ 	& 17.9 &  60$\%$  \\
 		& \Etmax\ 	& 21.6 &  36$\%$  \\
\cteqthreem\ 	& 0.75\Etmax\ 	& 23.1 &  28$\%$  \\
		& 0.5\Etmax\ 	& 20.5 &  43$\%$  \\
		& 0.25\Etmax\ 	& 15.1 &  77$\%$  \\
\hline                          	 	   
\cteqfourm\	& 0.5\Etmax\ 	& 22.4 &  32$\%$  \\
\cteqfourhj\ 	& 0.5\Etmax\ 	& 21.0 &  40$\%$  \\
\mrst\ 		& 0.5\Etmax\  	& 22.2 &  33$\%$  \\
\mrstgu\ 	& 0.5\Etmax\  	& 19.5 &  49$\%$  \\
\mrstgd\ 	& 0.5\Etmax\  	& 24.1 &  24$\%$  \\
\end{tabular}
\end{table}

\begin{figure}[htbp]
\begin{center}
\vbox{\centerline
{\psfig{figure=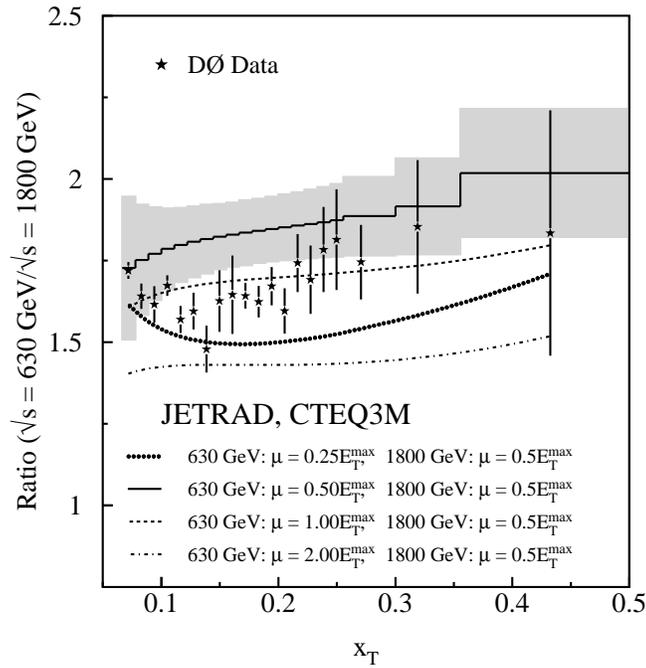,width=3.5in}}}
  \caption{The ratio of dimensionless cross sections for $\modeta <
  0.5$ compared with {\sc jetrad} predictions with $\mu = 0.5\Etmax$
  at $\sqrt{s}=1800$~GeV, $\mu = (0.25, 1.0, 2.0)\Etmax$ at
  $\sqrt{s}=630$~GeV, and the \cteqthreem\ PDF.  The shaded band
  represents the $\pm 1\sigma$ systematic uncertainty band about the
  prediction.  }
\label{compare_jetrad_ratio}
\end{center}
\end{figure}

 Different renormalization scales can be selected for the different CM
 energies since there is no explicit theoretical need for identical
 scales at $\sqrt{s}=630$ and 1800
 GeV. Figure~\ref{compare_jetrad_ratio} depicts a comparison between
 the ratio and theoretical predictions where the renormalization
 scales at the two CM energies are not equivalent. The resulting
 $\chi^{2}$ indicate good agreement between the data and the
 predictions (Table~\ref{t_rat_mu1mu2}).

\begin{table}[htbp] \centering
\caption{$\chi^{2}$ comparisons for the ratio of cross sections for
$\modeta < 0.5$ where the renormalization scale is mismatched between
CM energies.\label{t_rat_mu1mu2}}
\begin{tabular}{cccdd}
  PDF   	&  \multicolumn{2}{c}{Renormalization Scale} & {} &  \\ 
   		& 630~GeV		& 1800~GeV  & \chisq\   
 & \multicolumn{1}{c}{Prob.} \\ \hline\hline
   		& $2\Etmax$ 		& $0.5\Etmax$   & 14.9 &  78$\%$ \\ 
\cteqthreem\	& $\Etmax$ 		& $0.5\Etmax$   & 17.2 &  64$\%$ \\ 
   		& $0.25\Etmax$ 		& $0.5\Etmax$  	& 23.1 &  28$\%$ \\ 
\end{tabular}
\end{table}

 An additional analysis was carried out to measure the significance of
 the normalization difference between the data and the theoretical
 predictions. The data are reduced to a single value by fitting them
 to a constant (horizontal line), resulting in a value of $1.60 \pm
 0.08$. The uncertainty in this value is given by the uncertainty of
 the fit. Each of the theoretical predictions is also reduced to a
 single value. Each $x_T$ point of the prediction is assigned a weight
 given by the statistical uncertainty of the corresponding point in
 the data (Table~~\ref{Table:ratio_normalization}). The uncertainty in
 the value representing the theoretical prediction is assumed to be
 zero. The resulting \chisq\ values are given in
 Table~\ref{Table:ratio_normalization}, and lie in the range 1.4--13.2
 (corresponding to probabilities of $0.03\%$ to $23\%$). In every
 case, discarding the shape information in favor of a comparison of
 normalization results in poorer agreement between data and the
 theory.

\begin{table*}[htbp]
\caption{Normalization-only predictions for the ratio of cross
 sections and the \chisq\ comparison with the data ($1.60 \pm 0.08$
 for one degree of freedom).}
\label{Table:ratio_normalization}
\centering
\begin{tabular}{cccddd}
{PDF} & \multicolumn{2}{c}{Renormalization Scale} & Theory Normalization
 &  \chisq\ &  \multicolumn{1}{c}{Prob.} \\
\hline\hline
                & \multicolumn{2}{c}{$2\Etmax$}   & 1.75 &  3.3 &  6.8$\%$ \\
                & \multicolumn{2}{c}{\Etmax\ }     & 1.82 &  7.1 &  0.8$\%$ \\
\cteqthreem\    & \multicolumn{2}{c}{0.75\Etmax\ } & 1.87 & 10.7 &  0.1$\%$ \\
                & \multicolumn{2}{c}{0.5\Etmax\ }  & 1.85 &  9.5 &  0.2$\%$ \\ 
                & \multicolumn{2}{c}{0.25\Etmax\ } & 1.70 &  1.5 & 22.9$\%$ \\ 
\hline                           		 
\cteqfourm\     & \multicolumn{2}{c}{0.5\Etmax\ }  & 1.90 & 13.2 &  0.03$\%$\\
\cteqfourhj\    & \multicolumn{2}{c}{0.5\Etmax\ }  & 1.87 & 10.7 &  0.1$\%$ \\
\mrst\          & \multicolumn{2}{c}{0.5\Etmax\ }  & 1.89 & 12.6 &  0.04$\%$\\
\mrstgu\        & \multicolumn{2}{c}{0.5\Etmax\ }  & 1.87 & 11.1 &  0.09$\%$\\
\mrstgd\        & \multicolumn{2}{c}{0.5\Etmax\ }  & 1.90 & 12.9 &  0.03$\%$\\
\hline\hline
\\
\hline\hline
   		& 630~GeV	& 1800~GeV          & 	   &      &          \\
   		& $2\Etmax$ 	& $0.5\Etmax$       & 1.46 &  2.7 & 10.2$\%$ \\
\cteqthreem\	& $\Etmax$ 	& $0.5\Etmax$       & 1.71 &  1.8 & 18.1$\%$ \\
   		& $0.25\Etmax$ 	& $0.5\Etmax$  	    & 1.44 &  3.7 &  5.4$\%$ \\
\end{tabular}
\end{table*}

\subsection{Conclusions}

 We have made the most precise measurement to date of the ratio of the
 inclusive jet cross sections at $\sqrt{s}=630$ and 1800 GeV.  This
 measurement is nearly insensitive to the choice of parton
 distribution functions. The ratio of cross sections is therefore a
 more stringent test of QCD matrix elements.  The NLO QCD predictions
 yield satisfactory agreement with the observed data for standard
 choices of renormalization scale or PDF. In terms of the
 normalization however, the absolute values of the standard
 predictions lie consistently and significantly higher than the data.
\clearpage

\section{Dijet Angular Distribution}
\label{sec:angular_distribution}

 The dijet angular distribution is given by:
\begin{equation}
{1 \over {\sum \sigma}}
 {d^3 \sigma \over {dM \, d\chi \, d\eta_{\rm boost}}} =
 {1 \over {\sum N_i}}
 {N_i \over {\Delta M  \Delta \chi  \Delta\eta_{\rm boost}}} 
\end{equation}
 where the invariant mass is calculated assuming massless jets:
\begin{equation}
\jjmass^{2} = 2  \Etu{1}  \Etu{2}
  \left[ \cosh \left( \Delta \eta \right) - \cos \left(
 \Delta \phi \right) \right],
\label{eq:dijet_mass}
\end{equation}
 the pseudorapidity of the center-of-mass of the dijet system is given
 by $\eta^{\star} = {{{1} \over {2}}(\Delta \eta)}$; the
 pseudorapidity boost is given by {$\eta_{\rm boost} =
 {{1}\over{2}}(\eta_1 + \eta_2)$}; $\chi = \exp \left({\mid \!  \Delta
 \eta \!  \mid}\right) = \exp \left({2|\eta ^{\star}|} \right)$;
 \Etu{1}, $\eta_1$, and $\phi_1$ refer to the values associated with
 the jet with the largest \Et\ in an event; \Etu{2}, $\eta_2$, and
 $\phi_2$ refer to the values associated with the jet with the second
 largest \Et\ in an event; $\Delta \eta = \mid \! \eta_1 - \eta_2 \!
 \mid$ ; $\Delta \phi = \phi_1 - \phi_2$, and $N_i$ is the number of
 events in a given $\chi$ and mass bin.  If the individual jet masses
 are taken into account, the change in the dijet invariant mass is
 less than 1$\%$ for jets used in this analysis. Since the bins of
 $\Delta\eta_{\rm boost}$ are constant and we plot the angular
 distribution for a given mass bin, $\Delta M$, we choose to measure
 $d\sigma/ d\chi$ which is uniform for Rutherford scattering)
\begin{equation}
{1 \over {\sum \sigma}} {d \sigma \over {d\chi}} = 
{1 \over {\sum N_i}} {N_i \over {\Delta \chi }}.
\end{equation}

\subsection{Data Selection}

 The selected data are events with two or more jets which satisfy the
 set of inclusive jet triggers and pass the standard jet and event
 quality requirements (Section~\ref{sec:quality_cuts}). Events are
 removed unless both of the leading two jets pass the jet quality
 requirements.  The vertex of the event must be within 50 cm of $z=0$.
 The $\chi$ distributions were corrected for the efficiencies of the
 standard jet quality cuts and the \met\ cut.

 To ensure that the jet triggers did not introduce a bias, the trigger
 requirement was verified by comparing the $\chi$ distribution of a
 lower trigger threshold to the $\chi$ distribution of the desired
 trigger threshold. It is known that the lower threshold trigger is
 100$\%$ efficient in the desired region and thus a comparison would
 show an inefficiency in the desired trigger sample.  No differences
 were seen. The \Et 's of all second jets are well within the region of
 100$\%$ jet reconstruction efficiency, so an additional \Et\
 requirement on the second jet was not necessary. The final
 energy-scale-corrected \Et\ requirement placed on each trigger sample
 is summarized in Table~\ref{table:dijet_ang_et_limits}.

\begin{table}[htbp]
\begin{center}
\caption{The cut on the \Et\ of the leading jet to ensure 
 that the trigger is 100$\%$ efficient.}
\label{table:dijet_ang_et_limits}
\begin{tabular}{rc}
Trigger & Corrected \Et\ Limit on Leading Jet(GeV)\\
\hline
Jet\_30 & $55.0$ \\ 
Jet\_50 & $90.0$ \\ 
Jet\_85 & $120.0$ \\ 
Jet\_115 & $175.0$ \\ 
\end{tabular}
\end{center}
\end{table}

\subsection{Acceptance: Limits on Mass and $\chi$}

\begin{figure}[htbp]
   \centerline
   {\psfig{figure=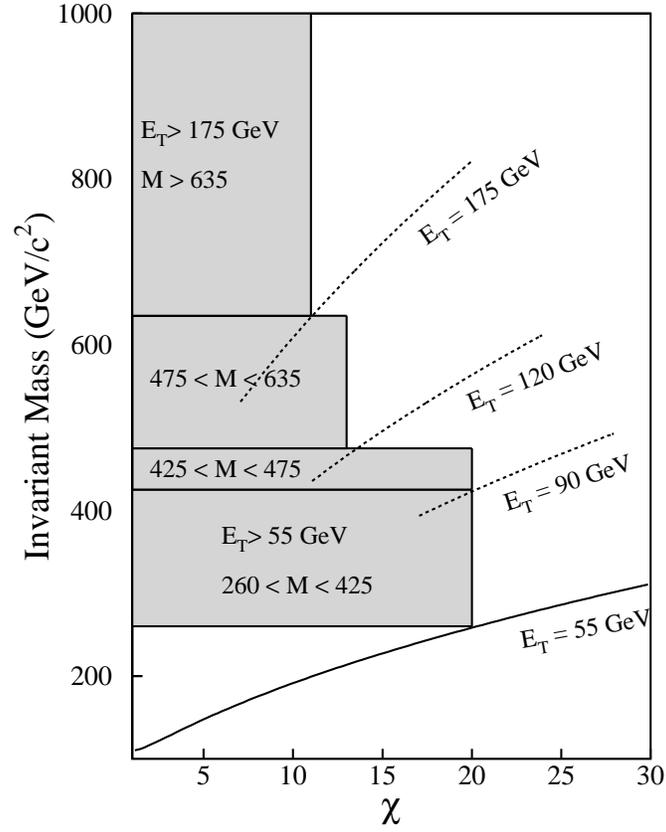,width=3.5in}}
   \caption{In the mass versus $\chi$ plane, the curves shown are
   contours of constant \Et .  The simplest form of uniform acceptance
   in this plane is a rectangle.  For a chosen mass region, the limit
   on $\chi$ corresponds to the intersection of the lower mass limit
   and the \Et\ contour.  The shaded regions shown are the mass bins
   chosen for this analysis.}  \label{FIG:mass_vs_chi}
\end{figure}

 Event acceptance is calculated using the kinematic relationships
 between mass, $\chi$, and $E_T$ shown in
 Fig.~\ref{FIG:mass_vs_chi}. Since an $E_T$ requirement is placed only
 on the leading jet, the maximum $\chi$ with 100$\%$ acceptance is
 determined from the $E_T$ requirement placed on the leading jet and
 the desired mass bin using the following formula:
\begin{equation}
{M^2 = 2E_{T1}^2\left[\cosh\left(\ln(\chi)\right)+1\right]}.
\end{equation}
 In this formula the $E_T$'s of the two leading jets are assumed to be
 identical.  Four mass bins were chosen in order to maximize the
 number of events per $\chi$ bin, and to attain a maximum $\chi$ of 20
 (corresponding to $\eta^{\star} = 1.5$). These mass bins are listed
 in Table~\ref{TABLE:range}.

\begin{table}[htbp]
\begin{center}
\caption{The average mass, maximum $\chi$ measured, and
the number of events after applying all kinematic cuts.}
\begin{tabular}{ccccc}
\multicolumn{1}{c}{Trigger $E_{T}$} &
\multicolumn{1}{c}{Mass} &
\multicolumn{1}{c}{Average} &
\multicolumn{1}{c}{$\chi_{\rm max}$} &
\multicolumn{1}{c}{Number} \\
\multicolumn{1}{c}{Threshold} &
\multicolumn{1}{c}{Range} &
\multicolumn{1}{c}{Mass} &
\multicolumn{1}{c}{ } &
\multicolumn{1}{c}{of Events} \\
\multicolumn{1}{c}{(GeV)} &
\multicolumn{1}{c}{(GeV/$c^{2}$)} &
\multicolumn{1}{c}{(GeV/$c^{2}$)} &
\multicolumn{1}{c}{ } &
\multicolumn{1}{c}{ } \\
\hline
55  & 260--425 & 302 & 20 & 4621 \\
90  & 425--475 & 447 & 20 & 1573 \\
120  & 475--635 & 524 & 13 & 8789 \\
175  & $>$635    & 700 & 11 & 1074 \\
\end{tabular}
\label{TABLE:range}
\end{center}
\end{table}

\begin{figure}[htbp]
   \centerline {\psfig{figure=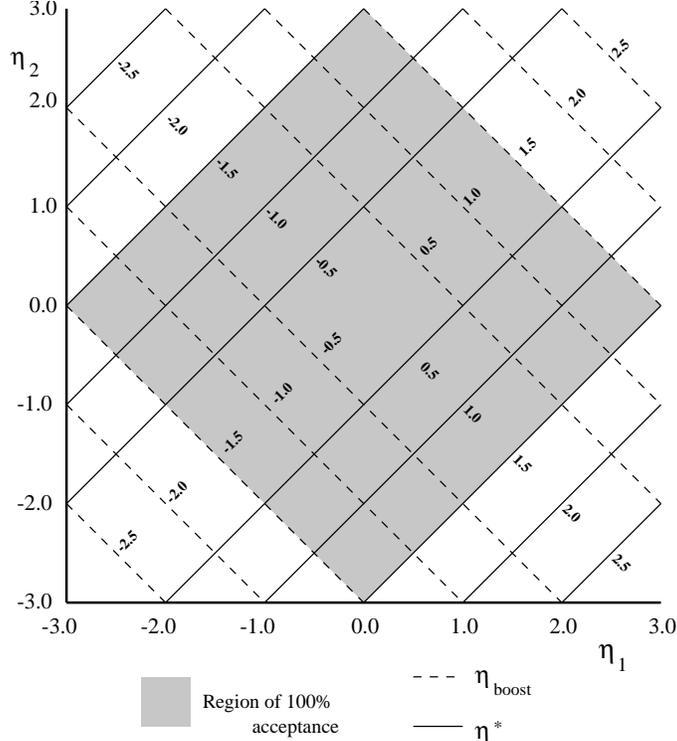,width=3.5in}}
   \caption{The $\modeta$'s of the two leading jets were required to
   be less than 3.0.  For a maximum $\eta^{\star}$ of 1.5, the $\mid
   \! \eta_{\rm boost} \! \mid$ was chosen to less than 1.5 to
   restrict the measurement to a region of 100$\%$ acceptance.}
   \label{FIG:etaboost}
\end{figure}

 Once the $\chi$ limit is known, a limit on  $\eta_{\rm boost}$ can
 be calculated. The $\eta_{\rm boost}$ parameter is used to restrict
 the $\chi$ distribution to the physical limits of the detector
 (Fig.~\ref{FIG:etaboost}). The $\eta_{\rm boost}$ limit is calculated
 using
\begin{eqnarray}
{|\eta_{\rm boost}|} & = & 
 \left\arrowvert |\eta^{\star}|-|\eta_{\rm max}|\right\arrowvert  \\
& = &  \left\arrowvert | 1.5 | - | 3.0 | \right\arrowvert = 1.5, \nonumber
\end{eqnarray}
 where $\mid \! \eta_{\rm max} \! \mid =3.0$ is the maximum $\eta$
 used for this analysis. The boost cut is chosen to be $\eta_{\rm
 boost} \le 1.5$.  For $\jjmass > 475$~\gevcc , $\mid \! \eta_{\rm
 boost} \! \mid$ is kinematically restricted to a value less than
 1.5. These mass bins are listed together with the average dijet
 invariant mass, the maximum $\chi$ measured, and the number of events
 for each of four mass ranges in Table~\ref{TABLE:range}.

\subsection{Systematic Studies}

 In order to study the systematic effects of the jet and event
 selection requirements, and the various corrections that are applied
 to the data, a series of systematic studies was performed.  For each
 requirement or correction, we measured the effect on the angular
 distribution by varying the requirement or correction by an
 appropriate amount.

\begin{figure}[htbp]
   \centerline
   {\psfig{figure=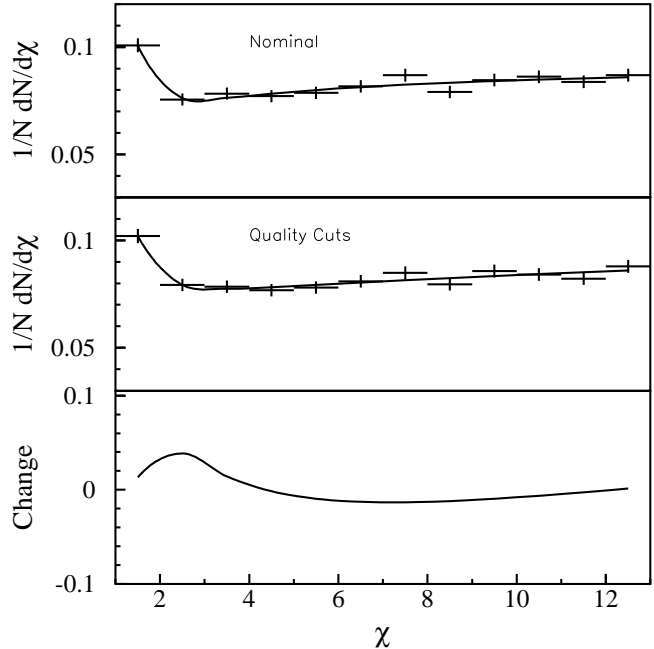,width=3.5in}}
   \caption{Technique to determine changes in angular distribution due
   to systematic uncertainties.  Top: Normalized angular distribution
   for the mass range 475--635 GeV/$c^{2}$ compared to a fit to the
   data.  Middle: Normalized angular distribution for the mass range
   475--635 GeV/$c^{2}$ after removing the jet quality cuts and a fit
   to the data.  Bottom: The (ratio$-$1) of the two fits shows the
   effects of the jet quality cuts on the shape of the angular
   distribution.}  \label{FIG:divfit}
\end{figure}

 To determine the systematic uncertainties on the shape of the angular
 distribution, each distribution is fit with a function:
 $\displaystyle{F = A + B\chi + C/ \chi + D/ \chi^2 +E/\chi^3}$.  The
 effect of varying each of the selection criteria or corrections is
 measured by taking the ratio of the distribution with the nominal
 selection criteria, and with the adjusted criteria
 (Fig.~\ref{FIG:divfit}), giving the size of the systematic
 uncertainty.

\begin{figure}[htbp]
   \centerline
   {\psfig{figure=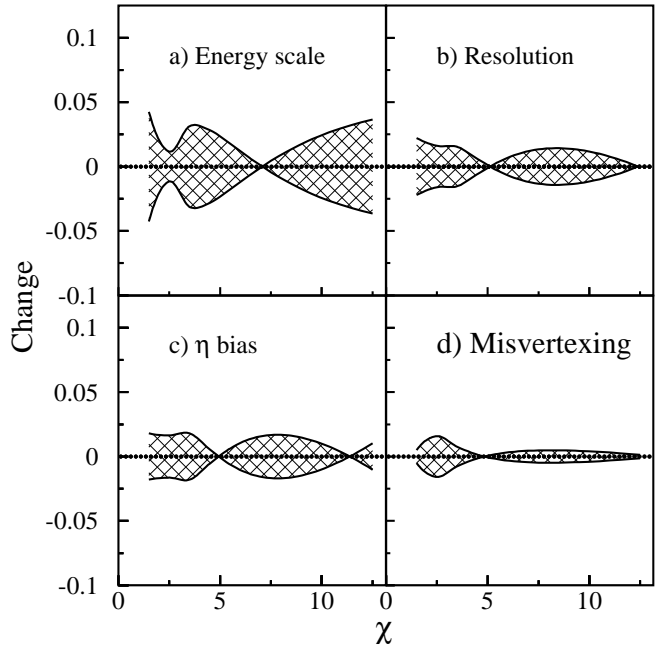,width=3.5in}}
   \caption{Ratios of parameterized curves showing the effects of 
    uncertainties on the shape of the angular distribution.
    Shown are the four largest uncertainties in the mass range 
    475--635 GeV/$c^{2}$.} 
   \label{FIG:rat1}
\end{figure}

 The largest source of uncertainty involves the $\eta$ dependence of
 the jet energy scale.  Small uncertainties in the relative response
 as a function of $\eta$ have large effects on the angular
 distribution.  The uncertainties in the jet energy scale are less
 than $2\%$ up to an $\modeta$ of 2.0 and become large near $\modeta
 \approx 3.0$.  The uncertainty in the showering correction is less
 than $2\%$ for $\modeta < $ 2.0 and becomes large at high
 $\modeta$. The effect of the $\eta$-dependent energy scale
 uncertainties are given in Fig.~\ref{FIG:rat1}(a).

 The resolution of our measurement of the jet energy can also affect
 the angular distribution. This was determined by measuring the
 difference between the smeared and unsmeared theory calculations.
 Since we are not unsmearing the data for the effects of $\eta$ and
 $E_{T}$ smearing, we apply this as an uncertainty in the measurement
 (Fig.~\ref{FIG:rat1}(b)).

 The effect on the angular distribution due to the $\eta$ bias in the
 jet reconstruction algorithm (Section~\ref{sec:eta_bias}) was studied
 by applying a correction for the bias. The difference between the
 corrected and uncorrected distributions was $1\%$ on average
 (Fig.~\ref{FIG:rat1}(c)).

 The overall energy scale does not affect the shape of the
 distribution, because a shift in the overall energy scale shifts the
 entire distribution in mass. The angular distribution changes very
 slowly with mass, so a small shift would not cause a significant
 change in the shape.

 For the Jet\_30 and Jet\_50 triggers, an online {\sc mitool} (see
 Section~\ref{sec:vertex}) requirement was used in the trigger for
 part of the run.  To determine if the {\sc mitool} requirement biased
 the angular distribution, runs with no {\sc mitool} requirement were
 compared to runs with the requirement.  A small shape difference was
 seen and an uncertainty equal to the difference between the two
 measurements was assigned.

 The effects of multiple interactions on the distributions were
 studied.  A secondary interaction adds approximately 0.6 GeV of $E_T$
 per unit $\Delta \eta \times \Delta \phi$ (Fig. \ref{fig:offset_1}).
 Since the angular distribution is measured in regions in which the
 $E_T$'s of the two leading jets are in excess of 50 GeV and are often
 above 100 GeV, the effect of this additional energy on the two
 leading jets is minimal. It is possible that a second interaction may
 produce a vertex which is incorrectly used as in the primary vertex
 for the leading two jets. This would cause an error in the measured
 $\eta$ positions of the jets as well as the measured $E_T$ of the
 jets. We studied the effect of not selecting the primary vertex by
 minimizing the \HT\ in the event (Section~\ref{sec:vertex}). This has
 a negligible effect on the angular distribution.

 It is possible that the vertex produced by a second interaction is
 the only vertex found in the event. This would also cause an error in
 the measured $\eta$ and $E_T$ values of the jets. We studied the
 possibility of multiple interactions affecting the angular
 distribution in this manner by the following method. For a determined
 percentage of events, we switched the vertex to a randomly chosen
 vertex. The new vertex was based on the measured vertex distribution,
 which has an approximate mean of $z=0$ and a $\sigma \approx
 30$~cm. We then recalculated the $\eta$ and $E_T$ of the two leading
 jets in the event and measured the angular distribution. The
 percentage of events with a new vertex was determined based on the
 efficiency of vertex reconstruction for events with large $E_T$ jets
 ($\approx 70\%$), and the percentage of multiple interactions in the
 data used for this analysis ($\approx 60\%$). The number of vertices
 switched was 20$\%$, which is an estimate of the number of times that
 the vertex reconstruction is incorrect. The size of the effect is
 less than $2\%$ and is dependent on the value of $\chi$ (see
 Fig.~\ref{FIG:rat1}(d)).

 The jet quality requirements and their corresponding efficiency
 corrections are necessary to remove noise from the event sample.
 Their effect on the shape of the angular distribution is minimal.

 The D\O\ jet algorithm allows for the splitting and merging of jets.
 This can cause a shift in the $\eta$ of the jet, and therefore affect
 the angular distribution. The effect on the shape of the distribution
 of removing those events in which either of the leading two jets were
 split or merged is minimal.  Since the theoretical predictions are
 expected to properly address merging and splitting, no uncertainty
 was assigned due to this effect.

\subsection{Results and Comparison to Theory}

\begin{figure}[htbp]
   \centerline
   {\psfig{figure=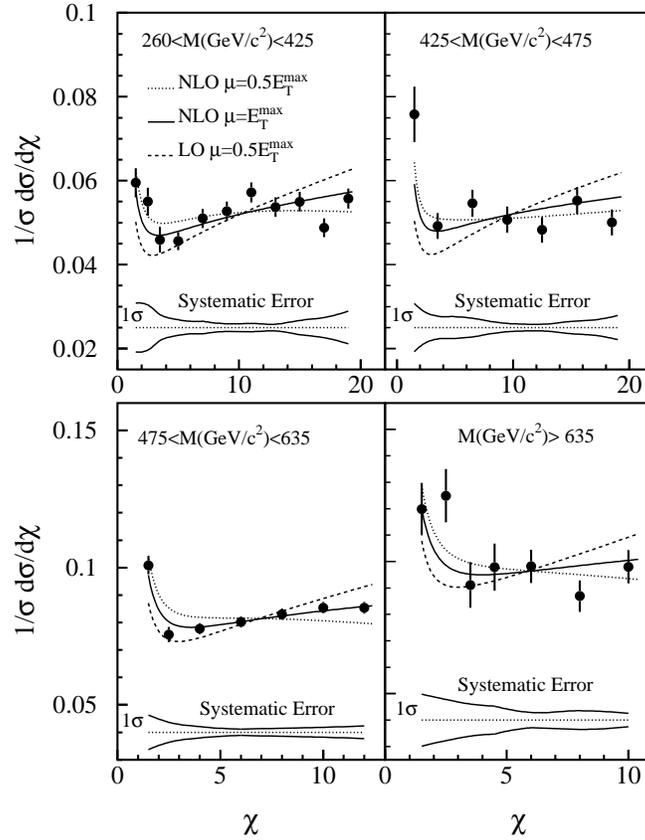,width=3.5in}} 
\caption{
   Dijet angular distributions for D\O\ data (points) compared to {\sc
   jetrad} for LO (dashed line) and NLO predictions with
   renormalization/factorization scale $\mu = 0.5\Etmax$ (dotted
   line).  The data are also compared to {\sc jetrad} NLO predictions
   with $\mu = \Etmax$ (solid line).  The errors on the data points
   are statistical only. The band at the bottom represents the $\pm$ 1
   $\sigma$ systematic uncertainty.}
\label{FIG:theo}
\end{figure}

 The measurement of the dijet angular cross section is given in
 Table~\ref{TABLE:numbers}.  The leading-order and
 next-to-leading-order theory predictions were obtained using the {\sc
 jetrad} parton-level event generator~\cite{jetrad} with CTEQ3M and
 $\mu = 0.5\Etmax$.  Four mass ranges are compared to the LO and NLO
 predictions of QCD in Fig.~\ref{FIG:theo}. The band at the bottom of
 each plot represents the $\pm$ 1 $\sigma$ systematic uncertainties.
 They are obtained by adding in quadrature all of the parameterized
 curves describing the shape uncertainties discussed earlier.

\begin{table*}[htbp]
\begin{center}
\caption{Dijet angular cross section 
 (100/$N$)(d$N$/d$\chi$) $\pm$ statistical $\pm$ systematic
 uncertainties for the four mass bins (GeV/$c^{2}$).}
\squeezetable
\begin{tabular}
{dd@{ $\pm$ }d@{ $\pm$ }dd@{ $\pm$ }d@{ $\pm$ }d
d@{ $\pm$ }d@{ $\pm$ }dd@{ $\pm$ }d@{ $\pm$ }d}
\multicolumn{1}{c}{$\chi$} &
\multicolumn{3}{c}{260$<M<$425} &
\multicolumn{3}{c}{425$<M<$475} &
\multicolumn{3}{c}{475$<M<$635} &
\multicolumn{3}{c}{$M>$635} \\
\cline{2-4}\cline{5-7}\cline{8-10}\cline{11-13}
\multicolumn{1}{c}{ } & 
\multicolumn{1}{c@{ $\pm$ }}{value} & 
\multicolumn{1}{r@{$\pm$ }}{stat.} & 
\multicolumn{1}{r}{sys.} &
\multicolumn{1}{c@{ $\pm$ }}{value} & 
\multicolumn{1}{r@{$\pm$ }}{stat.} & 
\multicolumn{1}{r}{sys.} & 
\multicolumn{1}{c@{ $\pm$ }}{value} & 
\multicolumn{1}{r@{$\pm$ }}{stat.} & 
\multicolumn{1}{r}{sys.} & 
\multicolumn{1}{c@{ $\pm$ }}{value} & 
\multicolumn{1}{r@{$\pm$ }}{stat.} & 
\multicolumn{1}{r}{sys.} \\
\hline
  1.5 & 5.95 & 0.35 & 0.58 & 7.58 & 0.66 & 2.08 & 10.08 & 0.33 & 0.63
& 11.98 & 0.99 & 0.49 \\ 2.5 & 5.50 & 0.33 & 0.54 & 4.26 & 0.50 & 0.75
& 7.56 & 0.28 & 0.36 & 12.49 & 1.01 & 0.78 \\ 3.5 & 4.59 & 0.30 & 0.31
& 4.96 & 0.53 & 0.67 & 7.83 & 0.29 & 0.33 & 9.11 & 0.86 & 0.61 \\ 4.5
& 4.57 & 0.30 & 0.28 & 5.54 & 0.56 & 1.04 & 7.71 & 0.28 & 0.25 & 9.79
& 0.89 & 0.23 \\ 5.5 & 4.56 & 0.30 & 0.25 & 5.29 & 0.55 & 0.86 & 7.87
& 0.29 & 0.17 & 10.06 & 0.91 & 0.26 \\ 6.5 & 5.10 & 0.32 & 0.23 & 6.26
& 0.60 & 0.73 & 8.17 & 0.29 & 0.16 & 9.58 & 0.88 & 0.51 \\ 7.5 & 5.10
& 0.32 & 0.19 & 4.83 & 0.53 & 0.33 & 8.70 & 0.30 & 0.20 & 9.30 & 0.87
& 0.57 \\ 8.5 & 5.61 & 0.34 & 0.15 & 4.40 & 0.50 & 0.16 & 7.91 & 0.29
& 0.21 & 8.08 & 0.81 & 0.42 \\ 9.5 & 4.93 & 0.32 & 0.09 & 5.60 & 0.57
& 0.25 & 8.46 & 0.30 & 0.24 & 8.96 & 0.85 & 0.30 \\ 10.5 & 6.04 & 0.35
& 0.06 & 5.22 & 0.55 & 0.37 & 8.62 & 0.30 & 0.27 & 10.65 & 0.93 & 0.41
\\ 11.5 & 5.40 & 0.33 & 0.04 & 4.30 & 0.50 & 0.40 & 8.38 & 0.30 & 0.29
\\ 12.5 & 5.33 & 0.33 & 0.08 & 4.75 & 0.52 & 0.52 & 8.69 & 0.30 & 0.36
\\ 13.5 & 5.41 & 0.33 & 0.14 & 5.43 & 0.56 & 0.65 \\ 14.5 & 5.40 &
0.33 & 0.20 & 5.69 & 0.57 & 0.70 \\ 15.5 & 5.60 & 0.34 & 0.28 & 6.18 &
0.60 & 0.76 \\ 16.5 & 4.81 & 0.31 & 0.30 & 4.70 & 0.52 & 0.57 \\ 17.5
& 4.95 & 0.32 & 0.38 & 4.83 & 0.53 & 0.56 \\ 18.5 & 5.78 & 0.34 & 0.53
& 5.01 & 0.54 & 0.55 \\ 19.5 & 5.37 & 0.33 & 0.57 & 5.17 & 0.55 & 0.55
\\
\end{tabular}
\label{TABLE:numbers}
\end{center}
\end{table*}

 Also shown in Fig.~\ref{FIG:theo}, are comparisons to NLO theory
 predictions calculated using $\mu = \Etmax$.  With the large
 angular reach measured, the angular distribution is sensitive to the
 choice of renormalization scale. The QCD theoretical predictions are
 in good agreement with the measured angular distributions.

\subsection{Compositeness Limits}
\label{sec:ang_compositeness}
\begin{figure}[htbp]
   \centerline
   {\psfig{figure=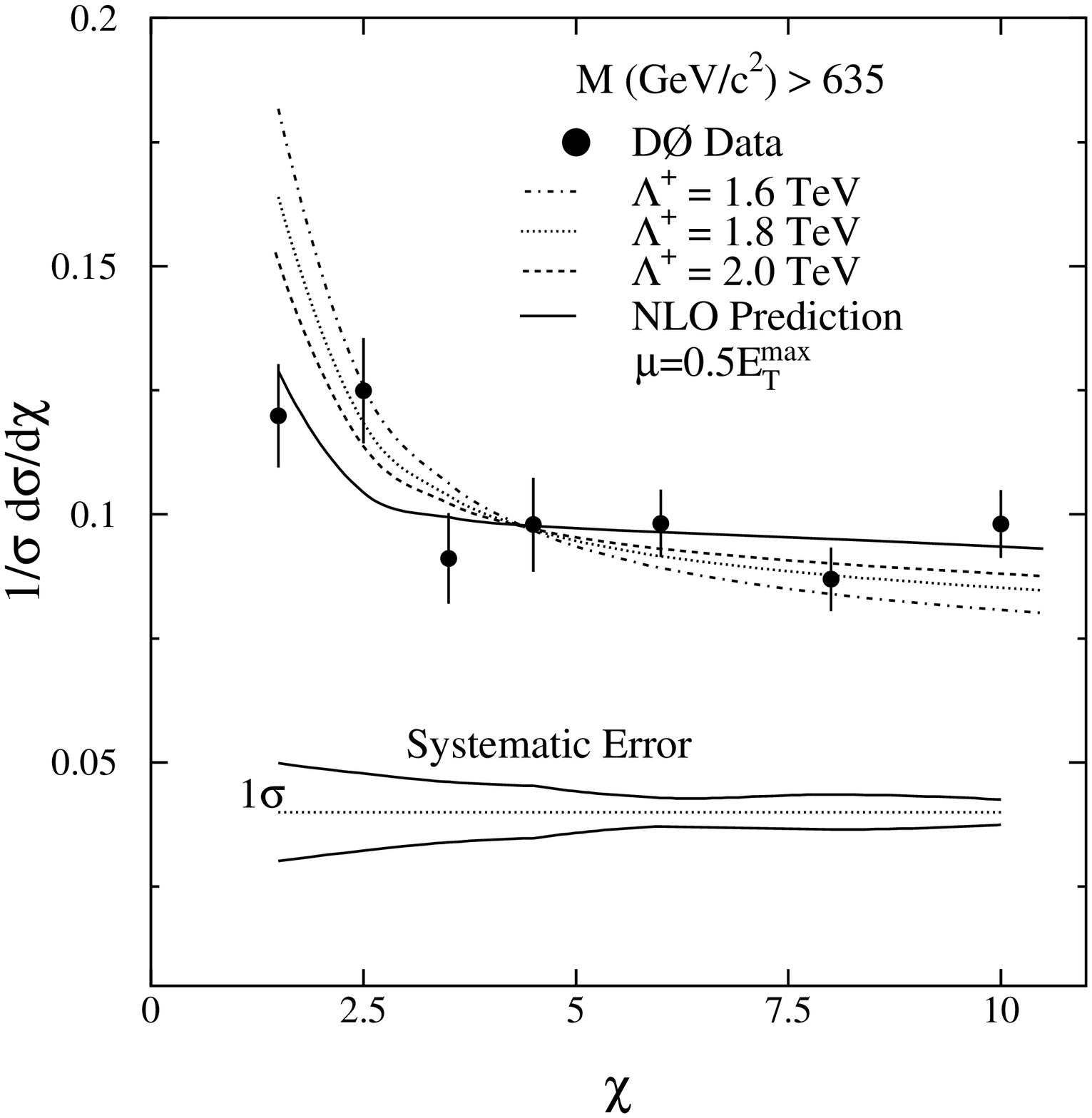,width=3.5in}}
   \caption{Data compared to theory for different compositeness
   scales.  See text for an explanation of the compositeness
   calculation.  The errors on the points are statistical and the band
   represents the $\pm$ 1 $\sigma$ systematic uncertainty.}
   \label{FIG:635}
\end{figure}

 A comparison to theory is made to test for quark compositeness
 (Section~\ref{sec:theory_compositeness}). Predictions of the theory
 of compositeness are available at LO. In order to simulate NLO
 prediction with compositeness, we generated LO curves at various
 values of $\Lambda$.  We measured the fractional differences between
 the LO angular distribution with $\Lambda = \infty$ and those with
 finite $\Lambda$ values.  We then multiplied the NLO prediction of
 the angular distribution by these fractional differences. The results
 are shown in Fig.~\ref{FIG:635} for the mass bin with $\jjmass >
 635$~\gevcc.

 To remove the point-to-point correlated uncertainties, the
 distribution can be characterized by a single number:
 $R_{\chi}=N({\chi < X })/N({ X < \chi < \chi_{\rm max} }$), the ratio
 of the number of events with $\chi < X$ to the number of events
 between $X < \chi < \chi_{\rm max}$ (where $\chi_{\rm max}$ is given
 in Table~\ref{TABLE:range}). The choice of the value of $X$ in the
 definition of $R_{\chi}$ is arbitrary.  To optimize the choice of
 $X$, the following study was performed. The percentage change in the
 largest mass bin between NLO QCD and NLO QCD with a contact term of
 $\Lambda_{LL}^{+} = 2.0$~TeV was measured as a function of the
 definition of $R_{\chi}$.  The change due to compositeness increases
 as one chooses smaller values of $X$.  However, the measurement error
 also increases for smaller values of $X$.  Only the statistical error
 was used to optimize the choice of $X$ so as not to bias the
 optimization with the data.  The ratio of percentage change to
 percentage statistical error peaks at $X=4$; hence we chose
 $R_{\chi}=N({\chi < 4 })/N({ 4 < \chi < \chi_{\rm max} }$).

 To determine the errors on $R_{\chi}$, the nominal value was compared
 to the value after each systematic uncertainty was varied within
 error.  Table~\ref{TABLE:rchi_errors} shows the size of the
 systematic uncertainties for the smallest and largest mass ranges.

\begin{table}[htbp]
\caption{The systematic uncertainties on the measurement of $R_{\chi}$
for the smallest and largest mass ranges.}
\label{TABLE:rchi_errors}
\begin{center}
\begin{tabular}{ccc}
\multicolumn{1}{c}{       } &
\multicolumn{2}{c}{Mass Range}\\
\multicolumn{1}{c}{       } &
\multicolumn{1}{c}{260--425} \gevcc &
\multicolumn{1}{c}{$>$ 635}  \gevcc \\ \hline
Misvertexing 		& 0.24		& 0.001 \\
{\sc mitool} 		& 0.0076 	& 0.000 \\
Jet quality cuts 	& 0.002 	& 0.010 \\
$\eta$ bias 		& 0.007 	& 0.009 \\
Energy scale 		& 0.01 		& 0.023 \\
Resolution 		& 0.004 	& 0.010 \\ 
\end{tabular}
\end{center}
\end{table}

\begin{figure}[htbp]
   \centerline
   {\psfig{figure=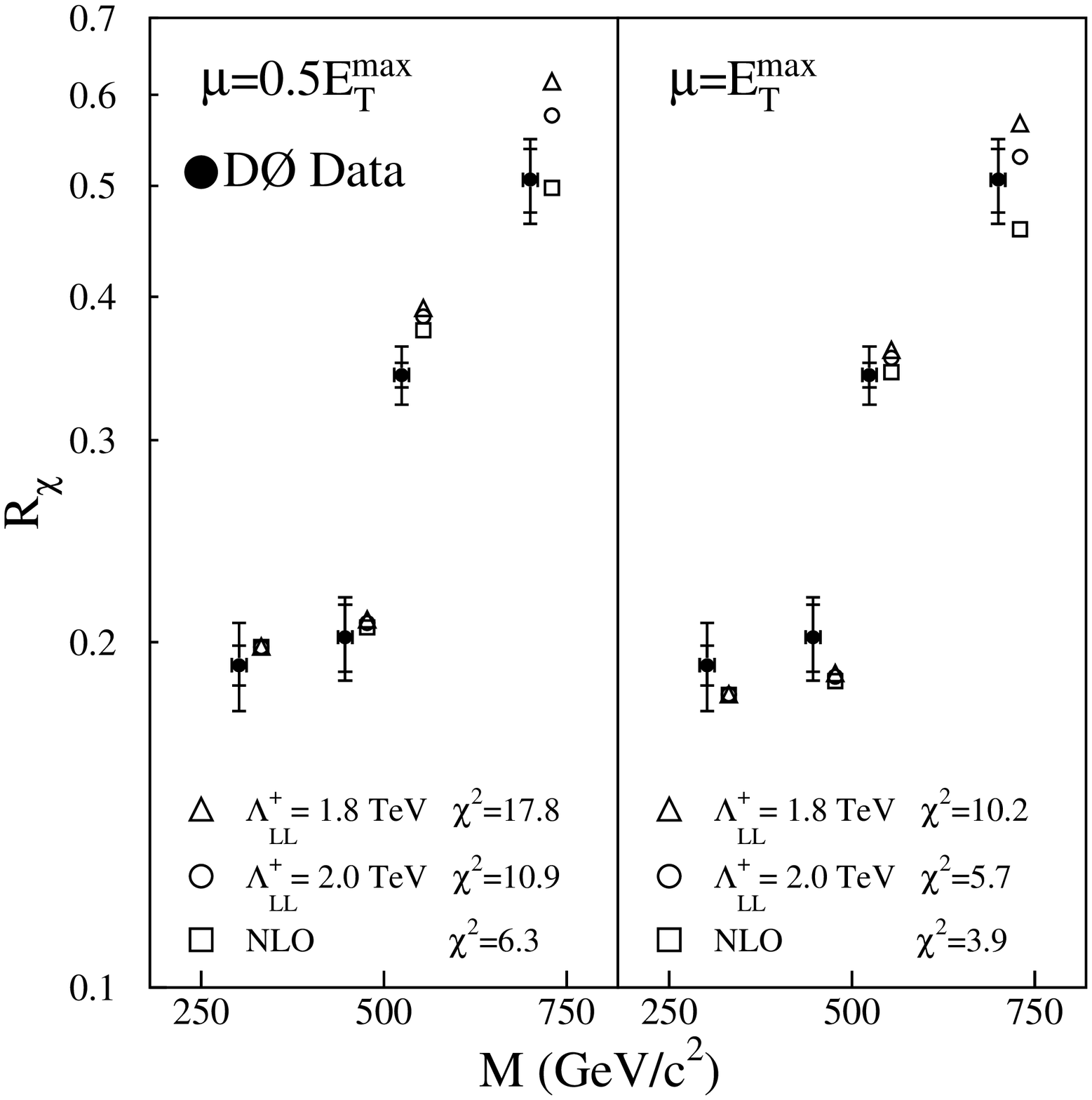,width=3.5in}}
   \caption{$R_{\rm \chi}$ as a function of dijet invariant mass for
   two different renormalization scales.  The inner error bars are the
   statistical uncertainties and the outer error bars include the
   statistical and systematic uncertainties added in quadrature.  The
   $\chi^{2}$ values for the four degrees of freedom are shown for the
   different values of the compositeness scale.  The data are plotted
   at the average mass for each mass range.  The NLO points are offset
   in mass to allow the data points to be seen.}  \label{FIG:Rchi}
\end{figure}

 Table~\ref{TABLE:R} shows the experimental ratio $R_{\rm \chi}$ for
 the different mass ranges with their statistical and their systematic
 uncertainties, which are fully correlated in
 mass. Figure~\ref{FIG:Rchi} shows $R_{\rm \chi}$ as a function of $M$
 for two different renormalization scales, along with the theoretical
 predictions for different compositeness scales. The effects of
 compositeness should be greatest at the highest masses. Note that the
 two largest dijet invariant mass bins have a lower $\chi_{\rm max}$
 value (Table~\ref{TABLE:range}), and thus a higher value of $R_{\rm
 \chi}$ is expected independent of compositeness assumptions.  Also
 shown in Fig. \ref{FIG:Rchi} are the $\chi^{2}$ values for the four
 degrees of freedom for different values of the compositeness scale.

\begin{table}[htbp]
\caption{The dijet angular ratio $R_{\chi}$ and its statistical and
systematic uncertainty. Also listed are the {\sc jetrad} predictions
 with ${\mathcal{R}}_{\rm sep} =$ 1.3, the CTEQ3M PDF, and $\mu =
 0.5\Etmax$ and $\Etmax$.}
\label{TABLE:R}
\begin{center}
\begin{tabular}{ccdd}
\multicolumn{1}{c}{Mass Range} & & \multicolumn{2}{c}{Theory}\\
\multicolumn{1}{c}{\gevcc\ } &
 \multicolumn{1}{c}{$R_{\chi} \pm \rm{Stat.} \pm \rm{Sys.}$ } &
\multicolumn{1}{c}{$\mu=0.5\Etmax$} &
\multicolumn{1}{c}{$\mu=\Etmax$} \\ \hline
260--425 & $0.191 \pm 0.0077 \pm 0.015$ & 0.198  & 0.180 \\
425--475 & $0.202 \pm 0.0136 \pm 0.010$ & 0.206  & 0.185 \\
475--635 & $0.342 \pm 0.0085 \pm 0.018$ & 0.342  & 0.344 \\
$>$635   & $0.506 \pm 0.0324 \pm 0.028$ & 0.506  & 0.458 \\
\end{tabular}
\end{center}
\end{table}

 The method chosen to obtain a compositeness limit uses Bayesian
 statistics~\cite{bayesian}.  The compositeness limit is determined
 using a Gaussian likelihood function for $R_{\chi}$ as a function of
 dijet mass. The likelihood function is defined as
\begin{equation}
 L(\xi) = {\frac{1}{|S|2 \pi}}
\exp\left\{-{\frac{1}{2}} [d-f(\xi)]^{T}V^{-1}[d-f(\xi)]\right\} P(\xi)
\end{equation}
 where $d$ is a 4 component vector of data points for the different
 mass bins, $f(\xi)$ is a 4 component vector of theory points for the
 different mass bins for different values of $\xi$ where $\xi$ is
 related to $\Lambda$ (see below) , $V^{-1}$ is the inverse of the
 covariance matrix, and $P(\xi)$ is the prior probability
 distribution, $P(\xi)$. The covariance matrix is defined so that the
 element $i,j$ of the covariance matrix, $V_{ij}$, is
\begin{equation}
 V_{ij} = \Delta \sigma_i \cdot \Delta \sigma_j ,
\end{equation}
 where $\Delta \sigma$ is the sum of the systematic and statistical
 uncertainties added in quadrature if $i=j$, and the systematic
 uncertainty only if $i \neq j$. The systematic uncertainties are
 assumed to be $100\%$ correlated as a function of mass.

 The compositeness limit depends on the choice of the prior
 probability distribution, $P(\xi)$.  Motivated by the form of the
 Lagrangian, $P(\xi)$ is assumed to be flat in $\xi=1/\Lambda^2$.
 Since the dijet angular distribution at NLO is sensitive to the
 renormalization scale, each renormalization scale is treated as a
 different theory.  To determine the $95\%$ confidence level (C.L.)
 limit in $\Lambda$, a limit in $\xi$ is first calculated by requiring
 that $Q(\xi) =\int_{0}^{\xi}L(R_{\rm \chi}|\xi') d\xi'=
 0.95Q(\infty)$.  The limit in $\xi$ is then transformed into a limit
 in $\Lambda$.  Table~\ref{TABLE:limits} shows the 95\% C.L. limit for
 the compositeness scale obtained for different choices of
 models. These results supersede those reported in~\cite{d0_angular}
 following the correction of an error in the program used to calculate
 the effects of compositeness in that paper. The resulting limits are
 reduced by approximately 150~GeV. If the prior distribution is
 assumed to be flat in $1/\Lambda^4$, the limits are slightly reduced,
 as shown in Table~\ref{TABLE:limits2}.

\begin{table}[htbp]
\begin{center}
\caption{The 95\% confidence level limits for the
left-handed contact compositeness scale for different
models.  The prior probability distribution is assumed
to be flat in 1/$\Lambda^2$.}
\begin{tabular}{ccc}
\multicolumn{1}{c}{Compositeness scale} &
\multicolumn{1}{c}{$\mu = \Etmax$} &
\multicolumn{1}{c}{$\mu = 0.5 \Etmax$} \\
\hline
$\Lambda_{LL}^{+}$     & 2.0 TeV & 2.1 TeV \\
$\Lambda_{LL}^{-}$     & 2.0 TeV & 2.2 TeV \\
$\Lambda_{LL(ud)}^{+}$ & 1.8 TeV & 1.8 TeV \\
$\Lambda_{LL(ud)}^{-}$ & 1.8 TeV & 2.0 TeV \\
\end{tabular}
\label{TABLE:limits}
\end{center}
\end{table}

\begin{table}[htbp]
\begin{center} 
\caption{The 95\% confidence level limits for the
left-handed contact compositeness scale for different models.  The
prior probability distribution is assumed to be flat in
1/$\Lambda^4$.}
\begin{tabular}{ccc}
\multicolumn{1}{c}{Compositeness scale} &
\multicolumn{1}{c}{$\mu = \Etmax$} &
\multicolumn{1}{c}{$\mu = 0.5\Etmax$} \\
\hline
$\Lambda_{LL}^{+}$     & 2.0 TeV & 2.0 TeV \\
$\Lambda_{LL}^{-}$     & 1.9 TeV & 2.1 TeV \\
$\Lambda_{LL(ud)}^{+}$ & 1.7 TeV & 1.7 TeV \\
$\Lambda_{LL(ud)}^{-}$ & 1.7 TeV & 1.9 TeV \\
\end{tabular}
\label{TABLE:limits2}
\end{center}
\end{table}

 Recently published results from CDF~\cite{cdf_angular} on dijet
 angular distributions compare to the model in which all quarks are
 composite, yielding 95$\%$ confidence limits $\Lambda_{LL}^{+}>1.8$
 TeV and $\Lambda_{LL}^{-}>1.6$ TeV.

\subsection{Coloron Limits}

\begin{figure}[hbtp]
\vbox{\centerline
{\psfig{figure=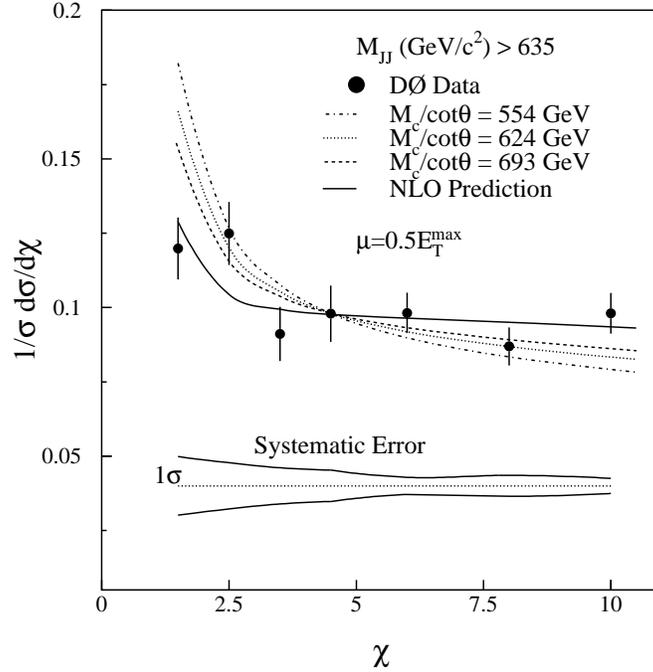,width=3.5in}}}
\caption{D\O\ data compared to theory for different values of 
 \mbox{$M_{c} / \cot{\theta}$} (see text for details of the 
 coloron distribution calculation). The errors on the points are
 statistical and the band represents the correlated $\pm 1\sigma$
 systematic uncertainty.}
\label{coloron_angular_fig_1}
\end{figure}

 Predictions of the dijet angular distribution with colorons are
 available at LO (Section~\ref{sec:theory_coloron}). To simulate NLO
 predictions, coloron LO predictions are generated for several values
 of \mbox{$M_{c} / \cot{\theta}$}. The fractional differences between
 the angular distribution with \mbox{$M_{c} / \cot{\theta} = \infty$}
 and the distributions with finite values of \mbox{$M_{c} /
 \cot{\theta}$} are measured. The coloron NLO predictions are then
 obtained by multiplying the NLO QCD prediction obtained using {\sc
 jetrad} by the LO fractional differences obtained above. The results
 are shown in Fig.~\ref{coloron_angular_fig_1}.

\begin{figure}[hbt]
\vbox{\centerline
{\psfig{figure=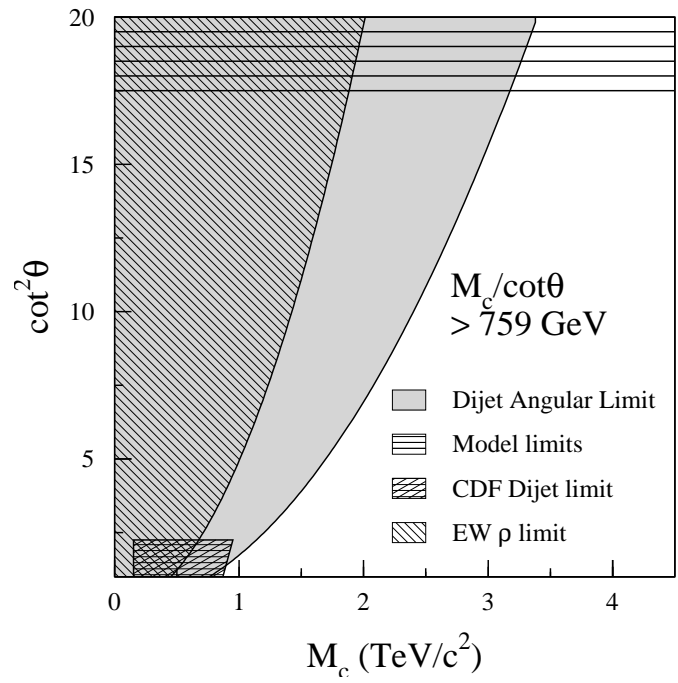,width=3.5in}}}
\caption{ Limits on the coloron parameter space: coloron mass $M_c$
vs.  mixing parameter $\cot{\theta}$.  }
\label{coloron_angular_fig_2}
\end{figure}

 Limits on the coloron mass are calculated using the same method as in
 the previous section.  For a renormalization scale of $\mu = \Etmax$,
 the 95$\%$ C.L. limit on the coloron mass is \mbox{$M_{c} /
 \cot{\theta} > 759$ GeV/$c^2$}. If $\mu = 0.5\Etmax$, the 95$\%$
 C.L. limit is \mbox{$M_{c} / \cot{\theta} > 786$ GeV/$c^2$}. The
 resulting limits are shown in Fig.~\ref{coloron_angular_fig_2}. The
 shaded region shows the 95$\%$ C.L.  exclusion region for the D\O\
 dijet angular distribution measurement (\mbox{$M_{c} / \cot{\theta} >
 759$ GeV/$c^2$}). The horizontally-hatched region at large
 $\cot{\theta}$ is excluded by the
 model~\cite{coloron_1,coloron_2}. The diagonally-hatched region is
 excluded by the value of the weak-interaction $\rho$ parameter
 (\mbox{$M_{c} / \cot{\theta} > 450$ GeV/$c^2$})~\cite{coloron_2}. The
 cross-hatched region is excluded by the CDF search for new particles
 decaying to dijets~\cite{cdfxjj}.  These limits are then converted
 into more general limits on color-octet vector current-current
 interactions: $\Lambda^{-}_{V_8} > 2.1$ TeV assuming $\als(M_Z) =
 0.12$ (Section~\ref{sec:theory_compositeness}).

\subsection{Conclusions}

 We have measured the dijet angular distribution over a large angular
 range.  The data distributions are in good agreement with NLO QCD
 predictions.  The compositeness limits depend on the choice of the
 renormalization/factorization scale, the model of compositeness, and
 the choice of the prior probability function.  Models of quark
 compositeness with a contact interaction scale of less than 2 TeV are
 ruled out at the $95\%$ C.L.
\clearpage

\section{The Inclusive Dijet Mass Spectrum}
\label{sec:dijet_mass}

 The dijet mass spectrum is calculated using the relation:
\begin{equation}
 \kappa \equiv \frac{d^{3} \sigma}{ d \jjmass d \eta_{1} d \eta_{2}} = 
 \frac { N_{i} C_{i} }
  {{\cal L}_{i} 
  \epsilon_{i} \Delta \jjmass  \Delta \eta_{1}  \Delta \eta_{2}},
 \label{EQ:dijet_cross_section}
\end{equation}
 where $N_i$ is the number of events in mass bin $i$; $C_i$ is the
 unsmearing correction; ${\cal L}_{i}$ is the integrated luminosity;
 $\epsilon_{i}$ is the efficiency of the trigger, vertex selection,
 and the jet quality cuts; $\Delta \jjmass$ is the width of the mass
 bin; and $\Delta \eta_{1,2}$ are the widths of the pseudorapidity
 bins for jets 1 and 2.  The dijet mass is calculated assuming
 massless jets using Eq.~\ref{eq:dijet_mass}. If we define the mass
 using four vectors, $m^2 = {(E_1+E_2)^2 - ({\vec{p_1}} +
 {\vec{p_2}})^2}$, the cross section changes by less than 2$\%$.

 \subsection{Data Selection}                                 

\begin{figure}[hbtp]
\vbox{\centerline
{\psfig{figure=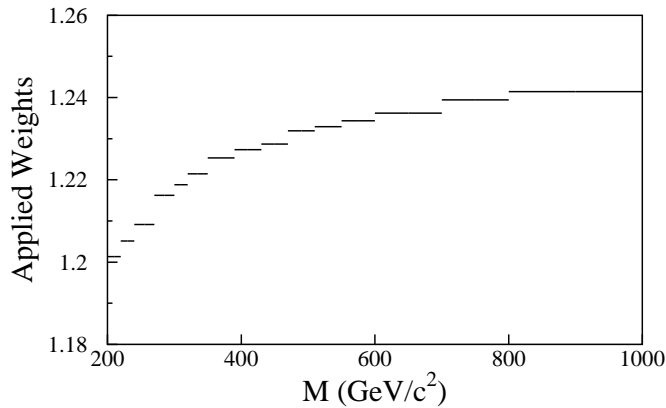,width=3.5in}}}
\caption{The reciprocal of the event efficiencies $\epsilon_i$ for each 
mass bin.}
\label{sec_40:fig_weight}
\end{figure}

 The selected data are events with two or more jets which satisfy the
 set of inclusive jet triggers and pass the standard jet and event
 quality requirements (see Section~\ref{sec:quality_cuts}).  Events
 are removed unless both of the two leading jets pass the jet quality
 requirements.  The vertex of the event must be within 50 cm of $z=0$.
 The efficiency for each event is then given by the product of the
 efficiencies ($\epsilon_{{\rm jet}i}$) of the jet quality cuts, the
 efficiency ($\epsilon_{\rm met}$) of the cut on $\met$, the
 efficiency ($\epsilon_{\rm trigger}$) for an event to pass the
 trigger, and the efficiency ($\epsilon_{\rm vertex}$) for passing the
 vertex cut.
 The reciprocal of the resulting efficiencies (the event weights) is
 plotted as a function of \jjmass\ in
 Fig.~\ref{sec_40:fig_weight}. The efficiency of the vertex
 requirement is $90 \pm 1\%$.  The data are used to select a sample
 where both jets have pseudorapidity $\modetajet < 1.0$. To examine
 the inclusive dijet cross section more closely, two sub-samples are
 created where both jets satisfy either $\modetajet < 0.5$ or $0.5 <
 \modetajet < 1.0$.

\begin{figure}[htbp]
\vbox{\centerline
{\psfig{figure=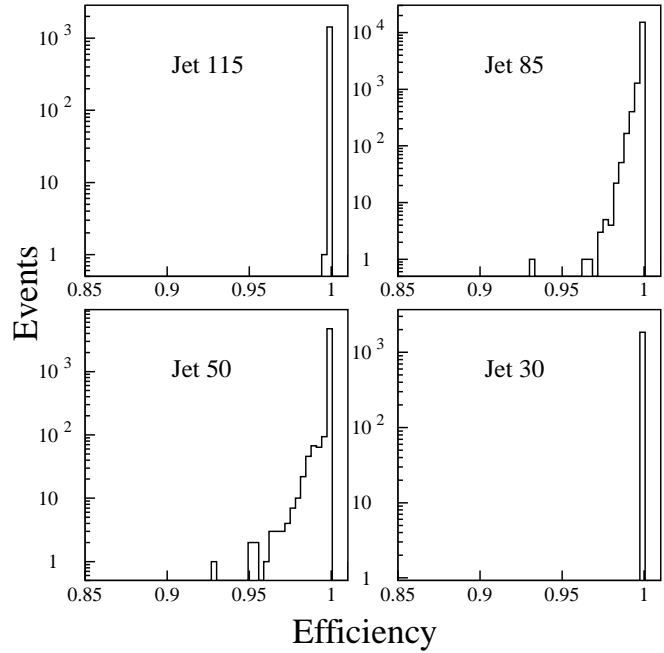,width=3.5in}}}
\caption{The trigger efficiencies of the events included in the dijet mass 
spectrum. Note that most events have an efficiency greater than
$99\%$.}
\label{FIG:mass-efficiencies}
\end{figure}

\begin{figure}[htbp]
\vbox{\centerline
{\psfig{figure=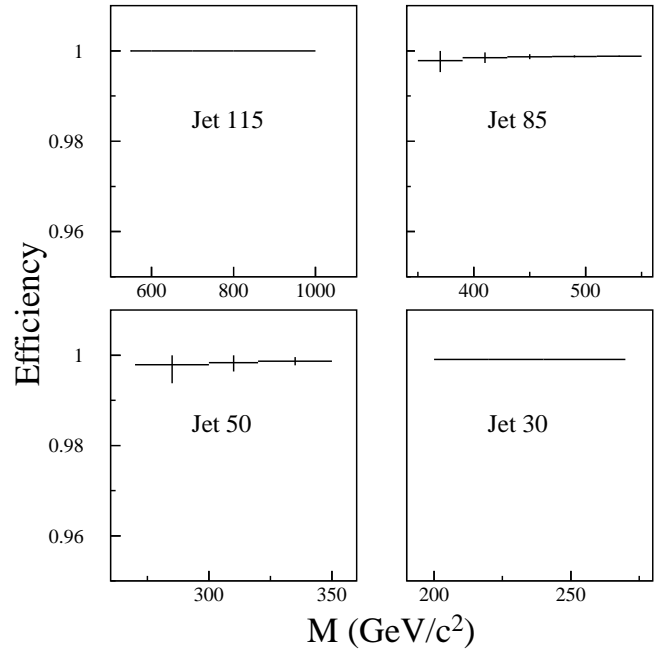,width=3.5in}}}
\caption{The average trigger efficiencies for each trigger
 as a function of mass. }
\label{FIG:mass-efficiencies_new}
\end{figure}

 To determine the mass at which a given trigger (Jet\_XX) becomes
 fully efficient, the event efficiencies are plotted for each of the
 triggers in the chosen mass ranges (Table \ref{TABLE:mass_table1})
 in Figs.~\ref{FIG:mass-efficiencies} and
 ~\ref{FIG:mass-efficiencies_new}. This plot shows that the triggers
 are $ > 95\%$ efficient for most of the data.
 
\begin{figure}[hbtp]
\vbox{\centerline
{\psfig{figure=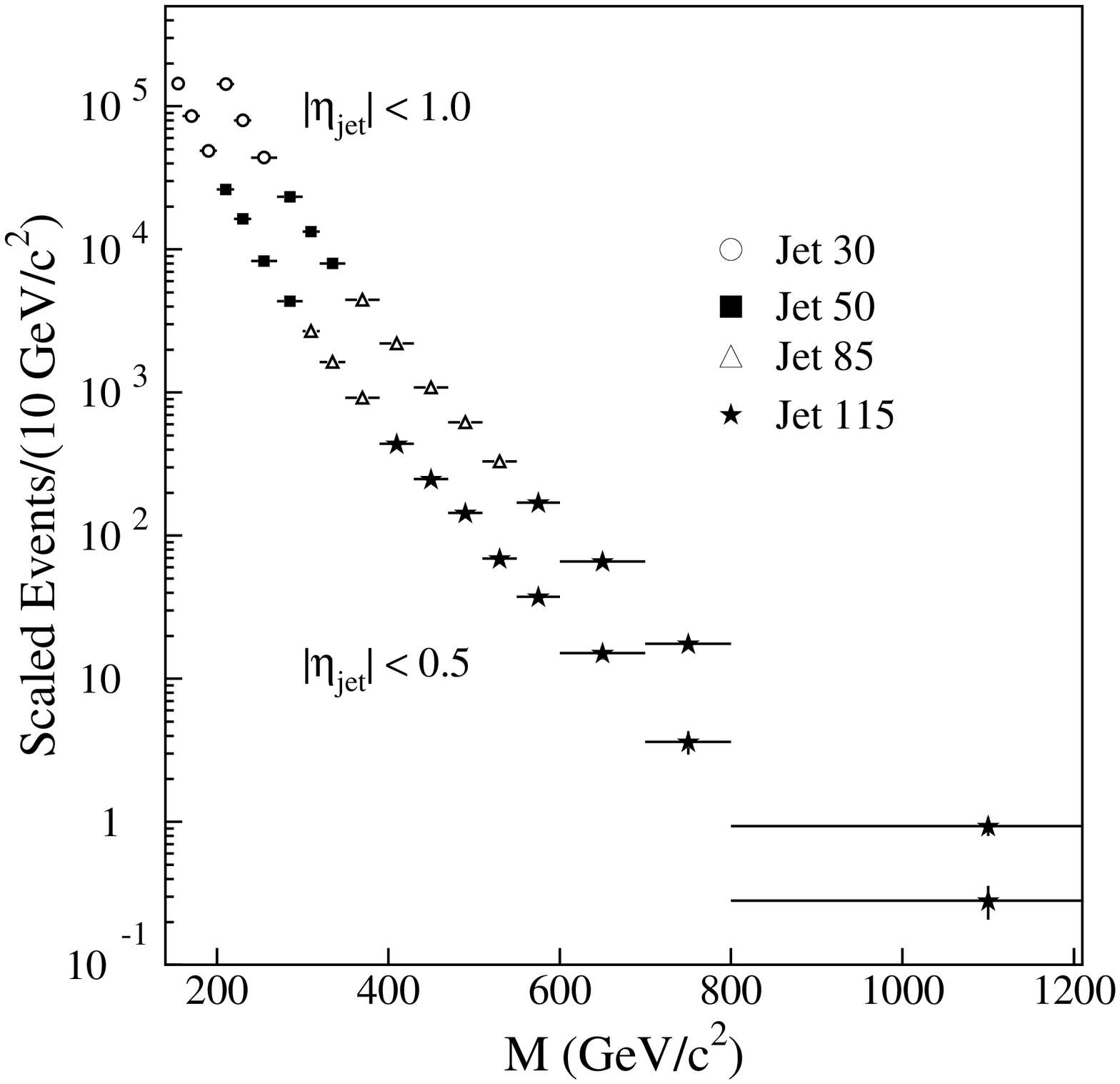,width=3.5in}}}
\caption{The scaled dijet mass spectrum for $\modetajet < 1.0$ 
 and $\modetajet < 0.5$.}
\label{FIG:mass_spectrum}
\end{figure}

 The mass spectra obtained from the triggers Jet\_30, Jet\_50 and
 Jet\_85 are then scaled to match the Jet\_115 mass spectrum by the
 scale factor $\displaystyle{S_{\rm Filt}}$ which is given by
\begin{equation}
\label{scale_factor}
S_{{{\rm{Filt}}}_i} = 
\left( \frac
{{\cal L}_{{{\rm{Filt}}}_i} }
{{\cal L}_{{{\rm{Jet\_115}}}} }
\right)
\times
\frac{1}{\epsilon_{{\rm{vertex}}_{i}}}.
\end{equation}
 The values of $S_{\rm Filt}$ used to scale the data in this analysis
 are given in Table~\ref{TABLE:mass_table1}. These scales (and the
 event weights) are then applied to the data to produce the mass
 spectra (two such spectra are depicted in
 Fig.~\ref{FIG:mass_spectrum}). The error plotted for each point is
 given by the statistical errors for that bin.

\begin{table}[htbp]
 \begin{center} \caption{The mass ranges and scale factors for the
 triggers used in this analysis.} 
 \label{TABLE:mass_table1}
 \begin{tabular}{ccc} 
 \multicolumn{3}{c}{Data Sets satisfying $\modetajet < 1.0$,
  and $0.5 < \modetajet < 1.0$. }\\ 
 \hline 
 Trigger Name & Mass Range & \multicolumn{1}{c}{Scaling Factor} \\ 
	      & (\gevcc )  & \multicolumn{1}{c}{$S_{\rm Filt}$} \\
 \hline \hline
 Jet\_30  & 200--270  &  $289.3   \pm 14.4 $\\
 Jet\_50  & 270--350  &  $ 21.7   \pm 1.1  $\\
 Jet\_85  & 350--550  &  $  1.845 \pm 0.005$\\
 Jet\_115 & 550--1400 &  $  1.095 \pm 0.009$ \\
 \hline \hline
 \multicolumn{3}{c}{}\\
 \hline\hline
 \multicolumn{3}{c}{Data Set satisfying $\modetajet < 0.5$.}\\
 \hline
 Trigger Name     &   Mass Range   & \multicolumn{1}{c}{Scaling Factor} \\
                 &    (\gevcc )       & \multicolumn{1}{c}{$S_{\rm Filt}$} \\
 \hline \hline 
 Jet\_30  & 150--200  & $ 289.3   \pm 14.4 $ \\
 Jet\_50  & 200--300  & $  21.7   \pm 1.1  $ \\
 Jet\_85  & 300--390  & $   1.845 \pm 0.005$ \\
 Jet\_115 & 390--1400 & $   1.095 \pm 0.009$  \\
\end{tabular}
\end{center}
\end{table}

\subsection{Vertex Selection Biases}

 The vertex selection procedure chooses the vertex with the smallest
 value of \HT\ (Section~\ref{sec:vertex}). This selection criterion
 may be biased for events where both of the two leading jets have the
 same absolute rapidity.  In this case the vertex chosen would be the
 one that minimizes the \Et\ for both of the leading jets and not
 necessarily the correct one. This bias was studied using the {\sc
 pythia}~\cite{pythia} MC event generator to generate events with
 multiple vertices at the same rate as the Jet\_85 and Jet\_115
 triggers. For dijet events with $\modetajet < 1.0$ the number of
 incorrectly chosen vertices is 5$\%$.

\begin{figure}[hbtp]
\vbox{\centerline
{\psfig{figure=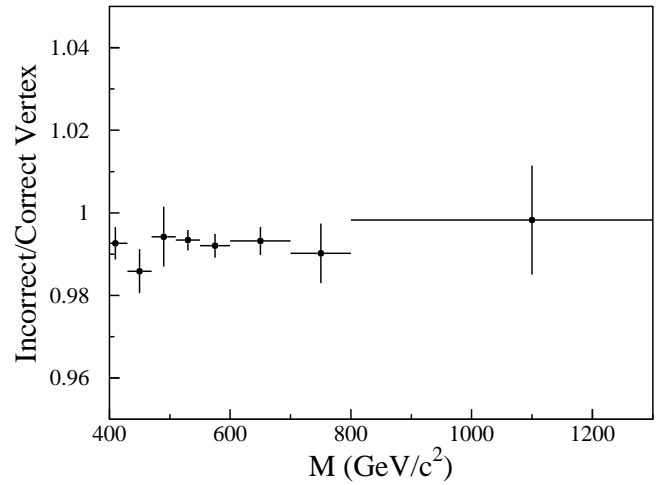,width=3.5in}}}
  \caption{The effect on the mass spectrum of incorrectly identified
  vertices for $\modetajet < 1.0$. }
  \label{sec2:fig0b}
\end{figure}

 The effect on the dijet mass cross section is measured by calculating
 the ratio of the mass spectrum produced using the selected vertex to
 that of the correct vertex. The result of this calculation is given
 in Fig.~\ref{sec2:fig0b} for $\modetajet < 1.0$ and shows that the
 effect is of the order of $1\%$ and that it is reasonably uniform as
 a function of mass. A $2\%$ uncertainty in the cross section,
 uncorrelated as a function of mass, was assumed.

\subsection{Dijet Mass Resolution}

\begin{figure}[hbtp]
\vbox{\centerline
{\psfig{figure=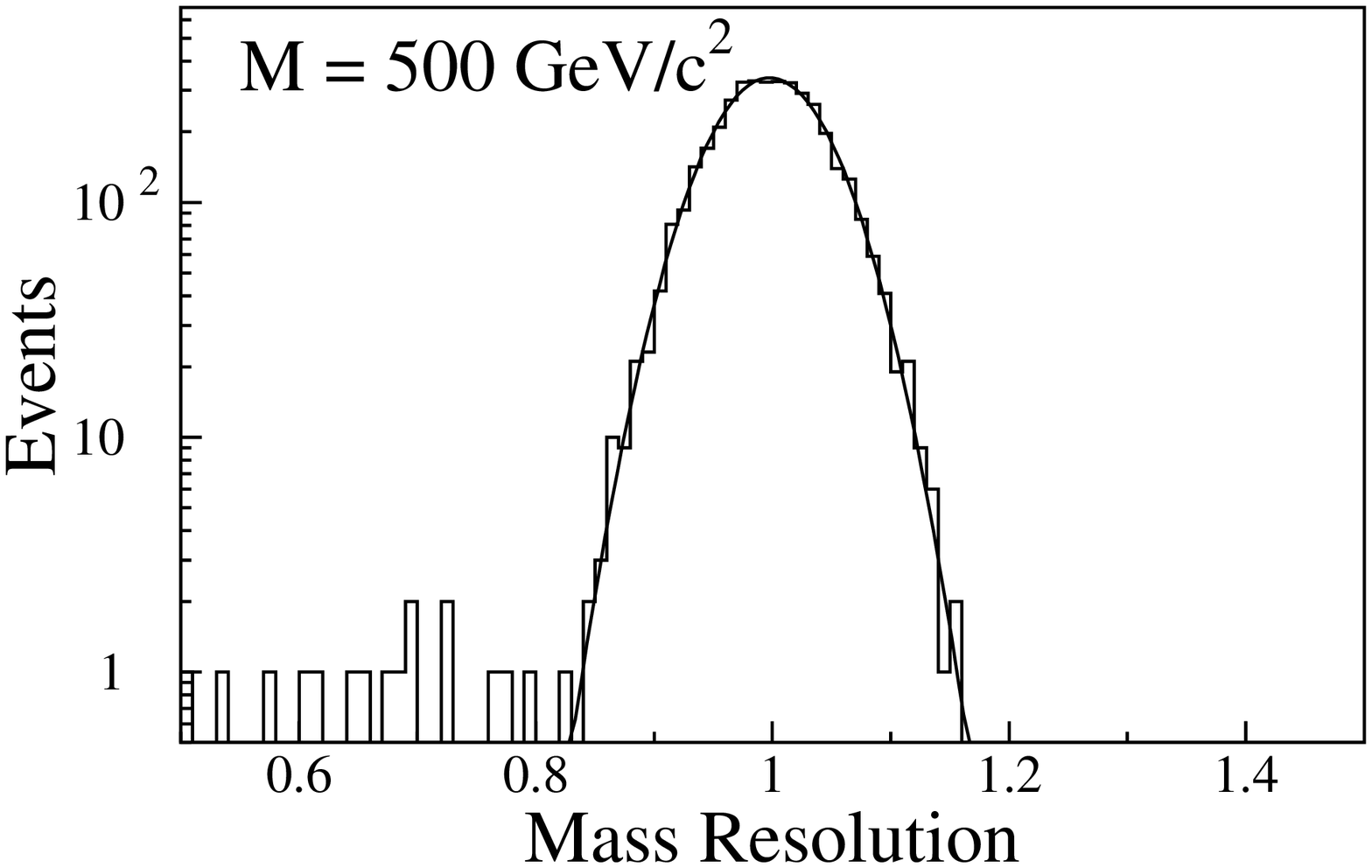,width=3.5in}}}
\caption{{{The distribution of $\displaystyle{\jjmass^{{\rm
 smeared}}/\jjmass^{{\rm unsmeared}}}$ for} {\sc
pythia}-generated events with {$\modetajet < 1.0$ and}
$\displaystyle{490 < \jjmass^{{\rm unsmeared}} < 510}$ GeV/$c^2$.}}
\label{FIG:dijet_mass_reolution_fit}
\end{figure}

\begin{figure}[htbp]
\vbox{\centerline
{\psfig{figure=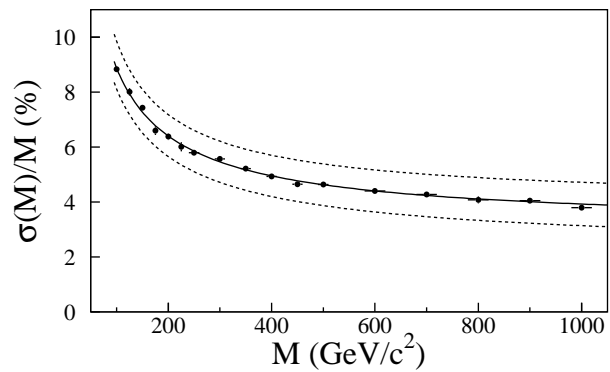,width=3.2in}}}
\caption{The mass resolutions for $\modetajet < 1.0$ generated using
the {\sc pythia} MC.  The solid curve and the MC data points show the
resolutions determined using the nominal jet energy resolutions; the
dashed lines show the resolutions determined with the $\pm1\sigma$ jet
energy resolution uncertainties.  }
\label{FIG:dijet_mass_reolution_fit_2}
\end{figure}

 The dijet mass resolutions were calculated using the measured single
 jet resolutions (Section \ref{SEC:jet_resolutions}). The dijet mass
 resolutions depend on the \Et\ and $\eta$ values of the two leading
 \Et\ jets in each event. Hence the mass resolutions are determined by
 using a Monte Carlo (MC) event generator to convolute the measured
 single jet resolutions (Table~\ref{Table:resolution_parameters}).
 For each MC event generated, the individual particle jets are smeared
 by the measured single jet resolutions. The unsmeared and smeared
 dijet masses are calculated and used to determine the mass smearing.
 The values of the mass smearing are plotted in discrete mass bins and
 fitted to a Gaussian ansatz (see
 Fig.~\ref{FIG:dijet_mass_reolution_fit} for an example). The
 distribution is well-represented by a Gaussian with only a small
 fraction ($\ll 1\%$) of events forming a tail (due to events where
 the jets are reordered after smearing). The resolution at each of
 these masses is given by the width of the Gaussian. The results
 obtained for $\modetajet < 1.0$ using the {\sc pythia} MC
 \cite{pythia} are plotted in
 Fig.~\ref{FIG:dijet_mass_reolution_fit_2}.

 The mass resolutions are then fitted using the functional form:
\begin{equation}
\sigma(M)/M (\%) =  {\sqrt{A + B M
 + C{M}^{2}
 + D{M}^{3}}} .
\end{equation}
 The results are depicted in
 Fig.~\ref{FIG:dijet_mass_reolution_fit_2}. The resulting fit
 parameters for all $\eta$ regions considered in this analysis are
 given in Table~\ref{TABLE:mass_resolutions_table_2}.

\begin{table*}[htbp]
 \begin{center}
  \caption{ Parameterizations of the mass resolutions (in percent)
    generated using  the {\sc pythia} MC.}
   \label{TABLE:mass_resolutions_table_2}
   \begin{tabular}{lr@{ $\pm$ }rr@{ $\pm$ }rr@{ $\pm$ }rr@{ $\pm$ }r}
   Data Set & \multicolumn{2}{c}{$A$} & \multicolumn{2}{c}{$B$} &
   \multicolumn{2}{c}{$C$} & \multicolumn{2}{c}{$D$} \\
   \hline\hline
 $\displaystyle{\modetajet < 1.0}$                                      
& 3.40 & 1.01 & 0.761 & 0.045 & 0.0302 & 0.0032 & 0.0002 & 0.0005 \\
 $\displaystyle{\modetajet < 0.5}$
& 3.78 & 0.94 & 0.701 & 0.041 & 0.0231 & 0.0025 & 0.0 & 0.0003 \\
 $\displaystyle{0.5 < \modetajet < 1.0}$
& 5.24 & 0.83 & 0.709 & 0.058 & 0.0389 & 0.0022 & 0.0 & 0.0004 \\
   \end{tabular}
 \end{center}
\end{table*}

 The mass resolution dependence on the MC generator used to convolute
 the single jet resolutions has been estimated by using the {\sc
 herwig}~\cite{herwig} and {\sc jetrad} event generators. The {\sc
 jetrad} program is used at LO with renormalization scales of $\mu =
 (0.25, 0.5, 1.0)\Etmax$ and $\mu = (0.25, 0.5, 1.0)\sqrt{\hat{s}}$
 with \cteqthreem~\cite{cteq3m}. To ensure that the choice of PDF does
 not affect the resolutions, {\sc jetrad} was run with the
 \mbox{CTEQ3L} and \mrsap~\cite{mrsap} PDFs.

\begin{figure}[htbp]
\vbox{\centerline
{\psfig{figure=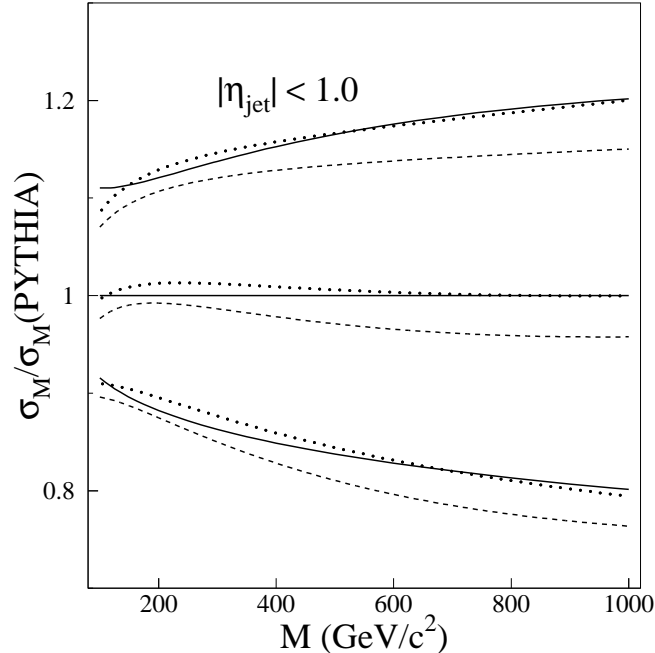,width=3.5in}}}
\caption{ A comparison of the mass resolutions for $\modetajet < 1.0$
obtained by running {\sc pythia} (solid line), {\sc herwig} (dashed
line), and {\sc jetrad} (dotted line) with $\mu = 0.5\Etmax$ and the
\cteqthreem\ PDF. The resolutions are divided by the nominal {\sc
pythia} resolutions. The upper and lower curves show the effect of the
$\pm1\sigma$ uncertainties on the measured jet resolutions.  }
\label{FIG:compare_mass_resolutions}
\end{figure}

\subsection{Data Unsmearing}

 The jet energy scale corrects only the average response to a jet. The
 steeply falling dijet mass spectrum is distorted by the jet energy
 resolution and, to a negligible extent, by the $\eta$ resolution.
 The observed mass spectrum is corrected for resolution smearing by
 assuming a trial unsmeared spectrum
\begin{eqnarray}
F(M^{\prime}) & = & 
 N \cdot {M^{\prime}}^{\displaystyle - \alpha}
 \left( 1 - \frac{M^{\prime}}{\sqrt{s}}
                \right)^{\displaystyle -\beta},
\label{param_1}
\end{eqnarray}
 which is convoluted with the measured mass resolutions
\begin{eqnarray}
f(M) & = & 
\int F(M^{\prime})\sigma\left(M^{\prime} - M, M^{\prime}\right) dM^{\prime},
\end{eqnarray}
 such that the number of events in any mass bin $i$ is given by
 integrating $f$ over that bin.  The data were fitted using a binned
 maximum likelihood method and the {\sc minuit} package~\cite{minuit}
 to determine the values of $N$, $\alpha$, and $\beta$. The smearing
 correction is given by
\begin{equation}
C_i = { { \int F(M) dM } \over { \int f dM} }.
\end{equation}
 To account for any uncertainties in the choice of trial function the
 data are fitted with two additional functions:
\begin{eqnarray}
F(M) & = &  N \cdot M^{\displaystyle -\alpha} \left[ 1 - \frac{M}{\sqrt{s}} 
       + \gamma \left( \frac{M}{\sqrt{s}} \right)^{2} 
 \right]^{\displaystyle -\beta}, \label{param_2} \\
\nonumber \\
F(M) & = & N \cdot M^{\displaystyle -\alpha}
	 \exp{\left[ -\beta \left(\frac{M}{100}\right) 
         -\gamma  \left(\frac{M}{100}\right)^{2} \right]}.
	 \label{param_3} \
\end{eqnarray}

\begin{figure}[htbp]
\vbox{\centerline
{\psfig{figure=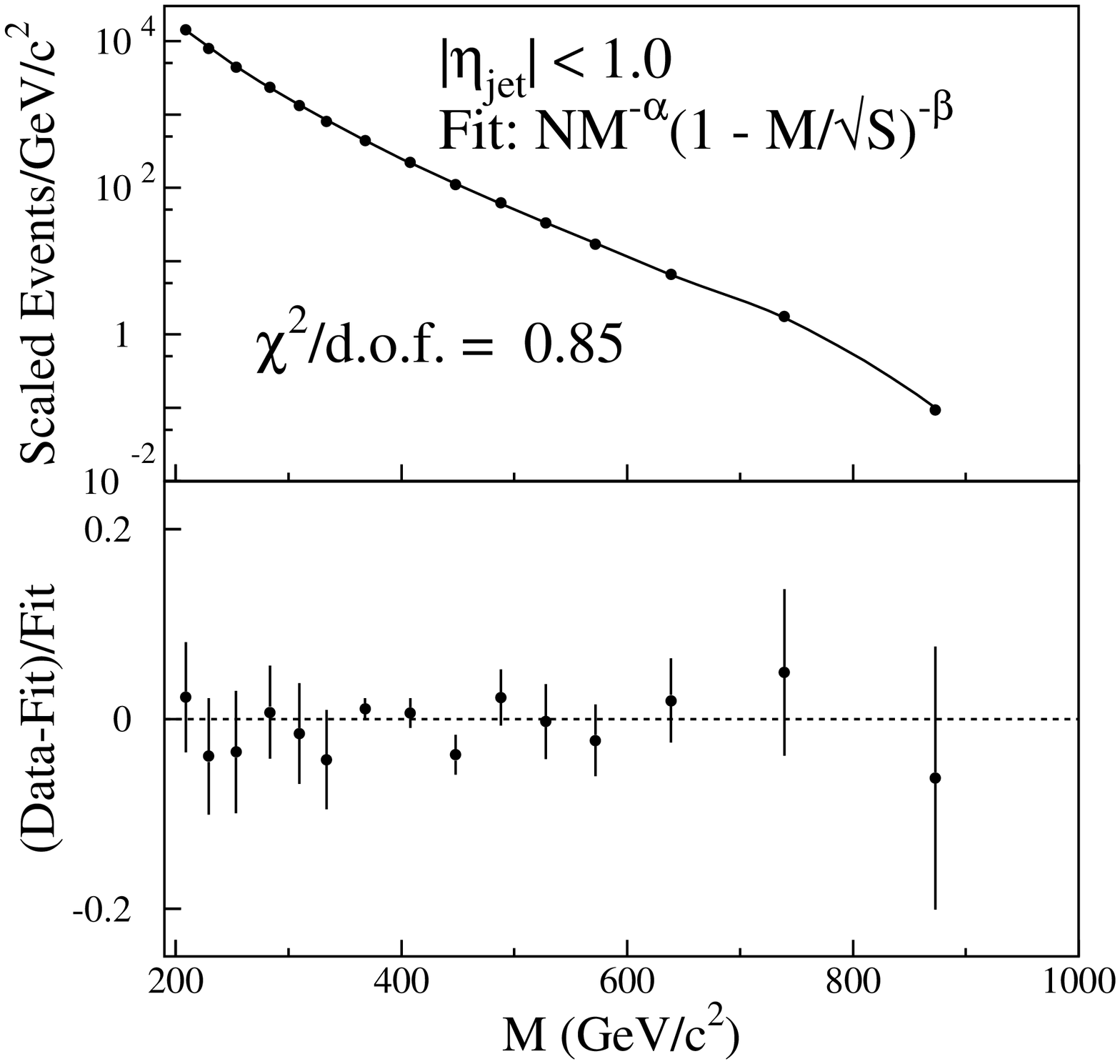,width=3.5in}}}
\caption{Top: The fit to the data for $\modetajet < 1.0$ using
Eq.~\protect{\ref{param_1}}. Bottom: The residuals of the fit are
plotted, (Data $-$ Fit)/Fit.}
\label{sec_2:fig_14}
\end{figure}

\begin{figure}[htbp]
\vbox{\centerline
{\psfig{figure=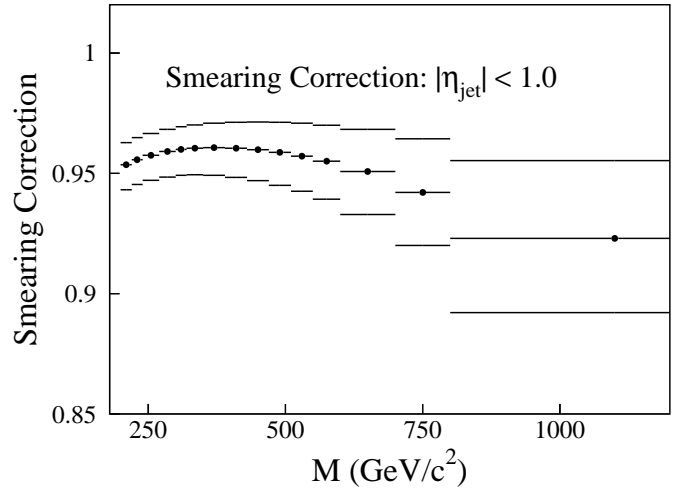,width=3.5in}}}
\caption{The smearing correction factor to be applied to the data for
 $\modetajet < 1.0$. The upper and lower curves represent the $\pm
 1\sigma$ uncertainties of the smearing correction. }
\label{sec_2:fig_15}
\end{figure}

 The nominal smearing correction is given by the fit to the data using
 the trial function given in Eq.~\ref{param_1} and the obtained mass
 resolutions (Table~\ref{TABLE:mass_resolutions_table_2}). The
 resulting fit for $\modetajet < 1.0$ has a $\chisq\ = 10.3$ for 13
 degrees of freedom and is given in Fig.~\ref{sec_2:fig_14}. The
 magnitude of the correction is approximately 5$\%$ at 209 \gevcc\ and
 drops to approximately 2$\%$ at 500 \gevcc , and then rises to 8$\%$
 at 1 \tevcc . The uncertainty in the smearing correction is obtained
 by fitting the data with each of the trial functions and all of the
 mass resolutions generated with the different MC generators. The
 error is given by the maximum and minimum corrections obtained for
 each mass bin and is approximately 2$\%$. The resulting smearing
 corrections are shown in Fig.~\ref{sec_2:fig_15}.

\subsection{Energy Scale Corrections}

\begin{figure}[htb]
\vbox{\centerline
{\psfig{figure=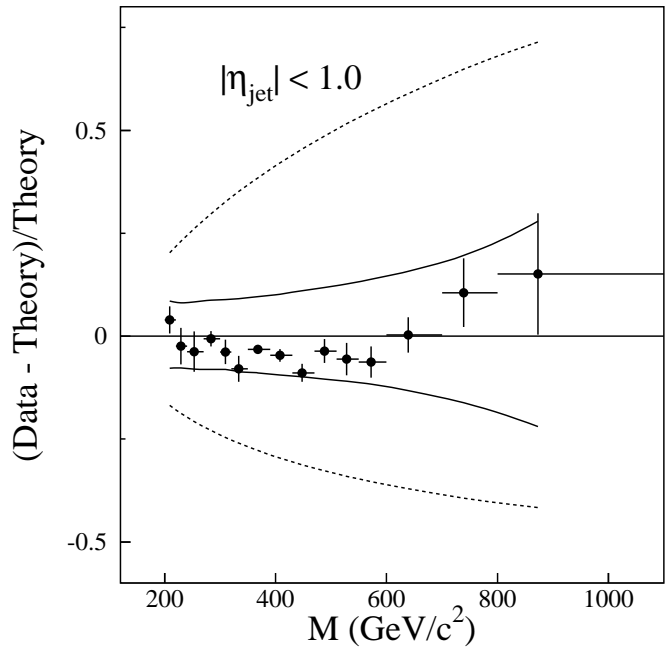,width=3.5in}}}
 \caption{ A comparison of the data and the mass spectrum produced by
 the Monte Carlo to study the energy scale uncertainties. The solid
 curves show the $\pm1\sigma$ energy scale uncertainties. The dashed
 curves show the MC predictions where the weights are set to
 $\jjmass^{-4.5}$ (upper) and $\jjmass^{-5.0}$ (lower).}
 \label{sec_2:fig_18}
\end{figure}

 The uncertainties in the dijet mass cross section due to the energy
 scale have also been determined using a Monte Carlo program. The MC
 program generates two initial state partons each with a uniform
 distribution in $x$ (the fraction of the proton momentum carried by
 the parton). The kinematic quantities of the two jets that result
 from this interaction are determined by generating a random value of
 $\chi$ from a uniform distribution. The jet \Et\ and $\eta$ are
 uniquely determined for the event; it is assumed that the two jets
 are back-to-back in $\phi$. The event is accepted if it satisfies the
 requirement $\modetajet < 1.0$. Each event is weighted by
 $\jjmass^{-4.75} \times \Pr(x_{1}) \times \Pr(x_{2})$ where
 $\Pr(x_{1,2})$ is the probability of finding a parton with momentum
 fraction $x$ in the \cteqfourm\ PDF~\cite{cteq4m}. The exponent,
 -4.75, was chosen to obtain a dijet mass spectrum with similar
 normalization and shape as the data. Finally, each of the resulting
 jets has its \Et\ smeared by the measured single jet
 resolutions. Figure~\ref{sec_2:fig_18} shows a comparison of the mass
 spectrum produced by the Monte Carlo and the data; the two are in
 reasonable agreement.  The effect of changing the weight applied in
 the Monte Carlo has been studied. If the weight is changed to
 $\jjmass^{-4.5}$ or $\jjmass^{-5.0}$ the resulting energy scale error
 changes by less than 1$\%$.

\begin{figure}[htb]
\vbox{\centerline
{\psfig{figure=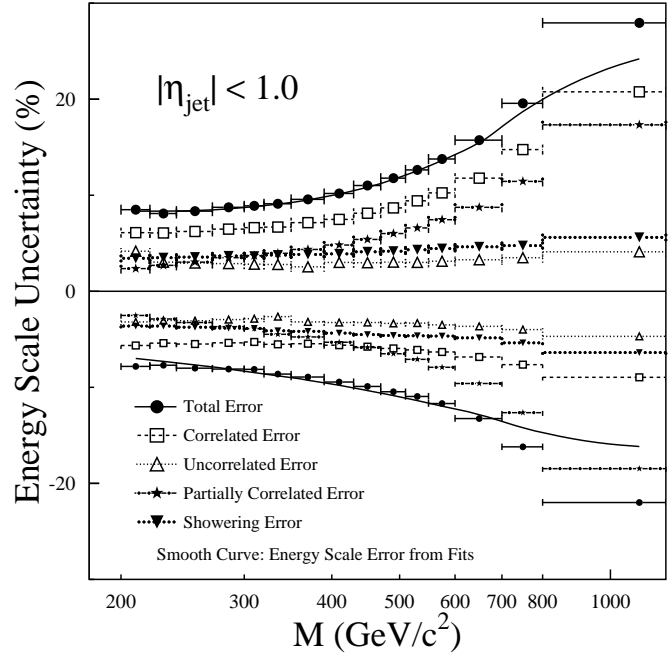,width=3.5in}}}
 \caption{ The energy scale errors obtained from the Monte Carlo. The
 full circles show the total energy scale uncertainties, the open
 squares show the correlated error, the open triangles show the
 uncorrelated error, and the stars show the partially correlated
 error. The curve shows the energy scale uncertainties obtained from
 fitting the data with the high and low energy scale corrections
 applied.} \label{sec_2:fig_19}
\end{figure}

 The energy scale uncertainties are calculated by generating a sample
 of MC jet events (in which the jets are fully corrected) and applying
 the inverse of jet energy scale correction. This sample of
 uncorrected jets then have the nominal, high (nominal$+1\sigma$), and
 low (nominal$-1\sigma$) energy scale corrections applied. The error
 due to the energy scale is split into components: the uncorrelated
 error, the fully correlated error, a partially correlated error, and
 the error due to the showering correction. The resulting errors are
 plotted in Fig.~\ref{sec_2:fig_19} along with errors obtained from
 fitting the data with the high and low energy scale corrections
 applied. The errors obtained by the two methods are in agreement.

\label{mass_correlations}

\begin{figure}[htb]
\vbox{\centerline
{\psfig{figure=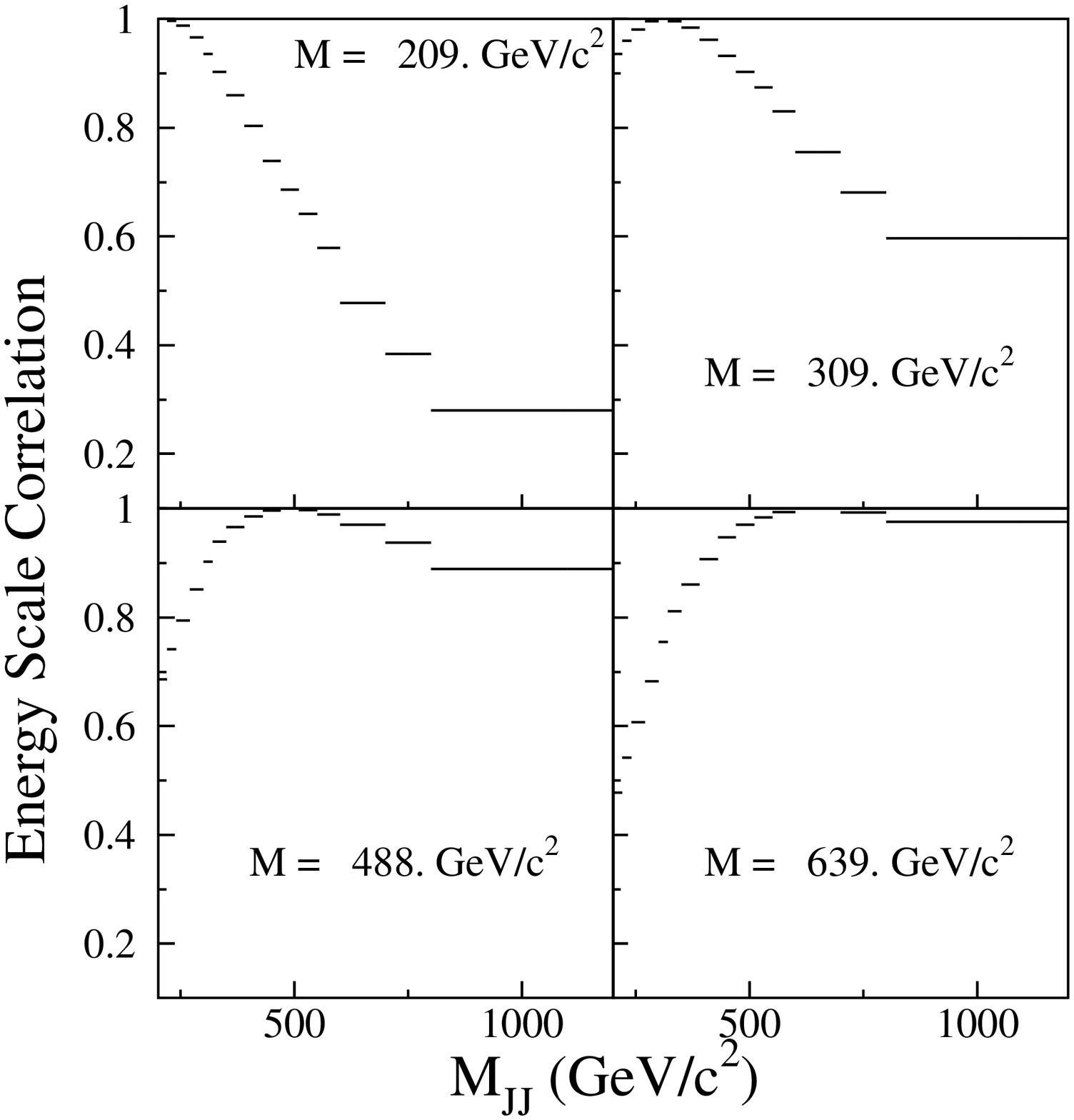,width=3.5in}}}
 \caption{The correlations between mass bins for $\modetajet <
 1.0$. The four plots show the mass correlations relative to four
 different mass bins: 200--220 \gevcc\ (209 \gevcc\ weighted average),
 300--320 \gevcc\ (309 \gevcc ), 470--510 \gevcc\ (488 \gevcc ), and
 600--700 \gevcc\ (639 \gevcc ).}
\label{sec_2:fig_20}
\end{figure}

 We calculated the correlation matrix for the partially correlated
 component of the energy scale uncertainty. The correlations have been
 calculated as a function of the jet energy
 (Section~\ref{sec:energy_scale_correlations}); hence the relationship
 between the jet energy and the dijet mass spectrum needs to be
 determined. The correlation matrix for the dijet mass spectrum can be
 written as
\begin{equation}
\langle \delta\sigma_i \delta\sigma_j\rangle =
{\partial{\sigma}\over {\partial M_i}} {\partial{\sigma}\over {\partial M_j}} 
\langle\delta M_i \delta M_j\rangle ,
\end{equation} 
 where $\delta X$ represents the shift in variable $X$ due to a
 systematic error parameter $\alpha$; $i$ and $j$ denote mass bins and
 $\sigma$ is the cross section. In the limit of massless jets, the
 dijet effective mass can be approximated by:
\begin{equation} 
 M_i = \sqrt{2 E_{i1} E_{i2} (1 - \cos\theta_{i12})}
\end{equation}
 where $E_{i1}$ and $E_{i2}$ are the energies, and $\theta_{i12}$ is
 the angle between the jets for event $i$. Hence:
\begin{equation}
  \delta M_i = 
  {{\partial{M_i}}\over{\partial{E_{i1}}}} \delta E_{i1} 
+ {{\partial{M_i}}\over{\partial{E_{i2}}}} \delta E_{i2} 
+ {{\partial{M_i} \over {\partial{\cos\theta_{i12}}}}} \delta
{\cos\theta_{i12}} ;
\end{equation}
 as we are only concerned with the energy scale, the angle error is
 ignored. Therefore:
\begin{equation}
 {{\partial{M_i}}\over{\partial{E_{i1}}}}  = 
 {{ E_{i2}}\over{M_{i}}}  (1 - \cos\theta_{i12}),
\end{equation}
\begin{equation}
  \delta M_i  = {1 \over M_{i}}
  ( E_{i1} \delta E_{i2} + E_{i2} \delta E_{i1})
  (1 - \cos\theta_{i12})
\end{equation}
 and
\begin{eqnarray}
\langle \delta M_i \delta M_j \rangle  = 
{{1}\over{M_{i}M_{j}}} (1- \cos\theta_{i12}) 
(1 - \cos\theta_{j12}) \times \nonumber \\
   \biggl( E_{i1} E_{j1} \langle \delta  E_{i2}  E_{j2}  \rangle 
       +  E_{i2} E_{j1} \langle \delta  E_{i1}  E_{j2}  \rangle + 
  \nonumber \\
  E_{i1} E_{j2} \langle \delta  E_{i2}  E_{j1}  \rangle 
    +  E_{i2} E_{j2} \langle \delta  E_{i1}  E_{j1}  \rangle \biggr) .
\end{eqnarray}
 Using this relationship, the correlations between jets due to the
 uncertainties in the jet energy scale can be translated into
 correlations between mass bins for the dijet mass cross section using
 the Monte Carlo. The resulting correlations are plotted for a
 selection of mass bins in Fig.~\ref{sec_2:fig_20}.

\subsection{Summary of Systematic Uncertainties}
\label{mass_sys_error_summary}

 In addition to the uncertainties in the luminosity, smearing
 correction, and the energy scale, there are uncertainties associated
 with the selection of the events that contribute to the data
 sample. These uncertainties are due to the jet quality cuts (see
 Eqs.~\ref{EMFRcut}, \ref{CHFRcut}, \ref{CELLcut}, and
 \ref{MISSETcut}) as well as the procedure used to add hot cells back
 into jets (Section~\ref{section06:hot_cells}). These uncertainties
 contribute a 1$\%$ uncorrelated uncertainty to the cross section. In
 addition, the uncertainty due to the unsmearing is assumed to be
 fully correlated as a function of the dijet mass in each event.

 A complete description of the systematic uncertainties in the dijet
 mass cross section is given in Table~\ref{error_cross}. The total
 systematic error in each mass bin is given by the sum in quadrature
 of these errors. The uncertainties are combined appropriately to
 obtain an overall correlation matrix for the bin-to-bin systematic
 uncertainties in the dijet mass spectrum.

\begin{table}[htbp]
\begin{center}
 \caption{Common systematic errors on the cross section.}
  \label{error_cross}
  \begin{tabular}{lrr}
   Source   &  Percentage Error & Comment  \\
   \hline \hline
   Jet Selection         & 1.  & Statistical  \\
			 &     & Uncorrelated  \\ \hline
   Vertex Selection      & 2.  & Systematic \\
			 &     & Uncorrelated \\ \hline
   Luminosity Scale      & 5.8 & Systematic  \\ 
			 &     & Fully Correlated  \\ \hline
   Luminosity Match      &     & Systematic  \\
   Jet\_30               & 4.9 & Statistical \\
   Jet\_50               &     & Correlated for \\
			 &     & Jet\_30 and 50 \\ \hline
   Unsmearing Correction & 0.5--3.0 & Systematic       \\
                         &     &  Fully Correlated \\ \hline
   Energy Scale          & 7.0--30.0 & Systematic \\
                         &     & Mixture\\ 
  \end{tabular}
\end{center}
\end{table}

\subsection{Cross Section}

 The dijet mass cross section is calculated using
 Eq.~\ref{EQ:dijet_cross_section} for the pseudorapidity range
 $\modetajet < 1.0$ , in mass ranges starting at 200, 270, 350, and
 550 \gevcc , corresponding to the jet \Et\ thresholds of 30, 50, 85,
 and 115 GeV.

\begin{figure}[htbp]
\vbox{\centerline
{\psfig{figure=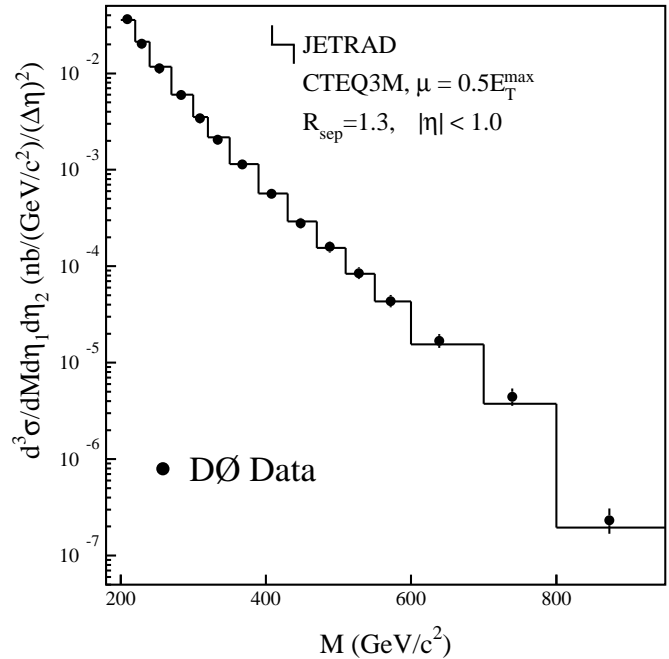,width=3.5in}}}
 \caption{Dijet mass cross section $d^{3} \sigma / d \jjmass d
 \eta_{1} d \eta_{2}$ for $\modetajet<1.0$. The D\O\ data are shown by
 the solid circles, with error bars representing the $\pm$$1\sigma$
 statistical and systematic uncertainties added in quadrature (in most
 cases smaller than the symbol). The histogram represents the {\sc
 jetrad} prediction.}
\label{sec_2:fig_21}
\end{figure}

\begin{figure}[htbp]
\vbox{\centerline{\psfig{figure=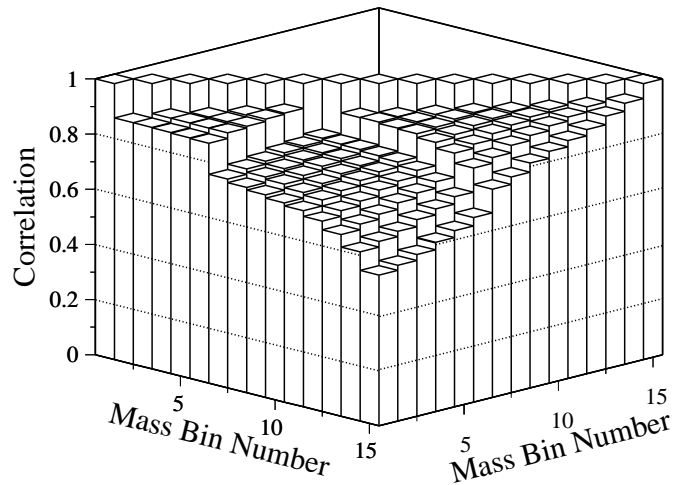,width=3.5in}}}
 \caption{The correlations between systematic uncertainties in bins of
 dijet mass (see Tables~\ref{cross_10} and
 ~\ref{high_correlations_table}) for $\modetajet<1.0$. The
 correlations are calculated using the average systematic
 uncertainty. The discontinuities arise from the uncorrelated errors
 (adjacent to correlations of 1.0) and to luminosity matching.}
\label{fig_2}
\end{figure}

 The cross section for the mass spectrum is plotted in
 Fig.~\ref{sec_2:fig_21}, and given in Table~\ref{cross_10}.  The data
 are plotted at the mass-weighted average of the fit function for each
 bin (${ \int M F dM} / {\int F dM }$). The systematic uncertainties
 are dominated by the uncertainties in the jet energy scale, which are
 7$\%$ (30$\%$) for the 209 (873) \gevcc\ mass bins.  The bin-to-bin
 correlations of the uncertainties are shown in Fig.~\ref{fig_2} and
 are given in
 Table~\ref{high_correlations_table}~\cite{inclusive_matrix}.

\begin{table*}[htbp]
\begin{center}
\caption{Dijet mass cross section for $\modetajet < 1.0$. High (low) 
 systematic uncertainties are the sum in quadrature of the
 uncertainties from the $\pm1\sigma$ variations in the energy
 calibration, the unsmearing, the vertex corrections, luminosity
 matching, jet selection, and the uncertainty in the luminosity.  Also
 included is the {\sc jetrad} prediction with $\mu = 0.5\Etmax$,
 ${\mathcal{R}}_{\rm sep} = 1.3$, and the \cteqthreem\ PDF. }
\label{cross_10}
\begin{tabular}{dddrcddc}
\multicolumn{3}{c}{Mass Bin (\gevcc )} &\multicolumn{1}{c}{$N_{i}$} &
  \multicolumn{1}{c}{Cross Section}  &  
\multicolumn{2}{c}{Systematic Error} & \multicolumn{1}{c}{Theoretical}  \\ 
\cline{1-3}\cline{6-7}
\multicolumn{1}{c}{Min.} & \multicolumn{1}{c}{Max.} &
  \multicolumn{1}{c}{Weighted}   & \multicolumn{1}{c}{} &
\multicolumn{1}{c}{$\pm$ Statistical Error} & 
 \multicolumn{1}{c}{~\hspace*{2.25mm}Low~\hspace*{2.25mm}} & 
\multicolumn{1}{c}{High} & \multicolumn{1}{c}{ Prediction} \\
\multicolumn{1}{c}{}   & \multicolumn{1}{c}{}   &
 \multicolumn{1}{c}{Center} &   \multicolumn{1}{c}{} &
\multicolumn{1}{c}{$\left({\rm nb}/(\gevcc )/(\Delta\eta)^2\right)$}  & 
\multicolumn{1}{c}{($\%$)}  & \multicolumn{1}{c}{($\%$)} &
\multicolumn{1}{l}{$\left({\rm nb}/(\gevcc )/{( \Delta\eta )^2}\right)$} \\
\hline \hline
  200. & 220. & 209.1 & 918 &$(3.66 \pm 0.12)\times 10^{-2}$& 11.2 &
11.7 &$3.57\times 10^{-2}$\\ 220. & 240. & 229.2 & 507 &$(2.03 \pm
0.09)\times 10^{-2}$& 11.1 & 11.4 &$2.12\times 10^{-2}$\\ 240. &
270. & 253.3 & 419 &$(1.13 \pm 0.06)\times 10^{-2}$& 11.3 & 11.6
&$1.17\times 10^{-2}$\\ 270. & 300. & 283.4 & 2944 &$(5.98 \pm
0.11)\times 10^{-3}$& 11.4 & 11.8 &$6.00\times 10^{-3}$\\ 300. &
320. & 309.3 & 1123 &$(3.43 \pm 0.10)\times 10^{-3}$& 11.4 & 12.0
&$3.53\times 10^{-3}$\\ 320. & 350. & 333.6 & 1006 &$(2.06 \pm
0.06)\times 10^{-3}$& 11.8 & 12.1 &$2.17\times 10^{-3}$\\ 350. &
390. & 367.6 & 8749 &$(1.14 \pm 0.01)\times 10^{-3}$& 10.9 & 11.5
&$1.15\times 10^{-3}$\\ 390. & 430. & 407.8 & 4323 &$(5.66 \pm
0.09)\times 10^{-4}$& 11.4 & 12.0 &$5.67\times 10^{-4}$\\ 430. &
470. & 447.9 & 2137 &$(2.80 \pm 0.06)\times 10^{-4}$& 11.8 & 12.7
&$2.92\times 10^{-4}$\\ 470. & 510. & 488.0 & 1210 &$(1.59 \pm
0.05)\times 10^{-4}$& 12.3 & 13.4 &$1.54\times 10^{-4}$\\ 510. &
550. & 528.0 & 646 &$(8.47 \pm 0.33)\times 10^{-5}$& 12.7 & 14.2
&$8.36\times 10^{-5}$\\ 550. & 600. & 572.0 & 699 &$(4.35 \pm
0.16)\times 10^{-5}$& 13.3 & 15.2 &$4.31\times 10^{-5}$\\ 600. &
700. & 638.9 & 542 &$(1.68 \pm 0.07)\times 10^{-5}$& 14.8 & 17.1
&$1.55\times 10^{-5}$\\ 700. & 800. & 739.2 & 144 &$(4.43 \pm
0.37)\times 10^{-6}$& 17.5 & 20.7 &$3.75\times 10^{-6}$\\ 800. &
1400. & 873.2 & 46 &$(2.32 \pm 0.34)\times 10^{-7}$& 23.1 & 28.9
&$1.95\times 10^{-7}$\\
\end{tabular}
\end{center}
\end{table*}

\begin{table*}[htbp]
 \caption{The systematic error correlations for the dijet cross
 section for $\modetajet < 1.0$, and the ratio $\kappa(\modetajet <
 0.5)/ \kappa(0.5 < \modetajet < 1.0)$. The correlation values above
 the diagonal are the correlations corresponding to the cross section
 and the correlations below the diagonal correspond to the ratio.  In
 both cases the correlation matrices are symmetric.}
 \label{high_correlations_table} 
 \begin{tabular}{rcccccccccccccccl} 
 & 1 & 2 & 3 & 4 & 5 & 6 & 7 & 8 & 9 & 10 & 11 & 12 & 13 & 14 & 15 & \\
 \hline\hline 
   & 1.00 & 0.88 & 0.88 & 0.87 & 0.87 & 0.87 & 0.76 & 0.74 & 0.73 &
   0.72 & 0.71 & 0.69 & 0.66 & 0.61 & 0.55 & 1 \\ & & 1.00 & 0.89 &
   0.89 & 0.89 & 0.89 & 0.78 & 0.77 & 0.76 & 0.74 & 0.73 & 0.71 & 0.68
   & 0.63 & 0.56 & 2 \\ 1 & 1.00 & & 1.00 & 0.90 & 0.90 & 0.90 & 0.79
   & 0.78 & 0.77 & 0.76 & 0.75 & 0.73 & 0.70 & 0.65 & 0.59 & 3 \\ 2 &
   0.58 & 1.00 & & 1.00 & 0.90 & 0.91 & 0.80 & 0.79 & 0.79 & 0.78 &
   0.77 & 0.75 & 0.72 & 0.68 & 0.61 & 4 \\ 3 & 0.61 & 0.60 & 1.00 & &
   1.00 & 0.91 & 0.81 & 0.80 & 0.80 & 0.79 & 0.78 & 0.77 & 0.74 & 0.70
   & 0.64 & 5 \\ 4 & 0.61 & 0.59 & 0.63 & 1.00 & & 1.00 & 0.82 & 0.82
   & 0.81 & 0.81 & 0.80 & 0.79 & 0.76 & 0.73 & 0.67 & 6 \\ 5 & 0.52 &
   0.51 & 0.54 & 0.54 & 1.00 & & 1.00 & 0.89 & 0.90 & 0.89 & 0.89 &
   0.87 & 0.85 & 0.81 & 0.76 & 7 \\ 6 & 0.56 & 0.54 & 0.57 & 0.58 &
   0.87 & 1.00 & & 1.00 & 0.90 & 0.90 & 0.89 & 0.88 & 0.87 & 0.84 &
   0.78 & 8 \\ 7 & 0.60 & 0.58 & 0.62 & 0.63 & 0.54 & 0.58 & 1.00 & &
   1.00 & 0.91 & 0.91 & 0.90 & 0.89 & 0.86 & 0.82 & 9 \\ 8 & 0.60 &
   0.58 & 0.61 & 0.62 & 0.53 & 0.57 & 0.63 & 1.00 & & 1.00 & 0.91 &
   0.91 & 0.90 & 0.88 & 0.84 & 10 \\ 9 & 0.59 & 0.58 & 0.61 & 0.62 &
   0.53 & 0.57 & 0.62 & 0.63 & 1.00 & & 1.00 & 0.92 & 0.91 & 0.90 &
   0.86 & 11 \\ 10 & 0.58 & 0.57 & 0.60 & 0.61 & 0.53 & 0.56 & 0.61 &
   0.62 & 0.62 & 1.00 & & 1.00 & 0.92 & 0.91 & 0.89 & 12 \\ 11 & 0.59
   & 0.58 & 0.61 & 0.63 & 0.53 & 0.58 & 0.62 & 0.63 & 0.63 & 0.62 &
   1.00 & & 1.00 & 0.93 & 0.92 & 13 \\ 12 & 0.60 & 0.58 & 0.62 & 0.63
   & 0.54 & 0.58 & 0.63 & 0.62 & 0.62 & 0.61 & 0.63 & 1.00 & & 1.00 &
   0.95 & 14 \\ 13 & 0.58 & 0.56 & 0.59 & 0.61 & 0.53 & 0.56 & 0.61 &
   0.60 & 0.60 & 0.59 & 0.61 & 0.62 & 1.00 & & 1.00 & 15 \\ 14 & 0.53
   & 0.52 & 0.55 & 0.57 & 0.50 & 0.53 & 0.56 & 0.56 & 0.56 & 0.55 &
   0.57 & 0.58 & 0.58 & 1.00 & & \\ 15 & 0.51 & 0.50 & 0.53 & 0.55 &
   0.48 & 0.52 & 0.54 & 0.54 & 0.54 & 0.53 & 0.55 & 0.57 & 0.57 & 0.55
   & 1.00 & \\
\end{tabular}
\end{table*}

 The dijet mass cross section measurement was then repeated for
 $\modetajet < 0.5$, and $0.5 < \modetajet < 1.0$. The resulting cross
 sections are given in Tables~\ref{cross_05} and \ref{cross_05_10}.

\begin{table*}[htbp]
\begin{center}
\caption{Dijet mass cross section for $\modetajet < 0.5$. High (low) 
 systematic uncertainties are the sum in quadrature of the
 uncertainties from the $\pm1\sigma$ variations in the energy
 calibration, the unsmearing, the vertex corrections, luminosity
 matching, jet selection, and the uncertainty in the luminosity.  Also
 included is the {\sc jetrad} prediction with $\mu = 0.5\Etmax$,
 ${\mathcal{R}}_{\rm sep} = 1.3$, and the \cteqthreem\ PDF. }
\label{cross_05}
\begin{tabular}{dddrcddc}
\multicolumn{3}{c}{Mass Bin (\gevcc )} &
\multicolumn{1}{c}{$N_{i}$} &  
\multicolumn{1}{c}{Cross Section}  &  
\multicolumn{2}{c}{Systematic Error} & 
\multicolumn{1}{c}{Theoretical}  \\ 
\cline{1-3}\cline{6-7}
\multicolumn{1}{c}{Min.} & 
\multicolumn{1}{c}{Max.} &  
\multicolumn{1}{c}{Weighted}   & 
\multicolumn{1}{c}{} &
\multicolumn{1}{c}{$\pm$ Statistical Error} & 
 \multicolumn{1}{c}{~\hspace*{2.25mm}Low~\hspace*{2.25mm}} & 
\multicolumn{1}{c}{High} &
 \multicolumn{1}{c}{ Prediction} \\
\multicolumn{1}{c}{}   & 
\multicolumn{1}{c}{}   & 
\multicolumn{1}{c}{Center} &   
\multicolumn{1}{c}{} &
\multicolumn{1}{c}{$\left({\rm nb}/(\gevcc )/(\Delta\eta)^2\right)$}  & 
\multicolumn{1}{c}{($\%$)}  & 
\multicolumn{1}{c}{($\%$)} &
\multicolumn{1}{l}{$\left({\rm nb}/(\gevcc )/{( \Delta\eta )^2}\right)$}\\
\hline \hline
  150. & 160. & 154.7 & 467 &$(1.46 \pm 0.07)\times 10^{-1}$& 10.8 &
11.0 &$1.46\times 10^{-1}$\\ 160. & 180. & 168.9 & 552 &$(8.69 \pm
0.37)\times 10^{-2}$& 10.5 & 10.5 &$9.03\times 10^{-2}$\\ 180. &
200. & 189.0 & 315 &$(4.99 \pm 0.28)\times 10^{-2}$& 10.5 & 10.9
&$4.91\times 10^{-2}$\\ 200. & 220. & 209.1 & 2243 &$(2.69 \pm
0.06)\times 10^{-2}$& 10.3 & 10.5 &$2.81\times 10^{-2}$\\ 220. &
240. & 229.2 & 1390 &$(1.67 \pm 0.04)\times 10^{-2}$& 10.2 & 10.7
&$1.68\times 10^{-2}$\\ 240. & 270. & 253.3 & 1055 &$(8.52 \pm
0.26)\times 10^{-3}$& 10.5 & 10.5 &$9.35\times 10^{-3}$\\ 270. &
300. & 283.4 & 550 &$(4.47 \pm 0.19)\times 10^{-3}$& 10.7 & 10.8
&$4.82\times 10^{-3}$\\ 300. & 320. & 309.3 & 2671 &$(2.78 \pm
0.05)\times 10^{-3}$& 9.1 & 10.0 &$2.86\times 10^{-3}$\\ 320. & 350. &
333.6 & 2434 &$(1.69 \pm 0.03)\times 10^{-3}$& 9.9 & 9.9 &$1.78\times
10^{-3}$\\ 350. & 390. & 367.7 & 1823 &$(9.50 \pm 0.22)\times
10^{-4}$& 9.9 & 10.3 &$9.49\times 10^{-4}$\\ 390. & 430. & 407.8 &
1459 &$(4.52 \pm 0.12)\times 10^{-4}$& 10.3 & 10.8 &$4.78\times
10^{-4}$\\ 430. & 470. & 448.0 & 831 &$(2.58 \pm 0.09)\times 10^{-4}$&
10.7 & 11.4 &$2.50\times 10^{-4}$\\ 470. & 510. & 488.0 & 480 &$(1.49
\pm 0.07)\times 10^{-4}$& 11.1 & 12.1 &$1.35\times 10^{-4}$\\ 510. &
550. & 528.1 & 231 &$(7.17 \pm 0.47)\times 10^{-5}$& 11.7 & 12.8
&$7.43\times 10^{-5}$\\ 550. & 600. & 572.2 & 156 &$(3.87 \pm
0.31)\times 10^{-5}$& 12.4 & 13.7 &$3.91\times 10^{-5}$\\ 600. &
700. & 639.4 & 125 &$(1.55 \pm 0.14)\times 10^{-5}$& 13.8 & 15.5
&$1.46\times 10^{-5}$\\ 700. & 800. & 739.8 & 30 &$(3.71 \pm
0.68)\times 10^{-6}$& 16.5 & 19.3 &$3.72\times 10^{-6}$\\ 800. &
1400. & 878.1 & 14 &$(2.86 \pm 0.77)\times 10^{-7}$& 22.0 & 27.8
&$2.08\times 10^{-7}$\\
\end{tabular}
\end{center}
\end{table*}

\begin{table*}[htbp]
\begin{center}
\caption{Dijet mass cross section for $0.5 < \modetajet < 1.0$. High (low) 
 systematic uncertainties are the sum in quadrature of the
 uncertainties from the $\pm1\sigma$ variations in the energy
 calibration, the unsmearing, the vertex corrections, luminosity
 matching, jet selection, and the uncertainty in the luminosity.  Also
 included is the {\sc jetrad} prediction with $\mu = 0.5\Etmax$,
 ${\mathcal{R}}_{\rm sep} =$ 1.3, and the \cteqthreem\ PDF. }
\label{cross_05_10}
\begin{tabular}{dddrcddc}
\multicolumn{3}{c}{Mass Bin (\gevcc )} &
\multicolumn{1}{c}{$N_{i}$} &  
\multicolumn{1}{c}{Cross Section}  &  
\multicolumn{2}{c}{Systematic Error} & 
\multicolumn{1}{c}{Theoretical}  \\ 
\cline{1-3}\cline{6-7}
\multicolumn{1}{c}{Min.} & 
\multicolumn{1}{c}{Max.} &  
\multicolumn{1}{c}{Weighted}   & 
\multicolumn{1}{c}{} &
\multicolumn{1}{c}{$\pm$ Statistical Error} & 
\multicolumn{1}{c}{~\hspace*{2.25mm}Low~\hspace*{2.25mm}} & 
\multicolumn{1}{c}{High} & 
\multicolumn{1}{c}{ Prediction} \\
\multicolumn{1}{c}{}   & 
\multicolumn{1}{c}{}   & 
\multicolumn{1}{c}{Center} &   
\multicolumn{1}{c}{} &
\multicolumn{1}{c}{$\left({\rm nb}/(\gevcc )/(\Delta\eta)^2\right)$}  & 
\multicolumn{1}{c}{($\%$)}  & 
\multicolumn{1}{c}{($\%$)} &
\multicolumn{1}{l}{$\left({\rm nb}/(\gevcc )/{( \Delta\eta )^2}\right)$} \\
\hline \hline
  200. & 220. & 209.1 & 275 &$(4.39 \pm 0.26) \times 10^{-2}$& 12.3 &
12.7 &$4.56\times 10^{-2}$\\ 220. & 240. & 229.1 & 170 &$(2.73 \pm
0.21) \times 10^{-2}$& 11.7 & 12.1 &$2.70\times 10^{-2}$\\ 240. &
270. & 253.2 & 139 &$(1.49 \pm 0.13) \times 10^{-2}$& 12.0 & 12.4
&$1.49\times 10^{-2}$\\ 270. & 300. & 283.4 & 964 &$(7.87 \pm 0.25)
\times 10^{-3}$& 11.9 & 12.7 &$7.60\times 10^{-3}$\\ 300. & 320. &
309.3 & 371 &$(4.55 \pm 0.24) \times 10^{-3}$& 12.1 & 12.3
&$4.46\times 10^{-3}$\\ 320. & 350. & 333.6 & 292 &$(2.40 \pm 0.14)
\times 10^{-3}$& 12.5 & 13.0 &$2.75\times 10^{-3}$\\ 350. & 390. &
367.6 & 2682 &$(1.41 \pm 0.03) \times 10^{-3}$& 11.6 & 12.4
&$1.44\times 10^{-3}$\\ 390. & 430. & 407.8 & 1445 &$(7.62 \pm 0.20)
\times 10^{-4}$& 12.0 & 13.0 &$7.16\times 10^{-4}$\\ 430. & 470. &
447.9 & 689 &$(3.64 \pm 0.14) \times 10^{-4}$& 12.6 & 13.6
&$3.70\times 10^{-4}$\\ 470. & 510. & 488.0 & 408 &$(2.16 \pm 0.11)
\times 10^{-4}$& 13.1 & 14.1 &$1.97\times 10^{-4}$\\ 510. & 550. &
528.1 & 219 &$(1.16 \pm 0.08) \times 10^{-4}$& 13.4 & 15.0
&$1.07\times 10^{-4}$\\ 550. & 600. & 572.2 & 244 &$(6.11 \pm 0.39)
\times 10^{-5}$& 13.6 & 16.0 &$5.59\times 10^{-5}$\\ 600. & 700. &
639.4 & 192 &$(2.40 \pm 0.17) \times 10^{-5}$& 14.7 & 17.5
&$2.05\times 10^{-5}$\\ 700. & 800. & 739.8 & 49 &$(6.10 \pm 0.87)
\times 10^{-6}$& 17.1 & 20.4 &$5.19\times 10^{-6}$\\ 800. & 1400. &
878.8 & 20 &$(4.06 \pm 0.91) \times 10^{-7}$& 22.4 & 28.0 &$2.92\times
10^{-7}$\\
\end{tabular}
\end{center}
\end{table*}

 Most of the systematic uncertainties in the measurement of the
 inclusive dijet mass spectrum are highly correlated as a function of
 dijet mass and $\eta$ and to a good approximation, cancel when a
 ratio of two cross sections is made. For this reason the cross
 section ratio for the rapidity ranges $\modetajet < 0.5$ and $0.5 <
 \modetajet < 1.0$ will be calculated:
\begin{equation}
\frac{\kappa\left(\modetajet < 0.5\right)}
{\kappa\left(0.5 < \modetajet < 1.0\right)}.
\end{equation}
 The uncertainty in the theoretical prediction of this ratio due to
 the choice of PDF is less than $3\%$, and $6\%$ from the choice of
 renormalization and factorization scale (excluding $\mu = 0.25
 \Etmax$).  The luminosity matching error only contributes to those
 bins where the data from triggers Jet\_30 and Jet\_50 overlaps with
 the data from triggers Jet\_85 and Jet\_115 ({\rm i.e.} for masses
 between 300 and 350 \gevcc ). The errors from the vertex selection
 cancel when the data in a bin come from the same trigger for each of
 the cross sections. The errors due to the unsmearing and the
 (partially) correlated part of the energy scale are assumed to be
 correlated for the two cross sections and mostly cancel out leaving
 small errors ($\ll1\%$). In addition the uncertainty due to the hot
 cell restoration is assumed to be correlated between the two
 cross-sections. All other errors are assumed to be uncorrelated
 between the two measurements. For the purposes of calculating a
 covariance matrix, the correlated energy scale and unsmearing errors
 are assumed to be fully correlated as a function of mass.

\begin{figure}[hbt]
\vbox{\centerline
{\psfig{figure=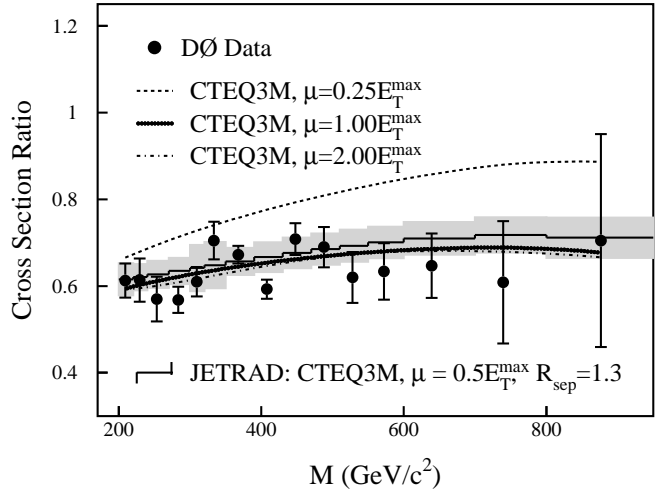,width=3.5in}}}
 \caption{The ratio of cross sections for $\modetajet<0.5$ and
 $0.5<\modetajet <1.0$ for data (solid circles) and theory (various
 lines).  The error bars show the statistical uncertainties.  The
 shaded region represents the $\pm1\sigma$ systematic uncertainties
 about the prediction. The effects on the prediction of changing the
 renormalization scale are also shown.}
\label{sec_2:fig_22}
\end{figure}

 The resulting cross section ratios are given in
 Table~\ref{cross_ratio_table} and plotted in
 Fig.~\ref{sec_2:fig_22}. Taking the ratio of the cross sections
 reduces the systematic uncertainties to less than 10$\%$.  The
 correlations of the systematic uncertainties are given in
 Table~\ref{high_correlations_table}~\cite{inclusive_matrix}.

\begin{table*}[htbp]
\begin{center}
\caption{The ratio $\kappa(\modetajet < 0.5)/ \kappa(0.5 < \modetajet
< 1.0)$. The systematic uncertainties are the sum in quadrature of the
uncertainties from the $\pm1\sigma$ variations in the energy
calibration, the unsmearing, the vertex corrections, luminosity
matching, jet selection, and the uncertainty in the luminosity.  Also
shown is the {\sc jetrad} prediction with $\mu = 0.5\Etmax$,
${\mathcal{R}}_{\rm sep} =$ 1.3, and the \cteqthreem\ PDF. }
\label{cross_ratio_table}
\begin{tabular}{ddr@{ $\pm$ }l@{ $\pm$ }ld}
\multicolumn{2}{c}{Mass Bin} & 
\multicolumn{3}{c}{Ratio of Mass Spectra} &   
\multicolumn{1}{c}{Theoretical}\\ 
\multicolumn{2}{c}{(\gevcc )}&
 \multicolumn{3}{c}{$\kappa(\modetajet < 0.5)/\kappa(0.5 <\modetajet < 1.0)$} 
& \multicolumn{1}{c}{Prediction}\\
\cline{1-2}
\multicolumn{1}{c}{Min.} & 
\multicolumn{1}{c}{Max.}&
\multicolumn{3}{c}{($\pm$ stat. error $\pm$ syst. errror)}  &  
\multicolumn{1}{c}{} \\
\hline \hline
   200. &   220. &  \hspace*{10mm}0.613 &  0.039 &  0.037 &  0.616 \\ 
   220. &   240. &  0.614 &  0.050 &  0.030 &  0.621 \\ 
   240. &   270. &  0.570 &  0.051 &  0.029 &  0.627 \\ 
   270. &   300. &  0.568 &  0.030 &  0.027 &  0.635 \\ 
   300. &   320. &  0.610 &  0.034 &  0.050 &  0.642 \\ 
   320. &   350. &  0.705 &  0.044 &  0.058 &  0.648 \\ 
   350. &   390. &  0.672 &  0.020 &  0.032 &  0.657 \\ 
   390. &   430. &  0.593 &  0.022 &  0.030 &  0.667 \\ 
   430. &   470. &  0.708 &  0.036 &  0.037 &  0.676 \\ 
   470. &   510. &  0.690 &  0.046 &  0.036 &  0.685 \\ 
   510. &   550. &  0.620 &  0.058 &  0.033 &  0.693 \\ 
   550. &   600. &  0.634 &  0.065 &  0.033 &  0.701 \\ 
   600. &   700. &  0.647 &  0.074 &  0.034 &  0.710 \\ 
   700. &   800. &  0.608 &  0.141 &  0.035 &  0.718 \\ 
   800. &  1400. &  0.705 &  0.246 &  0.046 &  0.711 \\ 
\end{tabular}
\end{center}
\end{table*}

\subsection{Comparison of Data with Theory}
\label{sec_comp_data_theory}

\begin{figure}[htb]
\vbox{\centerline
{\psfig{figure=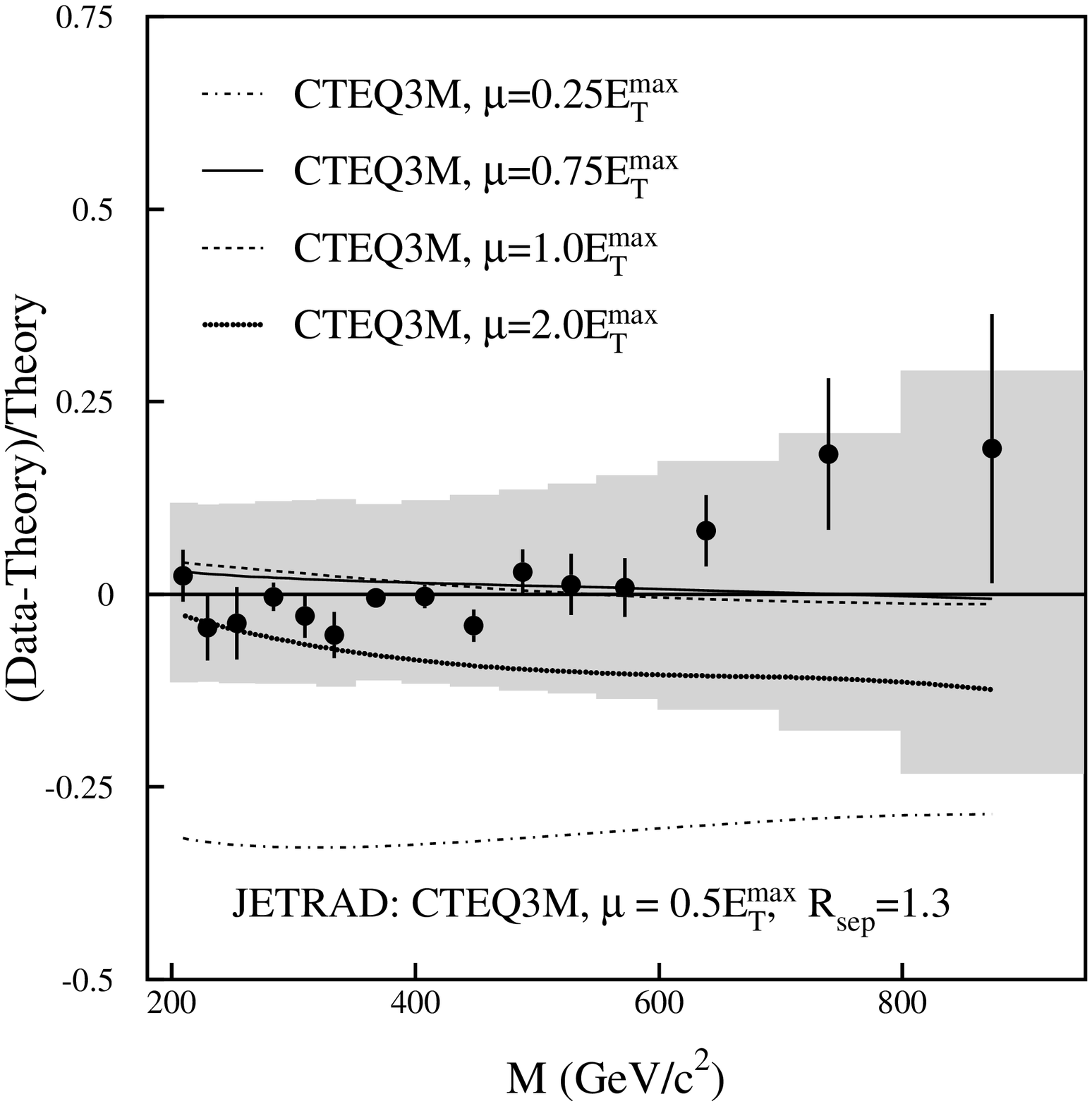,width=3.5in}}}
\caption{The difference between the data and the prediction ({\sc
jetrad}) divided by the prediction for $\modetajet < 1.0$. The solid
circles represent the comparison to the calculation using \cteqthreem\
with $\mu = 0.5 \Etmax$. The shaded region represents the $\pm1\sigma$
systematic uncertainties about the prediction. The effects of changing
the renormalization scale are also shown (each curve shows the
difference between the alternative prediction and the prediction using
\cteqthreem\ with $\mu = 0.5 \Etmax$).}
\label{sec_4:fig_1}
\end{figure}

\begin{figure}[htbp]
\vbox{\centerline
{\psfig{figure=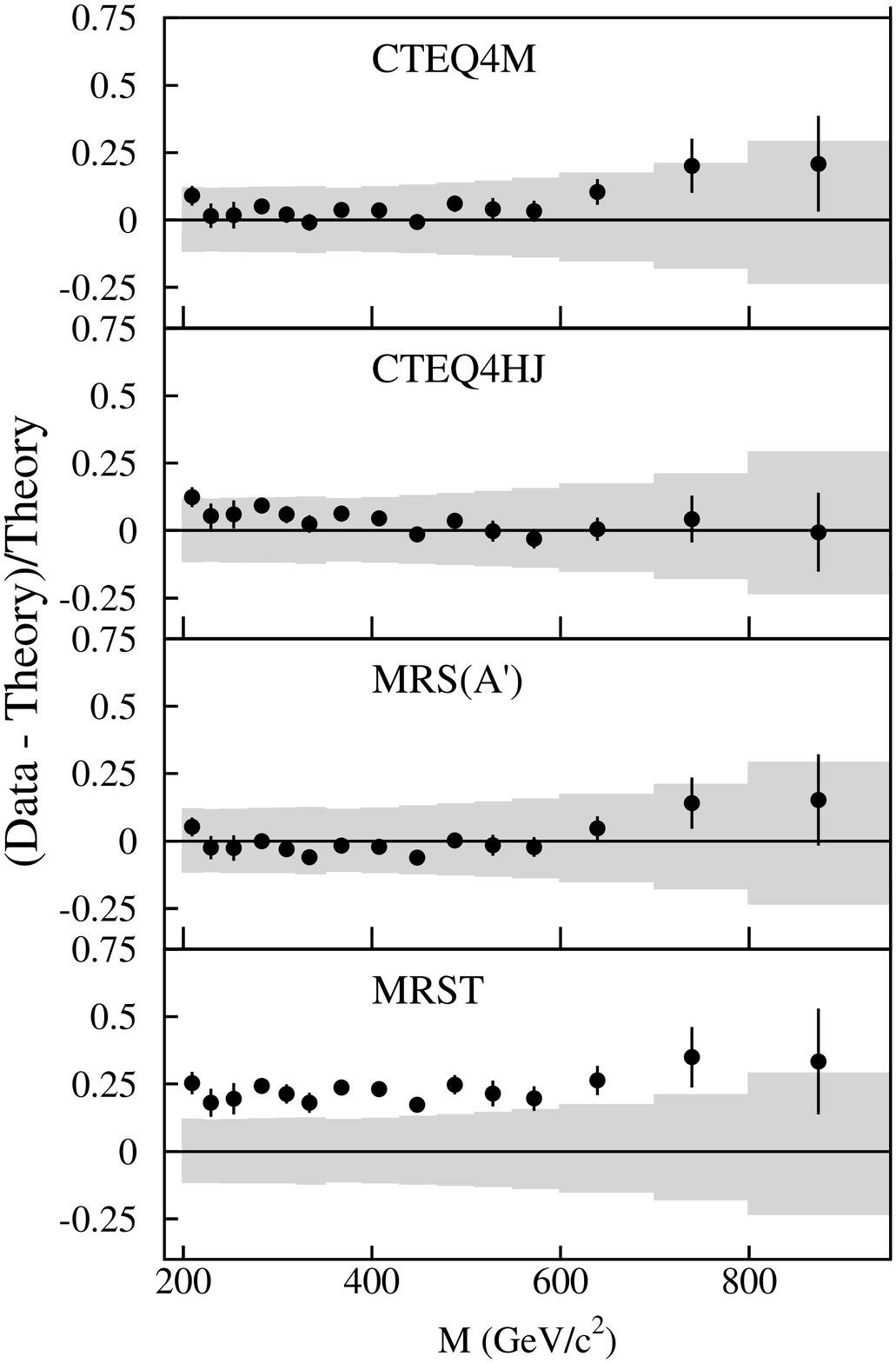,width=3.5in}}}
\caption{The difference between the data and the prediction ({\sc
jetrad}) divided by the prediction for $\modetajet < 1.0$. The solid
circles represent the comparison to the calculation using $\mu = 0.5
\Etmax$ and the PDFs \cteqfourm , \cteqfourhj , \mrsap , and \mrst
. The shaded region represents the $\pm1\sigma$ systematic
uncertainties about the prediction.}
\label{sec_4:vary_pdf}
\end{figure}

\begin{figure}[hbtp]
\vbox{\centerline
{\psfig{figure=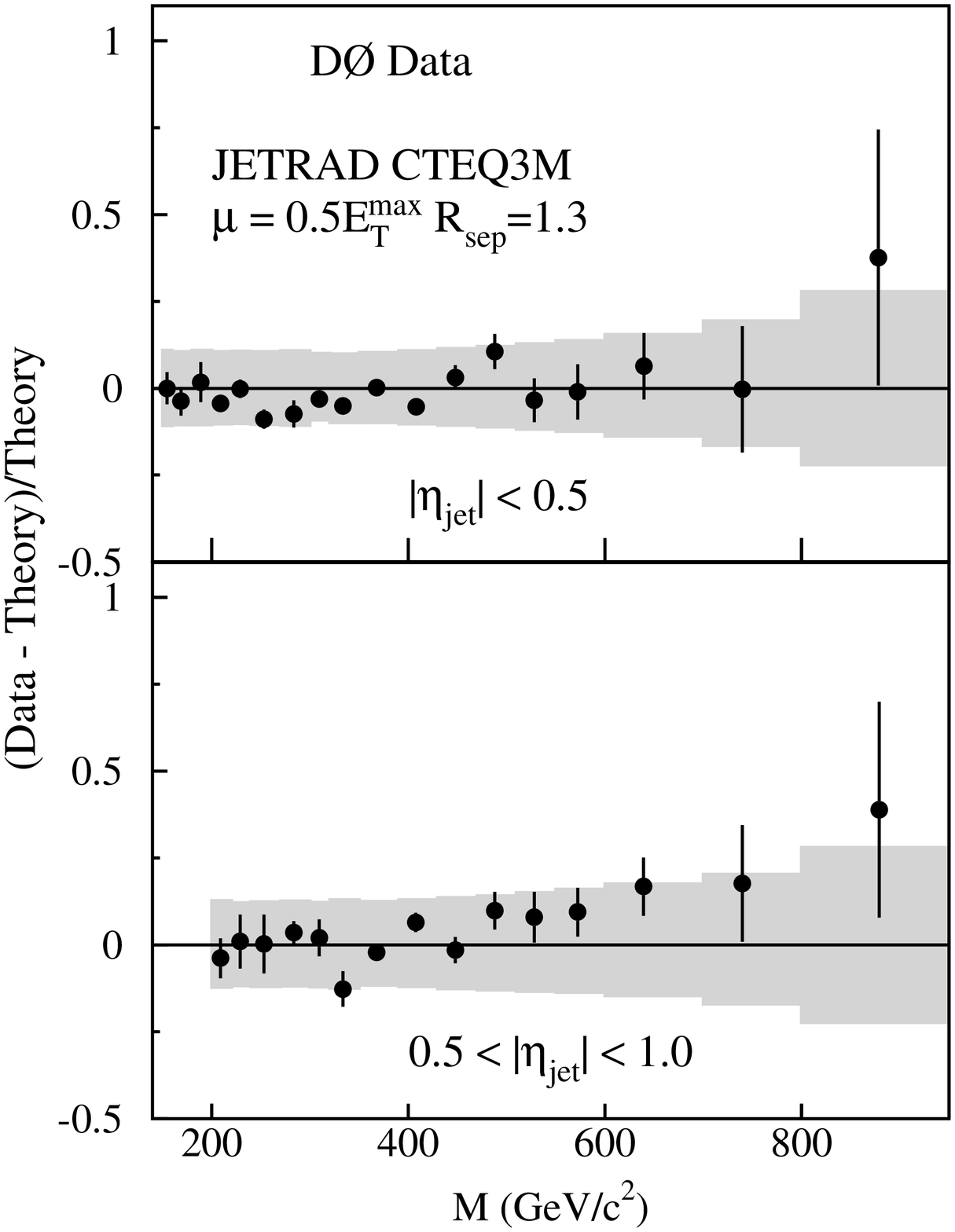
,width=3.5in}}}
\caption{The difference between the data and the  prediction ({\sc jetrad})
 divided by the prediction for $\modetajet < 0.5$ and $0.5 <
 \modetajet < 1.0$. The solid circles represent the comparison to the
 calculation using \cteqthreem\ with $\mu = 0.5 \Etmax$. The shaded region
 represents the $\pm1\sigma$ systematic uncertainties about the prediction.}
\label{sec_4:fig_1a}
\end{figure}

 Figure~\ref{sec_4:fig_1} shows the ratio $({\rm Data} - {\rm
 Theory})/{\rm Theory}$ for $\modetajet < 1.0$ and the {\sc jetrad}
 prediction using \cteqthreem\ with $\mu = 0.5\Etmax$. The effect of
 varying the renormalization scale in the prediction is also shown;
 all are in good agreement except for $\mu = 0.25\Etmax$ which lies
 approximately $30\%$ below the data.  Figure~\ref{sec_4:vary_pdf}
 shows $({\rm Data} - {\rm Theory})/{\rm Theory}$ for {\sc jetrad}
 predictions with different choices of PDFs. Given the experimental
 and theoretical uncertainties, the predictions can be regarded as
 being in good agreement with the data. Figure~\ref{sec_4:fig_1a}
 shows that the data and {\sc jetrad} predictions are in agreement for
 $\modetajet < 0.5$ and $0.5 < \modetajet < 1.0$.  The data are also
 in agreement, within the uncertainties, with the cross section
 measured by \mbox{CDF}~\cite{cdf_dijet_mass}.

 A \chisq\ can be calculated for each of the comparisons between the
 data (cross sections and ratio of cross sections) and the
 predictions. The \chisq\ is given by:
\begin{equation}
\chisq = \sum_{i,j} \delta_{i} V_{ij}^{-1} \delta_{j},
\end{equation}
 where $\displaystyle{\delta_{i}}$ is the difference between the data
 and theory for mass bin $i$, and $\displaystyle{V_{ij}}$ is
 element $i,j$ of the covariance matrix:
\begin{equation}
V_{ij} = \rho_{ij} \cdot \Delta \sigma_{i} \cdot \Delta \sigma_{j},
\end{equation} 
 where $\displaystyle{\Delta \sigma}$ is the sum of the systematic
 error and the statistical error added in quadrature if $i=j$ and the
 systematic error if $i \neq j$, and $\displaystyle{\rho_{ij}}$ is the
 correlation between the systematic uncertainties of mass bins $i$ and
 $j$ as given in Table~\ref{high_correlations_table}.  The systematic
 uncertainty is given by the fractional uncertainty times the
 theoretical prediction. The resulting \chisq\ values are given in
 Table~\ref{theory_data_chisquares} for all of the theoretical choices
 described above. The choice of PDF and renormalization scale is
 varied; all choices are in good agreement with the data, except for
 $\mu = 0.25 \Etmax$ which is excluded by the data.

\begin{table*}[htbp] \caption{The calculated \chisq\ for
 $\kappa(\modetajet < 1.0)$ (15 degrees of freedom),
 $\kappa(\modetajet < 0.5)$ (18 d.o.f.), and $\kappa(0.5 <
 \modetajet < 1.0)$ (15 d.o.f.) and for the ratio
 $\kappa(\modetajet < 0.5)/ \kappa(0.5 < \modetajet < 1.0)$. The
 probability of obtaining a larger  \chisq\ is also given.}
  \label{theory_data_chisquares}
 \begin{center}
 \begin{tabular}{lldddddddd}
PDF       & $\displaystyle{\mu}$            & 
\multicolumn{2}{c}{$\kappa(\modetajet < 1.0)$} & 
 \multicolumn{2}{c}{$\kappa(\modetajet < 0.5)$}& 
\multicolumn{2}{c}{$\kappa(0.5 < \modetajet < 1.0)$} &
\multicolumn{2}{c}{Ratio}  \\
 \cline{3-10}
  &        & \chisq  & \multicolumn{1}{c}{Prob.} &  
	     \chisq  & \multicolumn{1}{c}{Prob.} 
           & \chisq  & \multicolumn{1}{c}{Prob.} &  
	     \chisq  & \multicolumn{1}{c}{Prob.}  \\
 \hline\hline
 \cteqthreem\ & $0.25 \Etmax$ & 24.7 & 0.05 &26.4 & 0.09 & 38.3 &0.001
 & 29.1 & 0.02 \\ \cteqthreem\ & $0.50 \Etmax$ & 5.7 & 0.98 &11.2 &
 0.89 & 8.9 &0.88 & 14.1 & 0.52 \\ \cteqthreem\ & $0.75 \Etmax$ & 6.1
 & 0.98 &11.2 & 0.89 & 9.1 &0.87 & 13.6 & 0.56 \\ \cteqthreem\ & $1.00
 \Etmax$ & 6.3 & 0.97 &12.1 & 0.84 & 9.2 &0.87 & 13.3 & 0.58 \\
 \cteqthreem\ & $2.00 \Etmax$ & 6.0 & 0.98 &12.5 & 0.82 & 11.5 &0.71 &
 13.0 & 0.60 \\ \cteqthreem\ & $0.25\sqrt{x_{1}x_{2}s}$ & 12.7 & 0.63
 &28.7 & 0.05 & 10.2 &0.81 & 14.9 & 0.46 \\ \cteqthreem\ &
 $0.50\sqrt{x_{1}x_{2}s}$ & 6.1 & 0.98 &14.5 & 0.70 & 8.8 &0.89 & 13.8
 & 0.54 \\ \cteqthreem\ & $1.00\sqrt{x_{1}x_{2}s}$ & 7.7 & 0.93 &13.4
 & 0.77 & 13.3 &0.58 & 14.3 & 0.51 \\ \cteqfourm\ & $0.50 \Etmax$ &
 5.8 & 0.98 &11.5 & 0.87 & 8.3 &0.91 & 14.0 & 0.52 \\ CTEQ4A1 & $0.50
 \Etmax$ & 5.8 & 0.98 &13.1 & 0.79 & 8.1 &0.92 & 14.1 & 0.52 \\
 CTEQ4A2 & $0.50 \Etmax$ & 6.5 & 0.97 &12.4 & 0.83 & 8.0 &0.93 & 14.4
 & 0.50 \\ CTEQ4A4 & $0.50 \Etmax$ & 5.8 & 0.98 &11.7 & 0.86 & 8.5
 &0.90 & 14.5 & 0.49 \\ CTEQ4A5 & $0.50 \Etmax$ & 5.7 & 0.98 &11.4 &
 0.88 & 8.7 &0.89 & 14.9 & 0.46 \\ \cteqfourhj\ & $0.50 \Etmax$ & 5.6
 & 0.99 &11.4 & 0.88 & 6.8 &0.96 & 14.2 & 0.51 \\ \mrsap\ & $0.50
 \Etmax$ & 6.8 & 0.96 &11.0 & 0.89 & 8.3 &0.91 & 14.4 & 0.49 \\ \mrst\
 & $0.50 \Etmax$ & 8.8 & 0.89 &16.0 & 0.59 & 12.9 &0.61 & 14.5 & 0.49
 \\ \mrstgu\ & $0.50 \Etmax$ & 8.4 & 0.91 &16.7 & 0.54 & 10.2 &0.81 &
 14.2 & 0.51 \\ \mrstgd\ & $0.50 \Etmax$ & 13.9 & 0.54 &23.1 & 0.19 &
 19.6 &0.19 & 14.4 & 0.50 \\ 
\end{tabular} 
\end{center} 
\end{table*}

\subsection{Compositeness Limits}
 \label{sec:compositeness_limits}	

\begin{figure}[hbtp]
\vbox{\centerline
{\psfig{figure=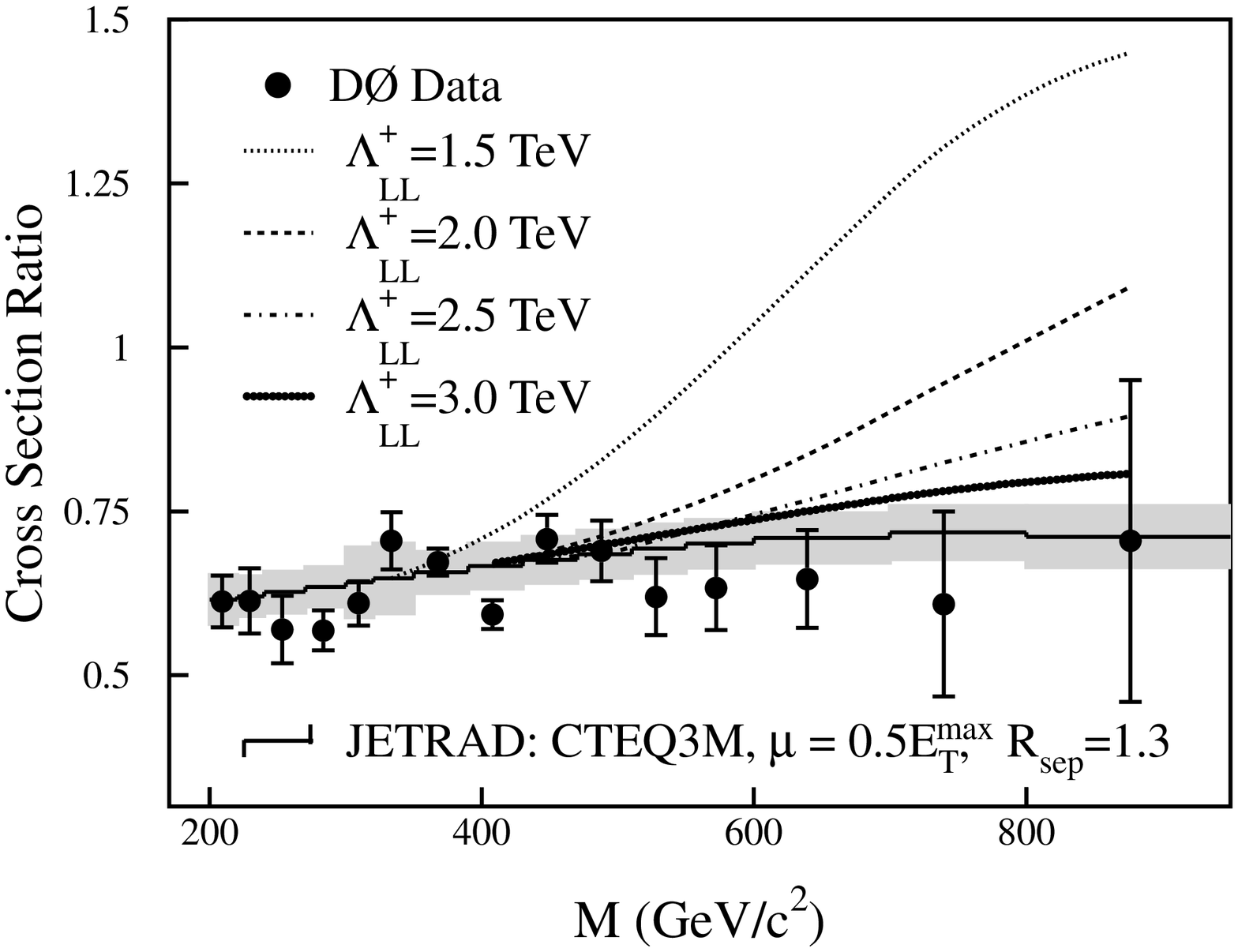,width=3.5in}}}
 \caption{The ratio of cross sections for $\modetajet<0.5$ and
 $0.5<\modetajet <1.0$ for data (solid circles) and theoretical
 predictions for compositeness models with various values of
 $\Lambda_{LL}^{+}$ (various lines, see
 Section~\ref{sec:theory_compositeness} for model details).  The error
 bars show the statistical uncertainties.  The shaded region
 represents the $\pm1\sigma$ systematic uncertainties about the {\sc
 jetrad} prediction. }
\label{fig:ratio_comp}
\end{figure}

 The ratio of the mass spectra can be used to place limits on quark
 compositeness (Section~\ref{sec:theory_compositeness}). Currently
 there are no NLO compositeness calculations available; therefore a LO
 event generator ({\sc pythia}) is used to simulate the effect of
 compositeness. The ratio of these LO predictions with compositeness,
 to the LO with no compositeness, is used to scale the {\sc jetrad}
 NLO prediction, shown in Fig.~\ref{fig:ratio_comp}.

\begin{figure}[htbp]
\vbox{\centerline
{\psfig{figure=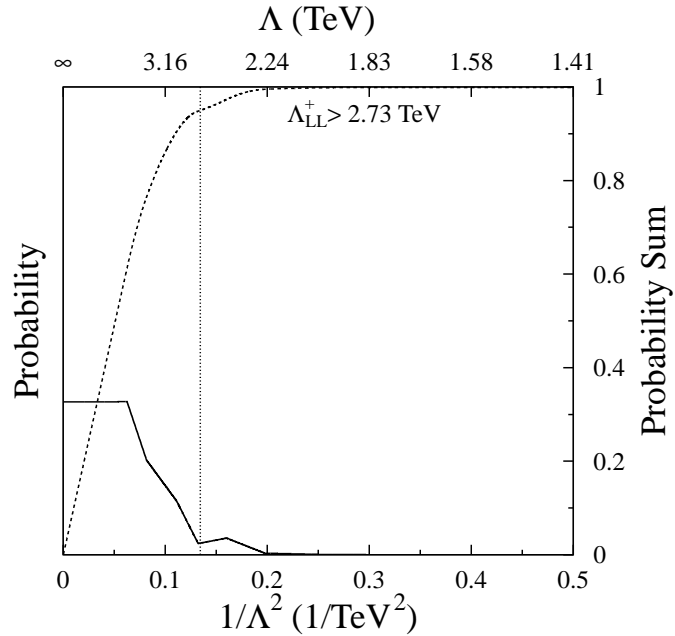,width=3.5in}}}
\caption{The probability distribution (solid curve) $P(\sigma|\xi')
P(\xi')/ Q(\infty)$ for the theoretical prediction {\sc jetrad} with
$\mu = \Etmax$. The dashed curve shows the integral of the probability
distribution and the dotted line shows the 95$\%$ C.L. limit on the
compositeness scale, 2.73 TeV. The most probable value for the
compositeness scale is $\Lambda_{LL}^{+} = \infty$.}
\label{sec_4:fig_5} \end{figure}

 The data show no evidence of compositeness and are used to set 95$\%$
 confidence level limits on $\Lambda_{LL}^{\pm}$. This was done using
 the same method that was used to extract compositeness limits from
 the dijet angular distribution
 (Section~\ref{sec:ang_compositeness}). Figure~\ref{sec_4:fig_5} shows
 the probability distribution for a theoretical prediction obtained
 using {\sc jetrad} with the \cteqthreem\ PDF and a renormalization
 scale of $\mu = \Etmax$. The 95$\%$ C.L. limit on the compositeness
 scale is $\Lambda_{LL}^{+} > 2.7$ TeV.

 Limits were also set for several different theoretical choices of PDF
 and renormalization scales for both the NLO {\sc jetrad} and LO
 compositeness predictions. The limits on the compositeness scale are
 summarized in Table~\ref{compositeness_limits}. The dijet mass
 spectrum rules out quark compositeness models at the 95$\%$
 confidence level where $\Lambda_{LL}^{+}$ is below 2.7~TeV and
 $\Lambda_{LL}^{-}$ is below 2.4~TeV.
\begin{table}[htbp]
 \caption{The 95$\%$ confidence level limits in TeV for the
 left-handed contact compositeness scales for different models.}
 \label{compositeness_limits}
 \begin{center}
 \begin{tabular}{ccdddd}
 {\sc PDF} & Renorm.      & \multicolumn{4}{c}{Compositeness Scale} \\
\cline{3-6}
           & Scale $\mu$ & \multicolumn{2}{c}{$\Lambda_{LL}^{+}$} &  
\multicolumn{2}{c}{$\Lambda_{LL}^{-}$} \\
\cline{3-6}
           &              & $1/\Lambda^2$ &  $1/\Lambda^4$ &
  $1/\Lambda^2$ &  $1/\Lambda^4$ \\
\hline\hline
 \cteqthreem\  	& $0.25  \Etmax$  & 3.51 & 3.21  &  2.87 &2.80\\
 \cteqthreem\  	& $0.50  \Etmax$  & 2.93 & 2.45  &  2.56 &2.38\\
 \cteqthreem\  	& $0.75  \Etmax$  & 2.88 & 2.43  &  2.52 &2.36\\
 \cteqthreem\  	& $1.00  \Etmax$  & 2.73 & 2.38  &  2.49 &2.35\\
 \cteqthreem\  	& $2.00  \Etmax$  & 2.84 & 2.39  &  2.48 &2.35\\
 \cteqfourm\   	& $0.50  \Etmax$  & 2.92 & 2.45  &  2.55 &2.38\\
 CTEQ4A1 	& $0.50  \Etmax$  & 2.96 & 2.47  &  2.55 &2.38\\
 CTEQ4A2 	& $0.50  \Etmax$  & 2.74 & 2.39  &  2.53 &2.36\\
 CTEQ4A4 	& $0.50  \Etmax$  & 2.76 & 2.40  &  2.54 &2.37\\
 CTEQ4A5 	& $0.50  \Etmax$  & 2.96 & 2.47  &  2.58 &2.39\\
 \cteqfourhj\ 	& $0.50  \Etmax$  & 2.87 & 2.42  &  2.58 &2.38\\
 \mrsap\	& $0.50  \Etmax$  & 2.97 & 2.47  &  2.59 &2.39\\
 \mrst\    	& $0.50  \Etmax$  & 3.00 & 2.50  &  2.58 &2.39\\
\mrstgu\ 	& $0.50  \Etmax$  & 3.00 & 2.50  &  2.57 &2.39\\
\mrstgd\ 	& $0.50  \Etmax$  & 2.93 & 2.45  &  2.57 &2.38\\
\end{tabular}
\end{center}
\end{table}

 Limits on models with color-singlet (octet) vector or axial contact
 interactions were also set using an analytic LO
 calculation~\cite{coloron_1} instead of the {\sc pythia} event
 generator. The resulting limits are given in
 Table~\ref{compositeness_limits_2}. The limits on the scale of
 $\displaystyle{\Lambda_{V_8}^{-}}$ can be converted into limits on a
 flavor-universal coloron~\cite{iab_coloron}, resulting in a 95$\%$
 C.L. limit of \mbox{$M_{c} / \cot{\theta} > 837$ GeV/$c^2$} (see
 Section~\ref{sec:theory_coloron} for a description of the theory).

\begin{table}[htbp]
\caption{95$\%$ confidence level limits in TeV for different contact
 compositeness scale for different models calculated using an analytic
 LO prediction~{\protect\cite{coloron_1}} (see
 Section~\ref{sec:theory_compositeness} for a description of the models). }
 \label{compositeness_limits_2} 
 \begin{center}
 \begin{tabular}{cdd}
 Model & \multicolumn{2}{c}{Interference Term X}\\
\cline{2-3}
 & $+1$ & $-1$\\ \hline\hline
 $\displaystyle{\Lambda_{LL}^{X}}$   & 2.2 & 2.2 \\
 $\displaystyle{\Lambda_{V}^{X}}$    & 3.2 & 3.1 \\
 $\displaystyle{\Lambda_{A}^{X}}$    & 3.2 & 3.1 \\
 $\displaystyle{\Lambda_{(V-A)}^{X}}$  & 2.7 & 2.7 \\
 $\displaystyle{\Lambda_{V_8}^{X}}$  & 2.0 & 2.3 \\
 $\displaystyle{\Lambda_{A_8}^{X}}$  & 2.1 & 2.1 \\
 $\displaystyle{\Lambda_{{(V-A)}_8}^{X}}$ & 1.7 & 1.9    \\
 \end{tabular}
 \end{center}
\end{table}

 The robustness of the confidence limits are tested in two ways. The
 first assumes that the systematic uncertainties are completely
 uncorrelated as a function of mass, which results in a degradation of
 the limit by 10~GeV (negligible compared to the scale of the
 limit). The second doubles the size of the systematic uncertainty,
 which results in a degradation of the limit by 20~GeV.

\subsection{Conclusions}

 We have measured the inclusive dijet mass spectrum for a
 pseudorapidity range of $\modetajet < 1.0$ and $200 < \jjmass < 1400$
 GeV at $\sqrt{s} = 1.8$ TeV to an accuracy of $10\%$ to $30\%$ as a
 function of mass. QCD NLO predictions, using several PDFs, show good
 agreement with the observed inclusive dijet mass spectrum.

 The ratio of the inclusive dijet mass cross sections for $\modetajet
 <0.5$ and $0.5 < \modetajet < 1.0$ has also been measured with a
 systematic uncertainty that is less than 10$\%$. The data
 distributions are in good agreement with NLO QCD predictions. Models
 of quark compositeness with a contact interaction scale of less than
 2.2~TeV are excluded at the 95$\%$ confidence level.
\clearpage

\section{Conclusions}

 We have presented a series of measurements of high energy jets at the
 Fermilab Tevatron which are sensitive to the various components of
 QCD predictions: the parton distributions, the matrix elements, and
 the scales. Measurements of the cross section as a function of jet
 \Et , and dijet invariant mass have been presented. By taking the
 ratio of the inclusive cross sections at two energies, both the
 experimental errors and the sensitivity to the parton distributions
 were reduced, providing a stringent test of the \Et\ dependence of
 the QCD matrix element at next-to-leading order. By looking at both
 the dijet angular distribution at fixed mass and the ratio of dijet
 invariant mass distributions in two different rapidity ranges, we
 have again minimized the experimental uncertainties and tested the
 angular dependence of the matrix element calculation.

 We have made the most precise measurement to date of the inclusive
 jet cross section for $\Et \geq 60$~GeV at $\sqrt{s}=1800$~GeV. No
 excess production of high-\Et\ jets is observed. QCD predictions are
 in good agreement with the observed cross section for standard parton
 distribution functions and different renormalization scales ($\mu =$
 0.25--2.00$\Et$ where \Et\ = \Etmax\ and \Etjet ). We have also made the
 most precise measurement to date of the ratio of the inclusive jet
 cross sections at $\sqrt{s}=630$ and 1800 GeV. The NLO QCD
 predictions yield satisfactory agreement with the observed data for
 standard choices of renormalization scale or PDF. In terms of the
 normalization however, the absolute values of the standard
 predictions lie consistently and significantly higher than the data.

 We have measured the dijet angular distribution over a large angular
 range and the inclusive dijet mass spectrum for a pseudorapidity
 range of $\modetajet < 1.0$. QCD NLO predictions, using several
 PDF's, show good agreement with the observed inclusive dijet mass
 spectrum. Since we found good agreement, the data have permitted us
 to provide sensitive limits on the existence of possible non-standard
 model phenomena.

\section*{Acknowledgements}
\addcontentsline{toc}{section}{Acknowledgements} 

We thank the staffs at Fermilab and collaborating institutions, and
acknowledge support from the Department of Energy and National Science
Foundation (USA), Commissariat \` a L'Energie Atomique and
CNRS/Institut National de Physique Nucl\'eaire et de Physique des
Particules (France), Ministry for Science and Technology and Ministry
for Atomic Energy (Russia), CAPES and CNPq (Brazil), Departments of
Atomic Energy and Science and Education (India), Colciencias
(Colombia), CONACyT (Mexico), Ministry of Education and KOSEF (Korea),
CONICET and UBACyT (Argentina), The Foundation for Fundamental
Research on Matter (The Netherlands), PPARC (United Kingdom),
A.P. Sloan Foundation, and the A. von Humboldt Foundation.

\clearpage
\appendix
\section*{$\protect{\bbox{\chisq}}$ Studies}%
\addcontentsline{toc}{section}{$\protect{\bbox{\chisq}}$ Studies}
\label{appendix_a}

 In this paper we have made quantitative \chisq\ comparisons between
 theoretical predictions and data to determine which predictions
 provide better agreement. The systematic uncertainties in the
 inclusive jet cross section (Sections~\ref{sec:inclusive_jet} and
 \ref{sec:inclusive_jet_ratio}) and the dijet mass spectrum
 (Section~\ref{sec:dijet_mass}) are highly correlated. An
 inappropriate definition of the uncertainties in \chisq\ analyses may
 result in theoretical predictions that have an average normalization
 below the data yielding a better fit (Peelle's Pertinent
 Puzzle~\cite{chisq}). The first section of this Appendix describes
 alternative methods for calculating the \chisq\ and our choice of an
 appropriate method. The second section describes studies of the
 probability distributions for the analyses presented in this paper.

 \subsection[Definition of \chisq ]{Definition of
 $\protect{\bbox{\chi^2}}$}

 The \chisq\ is given by 
\begin{equation}
\chisq = \sum_{i,j} \delta_{i} V_{ij}^{-1} \delta_{j},  \label{eq:chisq}
\end{equation}
 where $\displaystyle{\delta_{i}}$ is the difference between the data
 and the expected cross section for bin $i$, and
 $\displaystyle{V_{ij}}$ is element $i,j$ of the covariance matrix,
 with each element given by:
\begin{equation}
V_{ij} = \rho_{ij} \left( \Delta \sigma^{\rm stat}_{i} \Delta
 \sigma^{\rm stat}_{j} \delta_{ij} + \Delta \sigma^{\rm sys}_{i}
 \Delta \sigma^{\rm sys}_{j} \right) \label{eq:cov}
\end{equation} 
 where $\delta_{ij}$ is the Kronecker delta function, $\rho_{ij} = 1$
 for $i=j$, and $\displaystyle{\rho_{ij}}$ is the correlation of the
 systematic uncertainties between cross section bins $i$ and $j$.

 The analyses presented in this paper is based on using the fractional
 systematic uncertainties in each bin, but there are several ways of
 calculating the impact of the absolute systematic uncertainty on the
 \chisq\ values. We can use:
 \begin{enumerate}
  \item Fractional uncertainty multiplied by the observed cross section.
  \item The fractional uncertainty multiplied by a smooth fit to the
 observed cross section~\cite{kovacs} (which is normalized to the
 observed integrated cross section).
  \item The fractional uncertainty multiplied by a theoretical prediction.
 \end{enumerate} 
 This Appendix discusses these choices. In previous publications of
 the inclusive jet cross section~\cite{d0_inc} and the dijet mass
 spectrum~\cite{d0_dijet_mass} the \chisq\ values were calculated
 using the first option.

\begin{figure}[htbp]
\begin{center}
\vbox{\centerline
{\psfig{figure=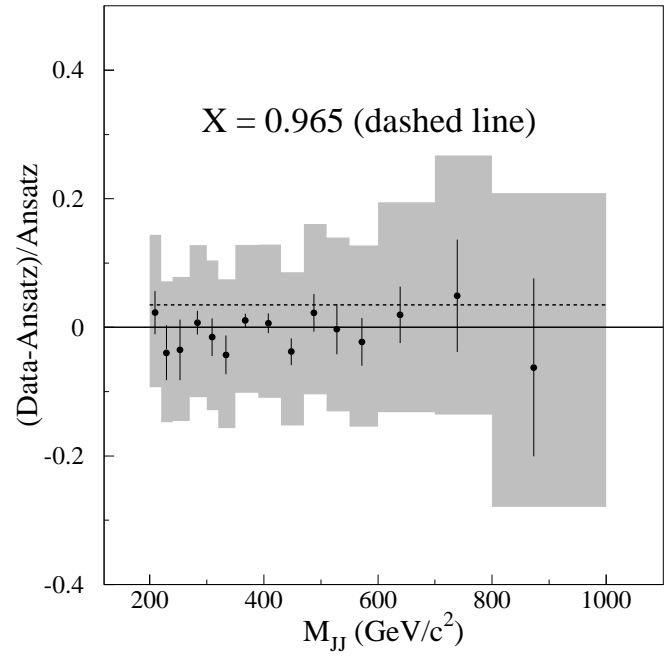,width=3.5in}}}
\caption{The difference between the dijet mass cross section for
$\modetaonetwo < 1.0$ and the Ansatz (see text). The dashed line shows
the best fit obtained by using the standard \chisq\ and absolute
systematic uncertainties obtained using the product of the fractional
systematic uncertainties and the measured cross section in each bin
(Option 1).}
\label{norm_chisquare}
\end{center}
\end{figure}

 The choice of calculation for the absolute systematic uncertainty
 used in the \chisq\ is investigated using the measurement of the
 dijet mass spectrum (Section~\ref{sec:dijet_mass}). A theoretical
 prediction, called the Ansatz (A), based on a fit to the observed
 cross section (Fig.~\ref{sec_2:fig_14}) is obtained by normalizing
 the fit to the observed integrated cross section (cf. Option 2). We
 also define a Floating Ansatz (FA) through a multiplicative factor
 $X$ that is used to change the normalization of the Ansatz (${\rm FA}
 = X{\rm A}$).  A comparison between the Ansatz and the data is given
 in Fig.~\ref{norm_chisquare}.

\begin{figure}[htb]
\begin{center}
\vbox{\centerline
{\psfig{figure=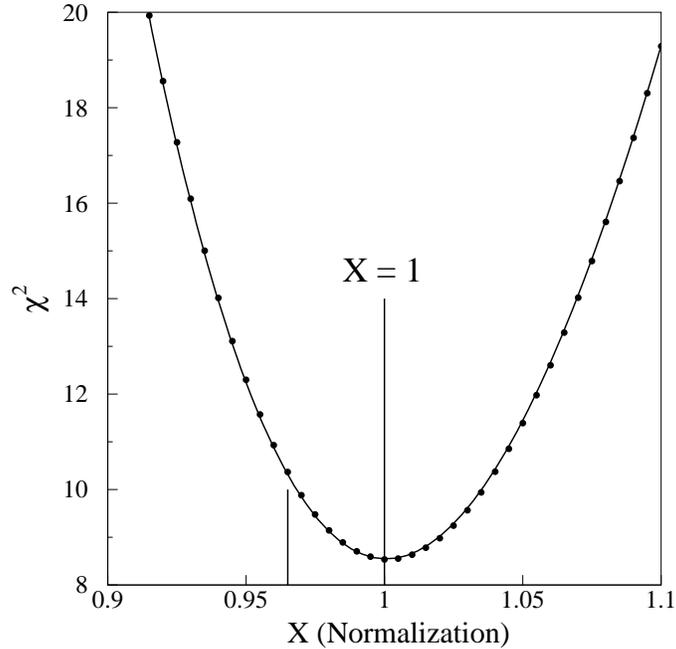,width=3.5in}}}
\caption{The \chisq\ for the Ansatz as a function of
 the floating normalization $X$ for Option 2 (see text). The minimum
 \chisq\ is obtained for a normalization of $X = 1.0$. The short
 vertical line indicates a normalization of $X = 0.965$, illustrating
 the bias of Option 1.}
\label{norm_chisquare_2}
\end{center}
\end{figure}

 If the systematic uncertainty is given by the product of the
 fractional uncertainty and the observed cross section in each bin
 (Option 1), the minimum value of the \chisq\ of the Floating Ansatz
 is obtained for a normalization of $X = 0.965$ (the dashed line in
 Fig.~\ref{norm_chisquare}).  This is clearly not the best visual fit
 to the observed cross section.  When this test is repeated using
 Option 2, the preferred normalization is $X = 1.0$
 (Fig.~\ref{norm_chisquare_2}). Using several different predictions
 from {\sc jetrad} (Option 3) also yields $X = 1.0$ as a best fit (not
 shown).

 Calculating systematic uncertainties using the observed cross section
 per bin introduces a statistical component to the systematic
 uncertainty; i.e., when the cross section fluctuates to a small value
 in a given bin the absolute systematic uncertainty also fluctuates to
 a smaller value. The smaller values of cross section therefore appear
 to be more precise relative to any given theory. This bias has been
 called Peelle's Pertinent Puzzle~\cite{chisq}.

 We choose to rely on Options 2 and 3 for determining systematic
 uncertainties. The choice depends on the question that is posed. In
 our work we wish to
\begin{quote}
 ``Determine the probability that the theoretical prediction could
 have produced the observed number of events.''
\end{quote}
 This requires that we determine the systematic uncertainties using
 the theoretical predictions (Option 3). For example, if we
 underestimated the integrated luminosity the number of predicted
 events would also be underestimated.

 This choice of \chisq\ definition means that the current results
 differ from those published previously for the inclusive jet cross
 section~\cite{d0_inc} and the dijet mass
 spectrum~\cite{d0_dijet_mass}.
 Table~\ref{theory_data_chisquares_app} summarizes the differences in
 \chisq\ values for the dijet mass analysis. The \chisq\ values in
 Table~\ref{theory_data_chisquares_app} are calculated using the same
 luminosity definition as given in Ref.~\cite{d0_dijet_mass}, and
 differ from those given in Table~\ref{theory_data_chisquares}. The
 \chisq\ changes most for theoretical predictions with the largest
 normalization differences with the data. If the theoretical
 prediction has a smaller normalization than the data then the size of
 the systematic uncertainties are reduced, hence increasing the value
 of the \chisq .

\begin{table*}[htbp] 
\caption{The \chisq\ for the cross section in dijet mass  for $\modetajet
< 1.0$ (15 degrees of freedom).}
  \label{theory_data_chisquares_app}
 \begin{center}
 \begin{tabular}{lldddd}
PDF       & $\displaystyle{\mu}$            & 
\multicolumn{2}{c}{Published \chisq \cite{d0_dijet_mass} } &
 \multicolumn{2}{c}{Updated \chisq\ } \\
 \cline{3-4}\cline{5-6}
  &        & \chisq  & Probability &  \chisq  & Probability \\
 \hline\hline
 \cteqthreem\  & $0.25  \Etmax$             & 12.2 & 0.66 & 28.9    &  0.02 \\
 \cteqthreem\  & $0.50  \Etmax$             &  5.0 & 0.99 &  5.8    &  0.98 \\
 \cteqthreem\  & $0.75  \Etmax$             &  5.3 & 0.99 &  5.9    &  0.98 \\
 \cteqthreem\  & $1.00  \Etmax$             &  5.4 & 0.99 &  6.1    &  0.98 \\
 \cteqthreem\  & $2.00  \Etmax$             &  4.2 & 1.00 &  6.4    &  0.97 \\
 \cteqthreem\  & $0.25\sqrt{x_{1}x_{2}s}$   &  8.6 & 0.90 & 14.6    &  0.48 \\
 \cteqthreem\  & $0.50\sqrt{x_{1}x_{2}s}$   &  4.8 & 0.99 &  6.8    &  0.96 \\
 \cteqthreem\  & $1.00\sqrt{x_{1}x_{2}s}$   &  5.1 & 0.99 &  8.9    &  0.88 \\
 \cteqfourm\  & $0.50  \Etmax$             &  4.9 & 0.99  &  6.3    &  0.97 \\
 \cteqfoura 1 & $0.50  \Etmax$             &  5.0 & 0.99  &  6.5    &  0.97 \\
 \cteqfoura 2 & $0.50  \Etmax$             &  5.7 & 0.99  &  7.2    &  0.95 \\
 \cteqfoura 4 & $0.50  \Etmax$             &  4.9 & 0.99  &  6.4    &  0.97 \\
 \cteqfoura 5 & $0.50  \Etmax$             &  4.8 & 0.99  &  6.2    &  0.98 \\
 \cteqfourhj\ & $0.50  \Etmax$             &  5.4 & 0.99  &  6.8    &  0.96 \\
 \mrsap\ & $0.50  \Etmax$     &  6.3 & 0.97    &  6.9    &  0.96   \\
 \mrst\    & $0.50  \Etmax$             &  6.2 & 0.98    & 10.9    &  0.76  \\
 \mrstgu\ & $0.50  \Etmax$             &  6.3 & 0.97    &  9.6    &  0.84   \\
 \mrstgd\ & $0.50  \Etmax$             &  6.5 & 0.97    & 16.7    &  0.33   \\
 \end{tabular}
 \end{center}
 \end{table*}

\subsection{Probabilities}

 The probability that a given theoretical prediction agrees with the
 data for a given \chisq\ is calculated assuming that the \chisq\ is
 given by the standard distribution~\cite{ppr}:
\begin{equation}
f \left( x ; n\right) = \frac{ x^{\left( n/2 - 1
\right)} \exp\left( -x /2 \right) } {2^{n/2} \Gamma\left( n/2\right)},
\label{eq:chisq_1}
\end{equation}
 where $n$ is the number of degrees of freedom (d.o.f.) in the
 data. The probability of getting a value of \chisq\ larger than the
 one obtained is then given by
\begin{equation}
P\left(\chisq ; n \right) = \int^{\infty}_{\chisq}  f \left( x ; n\right)  dx.
\label{eq:chisq_2}
\end{equation}
 Hence, for the probabilities quoted in
 Sections~\ref{sec:inclusive_jet} and \ref{sec:dijet_mass} to be
 reliable, the \chisq\ distribution for comparisons between
 theoretical predictions and the data must follow
 Eq.~\ref{eq:chisq_1}.

\begin{figure}[htbp]
\begin{center}
\vbox{\centerline{\psfig{figure=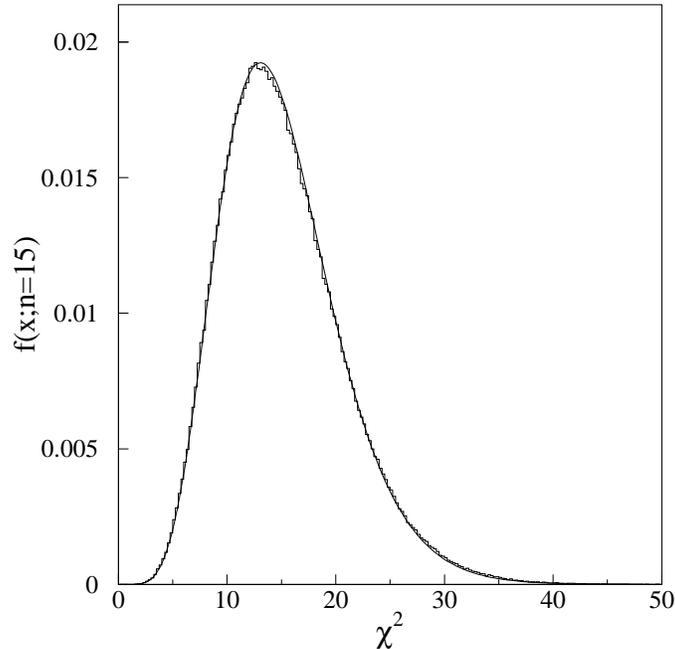,width=3.5in}}}
\caption{The \chisq\ distribution generated by sampling the Ansatz
cross section using only statistical fluctuations. The histogram shows
the generated distribution, and the curve is a fit to the histogram
using Eq.~\ref{eq:chisq_1}. The fitted number of degrees of freedom is
15.08. }
\label{fig:chi2_stat}
\end{center}
\end{figure}

 The distribution of \chisq\ for comparisons with the dijet mass
 spectrum was tested by developing a Monte Carlo program that
 generates many trial predictions based on the Ansatz (with a total of
 15 bins, or 15 degrees of freedom). The first step is to generate
 trials based on statistical fluctuations. The {\it true} number of
 events per bin is given by the Ansatz. The trial spectra are then
 generated for each bin by sampling a Poisson distribution with a mean
 defined by the {\it true} number of events. The \chisq\ for each of
 these trials is calculated using the difference between the {\it
 true} and the generated values. Figure~\ref{fig:chi2_stat} shows the
 \chisq\ distribution for all of the generated trials. The
 distribution is fitted to Eq.~\ref{eq:chisq_1}, with the best fit
 obtained for $n=15.08 \pm 0.20$, which is consistent with the
 expected value of $n=15$ for a normalized distribution of bins.

\begin{figure}[hbtp]
\begin{center}
\vbox{\centerline{\psfig{figure=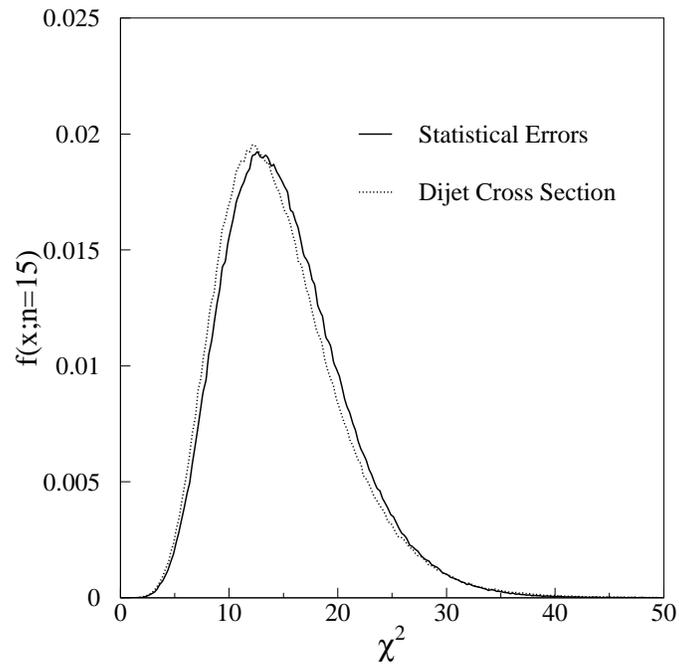,width=3.5in}}}
\caption{The \chisq\ distribution generated by sampling the Ansatz
cross section using only statistical fluctuations (solid curve), and
fluctuations based on the uncertainties in the dijet cross section as
a function of dijet mass (dotted curve). }
\label{fig:chi2_all}
\end{center}
\end{figure}

\begin{figure}[htbp]
\begin{center}
\vbox{\centerline{\psfig{figure=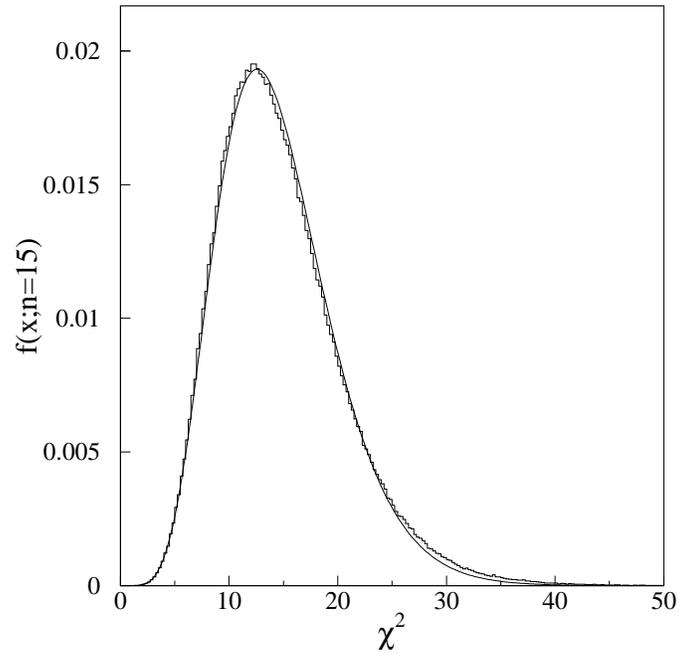,width=3.5in}}}
\caption{The \chisq\ distribution generated by sampling the Ansatz
cross section using all of the systematic uncertainties of the dijet
cross section. The histogram shows the generated distribution and the
curve is a fit to the histogram using Eq.~\ref{eq:chisq_1}. The fitted
number of degrees of freedom is $14.6 \pm 0.2$.}
\label{fig:chi2_dijet}
\end{center}
\end{figure}

 The final step is to assume that the uncertainties are the same as
 the uncertainties in the measurement of the dijet cross
 section. Trial spectra are generated using these uncertainties in
 order to obtain a \chisq\ distribution (see the dotted curve in
 Fig.~\ref{fig:chi2_all}). It is clear that the \chisq\ distribution
 is very similar to that predicted by Eq.~\ref{eq:chisq_1}; hence any
 probability generated using Eq.~\ref{eq:chisq_2} should be
 approximately right. The resulting \chisq\ distribution was fitted
 using Eq.~\ref{eq:chisq_1} (Fig.~\ref{fig:chi2_dijet}) and yielded
 $n= 14.6 \pm 0.2$.

\begin{figure}[hbtp]
\begin{center}
\vbox{\centerline{\psfig{figure=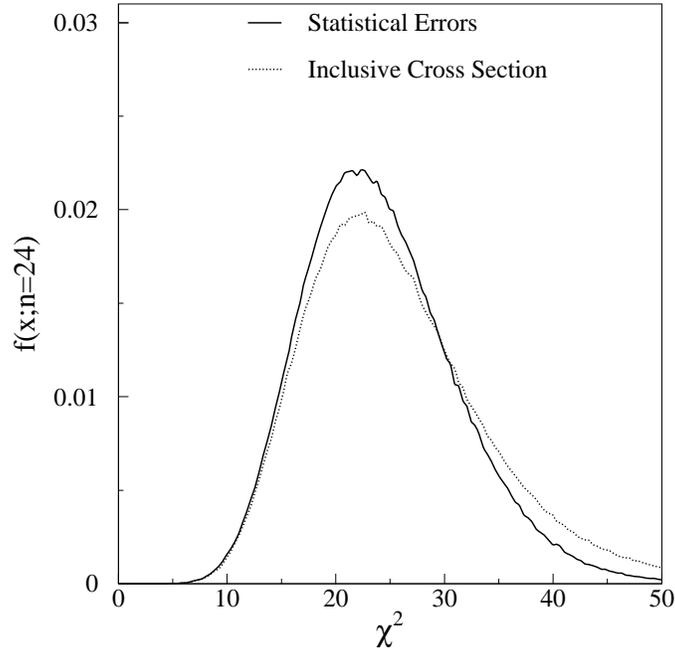,width=3.5in}}}
\caption{The  \chisq\ distribution
generated by sampling the inclusive jet cross section Ansatz using
only statistical fluctuations (solid curve), and fluctuations based on
the uncertainties in the inclusive jet cross section (dotted curve). }
\label{fig:chi2_all_inc}
\end{center}
\end{figure}

 The study of the \chisq\ distribution was repeated for the
 measurement of the inclusive jet cross section, which has 24 bins
 (d.o.f.). Figure~\ref{fig:chi2_all_inc} shows the resulting
 distributions for statistical fluctuations (solid curve) and the
 systematic uncertainties in the inclusive jet cross section (dotted
 curve). The two distributions agree for \chisq\ values below
 approximately 15, and then begin to diverge. The distribution based
 on the cross section uncertainties has a longer tail than the
 statistical \chisq\ distribution. This implies that all the
 probabilities quoted in Section~\ref{sec:inclusive_jet} are slightly
 underestimated.

\begin{figure}[hbtp]
\begin{center}
\vbox{\centerline{\psfig{figure=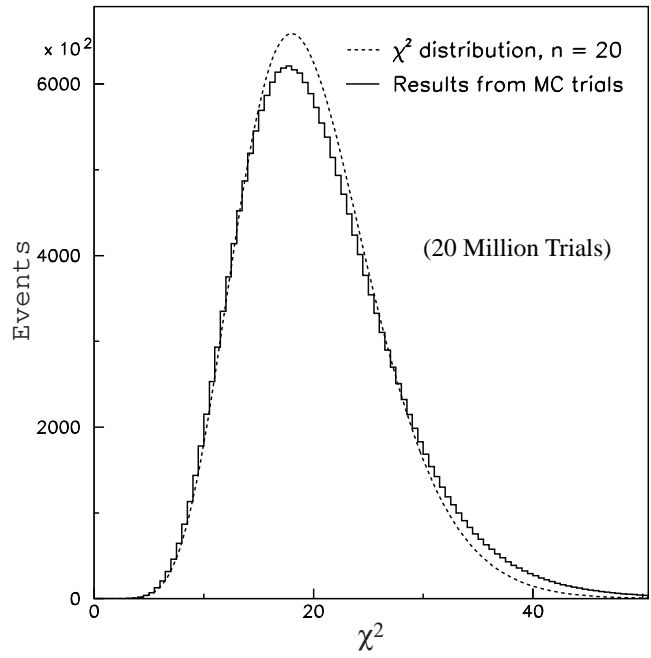,width=3.5in}}}
\caption{The  \chisq\ distribution
generated by sampling the ratio of inclusive jet cross sections Ansatz
using only statistical fluctuations (solid curve), and fluctuations
based on the uncertainties in the inclusive jet cross section (dotted
curve). }
\label{fig:chi2_all_ratio}
\end{center}
\end{figure}

 Finally, the ratio of inclusive jet cross sections
 (Section~\ref{sec:inclusive_jet_ratio}), was also examined with the
 results of the study given in Fig.~\ref{fig:chi2_all_ratio}.  The
 resulting distribution is similar to the one obtained for the
 inclusive jet cross section, with the distribution based on the
 uncertainties having a larger tail than the standard \chisq\
 distribution. The maximum deviation between the probability obtained
 assuming the standard distribution and the measured distribution is
 $2.9\%$, and probabilities quoted in
 Section~\ref{sec:inclusive_jet_ratio} will therefore be slight
 underestimates of the correct probabilities.

 The studies presented describe the \chisq\ comparisons made between
 the observed data and the theoretical predictions. We have
 demonstrated that they give an accurate representation of the
 probability of agreement between a given theoretical prediction and
 the data.


\begin{references}
%
%


\bibitem{first_theory_quark}
M.~Gell-Mann, Phys. Lett. {\bf 8}, 214 (1964);
G.~Zweig, CERN preprint 8182/Th~401 January 1964 (unpublished);
G.~Zweig, CERN preprint 8419/Th~412, February 1964 (unpublished),
reprinted in {\it Developments in quark theories of hadrons}, edited
by D.B.~Lichtenberg and S.P.~Rosen (Hadronic Press, Nonamtum, Mass.,
1980), Vol. 1, p. 22.


\bibitem{SLAC_quarks}
J.I. Friedman, H.W. Kendall, and R.E. Taylor,
Rev. Mod. Phys. {\bf 63}, 629 (1991).

\bibitem{jetsreview} 
R.K.~Ellis, W.J.~Stirling, and B.R.~Webber, 
{\it QCD and Collider Physics}, 
Cambridge University Press, UK: Univ. Pr. (1996) 435 p. (
Cambridge monographs on particle physics, nuclear physics and
cosmology: 8) and references therein.

\bibitem{ua1_inc}
UA1 Collaboration, G.\ Arnison {\em et al.},
Phys.\ Lett.\ B {\bf 172}, 461 (1986);

\bibitem{ua2_inc}
UA2 Collaboration, J.\ Alitti {\em et al.},
Phys.\ Lett. B {\bf 257}, 232 (1990); 
UA2 Collaboration, J.\ Alitti {\em et al.},
Z.\ Phys.\ C {\bf 49}, 17 (1991).

\bibitem{CDF_1} 
CDF Collaboration,  F. Abe {\em et al.},
Phys. Rev. Lett. {\bf 70}, 1376 (1993).

\bibitem{CDF_2} CDF Collaboration, F. Abe {\em et al.},
Phys. Rev. Lett. {\bf 77}, 438 (1996). Preliminary results from an
extended data set have been shown by CDF; we compare to the CDF data
which have been published.

\bibitem{d0_inc}
D\O\ Collaboration, B. Abbott {\em et al.}, 
Phys. Rev. Lett.  {\bf 82}, 2451 (1999).

\bibitem{afs_jets}
AFS Collaboration, T.~Akesson {\em et al.}, 
Phys.\ Lett. {\bf 118B}, 185 (1982).

\bibitem{first_jets}
UA2 Collaboration, M.~Banner {\em et al.},  
Phys.\ Lett. {\bf 118B}, 203 (1982).

\bibitem{aversa}
F.~Aversa, M.~Greco,  P.~Chiappetta, and J.P.~Guillet,
Phys.~Rev.~Lett. {\bf 65}, 401 (1990).

\bibitem{eks}
S.D.~Ellis, Z.~Kunszt, and D.E.~Soper, 
Phys. Rev. Lett. {\bf 64}, 2121 (1990);
Z.~Kunszt and D.E.~Soper, Phys. Rev. D {\bf 46}, 192 (1992).

\bibitem{jetrad} 
 W.T.~Giele, E.W.N. Glover, and D.A. Kosower, 
 Nucl. Phys. {\bf B403}, 633 (1993). 
 We used {\sc jetrad} version 2.0.

\bibitem{cteq4m} 
H.L. Lai {\em et al.}, Phys. Rev. D {\bf 55}, 1280 (1997), 
and references therein.

\bibitem{mrst} A.D.~Martin, R.G.~Roberts, W.J.~Stirling, and R.S.~Thorne,
Eur.\ Phys.\ J. C {\bf 4}, 463 (1998), 
and references therein. 

\bibitem{inc_jet_theory_uncertainties}
B.~Abbott {\em et al.},  Eur.\ Phys.\ J. C {\bf 5}, 687 (1998).

\bibitem{compos}
 E. Eichten, K. Lane, and M.E. Peskin, Phys. Rev. Lett. {\bf 50}, 811 (1983);
 E. Eichten, I. Hinchcliffe, K. Lane, and C. Quigg, 
 Rev. Mod. Phys. {\bf 56}, 579 (1984),
 Addendum --- {\sl ibid.} {\bf 58}, 1065 (1986);
 K. Lane,  hep-ph/9605257 (1996).


\bibitem{cdf_dijet_mass}
CDF Collaboration, F. Abe {\em et al.}, 
Phys. Rev. D {\bf 48}, 998 (1993)

\bibitem{cdf_angular} 
CDF Collaboration, F. Abe {\em et al.}, 
Phys. Rev. Lett. {\bf 77}, 5336 (1996),
erratum --- {\sl ibid.} {\bf 78}, 4307 (1997);
{\sl ibid.} {\bf 69}, 2896 (1992); 
{\sl ibid.} {\bf 62}, 3020 (1989).

\bibitem{Elvira} 
V.D. Elvira, Ph.D.~thesis, Universidad de Buenos Aires, 1995 (unpublished).


\bibitem{d0_inc_630}
D\O\ Collaboration, B. Abbott {\em et al.}, 
hep-ex/0008072, submitted to Phys. Rev. Lett.  

\bibitem{jkrane} 
J. Krane, Ph.D.~thesis, University of Nebraska, 1998 (unpublished).

\bibitem{d0_angular}
D\O\ Collaboration, B. Abbott {\em et al.}, 
Phys. Rev. Lett. {\bf 80}, 666 (1998).

\bibitem{fatyga} 
K. Fatyga, Ph.D.~thesis, University of Rochester, 1996 (unpublished).

\bibitem{d0_dijet_mass}
D\O\ Collaboration, B. Abbott {\em et al.}, 
Phys. Rev. Lett.  {\bf 82}, 2457 (1999).

\bibitem{levan}
D\O\ Collaboration, B. Abbott {\em et al.}, 
FERMILAB-Pub-00/271-E, hep-ex/0011036,
submitted to Phys. Rev. Lett;
L.~Babukhadia, Ph.D. Dissertation, University of Arizona, 
Tucson, Arizona, USA (1999) (unpublished),\\
{\tt http://fnalpubs.fnal.gov/techpubs/theses.html}.\\
Data from this analysis will be available from the AIP E-PAPS service.


\bibitem{pdh} 
D.E. Groom {\em et al.},  Eur.\ Phys.\ J. C {\bf 15}, 1 (2000)
and references therein.

\bibitem{d0_detector}  
 D\O\ Collaboration, S. Abachi {\em et al.},
 Nucl. Instrum. Methods Phys. Res. A {\bf 338}, 185 (1994).

\bibitem{d0_cal_test_beam}
 M. Abolins {\em et al.}, 
 Nucl. Instrum. Methods Phys. Res. A {\bf 280}, 36 (1989); 
 D\O\ Collaboration, S. Abachi {\em et al.},
 Nucl. Instrum. Methods Phys. Res. A {\bf 324},  53 (1993); 
 D\O\ Collaboration, H. Aihara {\em et al.}, 
 Nucl. Instrum. Methods Phys. Res. A {\bf 325}, 393 (1993).

\bibitem{snowmass} 
J.~Huth {\em et al.}, in proceedings of {\it
Research Directions for the Decade, Snowmass 1990}, edited by
E.L.~Berger (World Scientific, Singapore, 1992).

\bibitem{ellis} 
S.D.~Ellis, Z.~Kunszt, and D.E.~Soper,  
Phys. Rev. Lett.  {\bf 69}, 3615 (1992).

\bibitem{rsep} 
B. Abbott {\em et al.}, Fermilab-Pub-97/242-E (unpublished).

\bibitem{herwig}
G~.Marchesini {\em et al.}, Comput. Phys. Commun. {\bf 67}, 465 (1992).
We used {\sc herwig} version 5.8.

\bibitem{geant} 
D\O\ detector simulation package based on {\sc geant}.
{\sc geant} by R.\ Brun and F.\ Carminati, 
CERN Program Library Long Writeup W5013, 1993 (unpublished).


\bibitem{cteq3m}
H.L. Lai {\em et al.}, Phys. Rev. D {\bf 51}, 4763 (1995).

\bibitem{mrsap} 
A.D.~Martin, R.G.~Roberts, and W.J.~Stirling, 
Phys. Lett. B {\bf 354}, 154 (1995).

\bibitem{pdf_problem}
 A.D.~Martin, R.G.~Roberts, W.J.~Stirling, and R.S.~Thorne,
 hep-ph/9907231 (unpublished);
 H.L. Lai {\em et al.}, Eur.\ Phys.\ J. C {\bf 12}, 375 (2000).


\bibitem{coloron_1}
R.S. Chivukula, A.G. Cohen and E.H. Simmons, 
Phys. Lett. B {\bf 380}, 92 (1996).

\bibitem{chosimm} 
E.~H.~Simmons, Phys. Lett. B {\bf 226}, 132 (1989); 
{\sl ibid.}  {\bf 246}, 471 (1990); 
P.~Cho and E.~H.~Simmons, {\sl ibid.} {\bf 323}, 401 (1994);
Phys. Rev. D {\bf 51}, 2360 (1995).

\bibitem{coloron_2}
E.H. Simmons, Phys. Rev. D {\bf 55}, 1678 (1997).

\bibitem{cdfxjj}
CDF Collaboration, F. Abe {\em et al.},
Phys. Rev. D {\bf 55}, 5263 (1997).

\bibitem{d0_wcross} 
D\O\ Collaboration, B. Abbott {\em et al.},
Phys. Rev. D {\bf 61}, 072001 (2000), VII, pp. 15--16.

\bibitem{luminosity_630} 
J. Bantly {\em et al.}, Fermilab-TM-2000, (1997) (unpublished).

\bibitem{lumin_e710}
N.~Amos  {\em et al.},  Phys. Lett. B {\bf 242}, 158 (1990).

\bibitem{lumin_cdf} 
CDF Collaboration, F.~Abe {\em et al.},
Phys. Rev. D {\bf 50}, 5550 (1994).

\bibitem{E811_lumin}%
E811 Collaboration, C. Avila {\em et al.}, Phys. Lett. B {\bf 445}, 419 (1999).

\bibitem{ua4_lumin}
UA4 Collaboration, M.~Bozzo {\em et al.}, Phys. Lett. B {\bf 147}, 4 (1984);
UA4 Collaboration, D.~Bernard {\em et al.}, Phys. Lett. B {\bf 186}, 2 (1987).

\bibitem{Mbr}
CDF Collaboration, F. Abe {\em et al.}, Phys. Rev. D {\bf50}, 5550 (1994).

\bibitem{DtuJet}
P. Aurenche {\em et al.,} Phys. Rev. D {\bf 45}, 92 (1992);
F.W.~Bopp, A.~Capella, J.~Ranft, and J.~Tran Thanh Van, Z. Phys. C {\bf 51}, 99 (1991).


\bibitem{mitool}
 T.L. Taylor Thomas, Ph.D.~thesis, Northwestern University, 1997 (unpublished).

\bibitem{energy_scale} 
D\O\ Collaboration, B. Abbott {\em et al.},
Nucl. Instrum. Methods Phys. Rev. A {\bf 424}, 352 (1999).

\bibitem{ferbel} 
R. Wigmans, {\sl Experimental Techniques in High-Energy Nuclear and
Particle Physics}, edited by T.\ Ferbel ( World Scientific, 1991).

\bibitem{snihur}
 R. Snihur, Ph.D.~thesis, Northwestern University, 2000 (unpublished).

\bibitem{Ppoint} G.D. Lafferty and T.R. Wyatt, 
CERN-PPE/94--72 (1994) (unpublished).

\bibitem{inclusive_matrix}
Cross Sections and covariance matrices will be made available from the AIP E-PAPS service.

\bibitem{bayesian}
 H.~Jeffreys, {\it Theory of Probability} 
(Clarendon Press, Oxford, 1939, revised 1988), p 94;
F.T. Solmitz,
Ann. Rev. Nucl. Sci.  {\bf 14}, 375 (1964).

\bibitem{pythia} 
 T. Sj\"ostrand, 
 Comput. Phys. Commun. {\bf 82}, 74 (1994).
 We used {\sc pythia} version 5.7.

\bibitem{minuit}
 F. James, CERN Program Library Entry D506 (unpublished).
 We used {\sc minuit} version 96.a.


 
\bibitem{iab_coloron}
I.A.~Bertram and E.H.~Simmons, Phys. Lett. B {\bf 443}, 347 (1998).

\bibitem{chisq}
G. D'Agostini,  hep-ph/9512295 v3 p88-91 (1995) (unpublished).

\bibitem{kovacs}
E. Kovacs (private communication).

\bibitem{ppr}
C. Caso et al, Eur.\ Phys.\ J. C {\bf 3}, 1 (1998).


\end{references}
\end{document}